The monograph summarizes the studies of the temperature limits for the existence of a supercooled liquid phase of the components of nanodisperse structures. Original in situ methods for studying supercooling during crystallization in nanodispersed systems are proposed. The results of long-term studies of the features of crystallization of particles on different substrates, or as part of layered film systems are presented. The influence of the interaction at the melt-substrate (matrix) interface, as well as the conditions of preparation of the samples on the values of the limiting supercooling are determined.

## Contents

# Chapter 1
# Supercooling of the liquid phase. General aspects

**Abstract** From the point of view of thermodynamics, the stable phase is the phase that has a minimum of free energy. Thus, the phase transition temperature must be determined by the point at which the temperature dependencies of the free energy of those phases that can be in equilibrium intersect. As follows from approved thermodynamic models, the melting of a free particle happens without nucleation at a temperature determined by the equality of the free energies of the solid and liquid phases. However, for crystallization to begin in the melt, the nucleus of the crystalline phase, separated from the melt by an interface, must appear in the melt. The presence of such a boundary contributes to the free energy of the system. The relative share of the boundary energy increases as the size of the crystal nucleus decreases. Thus, the beginning of crystallization requires overcoming the energy barrier connected with the need to form a nucleus of sufficiently large size. The appearance of crystallites smaller than a certain size proves to be energetically disadvantageous. As a result, crystallization as a rule starts below the melting temperature. Due to the presence of the energy barrier during thermal cycling of samples, a melting-crystallization hysteresis can be observed. Thus, the melt itself can remain in a supercooled liquid state for a long time.

The chapter deals with a theoretical description on the base of the phenomenological thermodynamics of the phenomenon of crystallization of supercooled melts. The processes corresponding to slow cooling are considered. The role of wettable impurities in the melt is determined. It is shown that melt particles located on a poorly wetted substrate can be considered as free in the context of supercooling. The chapter concludes with some analysis of the structural peculiarities of the melts, which are in the state of deep supercooling.

## 1.1 Introduction

The study of phase transitions is one of the main tasks of thermodynamics. Based on thermodynamic considerations, the temperature of a phase transition can be determined as the point at which the free energies of two phases become equal. It should be recollected that under the possibility of existence of several phases in a system, the one with the lower free energy at a given temperature will be stable. Thus, different phases are thermodynamically stable above and below this temperature (Fig. 1.1). Accordingly, the transition from one phase to another should happen at the point of phase equilibrium, where the temperature dependences of the free energies of both phases intersect.

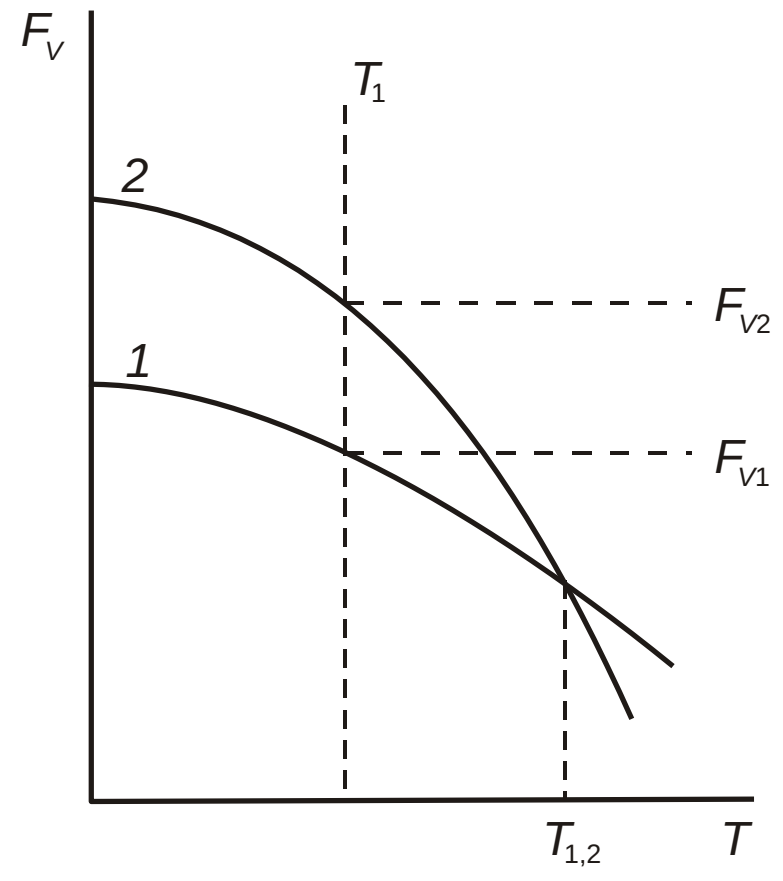

**Fig. 1.1** Free energy versus temperature for phases 1 and 2

However, this approach, which takes into account only the volumetric component of free energy, needs to be improved, since the appearance of a new phase in the system is accompanied by the appearance of an interface which some energy is connected with. Due to this, an energy barrier appears that prevents the formation of the equilibrium phase under these conditions. This, in turn, leads to the possibility of the existence of metastable states, in which a certain phase exists under conditions that correspond to the thermodynamic stability of another phase.

One of the most common phenomena connected with the presence of an energy barrier during the formation of a new phase is supercooling during the crystallization of melts. In everyday life, this phenomenon can be observed by the example of water during the crystallization of soap bubble shells (their crystallization happens at supercooling of 10 K or more), and in industry, it is widely used in the production of amorphous materials for which unique technological properties are possible.

It is worth noting that the creation of bulk metastable materials is usually based on the strategy of extremely rapid cooling. Due to the exponential nature of the temperature dependence of the diffusion rate, sufficiently rapid cooling allows the freezing of high-temperature phases that can remain in the metastable state indefinitely.

The tendency of recent decades has been the miniaturization of constructive elements, and the widespread industrial use of thin films, various composites, and other similar nanostructures. The physical and chemical peculiarities of the properties of such objects determine not only the manifestation of various size effects but also the ease of creation of metastable states. At the same time, metastable states happen to be possible not because of a decrease in the intensity of diffusion processes (as is observed during quenching), but due to the existence of an energy barrier connected with the appearance of additional interfaces during the formation of a nucleus of the new phase. The existence of nonequilibrium phases not only dramatically affects the peculiarities of the behavior of low-dimensional structures, but also allows the creation of various technological devices which operation is based on the presence of supercooled or overheated phases in the system. Such devices can be various kinds of keys, memory elements, sensors, and switches. In addition, the unique effects of inter-particle interaction that manifest in some supercooled liquids at a sufficient degree of supercooling can be very interesting from an applied point of view [1, 2].

It is equally important to study the phenomenon of supercooling of liquid metals and alloys in terms of developing lead-free solder technologies. For example, one way to overcome the relatively high melting point of many lead-free alloys is to use highly dispersed solder powders. The size effect makes it possible to control in a targeted way the melting point of particles by changing their size. This makes it possible to provide such values of processing temperatures for electronic circuit elements that are acceptable for practical reasons [3]. In addition, the size of the particles, due to which solder joints are made, is one of the factors that determine the possibility of further miniaturization of elements of modern electronics. Today, solder powders with particle sizes at the level of a few tens of nanometers are available on an industrial scale. However, precisely the small size of such particles is extremely favorable for the manifestation of the phenomenon of supercooling in them. This, together with the possible change in wetting in supercooled liquids, is of decisive importance for various technological processes.

To date, a significant amount of experimental data is available on supercooling in various thin film structures and nanocomposite materials. Many experimental methods have been developed to obtain samples suitable for studying supercooling, and the results of the studies are analyzed using various theoretical approaches.

The experimental study of supercooling involves two aspects. Firstly, it is necessary to purify the sample as much as possible from various impurities that change the properties of the system in a difficult-to-predict way. Second, it is necessary to register the moment of phase transition in the system under study.

For a long time, the need to obtain a sample sufficiently pure for research has limited progress in the study of metastable states. Usually, in addition to standard technological processes, multiple remelting and re-evaporations were used to purify the substances under study, combined with experiments in a protective environment. However, the extremely high sensitivity of metastable states to the presence of microparticles of insoluble impurities that provoke crystallization limits the application of these methods. A fairly effective approach to creating samples suitable for studying supercooling was proposed by Turnbull; [4, 5]. This method, which was later called the "method of micro-volumes", is based on the following. Since the number of insoluble impurities in the starting material is limited and, provided that under the condition of thorough pre-cleaning their proportion is small compared to the substance under test, it is possible to obtain some particles that are totally free of impurities if the sample is sufficiently dispersed.

The melt dispersion in [4, 5] was carried out using a powerful ultrasonic generator. Today, vacuum condensates and nanoporous structures saturated with the substance which supercooling is being studied are widely used to implement the micro-volumes method.

A separate task, which must be solved to obtain maximum supercoolings, is to reduce the influence of the crucible, which in this case acts as a foreign impurity. The most radical solution to this problem is the use of containerless methods of electrostatic or acoustic levitation, as well as conducting research in zero gravity or free fall. This task can be partially solved by using a crucible which material has a minimal impact on the behavior of the supercooled substance. For example, the results of a large set of studies indicate that the influence of the crucible on the value of supercooling decreases with the deterioration of wetting it by the melt. For contact pairs with a wetting edge angle of more than 120°, the solid phase has virtually no effect on melt supercooling. Carbon is a good material for the study of fusible metals, while for more refractory substances various oxides that are traditionally used as ceramic materials can be used as a crucible or substrate.

Different approaches can be applied to measure temperatures of phase transitions in pre-purified samples. For example, the crystallization and melting of a sample are accompanied by the appearance and disappearance of diffraction reflexes from its crystal structure. *In situ* electron diffraction studies have become widely used to implement the method of determining phase transition temperatures based on this phenomenon, especially in the application to nanostructures [3]. In bulk samples and some nanocomposites, XRD-methods are also effective. Melting and crystallization, as first-order phase transitions, can be registered by absorption or release of latent heat of the phase transition, which can be realized by the methods of differential scanning calorimetry.

In addition to the methods discussed above, which to some extent can be attributed to direct methods for measuring phase transition temperatures, indirect approaches are widely used. For example, melting or crystallization can be detected by changes in the optical properties of the medium. A large set of studies, in particular, those related to the study of supercooling of a low-melting component in porous materials, is based on the use of ultrasound, the peculiarities of spreading of which in solids and liquids are significantly different. A number of studies of supercooling in layered film systems have also been carried out by examining the temperature dependence of the electrical resistance of samples. These studies are based on a jump in the resistance of a multilayer film that accompanies the melting or crystallization of a low-melting component.

A special mention should be made of the method of changing the condensation mechanism, which has been used to obtain a large number of unique results concerning the thermal stability limits of the liquid phase in thin films. This method, which is designed exclusively for the use of vacuum condensates, is based on the experimentally and theoretically confirmed assumption that the temperature of the change in the condensation mechanism from "vapor-liquid" to "vapor-crystal" corresponds to the limiting temperature of the existence of the supercooled liquid phase. In turn, the measurement of the temperature of change of the condensation mechanism can be performed by the morphological features of the sample by electron microscopic examinations of the substrate with the deposited film.

Many studies have also been devoted to the theoretical analysis of the crystallization processes of supercooled melts [6-9]. When considering the peculiarities of supercooling in nanostructures, researchers focus on the reasons that ensure the liquid state of free particles remains below the equilibrium melting point and the size dependence of the supercooling value. The influence of the presence of a solid phase at temperature limits of stability of the liquid state, as well as thin effects at interfaces, are also studied.

In addition to the purely applied aspect, due to the unique properties of metastable melts, the study of supercooled liquids is also important from a general scientific point of view. For example, the structure of supercooled liquids is determined by the value of supercooling achieved [1]. According to [2], the value of supercooling at which crystallization happens can determine the morphology of the sample. In particular, during crystallization under maximum supercooling, bismuth particles that are located on an amorphous carbon substrate have an irregular shape, which is typical for a substance that was not in a liquid state. At the same time, the particles which were crystallized at lower supercooling are characterized by the expected shape of the spherical segment. According to [2], this behavior is explained by the near-order peculiarities in the most supercooled liquids. The results of many studies [10-14], according to which supercooling in systems of type "supercooled liquid - solid body" is determined by wetting of a solid surface with melt, indicate the importance of investigating the stability limits of the liquid phase in such systems to study the phenomena happening at "solid-liquid" interfaces.

This monograph, along with a brief review of the thermodynamic concepts used to quantitatively describe the supercooled state and crystallization of metastable melts, contains a detailed discussion and systematization of many experimental results concerning the peculiarities of crystallization of supercooled melts under various conditions. Without claiming to be general and to cover all the results available in the literature, the authors do not consider methods of creating supercooled melts based on freezing of crystallization processes (which are widely used to create amorphous materials and, in particular, metal glasses), but focus on supercooled systems that are in a metastable state due to the removal of potential crystallization centers. The main results presented in this work were obtained using vacuum condensates, which not only ensure high sample purity but also automatically implement the method of micro-volumes in most cases. Foremost, our own results concerning supercooling during the crystallization of a low-melting component in thin films obtained by vacuum condensation are discussed. Given that much of the experimental data were obtained using the author's unique methods, the presentation of the results is accompanied, if necessary, by a description of the specifics of the experiment.

The authors consider it their duty to express their sincere gratitude to O. O. Nevgasimov for his assistance in preparing the graphic material. It should also be noted that a significant contribution to the creation of this monograph was made by Nikolai T. Gladkikh, whose effective leadership resulted in the main set of results obtained by the method of changing the condensation mechanism and the beginning of research using the original *in situ* methods.

## 1.2 Thermodynamics of homogeneous and heterogeneous crystallization

In this subsection, being briefly reviewed the simplest concepts about the thermodynamics of the crystallization process, which are necessary for understanding the experimental results presented in the following sections. A detailed description of some of the crystallization theories developed on the basis of different approaches can be found in numerous reviews, monographs and original works [15-21].

Despite the fact that the ratio of the volume components of free energies of components is usually used to determine the thermodynamic stability of a particular phase, the actual formation of a new phase during changes in thermodynamic parameters is a more complex process. Current theories of phase transitions are based on the assumption that the formation of a new phase is carried out by a germinal mechanism. Let us consider the process of crystallization of liquid as an example. Due to the thermal motion of atoms in the melt, the fluctuation appearance of structures with a crystal lattice typical of the solid state of a given sample is always possible. Fig. 1.2 shows a schematic representation of such a nucleus in a volume of liquid. If the free energy of the crystalline state under these conditions is higher than that of the melt, the decay of such crystallites will be statistically natural. Further growth of crystal nuclei and subsequent crystallization of the entire melt will only be possible if the sample temperature is low enough so that the free energy of the new phase is lower than that of the initial.

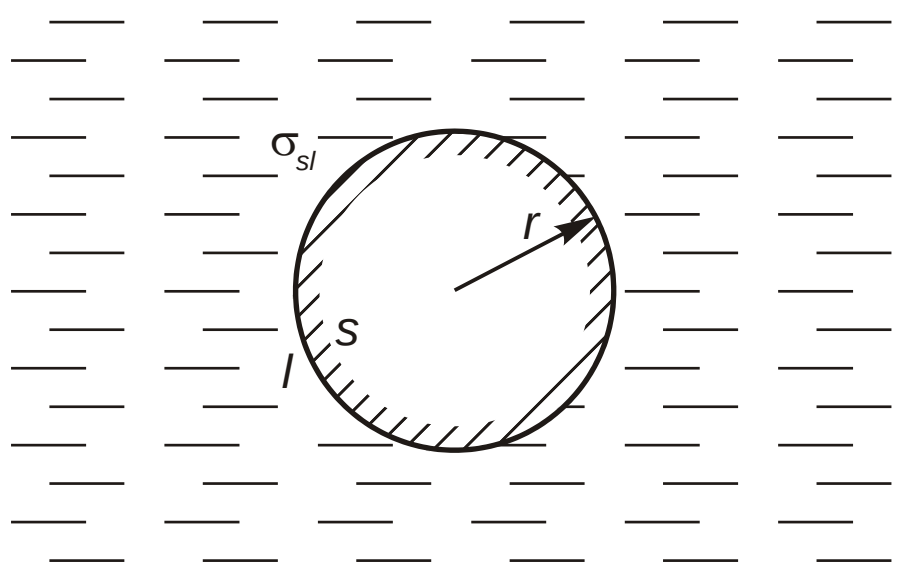

**Fig 1.2** Schematic representation of free crystallization nucleus in the melt

However, in order to describe the thermodynamic state of systems, especially in the case if the sample is small enough, in addition to the volumetric free energy term, it is often necessary to take into account the surface energy one, which is connected with the appearance of a crystal-melt interface at the initial stage of the phase transition. Thus, if the temperature of the sample is below the equilibrium melting point, the possibility of the start of a phase transition will be determined by two competing processes: a decrease in the free energy of the system (at this temperature, the change in the volumetric free energy is negative: $\Delta G_v = G_{sv} - G_{lv} < 0$) and a positive contribution to the total free energy, which is made to it by the surface of the growing nucleus (Fig. 1.3).

Based on this, it can be obtained that during the formation of a spherical nucleus of the crystalline phase with a radius *r*, the change in the free energy of the system will be determined by the expression:

$$\Delta G = \frac{4}{3}\pi r^3 \Delta G_v + 4\pi r^2 \sigma_{sl}\,, \tag{1.1}$$

where $\Delta G_v$ is the difference between the volumetric free energies of the liquid ($G_{lv}$) and solid ($G_{sv}$) phases; $\sigma_{sl}$ is the surface energy at the "crystal – supercooled melt" interface. The behavior of individual terms from formula (1.1) and their sum with increasing size of nucleus is shown in Fig. 1.3. In the graph, the lower curve represents a decrease in the volumetric term of the free energy, and the upper curve corresponds to its increase due to the appearance of a separating surface. Due to the different power law dependence of the terms, their sum (the middle curve) has a maximum at a certain value of the nucleus radius.

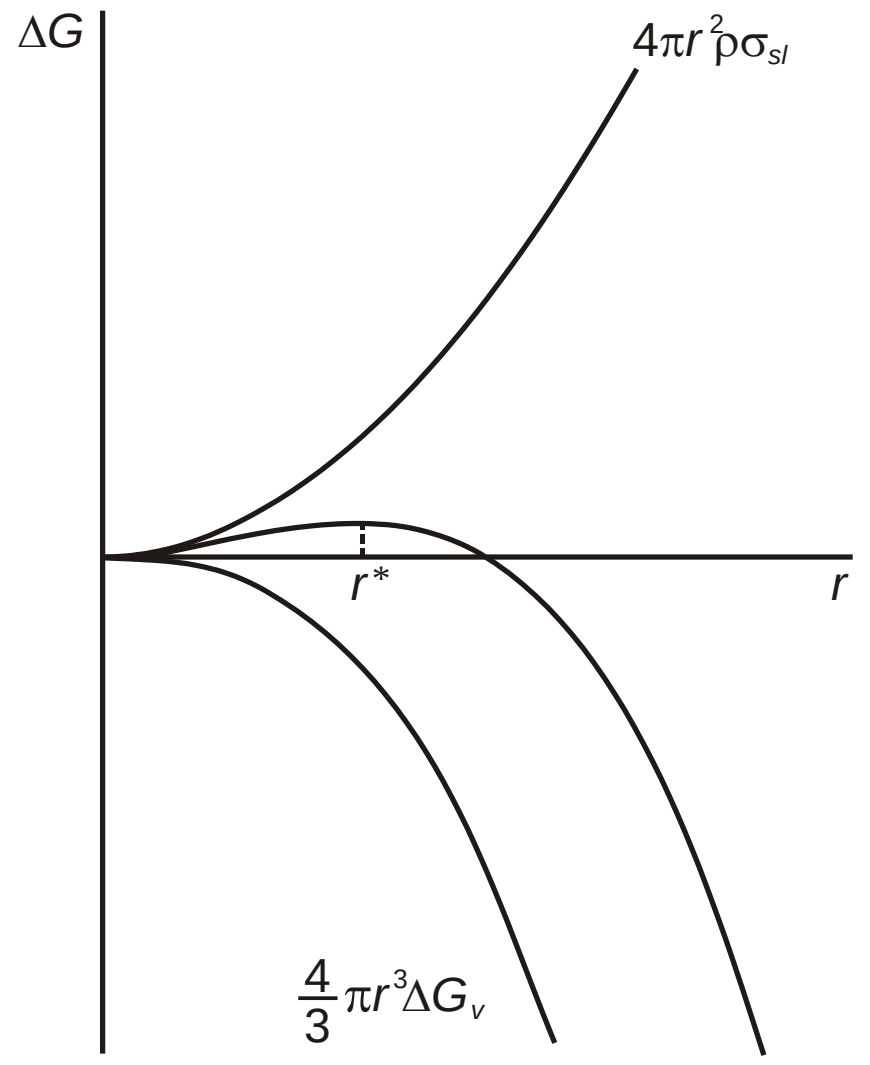

**Fig. 1.3** Size dependence of free energy change during the formation of new phase nucleus

Thus, from (1.1) and Fig. 1.3 it can be seen that under these thermodynamic conditions, there is a certain critical radius of the nucleus; adding atoms to such a nucleus makes it thermodynamically stable, which in turn will cause further growth. Below the critical size, the nucleus is not stable and therefore will decay. Thus, the radius of the critical nucleus, determined as the extremum of (1.1), will be equal to:

$$r^* = -\frac{2\sigma_{sl}}{\Delta G_v}\,. \tag{1.2}$$

Since the value of the difference in free energies in the liquid and solid state $\Delta G_v$ is determined by the temperature, the radius of the critical nucleus is a function of temperature. This temperature dependence can be obtained explicitly as follows. The partial derivative of the Gibbs potential with respect to temperature at constant pressure and number of particles is given by the expression:

$$\left(\frac{\partial G}{\partial T}\right)_{P,N} = -S\,.$$

By replacing the differentials in this equation with finite increments, for some temperature $T$ in the vicinity of the phase transition point $T_s$, we can obtain:

$$\Delta G_v = G_{sv}(T) - G_{lv}(T) =$$
$$= G_{sv}(T_s) - \frac{\partial G_{sv}}{\partial T}(T_s - T) - G_{lv}(T_s) + \frac{\partial G_{lv}}{\partial T}(T_s - T) = (T_s - T)(S_s - S_l).$$

Taking into account that the amount of heat released during the first-order phase transition is determined by the entropy jump: $q = T(S_2 - S_1)$, the expression for $\Delta G_v$ can be finally written in the following form:

$$\Delta G_v = \frac{T_s - T}{T_s} q = -\frac{\Delta T}{T_s}\lambda\,, \tag{1.3}$$

where $\lambda$ is the latent heat of fusion normalized to volume ($\lambda = -q$). Hence, assuming the independence of $\sigma_{sl}$ from temperature, the expression (1.2) can be written:

$$r^* = \frac{2\sigma_{sl}}{\Delta T}\frac{T_s}{\lambda}\,. \tag{1.4}$$

Using (1.4), it is possible to estimate the maximum value of supercooling $\Delta T = T_s - T$, which can be observed in thermodynamic systems. Assuming that the radius of the critical nucleus is equal to the size of the atom, which is obviously the smallest possible value for physical reasons, we obtain that the expected limiting supercooling in the case of free particles (i.e., under the condition of homogeneous crystallization) is $\Delta T \approx 0{,}3T_s$. The results obtained using levitation methods, as well as vacuum condensates on the surface of inert carbon or oxide layers, generally confirm the correctness of the estimation [22–26].

An explicit form for the height of the energy barrier $\Delta G^* = \Delta G(r^*)$ can be obtained by substituting the expressions for $r^*$ (1.2) and $\Delta G_v$ (1.3) into (1.1):

$$\Delta G^* = \frac{16\pi}{3}\left(\frac{T_s}{\Delta T}\right)^2 \frac{\sigma_{sl}^3}{\lambda^2}\,. \tag{1.5}$$

During homogeneous crystallization, the probability of the creation of a critical nucleus is the same at all points in the volume of the initial phase, and therefore any of its atoms can become the centre of nucleation. According to [27], the number of critical nuclei $n^*$ in a system from $N$ atoms is determined by the following expression:

$$n^* = N \exp\left(\frac{\Delta G^*}{kT}\right). \tag{1.6}$$

The formation of a single nucleus is sufficient to start the crystallization process, i.e. it can be put, $n^* = 1$, and then from (1.5) and (1.6) follows the equation that was used to analyze the results of studies of the limiting supercooling during the crystallization of many metals [10, 13]:

$$\left(\frac{\Delta T}{T_s}\right)^2 \frac{T}{T_s} = \frac{16\pi}{3k \ln N}\left(\frac{\sigma_{sl}}{\lambda}\right)^3\left(\frac{\lambda}{T_s}\right). \tag{1.7}$$

Expressions (1.4)–(1.7) were obtained for the case of homogeneous crystallization, i.e. they are applicable when the crystal phase nucleus is formed directly in the volume of supercooled melt and is in contact with it exclusively. At the same time, at the very beginning of the experimental study of supercooling, it became clear that the presence of foreign components (primarily solid impurities, the interphase energy on which differs from the value for the surface of free particles) affects the value of supercooling during crystallization.

Indeed, let us assume that a rather large (compared to the size of the critical nucleus) drop of supercooled melt is located on the flat surface of a particle of the solid impurity (Fig. 1.4).

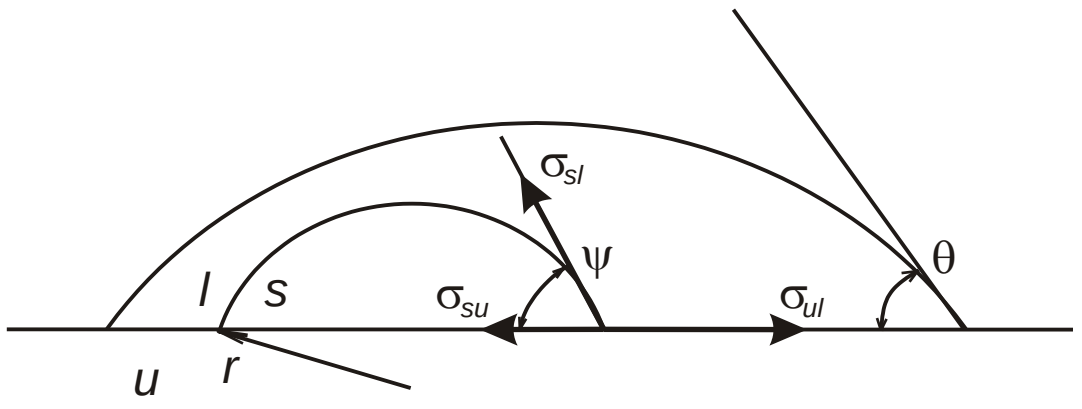

**Fig. 1.4** Scheme of the crystallization nucleus formation (subscript *s*) in a supercooled liquid (*l*) on a flat substrate surface (*u*)

Then, on the contact line of the three phases, where the surfaces of the interfaces "crystal nucleus – substrate" (interphase energy $\sigma_{su}$), "supercooled melt – substrate" ($\sigma_{lu}$) and "crystal nucleus – melt" ($\sigma_{sl}$) converge, the following relation will be fulfilled:

$$\sigma_{lu} = \sigma_{su} + \sigma_{sl} \cos\psi. \tag{1.8}$$

Experimental measurement of the contact angle $\psi$ is an extremely difficult, if not impossible, task. However, it should be noted that the influence of the solid surface on the energy balance during the phase transition is determined to a first approximation by the interaction between the melt and the substrate (quantified characteristics of that is the corresponding wetting angle θ). Therefore, instead of ψ, the value of θ will be used as a first approximation, which is quite simply determined in the experiment.

The change in free energy due to the formation of a nucleus on a flat surface during heterogeneous crystallization is measured by the following expression:

$$\Delta G_{het} = V\Delta G_v + S_{sl}\sigma_{sl} + S_{su}(\sigma_{su} - \sigma_{lu}). \tag{1.9}$$

Due to the fact that the presence of a solid substrate surface changes the shape of the nucleus, the volumetric component of the free energy also undergoes changes. Assuming that the nucleus on a flat surface has the shape of a spherical segment, its volume is reduced compared to the volume of a free spherical nucleus of the same radius:

$$V = \frac{4}{3}\pi r^3 \Phi(\theta),$$

where Φ(θ) is a geometric factor that determines the proportion of the volume of a full sphere that corresponds to a spherical segment with an angle of θ at the base:

$$\Phi(\theta) = \frac{1}{4}(2+\cos\theta)(1-\cos\theta)^2 = \frac{1}{4}\left(2 - 2\cos\theta - \sin^2\theta\cos\theta\right). \tag{1.10}$$

Using the obvious geometric relations for the surface areas of the interfaces that restrict the nucleus on a flat surface, namely $S_{sl} = 2\pi r^2(1-\cos\theta)$, and $S_{su} = \pi r^2 \sin^2\theta$ , we can write:

$$\Delta G_{het} = \frac{4}{3}\pi r^3 \Phi(\theta)\Delta G_v + 2\pi r^2(1-\cos\theta)\sigma_{sl} + \pi r^2 \sin^2\theta(\sigma_{su} - \sigma_{lu}),$$

and then, taking into account the balance of interphase energies on the line of triple contact (1.8), we obtain:

$$\Delta G_{het} = \frac{4}{3}\pi r^3 \Phi(\theta)\Delta G_v + 4\pi r^2 \sigma_{sl}\Phi(\theta). \tag{1.11}$$

The condition of the extremum (1.11) leads to the expression that determines the radius of the free surface of the critical nucleus during heterogeneous crystallization:

$$r^*_{het} = -\frac{2\sigma_{sl}}{\Delta G_v}. \tag{1.12}$$

The obtained expression is formally identical to (1.2), which determines the radius of the critical nucleus formed in the volume of the pure melt. At the same time, the actual size (the maximum distance between any two points) of the critical nucleus in the case of wetting (more precisely, starting from θ < 30) can be much smaller.

From the comparison of (1.1) and (1.11), it follows that the change in free energy due to nucleation during heterogeneous crystallization will be smaller than in comparison to the case of homogeneous crystallization (1 < Φ(θ) < 0):

$$\Delta G_{het} = \Phi(\theta)\Delta G .$$

In accordance with the factor Φ(θ), the value of the nucleation barrier will decrease. This circumstance allows us to extend expression (1.7) to the case of heterogeneous crystallization:

$$\left(\frac{\Delta T}{T_s}\right)^2 \frac{T}{T_s} = \frac{16\pi}{3k \ln N}\left(\frac{\sigma_{sl}}{\lambda}\right)^3 \left(\frac{\lambda}{T_s}\right)\Phi(\theta) . \quad (1.13)$$

The intensity of the influence of the wetting angle on the value of supercooling is shown in Fig. 1.5, which shows a graph of the function Φ(θ).

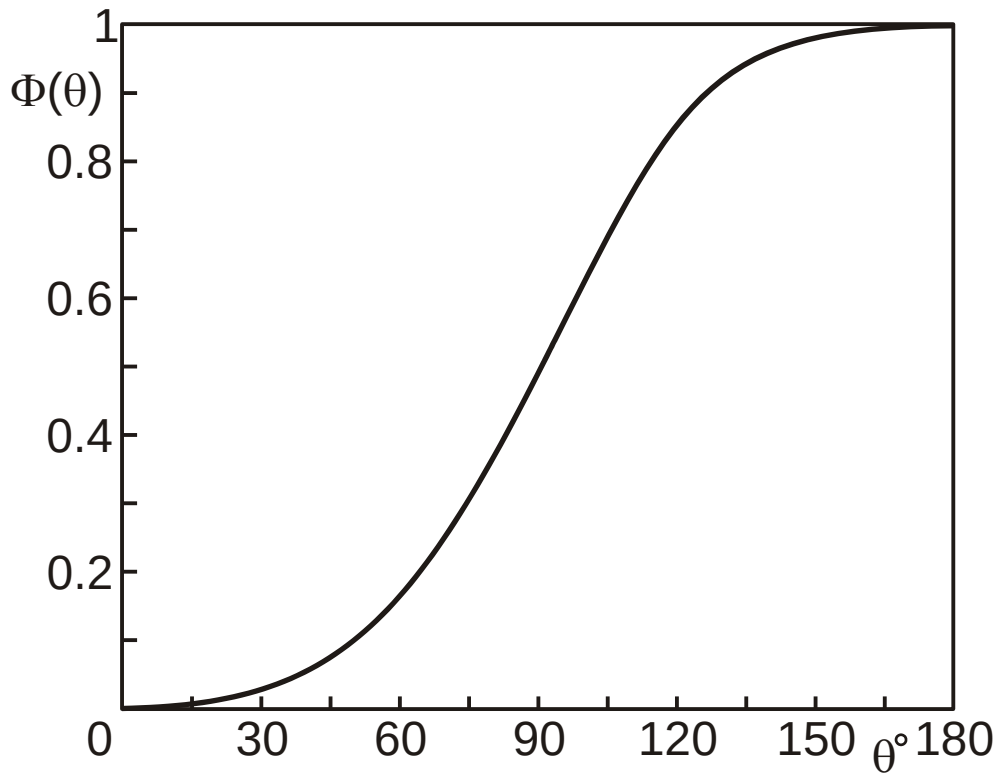

**Fig. 1.5** Function graph of Φ(θ)

It can be seen that the influence of the solid impurities (which in vacuum condensates are most often played by the substrate) on the value of supercooling decreases with the deterioration of their wetting by the melt. Starting at contact angles of approximately 130–140°, the solid particles hardly increase the crystallization temperature. Note that such values of θ are typical for fusible metals on carbon and some oxide substrates. This opens up the possibility of a fairly simple observation of crystallization, which happens under conditions as close as possible to homogeneous nucleation.

## 1.3 Structural peculiarities of melts in a supercooled state

The thermodynamic approach discussed in the previous section allows us to predict the very possibility of the existence of a supercooled state. However, simple thermodynamic considerations have limitations in describing the peculiarities that are specific precisely to a metastable melt. After all, as the temperature decreases, not only the free energy of a substance decreases, but also, for example, its diffusion activity. Under conditions of rapid cooling, the slowing down of diffusion makes it possible to form metallic glasses and other amorphous substances. Such materials combine the amorphous crystallographic state with the properties of a solid and are of great applied importance [28, 29, 30].

However, even in the case of slow cooling, which ensures thermodynamic equilibrium in the system, local atomic ordering and other peculiarities that are not typical of stable liquids can be observed in metastable melts. In particular, this can affect the crystallization process, resulting that the properties of the crystals may depend on the degree of supercooling at which they were obtained. A study of the relationship between the degree of supercooling and the morphology of crystals formed during the crystallization of melts was carried out in [2]. In this work, thin films of bismuth condensed on graphite substrates were studied. Unlike amorphous carbon, graphite has a significant orientation effect. At the same time, it retains the chemical inertness that amorphous carbon has in relation to fusible metals.

The authors of [2] deposited bismuth films on substrates at different temperatures and the phase state of the samples was determined via the electron-diffraction method. The electron-diffraction studies system was mounted directly in a vacuum chamber. According to [2], the supercooled melt is quite stable. Many hours of exposure to Bi/C films at a temperature 10-15 K above the temperature of maximum supercooling did not cause melt crystallization. On the other hand, it was found in [2] that an increase in the cooling rate of the metastable melt from 0.9 to 9.5 K/c causes an increase in supercooling by 6 K. At the same time, even this cooling rate is insignificant compared to the rates used to obtain solid amorphous structures [31, 32, 33]. The increase in the supercooling under the condition of a relatively small increase in the cooling rate indicates the formation of local partially ordered states in the metastable melt, which can be studied as an independent scientific field.

The particular structural state of a metastable melt becomes most important precisely in the vicinity of the crystallization temperature. As shown in [2], the shape of the particles obtained during the crystallization of melts, which are located on different degrees of supercooling, differs. Thus, under the condition of slight supercooling, the particles that form the Bi/C island films retain their spherical shape during crystallization. However, bismuth particles that have been condensed into a metastable liquid state near the crystallization temperature take on an irregular shape after cooling. They partially resemble the structures that form when deposited directly into the solid phase. However, according to electron diffraction patterns data, after condensation and before cooling of the substrate, such particles are in a liquid state. The authors of [2] explained the irregular shape of crystals that arise under the condition of maximum supercooling by the quasi-liquid state of melts. According to the authors of [2], despite the absence of a diffraction pattern, the substance under such supercooling conditions has a high degree of order, approaching the crystalline state in its properties.

In [34], it is shown that the value of the maximum supercooling is greater, the bigger the differences in the structure and density of the melt and its crystal are. Melting itself is described in [34] by introducing the fraction of delocalized atoms. In other words, according to [34], the formation of a liquid phase happens when the crystal lattice becomes unstable due to an increase in the number of delocalized atoms. Accordingly, crystallization, which is a reverse process, vice versa, requires an increase in the orderliness of the sample. This becomes possible when the relative number of localized atomic groups formed in the melt by fluctuations' way increases sufficiently to ensure the existence of a stable crystal lattice.

The complex nature of the structure of metastable states is also indicated by the results of [35]. In this work, the initial stages of metal-induced crystallization of germanium observed in the Ag/Ge system were investigated. The system under study is quite complex, since the germanium layers deposited by vacuum condensation on a substrate at room temperature are in an amorphous state. This state is metastable, and the films themselves can be converted to a crystalline state by annealing. This contact pair is of the eutectic type, with a significant reduction in melting point relative to the melting points of the pure components.

Thus, at a certain temperature, two metastable phases can simultaneously exist in Ag/Ge films: amorphous germanium layers and supercooled eutectic melt. According to the results of [36], the presence of a eutectic melt initiates the transformation of amorphous layers into a stable crystalline state. However, as established in [35], this process is complex and involves the appearance of intermediate phases which behavior resembles the liquid state of a substance.

Fig. 1.6 shows a sequential series of HRTEM images corresponding to the initial stages of metal-induced crystallization. The process of metal-induced crystallization starts under the supercooled eutectic particle with the formation of $Ag_{0.8}Ge_{0.2}$ structures. Such compounds show signs of crystallinity and coexist with a disordered phase.

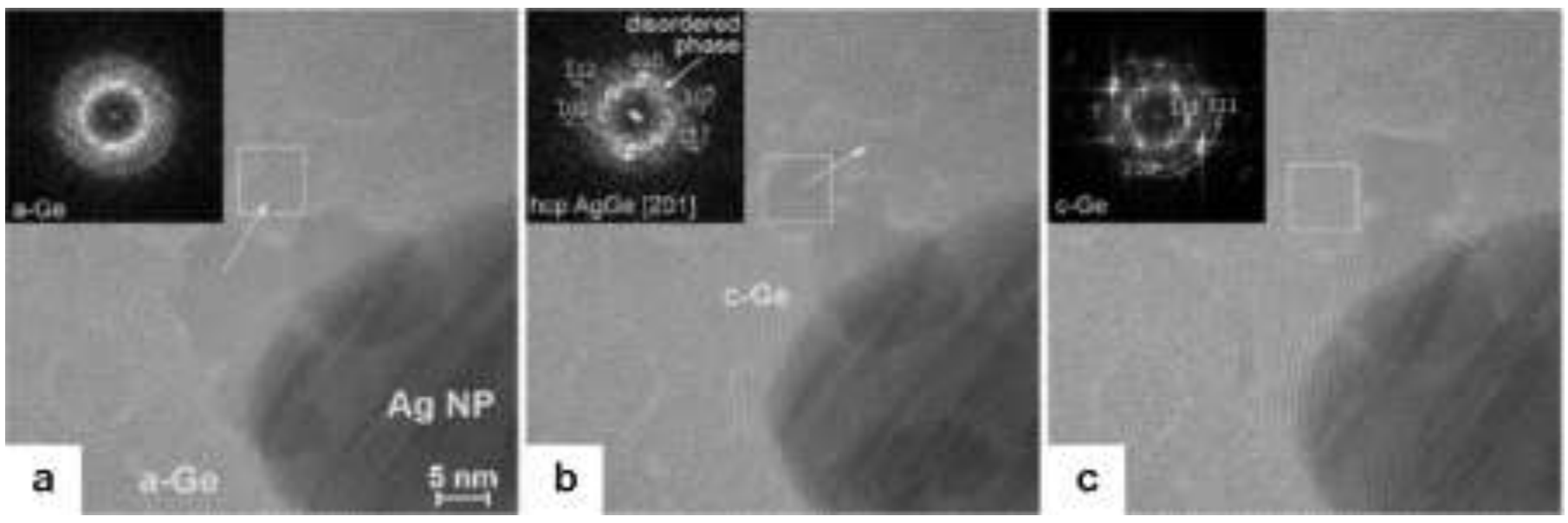

**Fig. 1.6** Time series TEM images of the same area of Ag/Ge film at 450 °C showing the propagation of a nucleus and crystallization of a-Ge film. FFT spectrum in the inset

corresponds to the rectangular region. Arrows in TEM images indicate the direction of the nucleus advancement. Dot arcs in the inset in (c) correspond to crystalline Ge [35]

The $Ag_{0.8}Ge_{0.2}$ structures, that have formed under a supercooled particle, flow to neighboring areas of the film (Fig. 1.6). This is what allows us to name them liquid-like particles to some extent. After the outflow, which happens in a jump, a stable crystalline phase is formed at the site of the particle. The silver that was involved in the formation of the intermediate phase moves into the neighbor area of the sample and begins its crystallization.

The structural peculiarities of the liquid phase are even more obvious in some multicomponent systems, in particular in Sb-based melts. In [37], the microscopic dynamics of $GeSb_2Te_4$ melts were studied using quasi-elastic neutron scattering. The quasi-elastic neutron scattering method allows independently determining the viscosity and diffusion coefficient of a material. On the other hand, the relationship between viscosity and diffusion coefficient can be obtained from the Stokes-Einstein equation:

$$D\eta_v = \frac{kT}{6\pi r_h} \quad , \qquad (1.14)$$

where $D$ is the diffusion coefficient, $\eta_v$ is the viscosity, and $r_h$ is the effective hydrodynamic radius.

It is generally believed that above the melting point, the dependence (1.14) is well fulfilled. However, in supercooled melts, deviations from (1.14) are observed, which is explained by the dynamic heterogeneity of metastable liquids [38, 39, 40].

In [37], it was found that for GeSb2Te4, the dependence (1.14) is violated even at temperatures well above the melting point. The authors determined the temperature of violation of dependence (1.14) to be 1050 K, which corresponds to $1.16T_s$. The high temperature at which deviations from (1.14) are observed does not allow the authors to use models of dynamic heterogeneity [38, 39, 40]. According to [37], at a temperature of 1050 K, a liquid-liquid phase transition (LLT) happens in the melt, i.e., the melt structure changes.

Violations of the dependence (1.14) were also observed in [41] for $Ge_2Sb_2Te_5$, GeTe, $Ag_4In_3Sb_{67}Te_{26}$, and $Ge_{15}Sb_{85}$ films. The deviation from the Stokes-Einstein relationship provides a crystallization rate that is approximately an order of magnitude higher than the expected values. The authors of [41] attribute the observed peculiarities to local ordering happening in the liquid phase.

The study of liquid-liquid structural transformations in $Ag_4In_3Sb_{67}Te_{26}$ and $Ge_{15}Sb_{85}$ films was performed in [42]. The combination of femtosecond heating and fast recording of diffraction patterns made it possible to achieve high time resolution in the study of processes in the liquid state. The authors of [42] used the ratio of the radii of the first ($r_1$) and second ($r_2$) coordination spheres as a parameter quantitatively characterizing the structural ordering. The criterion for the liquid phase transformation was the break at the temperature dependence of $r_1/r_2$. Such a break is observed at a temperature of 660 K for $Ag_4In_3Sb_{67}Te_{26}$ films. In $Ge_{15}Sb_{85}$ samples, the phase transformation in the liquid state happens at a temperature of 610 K. The typical size of the spatial ordering areas defined in [42] is 0.58-0.59 nm and is twice the radius of the first coordination sphere. The ordering observed by the authors of [42] is associated by them with Peierls distortion, which provides a gain in the chemical potential of the system. The liquid-liquid phase transition in the $Ge_3Sb_6Te_5$ melt was also observed in [43] by changing the temperature dependence of the melt viscosity.

Thus, the liquid phase is a complex structure. It cannot be fully described within the framework of the simple thermodynamic models presented in subsection 1.2. Due to the local ordering, the achieved value of supercooling can affect the properties of the obtained crystals, and liquid-liquid phase transitions can be observed directly in the liquid state. Given the high rate of phase transformations reported by many researchers [41, 43] and the stability of some metastable states [2], metastable structures are gaining prospects for applied application [44]. A positive feature of this approach is the significant difference between the physical properties of substances, which are in different phase states. For example, in addition to viscosity [41, 43], optical and structural characteristics, a change in the phase state can also change the type of conductivity, which seems to be beneficial for the development of modern electronics [45].

At the same time, the complexity of the liquid state, and especially the supercooled liquid state, makes it important to systematize data on the actual supercooling of melts in various systems. The monograph attempts to systematize the results on the existence of metastable states, which are caused actually by the presence of the crystallization energy barrier (subsection 1.1). The presented results were obtained using different and complementary experimental methods. They demonstrate the influence of interphase interaction and morphology on the temperature stability of the liquid phase in single-component and binary melts.

# Chapter 2
# Crystallization of free particles and island films

**Abstract** An experimental study of supercooling is carried out using various techniques. Due to the high sensitivity of the supercooled state to impurities, the most reliable results can be obtained using vacuum methods of studies. The use of vacuum deposition proves to be particularly effective in the case, when the study of supercooled liquid takes place immediately after it is obtained. This condition corresponds to the original method of changing the condensation mechanism, which is one of the variants of the technology of samples of variable content and variable state. It is with the use of this method that the main results presented in this chapter were obtained. The method of condensation mechanism change is based on the morphological criterion for determining the phase transition temperature and the assumption that particles that are liquid at the initial stages of obtaining remain liquid as long as the substrate temperature exceeds the temperature of maximum supercooling.

Using the method of condensation mechanism change, supercooling temperatures in a large number of contact pairs were determined and it was shown that the dependence of relative supercooling on the wetting angle is close to linear. The influence of the residual atmosphere on the supercooling value was established. It is worth noting that the results obtained using this method refer to the same vacuum cycle in which the samples were condensed. Moreover, they correspond to the supercooled state that exists directly during the deposition of the films.

## 2.1 Using the microvolumes method to study supercooling during metal crystallization

As an example of the first modern studies of the nucleation of the crystalline phase in a supercooled melt, we can mention the work [1], which was devoted to the study of the crystallization of droplets of liquid tin and water. Unlike his predecessors, the author focused not on the kinetics of crystal phase spread in the crystallizing melt, but on the process of its nucleation. For this purpose, an original methodology was developed in the work [1] based on the division of the melt into many separate spatially spread droplets, each of which should obviously crystallize independently of the others. The time of complete crystallization of each of these particles, in the framework of assuming no size dependence of the crystallization rate, is directly proportional to the radius of the particle. At the same time, according to the author's assumption, which was confirmed in further theoretical and experimental works, the probability of the formation of a crystal nucleus in one of the particles is proportional to the square or even the cube of its radius. Thus, in a particle of sufficiently small size, the time required for the formation of a nucleus will be significantly less than the time of its complete crystallization.

In the work [1], the behavior of rather large tin particles, ranging in size from 1 to 10 μm, was investigated. Two different methods were used to study the crystallization kinetics. In the first, tin particles were mixed with bakelite varnish and the resulting mixture was applied to a glass substrate. The sample was then examined by the XRD methods while being heated in a hydrogen atmosphere. The intensity of the diffraction reflexes normalized to the value that is to be for the initial sample was used as a criterion for the amount of the liquid phase.

The author found that after cooling the sample below the equilibrium melting point, the relative amount of particles that are in the liquid state decreases rather slowly and is determined by the value of supercooling. Thus, at a melt temperature of 130 °C, about 40 % of liquid particles remain in it after 20 minutes. It is only after about an hour that their number drops to 5–10%. It should be noted that in these studies, rather large relative supercoolings were obtained (the value of relative supercooling is determined by the expression $\eta = \Delta T/T_s$ , where $\Delta T = T_s - T_g$ – is the difference in melting and crystallization temperatures of the substance under study). The 130 °C crystallization temperature of tin melt obtained in the work [1] corresponds to $\eta = 0.2$.

The second technique, used by the author [1] and designed to overcome the high inertia of the furnace in the first method, was based on dilatometric measurements. The tin particle powder was in an ampule filled with liquid and connected to a capillary. The change in the level in the capillary was used to determine the proportion of the solid phase in the sample. It is important to note that before the powder was placed in the ampule, it was oxidized by heating in air. This was necessary so that the particles were covered with an oxide film that would prevent their coalescence. Thus, the studied particles of the supercooled melt were in contact with their own oxide. As a result of these studies, it was shown that the crystallization rate of the samples increases with their cooling and, upon reaching a temperature of 116.7 °C, is happening so rapidly, that it is impossible to speak with certainty about the establishment of thermal equilibrium in the experimental setup.

Similar and no less revealing results were obtained in the work [1] for supercooled particles of water. Thus, under the condition of supercooling by 16 K (relative supercooling η is approximately 0.06), after crystallization of 60 % of the droplets, which occurs within 10–15 minutes, further crystallization stops and supercooling by 18 K is required to the full completion of the process, which in this case takes about 10 minutes.

The study of supercooling features was further developed in the work of Turnbull [2, 3], who studied the crystallization of mercury, gallium and tin particles. As in the work [1], the main results were obtained for particles covered with a shell that prevented their coalescence. Despite the fact that crystallization in the samples under study also occurred gradually, unlike in [1], Turnbull did not study the kinetics of the phase transition but focused on the final moment of the process under study. With this approach, the last of the liquid particles should be considered as being in the conditions of the maximum possible supercooling under the given conditions. The key result of the work [3] was the establishment of the fact that the relative supercooling for the studied substances is practically constant and is in the range of 0.2-0.25. Such a large value of supercooling could not be explained, based on existing thermodynamic models, by a decrease in the probability of the formation of the crystal nucleus with a decrease in particle size. The extremely large supercooling values obtained at that time for individual particles were qualified in the work [3] as the result of homogeneous crystallization. Turnbull explained the presence of particles with a significantly higher crystallization temperature (i.e., a much lower supercooling value) by the influence of the surface and impurities that facilitate crystallization.

The peculiarity of the methodology of the studies carried out in [1, 3] is the coating of particles which were subjected to supercooling with a layer that prevents their coalescence. However, such a surface layer can itself stimulate crystallization. Probably, the effect of such a layer on the crystallization process was first discussed in the work [4]. In this work, in particular, it was found that lead sulfates are a more active catalyst for crystallization compared to its oxides. When choosing a material for an insulating layer, the author of [4] recommends, in addition to the obvious criterion of material insolubility in the melt under study, choosing those coatings that have a crystal lattice that differs as much as possible from the lattice of the substance being crystallized. It was also shown in the work [3] that for particles, which size ensures maximum supercooling, the thermal prehistory does not affect the temperature and nature of their crystallization. At the same time, according to the body of scientific data available at the time, the influence of thermal prehistory on the crystallization peculiarities of rather bulk samples of various materials was a reliably established phenomenon [5, 6, 7]. This effect consisted of the fact that the value of supercooling was found to depend on the temperature of the superheating that preceded the melt supercooling. According to [5, 6, 7, 8], for bulk samples with an increase in the overheating experienced by the melt after the end of the melting, the value of supercooling first increases rapidly, and then practically does not change. At the same time, the value of required overheating can vary from several degrees to 100–150 K [6]. Despite the many approaches that offer different interpretations of the effect of melt overheating on its supercooling, all of them [6, 7, 8, 9, 10, 11, 12] assumed the presence of impurities in the melt, i.e., the effect of thermal prehistory could be explained exclusively by heterogeneous crystallization in bulk samples. The absence of overheating and the high values of supercooling that occurred in works [1, 3, 4] allow us to suggest that in these experiments, conditions for the crystallization of supercooled melts were created that are close to those, under which homogeneous crystallization is observed.

A detailed theoretical consideration of the processes of heterogeneous crystallization, i.e. crystallization occurring on foreign impurity centers, is presented in the work [8]. The key peculiarity of this work, which is a development of the ideas of [2, 6], is the transition to more understandable terms, namely the use of interfacial energy instead of heat of absorption. It should be noted that this approach proved to be quite fruitful and was widely used in later works. It was precisely in the work [2] that it was proposed to use the value of the wetting angle by a supercooled melt of the surface of the given solid body to describe the activation ability of a solid catalyst.

Thus, the results of works [1, 3, 4, 8] indicate that homogeneous crystallization, which actually allows us to study the fundamental peculiarities of this phase transition, should be observed in that case, if the melt does not contain impurities. And such conditions can indeed be created during the cooling of sufficiently small particles. These considerations led to the creation of the microvolumes method [2], which is currently one of the main ways to study supercooling.

The essence of the microvolumes method according to [2] is that the initial sample containing $m_0$ possible centers of heterogeneous crystallization (usually such centers occur on insoluble impurities) is dispersed into $m$ particles. If the number of particles obtained is significantly higher than the number of centers for which catalytic activity is typical for ($m >> m_0$), then many of these particles will be free of this impurity and, in the absence of other factors facilitating the formation of nuclei, will be in conditions under which homogeneous crystallization can be realized. The effect of the degree of dispersion on the value of supercooling will vary and will be primarily determined by the composition of the impurities. Thus, if the sample contains only one class of impurities, which obviously have the same crystallization activity, the transition from heterogeneous to homogeneous crystallization will have an abrupt character and will be observed when a certain degree of dispersion is reached. At the same time, if the sample contains impurities of several classes with different catalytic activity, this transition is blurred and a blurred transition zone separating the regions of heterogeneous and homogeneous crystallization will be observed on the dependence of the supercooling value on the degree of dispersion.

This situation seems quite natural in cases when supercooling during crystallization is studied in samples with a typical size in the range of 10–1000 μm, obtained and studied in various, sometimes quite active, environments. At the same time, the use of methods of vacuum deposition in the study of supercooling not only allows a significant reduction of the minimum sample size but also holding of studies in conditions where the only significant impurity is actually the substrate, on the surface of which supercooled particles are located.

The study by Takagi, conducted in 1954, can be called one of the first works devoted to the study of supercooling in thin-film systems [13]. In that work, the size dependence of the melting point of thin films of lead, tin, and bismuth was experimentally revealed for the first time using in situ electron diffraction. At the same time, in the work [13] the dependence of the melting point on the thickness of the films is not only obtained, but it is also found that the temperature of the disappearance of diffraction reflexes (indicating the melting of the film) and their appearing, indicating the crystallization of the melt, do not coincide with each other. According to [13], during cooling below the melting point corresponding to this film thickness, the melt continues to be in a supercooled liquid state and crystallizes only when sufficient supercooling is achieved. For lead films, it is shown that their supercooling $\Delta T = T_s - T_g$, where $T_s$ is the melting point of the bulk sample and $T_g$ is the crystallization temperature of the studied films, depends on their thickness. For example, the supercooling of films with a thickness of 50 nm is 92 K, and when the thickness is reduced to 5 nm, it increases to 157 K. At the same time, the melting point of the films decreases by only 34 K.

## 2.2 Supercooling during the island vacuum condensates formation

### 2.2.1 Method of changing the condensation mechanism

The method of microvolumes has dramatically improved the purity of supercooled particles, but given the high sensitivity of melts to impurities, it is still insufficient to study the fundamental nature of this phenomenon. The next step in reducing the influence of external factors is the use of vacuum methods, especially those that allow combining obtaining of the sample and its examination within a single vacuum cycle. A set of systematic results concerning the supercooling of island films of various metals, which are located on a substrate of more refractory substances was obtained using vacuum condensates with the help of the method of changing the condensation mechanism [14, 15, 16, 17, 18, 19, 20, 21]. Interpretation of these results requires some consideration of the peculiarities of vacuum condensation of substances.

The growth of thin films during condensation in a vacuum from the vapor phase can be carried out due to several qualitatively different mechanisms, depending on the conditions [22, 23]. For metals on non-orienting substrates, the Volmer-Weber mechanism is usually implemented. Films that grow by this mechanism are island-like until the moment when islands, which become bigger by size begin to touch each other, which ultimately leads to the formation of a continuous polycrystalline film. At the same time, due to the size dependence of the melting point at the initial stage of condensation, particles with a size of less than 1 nm can be in a liquid state. This liquid state, due to its small size, will be thermodynamically stable, i.e. it can be kept unlimitedly long. However, as the mass thickness increases, the size of these islands will also increase. If the temperature of the substrate is lower than the melting point of the substance under study in the bulk, then, starting from a certain size, which is determined by the size dependence of the melting point, the crystalline state will become thermodynamically stable instead of the liquid state.

Experimental studies show that when condensing on a sufficiently cold substrate, vacuum condensates are usually crystalline and do not contain any signs of a liquid phase. In practice, this situation is realized when a metal is condensed onto a substrate with a temperature below two-thirds of its melting point.

At the same time, with increasing substrate temperature, condensed films show signs of the existence of liquid particles, the relative amount of which increases with temperature and decreases with increasing film thickness. For sufficiently thick films, it has been experimentally established that there is a limit temperature of the substrate during condensation $T_g$, above which the film deposition is carried out through the liquid phase. This temperature is significantly lower than the value expected based on the size dependence of the melting point. At the same time, this limiting temperature practically does not change with the film thickness, but significantly depends on the substrate material on which the substance under study is deposited.

The existence of such a limiting temperature can be explained by the phenomenon of supercooling, which is realized in island vacuum condensates. As noted above, condensation of films begins with the formation of liquid particles, and with an increase in their size, the crystalline state becomes thermodynamically stable. However, islanded vacuum condensates can be considered as a natural implementation of the method of microvolumes. Thus, despite the change in the thermodynamic conditions of the existence of phases, in the case, when the substrate temperature is higher than the supercooling value, possible under these conditions, the islands can remain in a supercooled liquid state. This allows us to state that the temperature $T_g$, which separates condensation by the vapor-liquid and vapor-crystal mechanisms, will be the temperature of maximum supercooling. At the same time, an important particularity of this approach to the study of supercoolings is that signs of the presence of the liquid phase (in particular, the close one to the spherical shape of the islands) remain in the sample even after it has been cooled to room temperature.

Electron microscopic studies show that films of the same thickness condensed by different mechanisms have significantly different microstructures (Fig. 2.1).

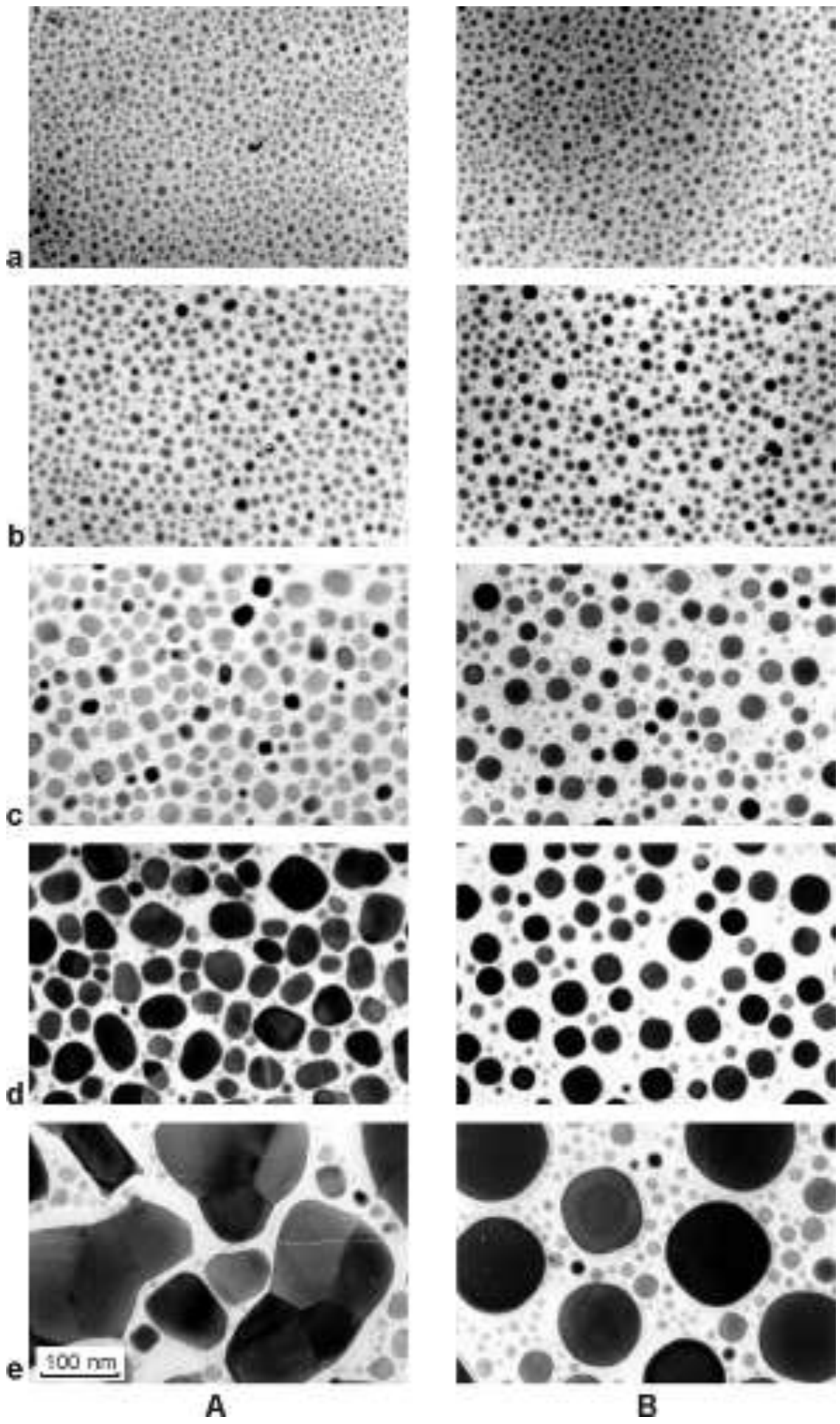

**Fig. 2.1** TEM images of tin films deposited on a carbon substrate at the temperature of 40 °C (column A – vapor → crystal mechanism) and 90 °C (column **B** – vapor → liquid mechanism). Mass thickness of the tin layer: **a** – 2 nm; **b** – 3.7 nm; **c** – 4.7 nm; **d** – 12.5 nm; **e** – 30 nm [10]

As it can be seen, when condensing on a non-wettable substrate using the vapor-liquid mechanism, the film, even after cooling to room temperature, consists of individual particles that have the shape of a spherical segment. At the same time, samples condensed by the vapor-crystal mechanism are characterized by a structure consisting of particles with a flat surface shape, and such a structure quickly turns into a continuous film with an increase in mass thickness. On non-wettable substrates, such structures differ significantly from those formed during condensation into the liquid phase. The reflection of visible light from the surface of the condensate is also significantly different so that the temperature limit $T_g$ can be visualized on a substrate with a temperature gradient [14, 17, 24]. At the same time, as the mass thickness of the sample decreases, the differences in the microstructure of films obtained by different mechanisms decrease and, accordingly, the identification of the phase transformation temperature becomes more difficult. For example, for tin films on an amorphous carbon substrate, a confident identification of the crystallization temperature of islands can be implemented at mass thicknesses of condensates of more than 5–10 nm.

More reliably, the temperature of change in the condensation mechanism can be recorded from the temperature dependence of the average particles size formed in the sample. Fig. 2.2 shows such dependencies for island condensates of tin and bismuth on an amorphous carbon substrate [14, 20]. It can be seen that on the temperature dependencies of the radius of the particles, which corresponds to the maximum in the size distribution histograms, there is a jump, which can be compared to the temperature of the change in the condensation mechanism. This method allows reliable recording of the value of supercooling for tin films with a thickness of more than 5 nm. For films with a smaller mass thickness, although this approach remains applicable, its reliability decreases.

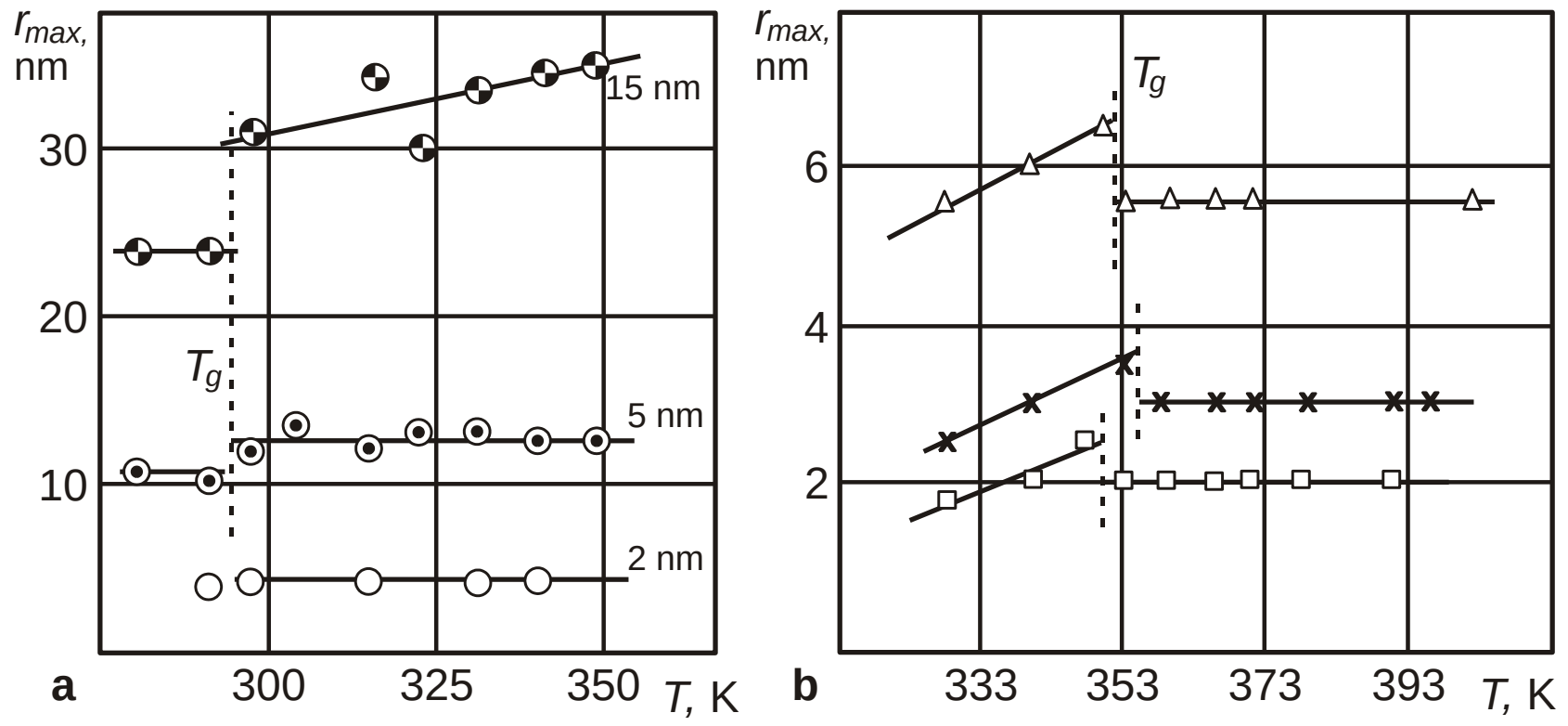

**Fig. 2.2** Island size dependence, which corresponds to the largest volume of the film substance, on the substrate temperature for the condensates of tin (**a**) and bismuth (**b**) of various mass thicknesses (shown in the figures) [10, 20]

Also, taking into account the above-mentioned differences in the morphological structure of condensates above and below $T_g$, the temperature of change in the condensation mechanism can be determined by studying the profile of islands on the substrate (Fig. 2.3). To implement this technique, one can use the methods of studying of wetting in condensed films (methods of cleavage, oblique observation, convolution [25, 26, 27]) or cross-sectional studies.

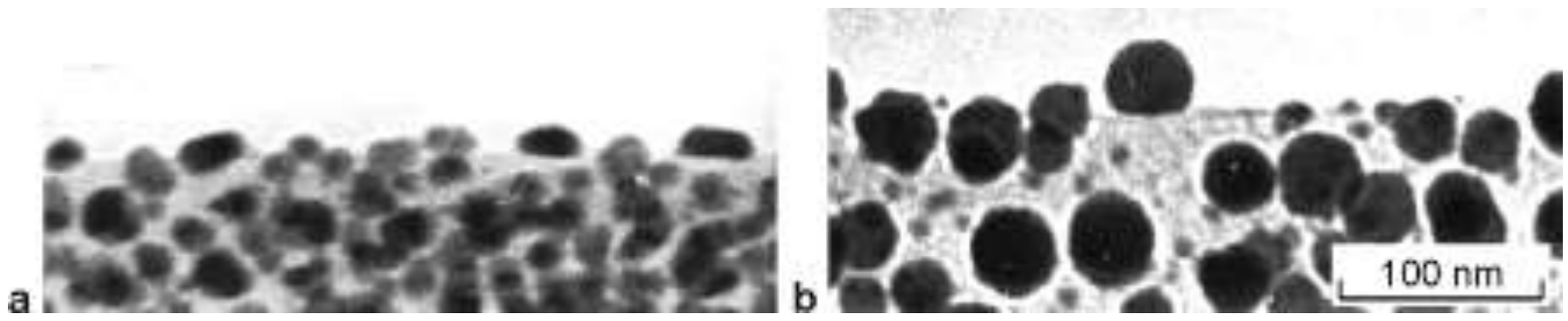

**Fig. 2.3** TEM images of bismuth particles condensed on a carbon substrate below (**a**) and above (**b**) of the temperature of changing the condensation mechanism [14]

Thus, the method of changing the condensation mechanism is a convenient means of recording supercooling values in cases where the only significant impurity is the substrate. Also, due to the possibility of varying vacuum conditions, this method allows studying the effect of different vacuum conditions on supercooling, which is of particular importance for practical application.

### 2.2.2 Effect of residual gases on supercooling

When studying the influence of the residual atmosphere on the peculiarities of supercooling of vacuum condensates, it was found that this parameter has a different effect on different metals [14]. Thus, the supercooling during the crystallization of gold films on amorphous carbon substrates obtained at a residual gas pressure of $10^{-3}$ mm Hg does not differ from the value observed in samples condensed in a vacuum of $10^{-9}$ mm Hg, which is provided by oil-free pumping systems.

However, for more active metals, such as lead, the effect of residual gases pressure is much more noticeable. Fig. 2.4 shows the results of studies of supercoolings during the crystallization of lead films condensed under different vacuum conditions. As we can see, there is a certain critical pressure of the residual atmosphere that delimits the regions where crystallization occurs with different supercoolings. It should be noted that at a pressure below the critical pressure, there is also a slight and almost linear increase in supercooling, which accompanies the improvement of vacuum conditions.

Also, from Fig. 2.4, it can be seen that the determining factor in the behavior of supercooled films is not so much the absolute value of the residual pressure as some function from the pressure and the rate of deposition of the sample. At that, the crystallization temperature $T_g$ increases with increasing residual pressure, but decreases in samples obtained with a higher condensation rate.

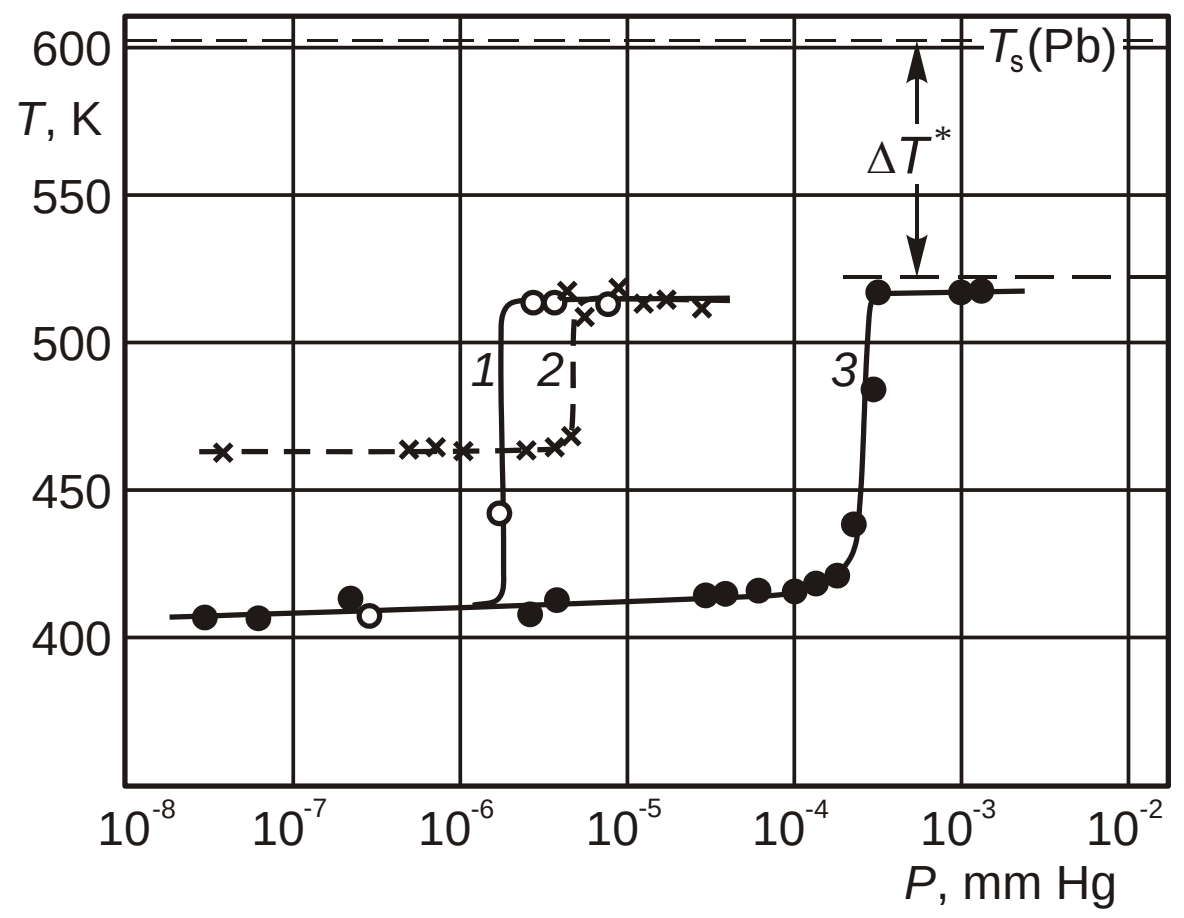

**Fig. 2.4** Boundary temperature $T_g$ dependence on residual gases pressure for lead on a carbon substrate at a condensation rate of 0.02 nm/s (1) and 0.5 nm/s (3) [10] and on a SiO substrate (2 – according to [28])

This behavior can be explained by the fact that during the process of depositions, samples are contaminated as a result of their interaction with the residual atmosphere. Impurities that got into the melt, can be divided into several classes according to their effect on supercooling. The first of these includes soluble impurities that increase surface tension. They tend to move away from the surfaces, where a new phase can be possibly formed in the sample, as much as possible. As a result of this, they do not actually facilitate phase formation in the melt. This turns out to be especially true for samples on fairly active, i.e., well-wettable substrates. At the same time, even low-active impurities that increase the surface tension can have a certain effect on practically non-wettable substrates that provide crystallization conditions that are close to homogeneous. In contrast to them, surfactants concentrate near the interfaces and improve wetting. This reduces the work of the formation of the crystal nucleus, which naturally leads to a reduction in supercooling. A separate class of impurities, which are typical for vacuum condensates, are oxides of the substances under study. Since most metal oxides are well wetted by their own pure metals, such impurities greatly facilitate the formation of crystal nuclei, thereby reducing the achievable supercoolings.

In view of this, the results presented in Fig. 2.4 can be explained by the formation of oxides in the films. In a low vacuum, virtually each of the islands, that make up the island films, will contain particles of oxides. Their size will be sufficient for the crystallization of an island at a lower supercooling than the one that is possible for the contact pair of the “film-substrate” under study. At the same time, starting from the moment when the particles of oxides, which are formed, will have a size insufficient to facilitate crystallization, or their catalytic effect will turn to be comparable to the influence of the substrate, the supercooling naturally increases and approaches the limit value. This situation probably occurs when the size of the emerging particles of oxides turns out to be comparable to or significantly smaller than the size of the critical nucleus characteristic of the given conditions. Since the formation and growth of an oxide particle demand a finite amount of time, the effect of the condensation rate on the value of supercooling becomes clear.

The significant role of gas impurities in the limitation of the maximum supercoolings is indicated by studies of supercooling during the crystallization of tin island films prepared under different vacuum and under conditions of oxygen, argon, and helium injection into the vacuum chamber [29]. It was found that the value of supercooling depends on the ratio of the number of deposited tin molecules ($n_m$) to the number of residual atmosphere molecules that fall on the substrate per unit time ($n_0$). Fig. 2.5 shows in log scale the results of such studies performed for tin condensates on a carbon substrate.

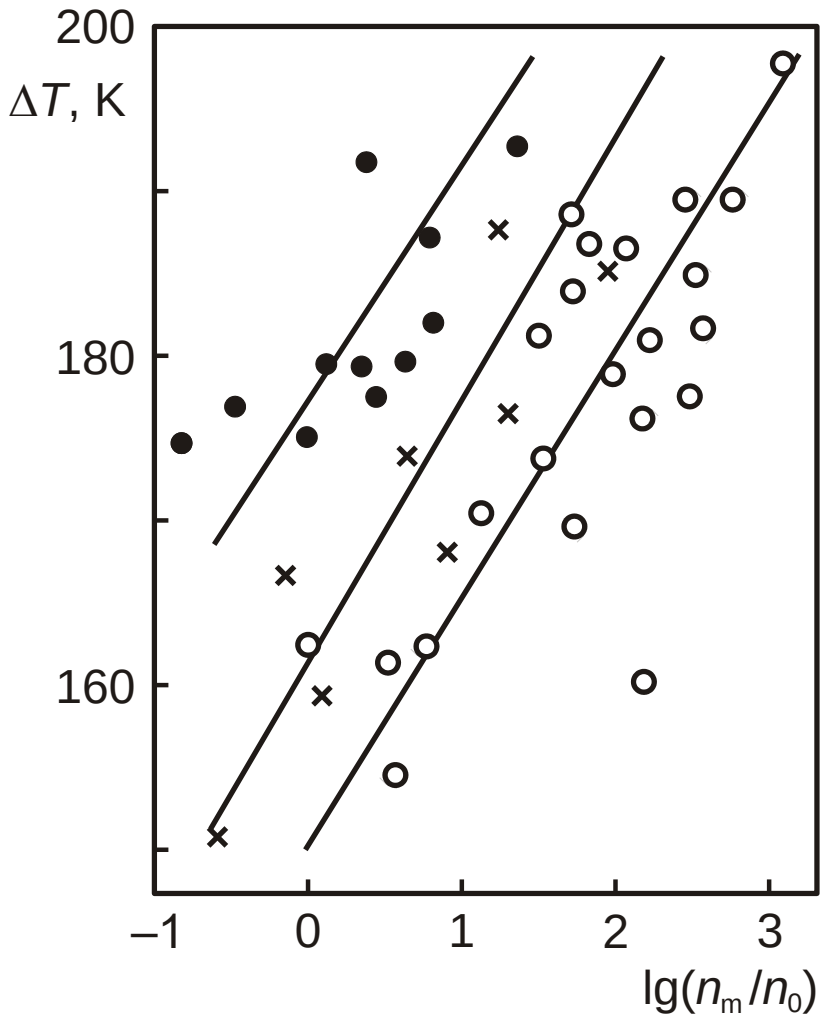

**Fig. 2.5** Boundary temperature $T_g$ dependence on $n_m/n_0$ for island tin films condensed on a carbon substrate in the initial residual atmosphere (◦) and enriched with oxygen (x) and inert gases (•) [29]

Figure 2.5 shows that it is the value of $n_m/n_0$ that is decisive in achieving the limit supercoolings. In addition, the activation capacity of different gases turns out to be different. The equilibrium residual atmosphere created by the getter ion pump of the orbitron type, which consists mainly of gases of the methane group and their decomposition products, has the highest crystallization activity among the studied environments. Oxygen is in second place, the effect of which is likely due to the formation of oxides, usually insoluble in the melt of the native metal. The most favorable environment for achieving deep supercoolings (at the same residual pressure) is, naturally, an inert gases-enriched atmosphere.

The results presented above made it possible to determine the conditions of sample preparation under which the values of supercooling, which are observed, will be close to the values that are limiting for the contact pair under study. Namely, to achieve maximum supercoolings during crystallization, i.e., those corresponding to homogeneous nucleation, it is necessary to obtain samples with a condensation rate of several nanometers per second at a residual gas pressure of $10^{-9}$–$10^{-8}$ mm Hg. At the same time, for substrates that are more active than amorphous carbon or refractory oxides (i.e., better wetted by the melt under study), the contribution of impurities from the residual atmosphere is overwhelmed by the influence of the substrate. Thus, supercoolings that are limiting for a particular contact pair can be observed under more modest vacuum conditions.

### 2.2.3 Limiting supercooling during the crystallization of metals

The above results indicate that the method of changing the condensation mechanism is a convenient means for studying the supercooling of metals that are in contact with various substrates. In addition, this method, on the condition of being supplemented by electron microscopic techniques, turns out to be effectively applicable in a very wide range of characteristic sizes of supercooled objects: from particles of the size of several nanometers to micron samples, that actually already have all the features and properties of the bulk state of the substance.

The wide possibilities, which are provided by the method of changing the condensation mechanism for contact pairs with different degrees of interaction, have made it possible to obtain a number of fundamental results on the particularities of supercooling in various systems. For example, in works [14, 17, 18], the mechanism of condensation of island films of indium, tin, bismuth, and lead on various substrates was studied: from almost inert non-wettable substances (C and $Al_2O_3$) to well-wettable metals (Cu, Ni, and W). Along with the determination of the crystallization temperature, the wetting angles in the studied systems were also measured in works [14, 17, 18]. Conducting experiments under high vacuum conditions and using sufficiently high condensation rates made it possible to obtain the limiting supercoolings and confirm the conclusion that it is the nature of the interaction with the substrate that determines the crystallization temperature in the studied contact pairs.

The analysis of the data obtained in works [14, 17, 18] made it possible to establish a connection between the maximum value of supercooling which is realized in this contact pair, and the wetting in the system. In full accordance with the theoretical consideration, presented above, with the improvement of wetting in the system, i.e., with a decrease in the wetting angle θ, the supercooling decreases.

The results of a large set of studies performed for various contact pairs "condensate-substrate" under conditions where the value of supercooling is limited by the substrate material are shown in Fig. 2.6 The figure shows the dependence of relative supercooling on the wetting angle in this contact pair.

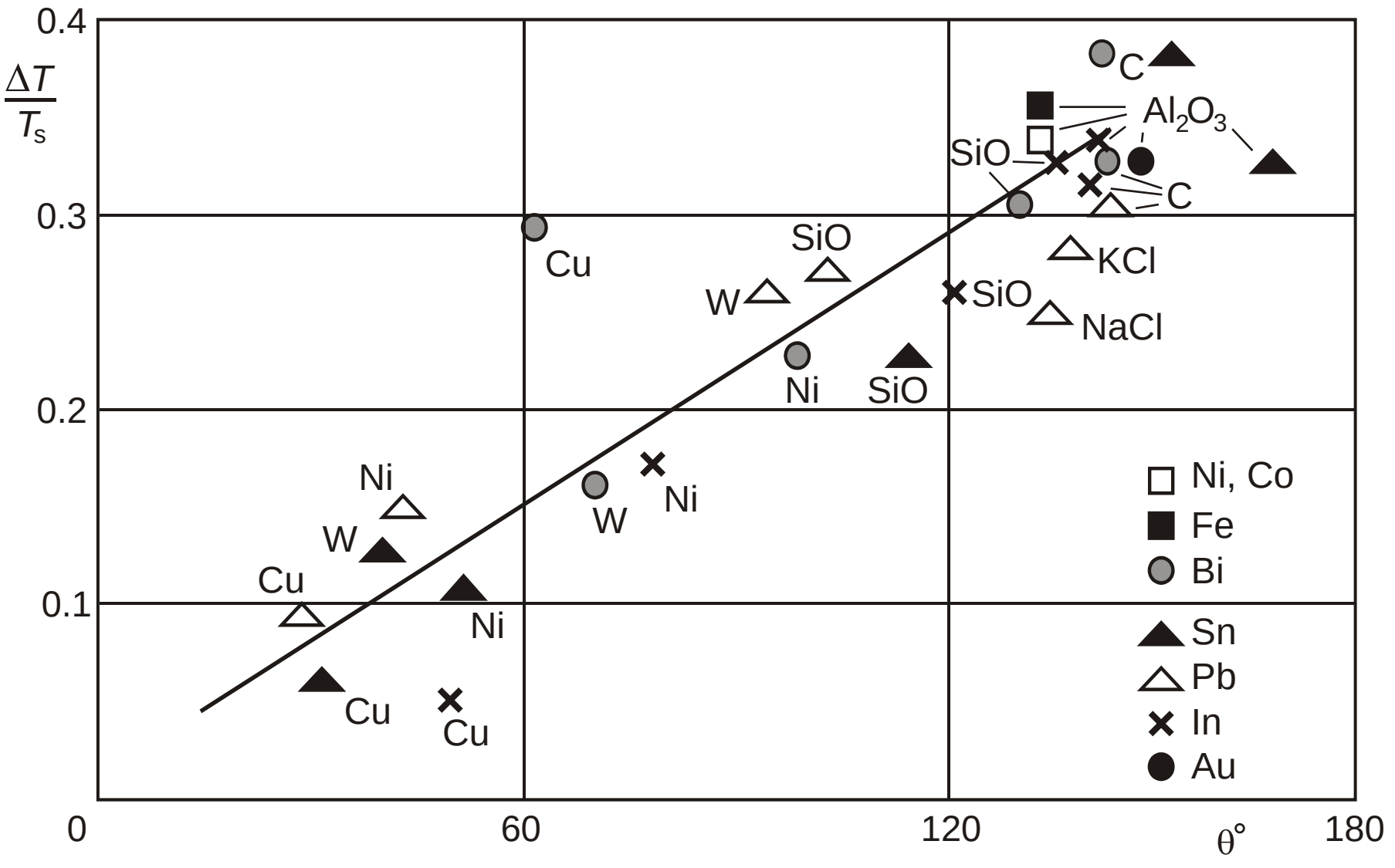

**Fig. 2.6** Dependence of relative supercooling during the crystallization on the contact angle wetting by the substrate metal. The substrate material is shown near the experimental points [14, 17, 18]

As can be seen, for wetting angles less than 130°, the supercooling almost linearly depends on the contact angle. At the same time, for larger angles, the supercooling reaches values that are probably typical for homogeneous crystallization and stop to depend on the contact angle and, therefore, on the substrate material. The Bi/Cu films that exhibit supercoolings, which are significantly exceeding the expected values, somehow fall out of the general map. The reasons for the violation of the observed empirical dependence in Bi/Cu films are not sufficiently understood. They are probably caused by some peculiarities of bismuth crystallization in layered film systems, which are discussed below.

The quantitative effect of wetting on the value of supercooling can be effectively described by expression (1.13). From this expression, as well as from Fig. 1.5, it is clearly seen that at contact angles greater than approximately 130°, the substrate practically stops to influence the process of crystallization, which can be considered homogeneous under these conditions. Experimental results (Fig. 2.6) confirm this conclusion. For many contact pairs with poor wetting, the stability of limiting supercoolings and the reproducibility of results are observed. This behavior is the main experimental criterion for crystallization homogeneity. This makes it possible to use the supercooling values obtained for the "metal-carbon" and "metal-refractory oxide" contact pairs to establish the physical laws of the "liquid-crystal" phase transition.

The values of supercooling determined in these experiments, which in the case of practically non-wettable substrates should be attributed to homogeneous crystallization, make it possible to estimate the interfacial energy of the "crystal-melt" interface of the metals under study using the equation (1.7). The results of such estimates performed in the work [14] are shown in Table 2.1. For comparison, the table also shows the values obtained from the measurement of dihedral angles and under the supercooling during crystallization observed in macroscopic systems.

**Table 2.1** Crystal-melt interfacial energy [10]

| Metal | $\lambda$, kJ/mol [30] | $\sigma_{sl}$, mJ /m$^2$ | | |
|---|---|---|---|---|
| | | Equation (1.7), [17] | Supercooling [2] | Dihedral angles |
| Au | 12,37 | 200 | 132 | - |
| In | 3,27 | 43,5 | - | - |
| Sn | 7,20 | 73 | 59 | - |
| Pb | 4,78 | 55 | 33,3 | 76 [31] |
| Bi | 10,89 | 85 | 54,4 | 82 [32]; 61,3 [33] |
| Fe | 15,37 | 330 | 204 | - |
| Co | 15,24 | 342 | 234 | - |
| Ni | 17,63 | 378 | 255 | - |

It is worth noting that the interfacial energies calculated on the basis of supercooling data during crystallization in island films deposited under high vacuum conditions are of the highest numerical value. This is a natural consequence of the fact that in the samples, obtained by the method of changing the condensation mechanism, the highest supercooling is achieved. Thus, these values should probably be considered as the best approximation to the actual values of the interfacial energy.

Supercooling during the crystallization of a low-melting Au-Ge eutectic deposited on a germanium substrate was studied in the work [34]. It is worth noting that metal-induced crystallization takes place in this contact pair [35], which should lead to a change in the crystal state of the germanium layers. The samples under study presented themselves as three-layer Au/Ge/C films obtained by the method of layer-by-layer vacuum condensation. The carbon and germanium layers were deposited on the substrate at room temperature, after which the required gradient of temperatures was created along the substrate, and only then the gold was condensed. The mass thicknesses of the germanium and gold layers were chosen to ensure a eutectic ratio of components in the films. Typical electron microscopic images of such samples are shown in Fig. 2.7.

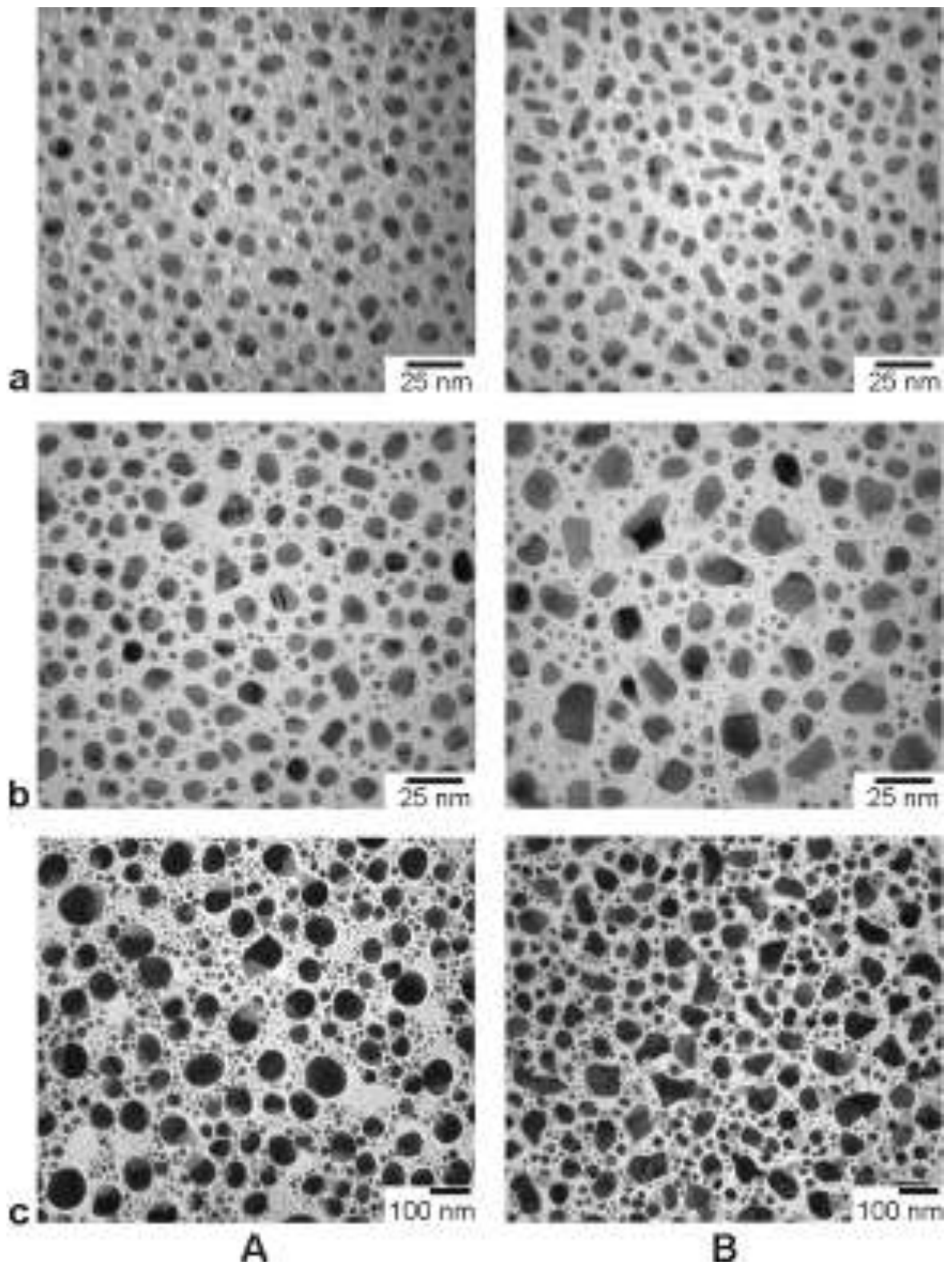

**Fig. 2.7** TEM images of Ge-Au film systems of eutectic composition with different mass thickness condensed by the vapor-liquid (**A**) and vapor-crystal (**B**) mechanisms. Ge/Au films thickness: a) 1.25/2.5 nm, b) 2.5/5 nm, c) 15/30 nm [34]

In the range of temperatures at which condensation by the "vapor-liquid" mechanism occurred (Fig. 2.7A), the images show only isolated islands with a shape close to spherical. This indicates the formation of a liquid phase in the Ge-Au system, which, without wetting the carbon sublayer, was collected in separate spherical drops. The morphology of the film systems, condensed on the substrate at temperatures below the temperature of change of the condensation mechanism, i.e., in the case when the "vapor-crystal" mechanism was realized, is sharply different (Fig. 2.7B): the islands have an irregular shape, far from spherical. The authors of [34] note that the smaller the mass thickness of the films is, the smaller the difference in their morphological structure above and below the critical temperature $T_g$. At the same time, it was shown in the work [36] that the formation of the liquid phase occurs in this system already at a film thickness of gold of ~0.2 nm. Thus, during the condensation of gold onto a germanium substrate with a temperature gradient, the formation of a liquid phase of eutectic composition above the critical temperature occurs on the surface of germanium. Thus, the gold-germanium eutectic is in contact with pure germanium, and the value of supercooling, as well as in other contact systems studied (Fig. 2.6), is set by the nature of the interaction between the melt and the substrate. The wetting angle by the Ge-Au eutectic of amorphous germanium, which was measured using the method of oblique observation (Fig. 2.8), was 57°.

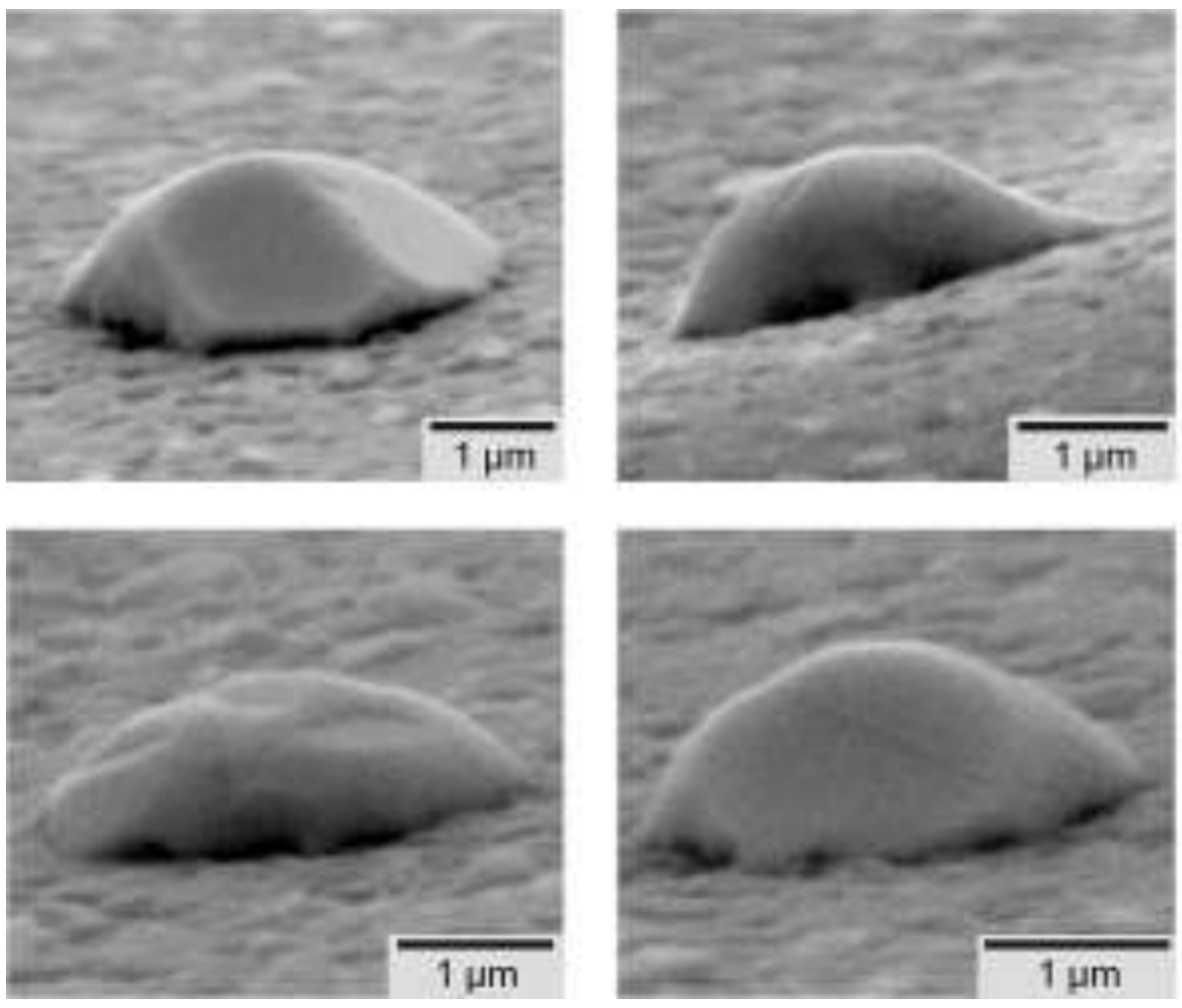

**Fig. 2.8** SEM images of Ge-Au eutectic droplets on an amorphous germanium substrate. The substrate is at the angle of 75° to the optical axis of the microscope [34]

As the result of the studies, the authors of [34], using the morphological criterion to determine the phase state of particles in the range of film system thicknesses of 2-60 nm, constructed a size dependence of the crystallization temperature of the Au-Ge eutectic, which is in contact with the amorphous film of germanium (Fig. 2.9).

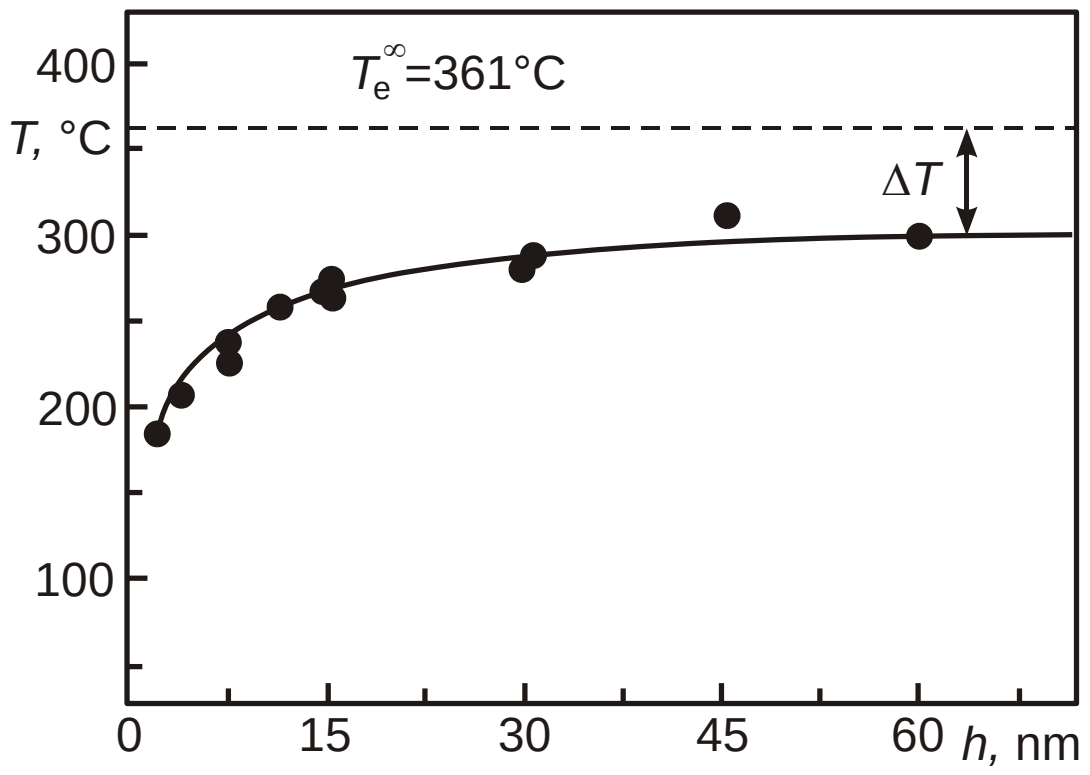

**Fig. 2.9** Temperature of condensation mechanism change during Au deposition onto amorphous Ge (•) layer depending on the thickness of the film system of the eutectic composition ($T_e^\infty$ is the eutectic temperature for the bulk samples [37], $\Delta T$ is the supercooling value) [34]

Discussing the results obtained, one cannot but note the rather strong dependence of the crystallization temperature $T_g$ on the thickness of the film system in the region of small sizes (Fig. 2.9), while for the previously studied single-component systems it was found that the crystallization temperature depends much less on the characteristic size [19]. This difference is probably due to the peculiarities of melting and crystallization in binary systems of eutectic type. Finally, it should be noted that the results obtained are in good agreement with the known empirical dependence of the supercooling value on the degree of interaction between the supercooled liquid and the substrate (Fig. 2.6).

### 2.2.4 Size effect during the crystallization of supercooled small particles

These results show that the use of island vacuum condensates makes it possible to obtain exceptionally deep and reproducible supercoolings during crystallization under controlled conditions. Such supercoolings can be considered close to the limiting, corresponding to homogeneous crystallization. The relative supercooling of islands of a number of metals (In, Sn, Bi, Pb, Au, Fe, Co, Ni) does not depend on their size in a wide range from about 40 nm to tens of microns and is $\eta$ = (0,3–0,4)$T_s$.

At the same time, in the result of numerous theoretical and experimental studies, is reliably established the existence of a size dependence of the melting point, due to which the melting point of sufficiently small particles can be lower than the crystallization temperature of a supercooled melt. The combination of these data with the results of studies of supercooling during crystallization in island films makes it possible to construct a diagram of "liquid-crystal" phase transitions that takes into account the presence of a size dependence of the melting point and the existence of a metastable supercooled state. A qualitative example of such a diagram in the "temperature-size" coordinates, proposed in the work [15], is shown in Fig. 2.10.

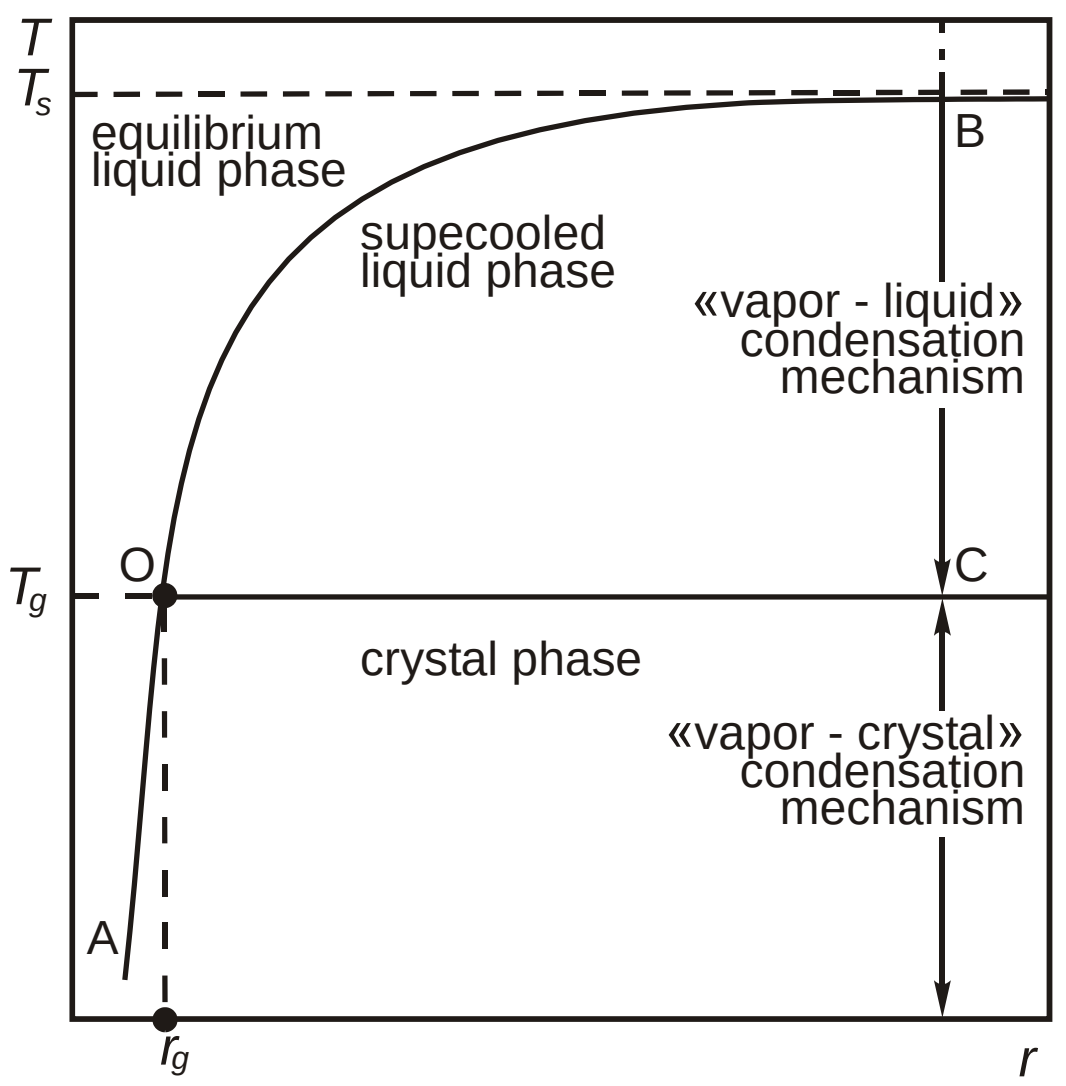

Fig. 2.10 Diagram of liquid-crystal phase transitions for condensed island films and small particles [10, 15, 16, 17, 14]

In Fig. 2.10, the solid line AB corresponds to the size dependence of the melting point and separates the region of existence of the equilibrium liquid phase (to the left of the AB curve). The horizontal straight drawn at temperature $T_g$ separates the regions of existence of the stable crystalline and metastable liquid phase. As can be seen from Fig. 2.10, the mentioned lines also delimit the realization zones of different condensation mechanisms. Thus, above the AOC line, condensation occurs by the "vapor-liquid" mechanism, and the "vapor-crystal" condensation mechanism is realized only within the region below the AOC line.

When constructing the diagram (Fig. 2.10), it was assumed that up to the particle size at which the melting point decreases so much that the formation of the crystalline phase turns out to be thermodynamically disadvantageous, the crystallization temperature has no size dependence. Under this assumption, state diagrams were constructed in the work [14] for island films of lead and tin on various substrates (Fig. 2.11).

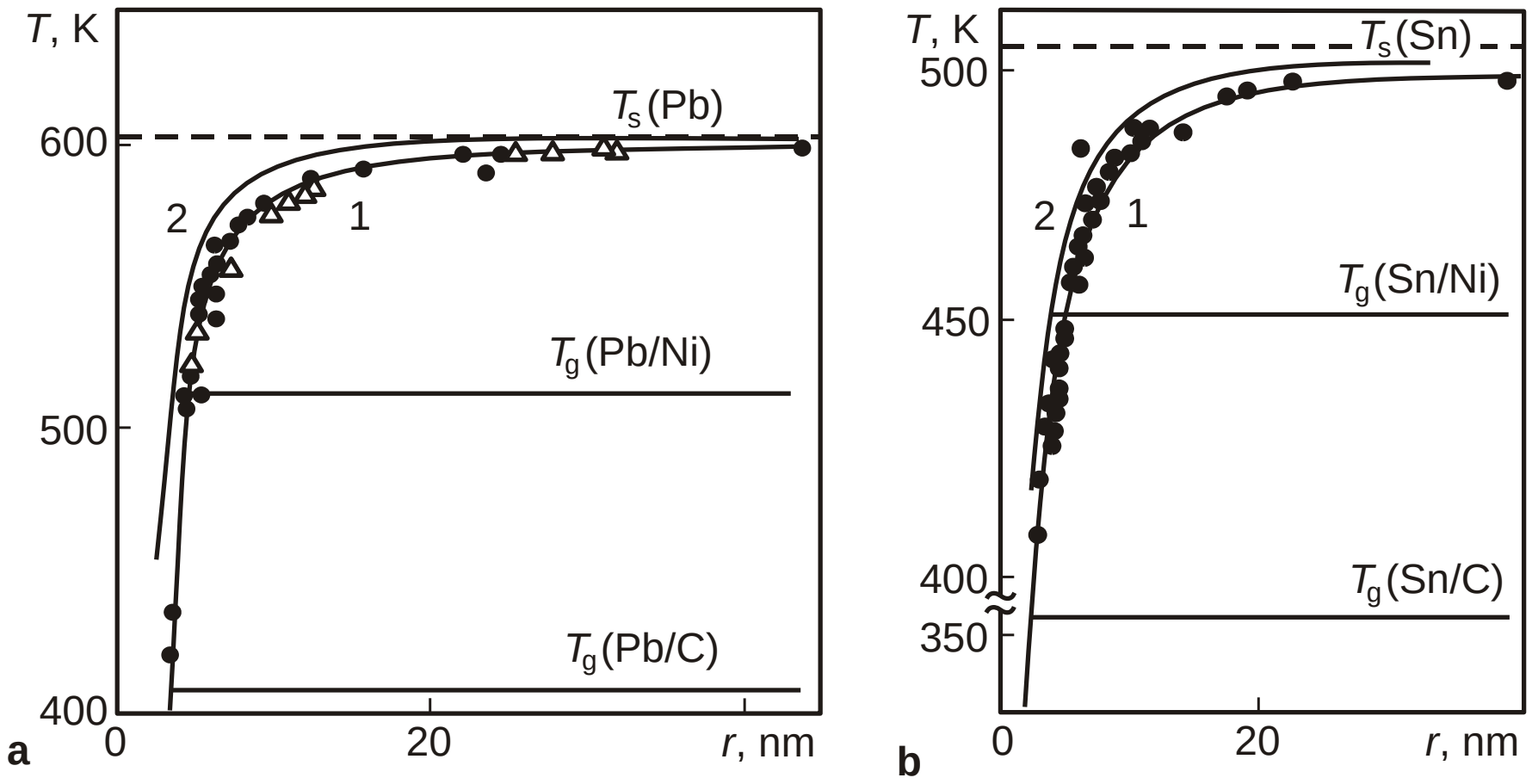

**Fig. 2.11** Diagrams of liquid-crystal phase transitions for lead (**a**) and tin (**b**) on carbon (1) and nickel (2) substrates [14]. The dots on the graph are experimental data on the size dependence of the melting temperature according to [38] (Δ) and [39] (●)

It is worth noting that the horizontal straights corresponding to the crystallization temperature were drawn due to the extrapolation of the supercooling values for particles larger than a few tens of nanometers to the intersection with the curve of the size dependence of the melting point. However, there are no grounds for such an extrapolation, generally speaking, and it can be expected that along with the size effect of the melting point, the size dependence of the crystallization temperature will be observed.

Indeed, in *in situ* experimental studies [40, 41, 42] using the method of dark-field electron microscopy it was revealed that the crystallization temperature of Hg, In, and Sn islands, as well as their melting point, decreases with decreasing size. Along with this, for Sn particles smaller than 10 nm, a decrease in the difference between the temperatures of their melting and crystallization was observed. At the same time, the rather high crystallization temperatures observed in these works suggest assuming that the detected values of supercooling correspond to heterogeneous nucleation on impurities. Similarly, using dark-field electron microscopy, similar results under high vacuum conditions were obtained in the work [43] for lead particles of 2–60 nm deposited on silicon oxide substrates. The size dependence of the melting point of Pb islands obtained in this work is consistent with the known literature data [44, 45]. For particles larger than 15 nm, the crystallization temperature of $T_g$ was almost constant and equal to approximately 490 K, which corresponds to 0,21$T_s$. The value of relative supercooling for lead particles, condensed under high vacuum conditions in the work [43], is so far from the limiting value probably due to the fact that the wetting angle for lead on a $SiO_2$ substrate is only $\theta \approx 110^\circ$, which, in accordance with expression (1.13), causes a significant decrease in supercooling. At the same time, according to data [43], with a decrease in particle size, the temperature of $T_g$ decreased and amounted to 410 K at a size of 2–3 nm. It is important to note that the melting point of such particles has the same value, i.e., when the particles reach a size of 2–3 nm, their supercooling actually turns to zero.

In the work [20], to study the size dependence of the crystallization temperature of small particles, the method of changing the condensation mechanism described above was used, based on the electron microscopic study of the morphology of island films that were condensed in a vacuum onto a substrate with a temperature gradient. Bismuth was chosen as the studied metal, in island condensates of which, due to the high anisotropy of the surface energy of different crystal faces, the "crystal-supercooled liquid" transition boundary is more clearly manifested in comparison with other fusible metals.

The samples in the work [20] were prepared by the method of thermal evaporation and condensation of bismuth in a vacuum of $10^{-5}$–$10^{-8}$ mm Hg, which was created by an oil-free system of pumping onto a substrate, along which a temperatures gradient in the range of 300–600 K was maintained. The substrate used was the chips of single crystals of NaCl, on which a layer of amorphous carbon was deposited immediately before bismuth deposition. In one experiment, several films of different mass thicknesses were obtained using mobile screens.

For films with an average island size of more than 10 nm, the crystallization temperature was determined based on the morphological criterion, i.e., under the change in the shape of the islands in electron microscopic images. For films with particles of smaller sizes, no significant difference in the shape of liquid and crystalline islands is observed. For such samples, histograms of the distribution of islands by size were constructed and the temperature of change in the condensation mechanism $T_g$ was determined as the one, which is corresponding to a jump in the graph of the temperature dependence of the average size of particles (Fig. 2.2).

Fig. 2.12 shows obtained in the work [20] size dependence of the crystallization temperature of bismuth particles, located on an amorphous carbon substrate. For particles with a diameter of more than 50 nm, the temperature of $T_g$ is about 380 K. This value is consistent with the results of previous studies by the authors [16, 17, 18] and with the literature data [46]. With a decrease in the size of bismuth islands, a decrease in the crystallization temperature is observed up to 325 K for particles with a diameter of less than 5 nm. Fig. 2.12 also shows the available literature data on the size dependence of the melting point of small bismuth particles [44, 45, 47].

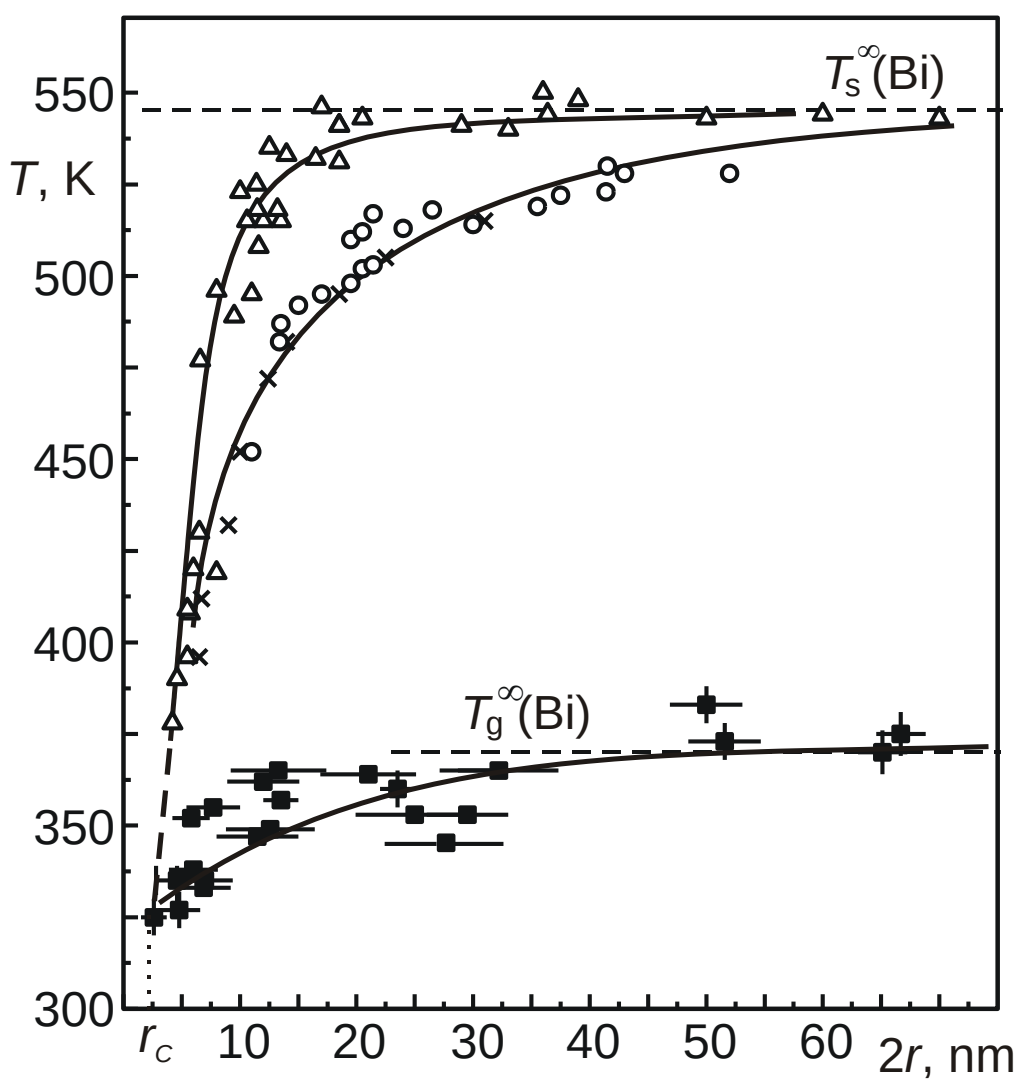

**Fig. 2.12** Size dependencies of melting (○ – [44], × – [45], Δ – [47]) and crystallization temperature of bismuth particles [20]

Thus, the data presented in this figure form a diagram of "liquid-crystal" phase transitions for small bismuth particles (on an amorphous carbon substrate) in the "temperature-size" coordinates. The graph of the size dependence of the melting point $T_r(r)$ delimits the regions of existence of the equilibrium liquid (left) and crystalline (right) phases. Below the $T_g(r)$ dependence, there will always be a crystalline phase, while above it, in the range between the lines $T_r(r)$ and $T_g(r)$ depending on the conditions and prehistory of the sample crystalline as well as supercooled liquid phase can exist. The difference between $T_r(r)$ and $T_g(r)$ values determines the temperature hysteresis of melting-crystallization for particles of radius $r$: $\Delta T(r) = T_r(r) - T_g(r)$.

Fig. 2.13 shows the dependence of the supercooling value on the characteristic particle size calculated using the averaged values of $T_r(r)$ [20]. The maximum supercooling $\Delta T = 0{,}33T_s$ corresponds to particles with a radius $r > 50$ nm. With a decrease in particle size, the value of $\Delta T(r)$ decreases and at $r_c$ = 2–3 nm it turns to zero. Thus, the "liquid-to-crystal" transition for particles with a size smaller than $r_c$ occurs without supercooling, i.e., it becomes continuous, as a second-order phase transition.

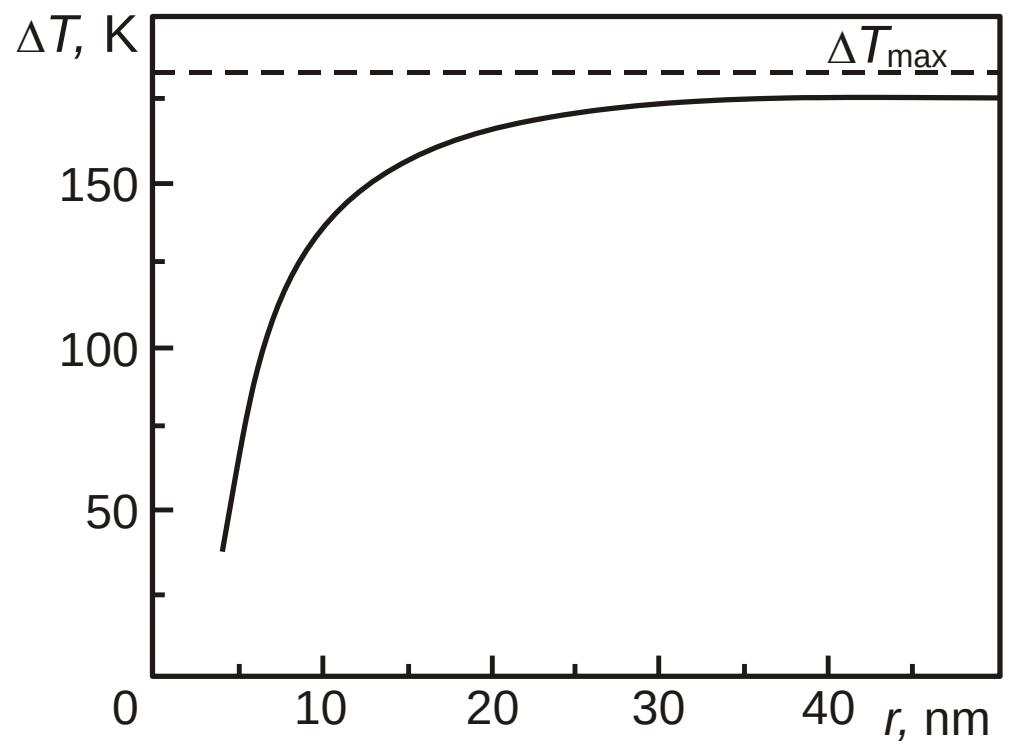

**Fig. 2.13** Size dependence of absolute supercooling of bismuth particles on an amorphous carbon substrate [20]

This fact cannot be interpreted in accordance with the classical theory of nucleation, which assumes the presence of an energy barrier between the solid and liquid states. At the same time, the dependence of the supercooling value on the particle size, which is observed experimentally, within this theory may be a consequence of the size dependence of the interfacial energy of the "crystal – own melt" interface $\sigma_{sl}$. The availability of experimental data concerning the size dependence of melting and crystallization temperatures makes it possible using the relation (1.7) to find the change $\sigma_{sl}$ with particle size. The dependence $\sigma_{sl}(r)$ obtained in the work [20] is shown in Fig. 2.14. As can be seen from the figure, along with a decrease of supercooling with a decrease in particle size, the interfacial energy of the "crystal – own melt" interface also decreases, which, in turn, indicates a decrease in the energy barrier between the liquid and crystalline phases.

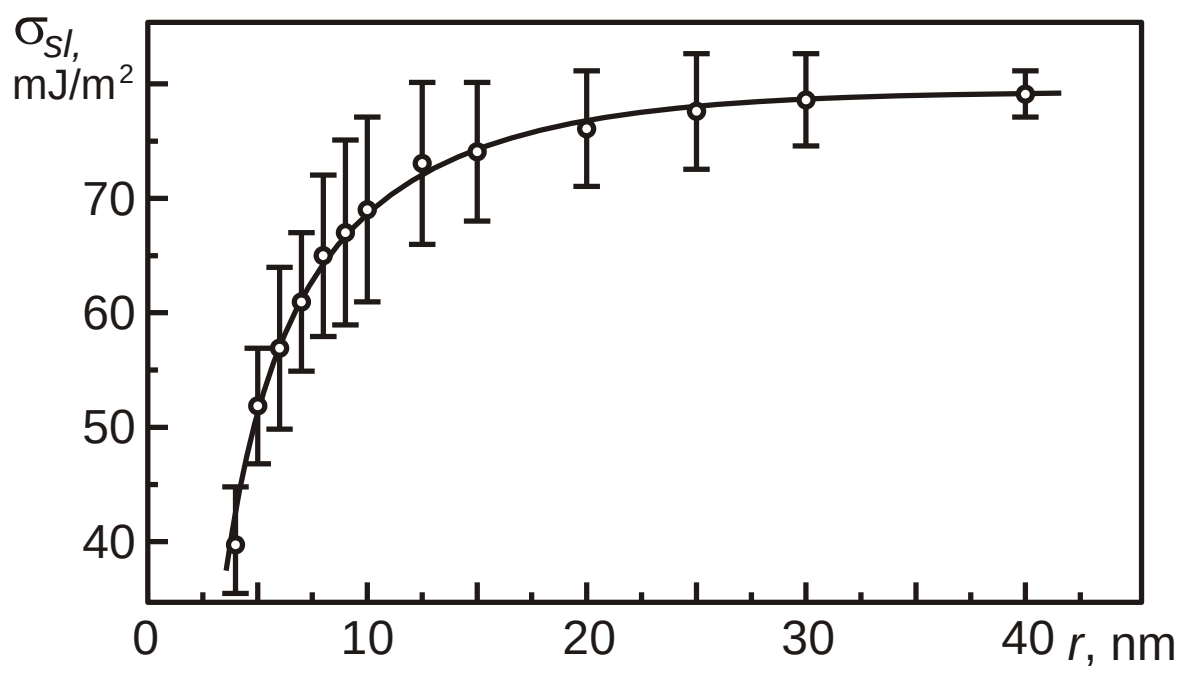

**Fig. 2.14** Size dependence of the interfacial energy of the "crystal – own melt" boundary for bismuth particles on an amorphous carbon substrate [20]

It should be noted that the conclusion that the interfacial energy of the "crystal – own melt" interface decreases with decreasing particle size follows from the statistical electron theory of the surface energy of metals [48]. According to [48], the dependence $\sigma_{sl}(r)$ is as follows:

$$\sigma_{sl}(r) = \sigma_{sl}^{\infty}(1 - \delta / r), \tag{2.1}$$

where $\delta$ is a parameter that has a physical meaning of the width of the transition zone between a crystal particle and its own melt.

The approximation of the data of the size dependence $\sigma_{sl}(r)$ for bismuth nanoparticles on an amorphous carbon substrate, shown in Fig. 2.14, gives the following parameter values: $\sigma_{sl}^{\infty} = 80$ mJ/m$^2$ and $\delta$ = 1.8 nm. The obtained value of the interfacial energy for bismuth in the bulk state is in good agreement with the data of the work [31] (82 mJ/m$^2$), in which this value was determined on the basis of measuring the dihedral angles of grain disorientation at the "crystal – own melt" interface. Estimation of the radius of the critical nucleus with the use of relations (1.4) and (2.1) gives a value of about 1 nm. The minimum radius of a particle in which a growth-capable nucleus can be formed will possibly be estimated as $r^* + \delta \approx 3$ nm, which corresponds to the value of the critical particle size, for which crystallization occurs without supercooling. The obtained value of the critical size of the particle is close to the size of the short-range order region in the liquid phase. Consequently, within the framework of the structure consideration for such particles, no difference between the liquid and crystalline states can be detected and such a particle has only one stable state, characterized by an ordering corresponding to the short-range order in a liquid. At the same time, large particles ($r \gg r_c$) near the melting point have two thermodynamically stable states with different degrees of ordering - liquid and crystalline. This circumstance is the physical reason why the phase transition in larger particles (and, of course, in the array) proceeds as a first-order transition, while in small particles it becomes of continuous nature.

### 2.2.5 Supercooling during crystallization of inorganic compounds

The results presented above correspond to the supercooling of metal particles on substrates of different nature. At the same time, the technique of changing the condensation mechanism can be effectively used to study supercooling during the crystallization of chemical compounds [21]. Fig. 2.15 shows a photograph of a nickel substrate on which a layer of NaCl is condensed. The substrate was heated by direct transmission of electric current without the use of an external heater, and its temperature was recorded at several points by K-type thermocouples welded to the back side. Due to the wedge shape, a temperature gradient within 400–900 K was established along the substrate during the heating process.

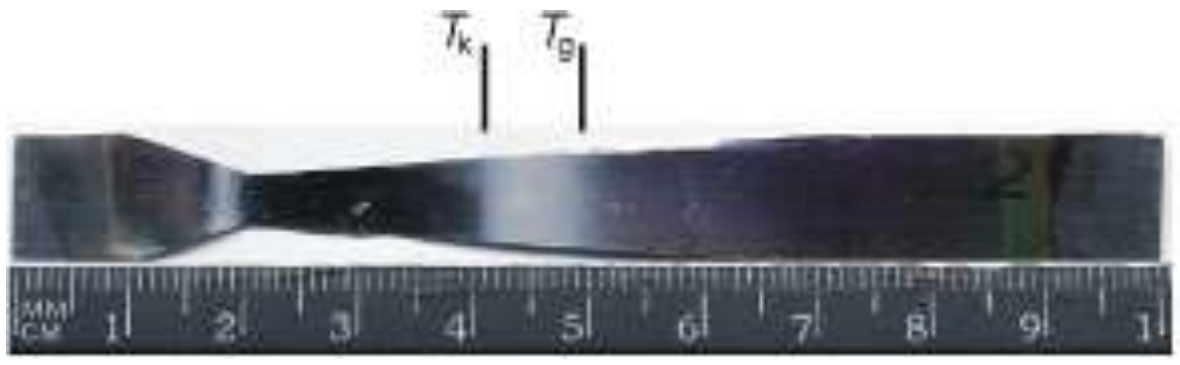

**Fig. 2.15** Photo of a nickel substrate with a NaCl film. The visible on the substrate boundaries correspond to the change of the condensation mechanism from vapor-liquid to vapor-crystal ($T_g$) and to the critical condensation temperature ($T_k$) [21]

Visually, the substrate reveals boundaries separating the regions of the substrate with different light scattering (Fig. 2.15). According to electron microscopic studies, the high-temperature line $T_k$ is the critical condensation temperature, and the low-temperature line corresponds to the temperature of change of the condensation mechanism in the studied contact pair**.** The effect of the substrate temperature on the morphological structure of sodium chloride condensates is illustrated by the micrographs shown in Fig. 2.16. It can be seen that up to a temperature of 705 K, condensation was carried out by the "vapor-crystal" mechanism, and above 710 K, the film under study contains obvious signs of liquid. The range of 705–710 K corresponds to the region of the sample in which the solid and liquid phases coexisted during condensation. The values of relative supercooling η obtained in the work [21] for AgCl and NaCl films on a nickel substrate are 0.13 and 0.34, respectively.

The use of the oblique observation method [25, 26, 27] made it possible to determine the values of the wetting angles in the studied contact pairs (Fig. 2.17). The measured values were 67° for NaCl films and 60° for AgCl samples. The ratio of the values of the wetting angle and relative supercooling obtained for AgCl is in good agreement with the known laws for metals and alloys [14, 15, 16, 17, 18] (Fig. 2.6).

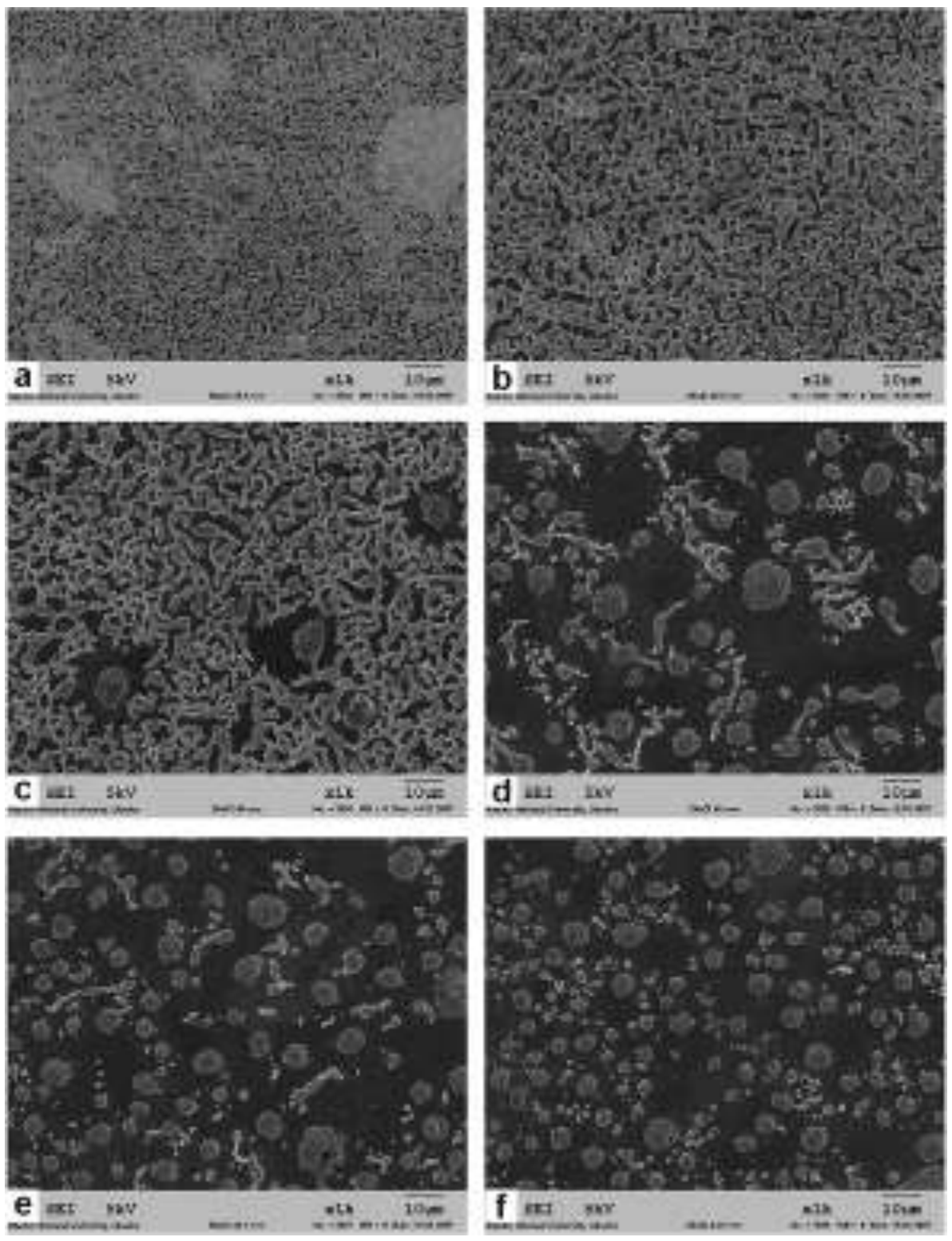

**Fig. 2.16** SEM images of NaCl films on a nickel substrate, which correspond to different condensation temperatures (**a** – 698, **b** – 703, **c** – 705, **d** – 710, **e** – 712, **f** – 713 K) [21]

It should be noted that the value of supercooling that occurs in NaCl films on a nickel substrate exceeds the value obtained in works [49, 50] for micron droplets. This is probably due to the small size of the particles studied in the work [21], as well as to the experiments conducted under high vacuum. Moreover, the supercooling of the NaCl particles significantly exceeds not only the previously obtained values but also the value expected based on the contact angle. The value of η = 0.34 seems to be close to the values typical for homogeneous crystallization.

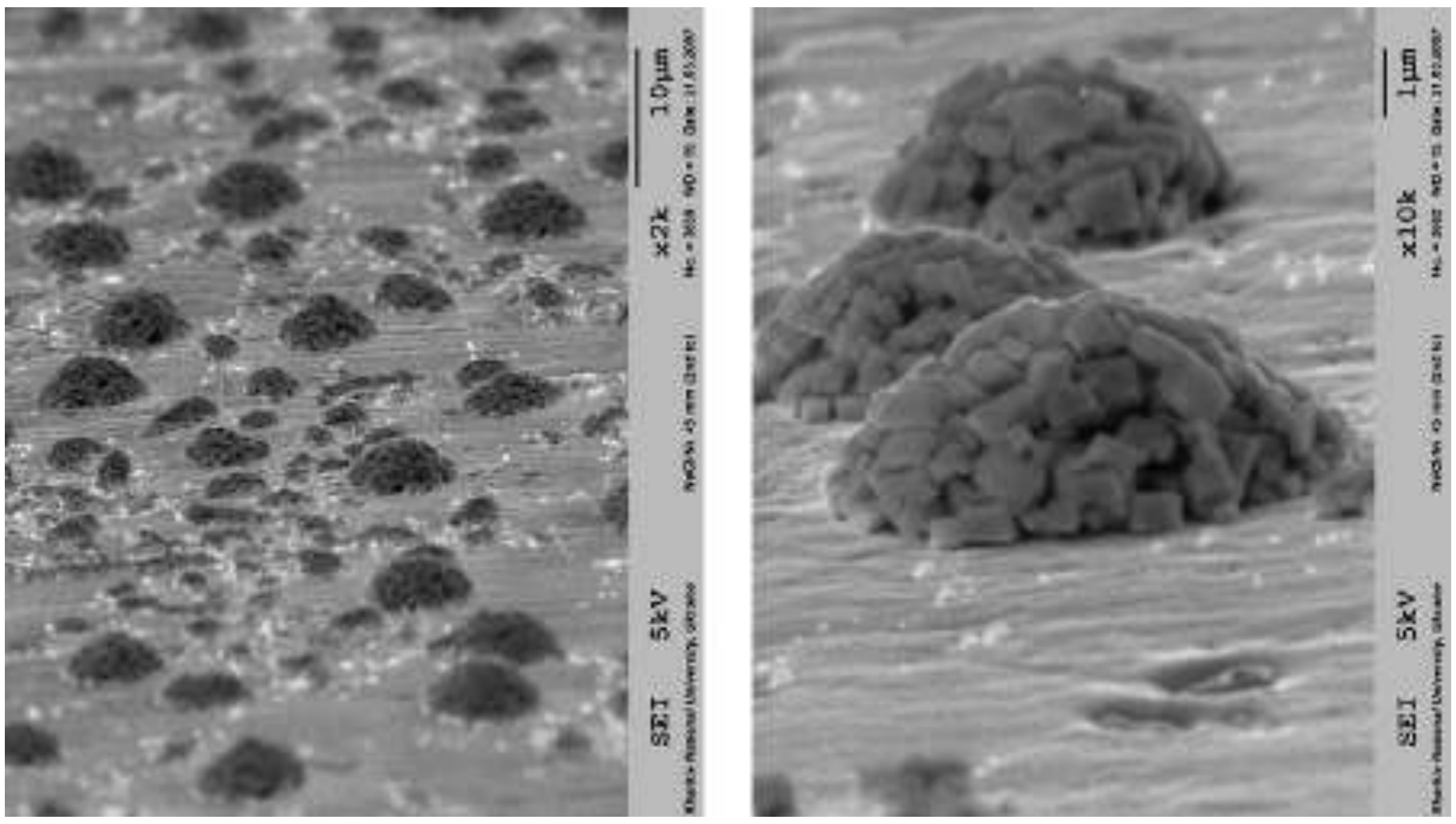

**Fig. 2.17** SEM images of NaCl films on a nickel substrate. The samples were condensed by the vapor-liquid mechanism. Images were obtained at an angle of 70° to the optical axis of the microscope [21]

A certain mismatch of the angle θ and the value of relative supercooling in the NaCl/Ni contact pair can be explained by the difficulty of measuring the wetting angles in this system. The wetting angles used earlier to construct the dependence of supercooling on the degree of contact interaction (Fig. 2.6) correspond to a melt located at or slightly above the melting point. At the same time, the critical condensation temperature (at which the rate of reverse evaporation of the substance from the substrate is equal to the rate of its deposition) in NaCl/Ni films is significantly lower than the melting point of sodium chloride. This does not allow us to determine the values of the wetting angles for a thermodynamically stable melt. The values of contact angles for the NaCl/Ni system were obtained for particles that are in a state of deep supercooling. However, for such particles, according to the results of works [24, 25, 51], a significant improvement in wetting can be observed compared to the values characteristic of an equilibrium melt.

It follows from the results of studies [21] that for chemical compounds, as well as for pure metals, the value of the limiting relative supercooling is determined only by the conditions of crystallization, and the method of studying supercooling during crystallization, based on the use of condensed in vacuum island films, is applicable to a wide class of substances. A natural limitation of the method of changing the condensation mechanism is the vapor pressure of the substance under study. In the case of high vapor pressure near the temperature of limiting supercooling, this temperature cannot be visualized on a substrate with a gradient of temperatures, because it will be higher than the critical condensation temperature.

### 2.3 Quartz resonator as an in situ method for determining the stability limits of the liquid phase in the "film-substrate" system

The method of changing the condensation mechanism is a powerful way to determine the value of supercooling during melts crystallization. However, it has some peculiarities, such as not allowing to implement of thermal cycling of samples, and is limited in its suitability for studying supercooling in systems with good wetting. Another peculiarity of this method is that it studies a substance that was initially obtained in a metastable state. On the one hand, this is a positive feature of the method, as it allows for the best possible reduction of the influence of extraneous factors. On the other hand, in complex systems, the behavior of the melt may depend on the prehistory, which should probably be taken into account for applied use.

The electron diffraction method [13] is an effective way to study the phase states of the substance. With the appropriate equipment, it can provide for the preparation and examination of samples in a single vacuum cycle and can be adapted to study both thin films and surface layers of bulk structures. However, the technological complexity of *in situ* electron diffraction studies (especially when it comes to working with reflected electrons, and samples obtained and examined in a single vacuum cycle) makes it advisable to develop additional research methods. One such method is the quartz resonator method.

The method is based on a combination of the piezoelectric effect and mechanical resonance in a quartz plate. The overlap between the mechanical oscillation of the quartz plate caused by an alternating voltage applied to it and the oscillation of the crystal lattice causes a resonance, which results in a standing transverse sound wave. Therefore, the conditions of sound propagation at the boundary of a quartz resonator and a condensed film significantly affect the resonance parameters for a piezoelectric crystal, such as its frequency of resonant oscillation, Q-factor, and amplitude of oscillation.

The use of a quartz resonator as a sensor for determining small amounts of condensed material was first proposed by Sauerbrey in 1959 [52]. In this work, it was shown that the change in the frequency of the fundamental harmonic of a quartz resonator is directly proportional to the mass of the condensed film. This property of the quartz resonator is widely used for *in situ* control of the thickness of deposited films. The quartz method of thickness control is currently the most sensitive. In the work [52], it was shown that the change in the resonant frequency $\Delta f$ of the quartz plate oscillation during the condensation of a metal layer with a mass $\Delta m$ on it is equal to:

$$\Delta f = -\frac{\Delta m}{N\rho_q A} \cdot f_0 (f_0 - 2\Delta f), \qquad (2.2)$$

where $f_0$ is the resonant frequency of quartz; $\rho_q$ is the density of quartz; $N$ is a frequency constant for the working crystallographic plane; $A$ is the electrode square of the excited part of quartz.

It follows from equation (2.2) that if we accept the condition of the uniformity of the distribution of the condensate layer over the surface of the quartz plate, then, having measured Δf and knowing the frequency constant, it is easy to calculate the increase of the mass $\Delta m$ and determine the thickness of the deposited layer. Assuming that $f_0 >> \Delta f$ (this condition is always satisfied for thin films), formula (2.2) will be transformed to the form commonly used in practice to measure the thicknesses of condensed layers:

$$h = \frac{N\rho_q A}{f_0^2 \rho_m S} \Delta f = \frac{1}{c} \Delta f \,,$$

where $c = f_0^2 \rho_m S / (N\rho_q A)$ is the sensitivity of the quartz sensor; $\rho_m$ is the density of the condensed metal; $S$ is the square of the electrode on which the metal is condensed.

It is worth noting that equation (2.2) was obtained under the assumption of a rigid connection between the infinitely elastic condensate and the substrate. That is, equation (2.2) allows quantitatively describing the effect on the natural frequency of a quartz resonator of condensation only of solid layers on it.

The particularities of the behavior of a quartz resonator with one of its surfaces in contact with a liquid were studied in works [53, 54, 55]. In these scientific works, a change in the resonant frequency of oscillation of a quartz crystal was observed, which the authors attributed either to a difference in the density of the contacting environments or to surface adsorption. These studies showed that oscillation in contact with a liquid can occur in a piezoelectric crystal. In addition, it was found that the presence of liquid causes a significant shift in the resonant frequency. A theoretical analysis based on the superposition of crystal motion and a shear wave in a liquid medium was carried out by the authors [53]. In this article, the distribution of the speeds of oscillation in the liquid was obtained, which made it possible to identify the effect of the viscosity and density of the liquid on the resonant phenomena in the system. Note that work [53] considers a quartz resonator, one side of which is in contact with a half-space of a viscous liquid. The resonance phenomenon is interpreted as the matching of shear waves at the "quartz-liquid" interface and in quartz. The equation for shear waves in contacting environments comes from a bulk response to shear stresses, and the resonance conditions were obtained by analyzing transverse waves. The following expression was obtained for the dependence of the resonator frequency on the fluid viscosity:

$$\Delta f = -f_0^{3/2}\sqrt{\frac{\rho_l \eta_l}{\pi \rho_q \mu_q}}, \tag{2.3}$$

where $f_0$ is the resonant frequency; $\rho_l$ is the density of the liquid layer; $\eta_l$ is its viscosity; $\rho_q$ is the density of the piezoelectric; $\mu_q$ is its shear modulus.

The given relations in work [53] were experimentally verified on the example of aqueous solutions of glucose and ethyl alcohol of various concentrations, and a good agreement with the experimental data was established. In [56, 57], the quartz resonator method was successfully used to measure the viscosity of liquids. At present, further development of the methodology for determining the viscosity of gases and liquids using piezoelectric sensors is observed. Theoretical foundations for the case of complete immersion of a quartz resonator into the fluid environment are being developed [58], as well as new technical solutions to increase the sensitivity of the method and miniaturize active elements [59].

The idea of using a quartz resonator to determine the temperature limits of the stability of the liquid phase, based on the ideas developed in works [52, 53], was realized in the work [60]. It turned out that the quartz resonator technique not only provides a convenient way to thermocycling samples but also allows us to study layers of considerable thickness. This is especially interesting for studying nanocomposite structures in which a fusible component is embedded in a refractory substance.

Taking into account equations (2.3) and (2.2), the authors of [60] suggested that quartz should be sensitive to "melting-crystallization" phase transitions, which occur in a pre-deposited solid film (Fig. 2.18).

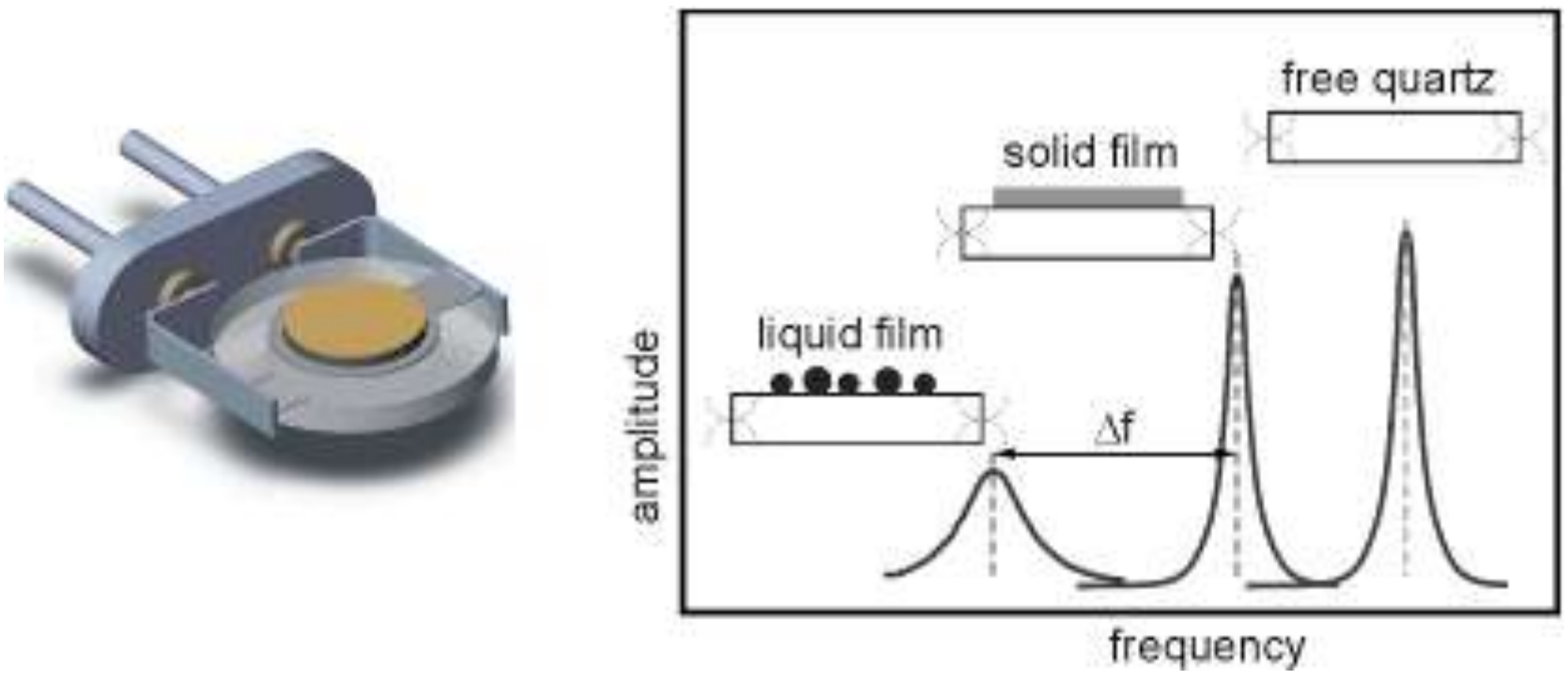

**Fig. 2.18** Typical quartz resonator (on the left) and the dependencies from amplitude frequency of its oscillation at the resonant frequency in a case of free quartz, of quartz with a layered solid film and with a film in the liquid state

For the practical implementation of the quartz resonator method, in the work [60], two identical piezoelectric plates (working and control) were used, which had been installed into a bulk copper block equipped with a resistive heater (Fig. 2.19). This block ensured the location of both sensors under the conditions of the same thermal influence.

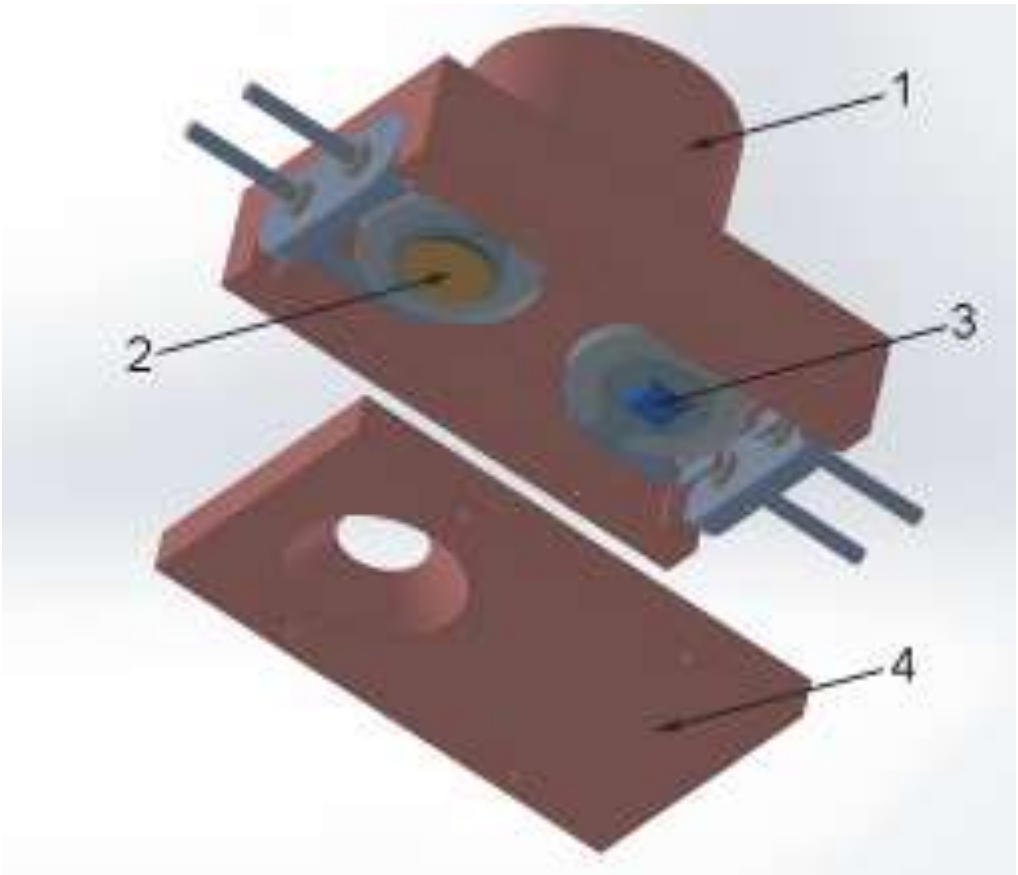

**Fig. 2.19** The scheme of using a quartz resonator for determining phase transition temperatures: 1 – a copper block with a resistive heater, 2 – a working quartz plate onto which the researched film system was condensed, 3 – a control quartz plate with a platinum temperature sensor, 4 – a copper mask, which covers the quartzes (is shown separately)

A special mask made it possible to realize the condensation of the investigated layered film system onto only one (working) resonator. Along with this, another (control) resonator was not coated and was used as a means of measuring the temperature. To do this, a miniature thin-film resistive platinum temperature sensor was fixed on a control quartz, which was located under the conditions of the same thermal influence as the working resonator with a condensed film.

It should be noted that vacuum condensates are extremely non-equilibrium structures, and during the annealing of as-deposited films, a number of processes are observed in them that bring the sample state closer to equilibrium. Such processes that occur during low-temperature annealing include recrystallization, diffusion, grain boundary migration, mass transfer, etc. At present, the theoretical understanding of the phenomena occurring at the interface between a piezoelectric and a condensed film is not complete enough to draw conclusions about the effect on the frequency of the resonator of processes, that aim to bring the deposited layer to an equilibrium state. Therefore, at the first stage of testing the quartz resonator method, the dependence of the frequency of the quartz resonator with the condensed Al/C system was studied in the temperature range from room temperature to 320 °C. Heating to this temperature is already sufficient to intensify the processes mentioned above. At the same time, this temperature is significantly lower than the values that can cause the appearance of a liquid phase in the studied contact pair. The resulting curves are shown in Fig. 2.20.

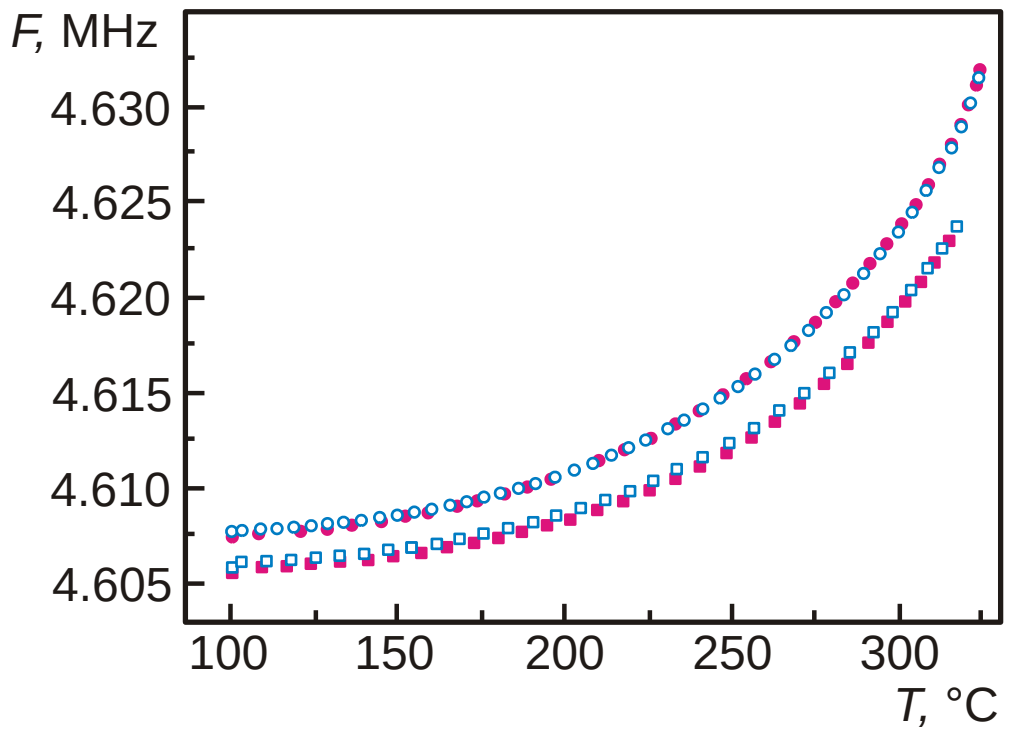

**Fig. 2.20** Resonance frequency of the quartz plate versus the temperature before (●○) and after (■□) the deposition of the Al/C system (filled points correspond to heating, unfilled – to cooling) [60]

It can be seen that the temperature dependence of pure quartz has no significant differences from the sensor on which the Al/C film system is condensed. The only difference observed in Fig. 2.20, is the vertical shift of the curves, which is caused by the change in the mass of the plate due to the condensation of the layered system.

Thus, it has been experimentally established that the quartz resonator turns out to be insensitive to recrystallization, vacancies diffusion, and other processes that homogenize the system without significantly changing its viscous properties.

At the same time, the jumps of the resonant frequency occurring during the heating and cooling of the sample are observed on the dependencies of the resonant frequency on temperature, obtained from the resonator, loaded with the deposited Bi/C film system (Fig. 2.21). The temperatures at which these peculiarities are observed are 271 and 175 °C, respectively.

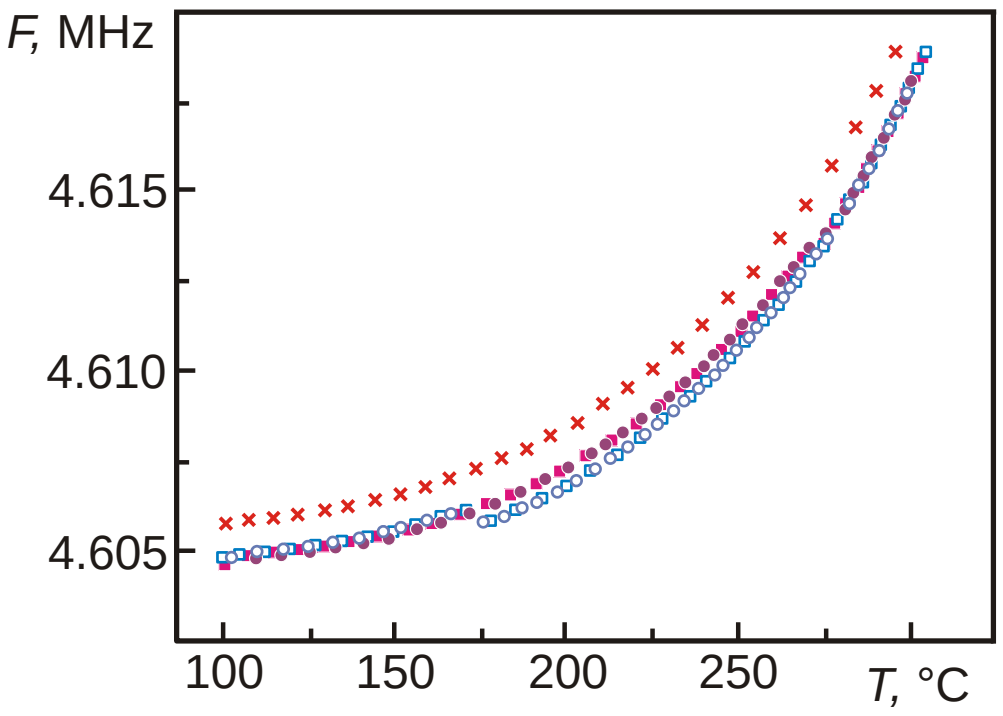

**Fig. 2.21** Resonance frequency temperature dependences of pure quartz and quartz with a Bi film: × is a pure quartz, ■□ is the first and ●○ is the second cycle of heating and cooling a quartz with a bismuth film of 30 nm thickness (filled points correspond to heating, unfilled – to cooling) [60]

The sharp change in the frequency of the quartz resonator can be associated with the melting and crystallization of bismuth. The spherical shape of the particles in the photographs (Fig. 2.22) clearly indicates the melting of a film of bismuth.

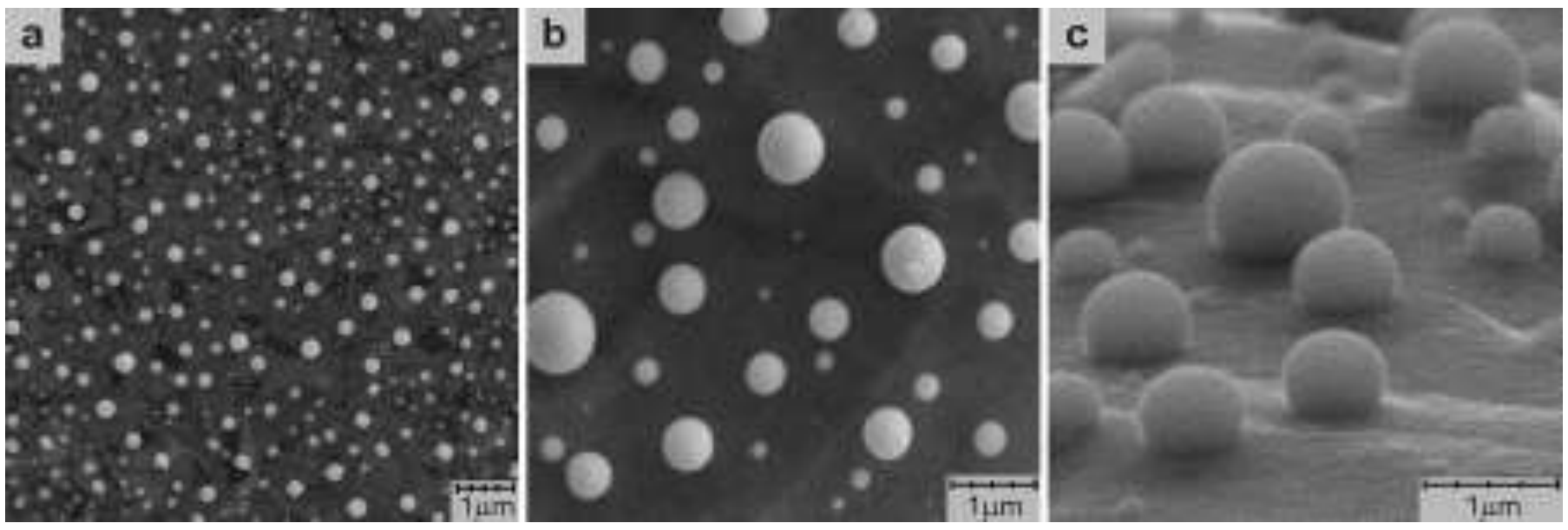

**Fig. 2.22** SEM images of the Bi/C system on a quartz plate with a bismuth mass thickness of 30 (**a**) and 100 nm (**b**, **c**) after melting. Image (**c**) was taken at an angle of 75° to the optical axis of the microscope [60]

This phenomenon can be registered more clearly if it is expressed in terms of the difference in quartz frequencies before and after condensation of the film system under study (Fig. 2.23). As can be seen, the jumps that identify melting and crystallization occur in a certain temperature range. This fact can be seen as an indication that both phase transitions for this binary system, in which bismuth is in contact with an amorphous carbon layer, are extended over a certain temperature range, i.e., they are of a diffuse nature. The resulting supercooling value is 96 K, which corresponds to $\eta = \Delta T/T_s = 0{,}18$.

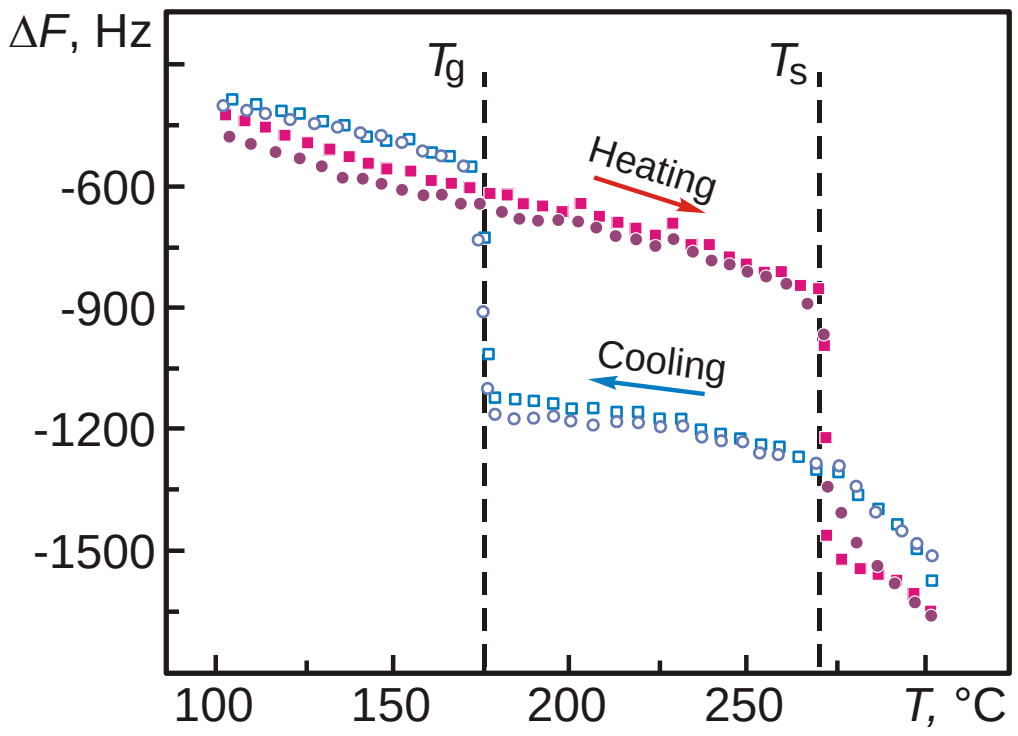

**Fig. 2.23** Frequency difference dependence of pure quartz and quartz with a condensed bismuth film on the temperature [60]

The strong dependence of the resonant frequency of a quartz plate on temperature complicates the research. It is worth noting that today there is no clear model that would describe the peculiarities of the behavior of a quartz resonator during a change in the phase state of a substance on its surface. Typically, during the theoretical analysis of the interaction of a quartz plate with a liquid, it is assumed, that the surface of the quartz is perfectly smooth, and the liquid is a Newtonian fluid. In a real experiment, the surface of quartz is quite rough and the interaction with the liquid is significantly complicated. In other words, under such conditions, the liquid can be captured by the cavities and pores of the surface, and turbulent flows can occur. As a result, the resonant frequency of quartz begins to depend on additional parameters [61, 62].

At the same time, it turned out that for the identification of phase transitions, a rather convenient characteristic of an oscillating system is its quality factor, which is essentially a value that numerically describes energy losses. Despite the fact that the quantitative description of the effect of viscosity on this value is rather complicated, the qualitative understanding looks simple and visually clear. It is well known that the propagation of elastic oscillation in a solid body (whose viscosity can be considered virtually infinite) occurs under significantly less energy dissipation than is the case during the propagation of oscillation in a viscous medium. In addition, extra energy losses can arise due to the spatial movement of the liquid that occurs in the sample after melting (Fig. 2.22). Being in the liquid state, such islands are able to move on the surface of the substrate, which causes additional energy losses.

Thus, the value of the quality factor seems to be an extremely convenient parameter for studying the melting and crystallization of vacuum condensates deposited on the surface of a quartz plate. Fig. 2.24 shows the temperature dependence of the quartz quality factor with the deposited Bi/C film. It can be seen that the heating and cooling graphs also show jumps in the recorded value, which correspond to the melting and crystallization of bismuth. It should be noted that both processes occur diffusely.

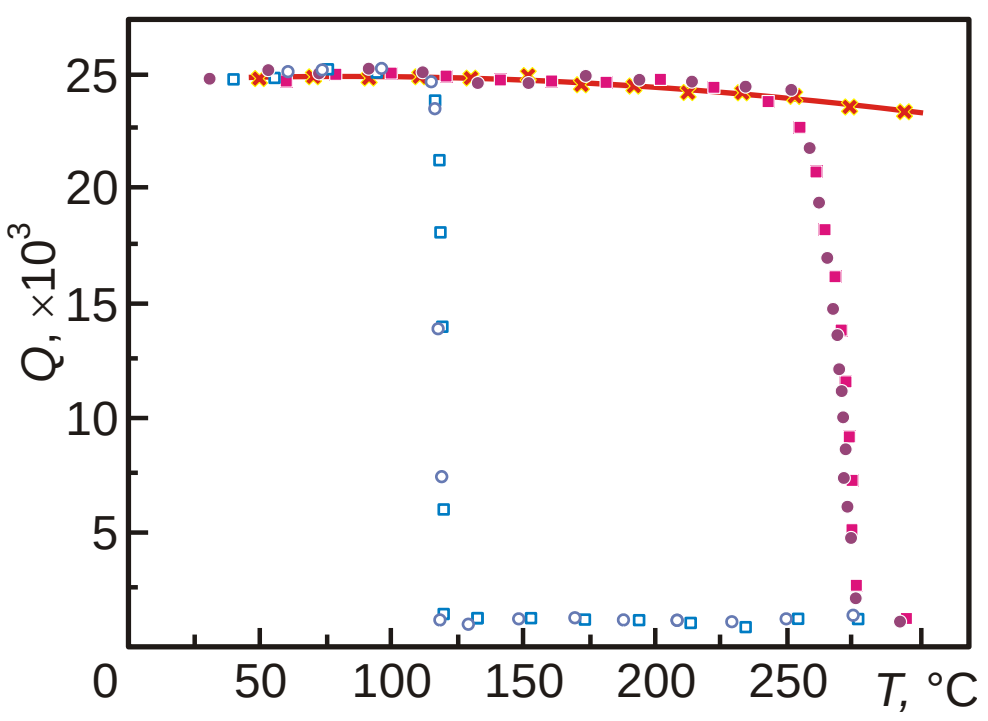

**Fig. 2.24** Quality factor dependence of a quartz resonator on the temperature of a pure quartz (×) and from the quartz with a bismuth film for the first (■, □) and second (●, ○) cycles (filled points correspond to heating, unfilled – to cooling) [60]

At the same time, the crystallization temperature of the samples by which these graphs were built, turns out to be significantly lower than in the case of the previously studied samples (Fig. 2.23). The crystallization is completed at a temperature of 118 °C, i.e., the supercooling in this case reaches 153 K, $\eta$ = 0,28.

Differences in the values of supercooling in the two experiments are explained by different condensation conditions. Thus, films for which the supercooling value is $\eta$ = 0.18 were obtained at a residual gas pressure of $10^{-5}$ mm Hg, which was created by a pumping system based on a diffusion vapor jet pump. At the same time, the films in which the supercooling $\eta$ = 0.28 was obtained were deposited at a pressure of the residual atmosphere of $5\cdot10^{-8}$ mm Hg in a vacuum chamber with an oil-free pumping system. This difference is consistent with the results presented above (subsection 2.2.2), according to which the change in the pressure and composition of the residual atmosphere affect the crystallization temperature.

Despite the fact that the supercooling values obtained by the authors of [60] do not reach the record values typical of Bi/C films, the main advantage of the study is the demonstration of the effectiveness of using a quartz resonator as a method for recording phase transitions in vacuum condensates. Subsequently, the quartz resonator method was used to determine the crystallization temperatures of supercooled melts in In/C, Sn/C, and Pb/C films. Fig. 2.25 shows the dependencies on the temperature of the quality factor of a quartz resonator with In/C, Sn/C, and Pb/C films, on which the temperature ranges corresponding to the melting and crystallization of the samples are well identified.

The phase transition temperatures obtained by this method are in good agreement with the values, which are determined using other techniques [14].

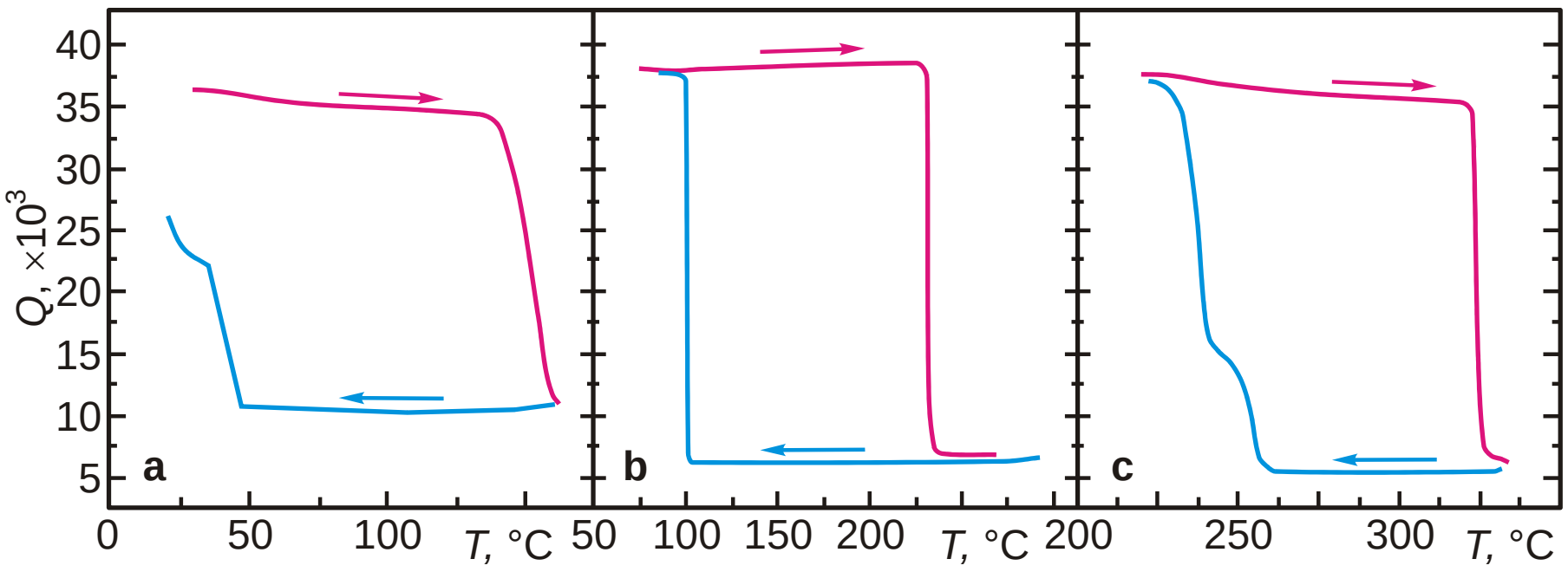

**Fig. 2.25** Temperature dependence of the quality factor of a quartz resonator with an In/C (**a**), Sn/C (**b**) and Pb/C (**c**) films

# Chapter 3
# Phase transitions in layered film systems

**Abstract** In addition to measuring the crystallization temperatures of supercooled melts, supercooling studies suppose obtaining information on the kinetics of phase transitions and the effect of successive heating and cooling series on the behaviour of melts. Nanocomposite systems, in particular, structures of the "small particles in a more refractory matrix" type, are currently attracting special interest. The study of such objects requires the development of special methods that provide data on the kinetics of phase transitions and allow the carrying out of multiple thermal cycling of nanocomposite samples. Effective techniques to investigate these aspects and systems are based on observing the resistance of the films or the resonant frequency of the piezoelectric crystal on which the films are deposited. In the first case, the criterion for melting and crystallization of the layer is a jump in the electrical resistance of the film. In the second case, it is a jump in the quartz resonator's quality factor. Using these techniques, studies of supercooling during crystallization of the fusible component in multilayer films have been carried out. It is shown that for contact pairs containing bismuth, the temperature and kinetics of crystallization depend on the condensation mechanism of the fusible metal. The observed differences are explained by the different morphology of samples obtained by condensation of bismuth into liquid and solid phases. Bismuth layers deposited into the solid phase form a connected system of inclusions, which is preserved even during thermal cycling. Such a system crystallizes as a single unit, and its formation is provided by grain boundaries of the more refractory layer.

## 3.1 Particularities of phase transitions in nanocomposite materials

The theoretical and experimental results discussed in Chapters 1 and 2 primarily concern particles, which are located on the surface of a solid substrate. At the same time, a recent trend is the growing interest in nanocomposite materials in which nanoparticles are embedded in a bulk matrix. The nanosized component in such structures can significantly modify the properties of the solid matrix, and the presence of the matrix, in turn, can change the properties of the embedded nanoparticles and, in particular, affect the temperatures of their melting and crystallization.

Since the value of supercooling is usually understood as the difference between the melting and crystallization temperatures, a shift in the melting point will also mean a change in the supercooling conditions in the system. The size effect of the melting point observed for free particles is a well-known phenomenon and consists of a decrease in the melting point of particles as their size decreases. However, due to the presence of a solid matrix, the interfacial energy of the nanoparticle surface changes and some overpressure may occur. This causes a change in the balance of free energy responsible for the phase transition. As a result, the system can realize the phenomenon of overheating [1, 2, 3, 4, 5, 6], in which the nanoparticle continues remaining in the solid state even after it is heated above the equilibrium melting point corresponding to the bulk state.

The question of the theoretical limiting value of the overheating temperature that can be obtained in systems with limited geometry looks quite interesting. In general, this question is similar to the question of the limiting value of supercooling, and different approaches are used to predict it. For example, Kauzmann [2] defined the minimum crystallization temperature as the temperature at which the entropies of the solid and liquid phases are equal to each other. A similar approach was used in the work [3], the authors of which showed that a similar particularity on the dependence of the difference of entropies on temperature is observed above the melting point. Based on these considerations, the overheating limiting value was estimated at $1{,}38T_s$. However, the estimate given in the work [3] seems to be somewhat overestimated and does not take into account the contribution of stresses arising under the conditions of the confined geometry. These factors were taken into account in the work [4], the authors of which showed that the overheating limiting value for aluminum particles should be $1{,}2T_s$. An even lower overheating value was obtained by the authors of [5], who considered heterogeneous melting occurring on dislocations and other defects. The overheating limit obtained within this model is about $1{,}1T_s$. Note that the above values correspond to a rather slow heating, which provides the equilibrium of the processes. At the same time, under conditions of ultra-fast heating, it is possible to observe an overheated state at higher temperatures [6].

Experimentally, the phenomenon of overheating of nanoparticles embedded in a solid matrix has been observed in many works. One of the first studies of overheating in composite systems is probably the work [7]. When studying gold-coated silver particles with a diameter of 120–160 µm, the authors of [7] found that silver in such a composite system is able to remain in a solid state when overheated by 25 K relative to the equilibrium melting point. The strategy proposed in the work [7] has been used in many subsequent studies devoted to overheating.

In particular, the authors of [8], while studying the Ag-Ni nanocomposite system, in which silver nanoparticles (with an average size of 30 nm) were embedded into a nickel matrix, found that the crystal matrix has a strong influence on the structure of the embedded nanoparticles. According to the conclusions of [8], silver has an orientation effect, which consists of the partial coherence of the crystal lattices of silver particles and nickel surrounding it. When the samples were analyzed by the method of differential scanning calorimetry (DSC), it was found that most of the particles melt below the equilibrium melting point. This is the result to be expected based on thermodynamic considerations used to describe the behavior of free particles. However, in addition to the main peaks corresponding to the melting of most particles, the authors of [8] observed peaks at higher temperatures on the DSC-curves, which are indicating that a small, but confidently recorded number of particles melt with an overheating, value of which reaches 70 K. Such peaks are discrete, which indicates the existence of particles of several classes in the matrix, and the orientation particularities of some of them provide the possibility of their overheating. At the same time, during thermal cycling, with an increase in the cycle number, the intensity of the peak corresponding to overheated particles decreases. Since during low-speed thermal cycling, the material is annealed rather than quenched, which partially eliminates the defective structure of the sample, this behaviour looks regular and confirms the connection between mechanical stresses and the possibility of overheating.

In addition to differential scanning calorimetry, the existence of the Ag crystalline phase above the equilibrium melting point was confirmed by the methods of *in situ* X-ray diffraction (XRD) in the work [8]. Thus, a clearly visible peak corresponding to the reflection from the [1 1 1] silver plane was observed up to a temperature of 1263 K, which corresponds to significant overheating. The overheating values obtained from XRD data do not coincide with those obtained using DSC. The authors attribute the indicated difference to the annealing that occurs during the study. For example, DSC-curves were obtained at a heating rate of 20 K/min, while each of the XRD spectra required at least 15 minutes to be recorded. This reduced the effective heating rate and caused the annealing of the samples directly during the XRD study. Thus, the difference in the overheating values obtained by the two methods under conditions of significantly different thermal effects again indicates the determining role of mechanical stresses in the overheating phenomenon.

The results of the study of overheating of lead embedded into various matrices are presented in the work [9]. It was found, that some part of the lead embedded into a silicon, copper, or aluminum matrix melts with overheating. Along with this, the maximum value of overheating depends on the matrix material and increases with decreasing particles size. In particular, a significant part of lead particles with an average size of 5.2 nm melts with an overheating of more than 100 K. Comparisons of the obtained data with the particularities of the lead crystal lattice, which are observed experimentally, led to the conclusion that, along with interface epitaxy, the pressure generated by the matrix plays an important role in overheating [9].

Thus, the phenomenon of overheating that causes a shift in the reference point makes the determination of the supercooling value somewhat ambiguous. Also, a certain amount of the fusible component, that may remain in the overheated crystalline state, can cause crystallization at a supercooling, which is less, than the value typical for this contact pair.

In addition to providing the possibility of overheating, the matrix into which more fusible inclusions are embedded can also affect their crystallization. It is obvious that supercooling of fusible particles, embedded in a more refractory matrix, is a much more complex phenomenon than the heterogeneous crystallization of particles located on the surface of the substrate, and even more complex than phase transition in free particles. At the same time, due to the high applied and fundamental importance that phase transition processes have in nanocomposite systems, numerous studies are currently devoted to the study of supercooling of a fusible component embedded into a refractory matrix.

In the work [10], an Al-Sn composite system containing 2-3 wt.% tin was created by the method of ion implantation, realized at a sample temperature of 425 K. According to electron microscopic studies, the size distribution of tin particles was close to normal with a maximum of about 5-6 nm. The melting and crystallization of the particles were studied *in situ* by heating the samples directly in an electron microscope. The temperature of the phase transitions was determined by the moment of appearance-disappearance of moiré patterns, the presence of which indicates the crystal structure of the sample.

It was found that the melting of tin particles occurs in a certain temperature range. Due to *in situ* studies, it was confirmed that the broadening of the melting interval is caused by the size effect. The melting of tin in the samples is completely finished at 460 K. Crystallization in the films also occurs diffusely and is most effectively carried out near 400 K.

The melting and crystallization of indium particles embedded into an aluminum matrix were studied in the work [11]. Samples for the study were obtained by the method of rapid cooling of the Al-In melt. They contained 4.7 at. % In. Melting and crystallization of the samples were studied using the differential scanning calorimetry technique. Using the methods of transmission electron microscopy (TEM), the authors of [11] found that the fusible particles formed in this contact system have a facet typical for bulk crystals and are characterized by a bimodal size distribution. The large fraction is approximately 200 nm in size, and the small-dispersed fraction is 10–20 nm. The aluminum matrix has a strong orienting influence on the nanoparticles embedded into it. During the DSC-experiments, it was found that the melting in the studied system occurs in two stages. The temperature of the first particularity on the DSC-curves corresponds to the tabulated melting point of bulk samples (429 K). In addition, in the process of heating nanocomposites, the authors of [11] observed another peak at a temperature of 435–440 K, which was matched to the melting of particles under overheating conditions.

Crystallization in the studied contact pair is also a multistage process, for which three characteristic peaks can be distinguished in the DSC-graphs, observed at temperatures of 428, 418, and 410 K. Despite the fact, that the presence of overheating and the multistage nature of crystallization make it somewhat ambiguous to determine quantitatively the value of supercooling in this system, the authors give a value of 19 K as the value of supercooling, which corresponds to a relative supercooling of η = 0.04. The growth rate of the crystalline phase obtained in the work [11] allowed us to estimate the value of the thermodynamic barrier to crystallization in this system, which turned out to be approximately 2.7 kcal/mol.

Supercooling during the crystallization of indium particles embedded into an aluminum matrix was also studied in the work [12]. Using the method of channelling, it was confirmed that the aluminum matrix has an orientation influence on the embedded nanoparticles. Two series of samples were obtained: in one of them, the average size of indium particles was 4 nm, and in the second, 40 nm. The authors found that the melting point of indium in the first series is 452 K, which corresponds to an overheating of 23 K. Crystallization in these samples is observed at a temperature of approximately 408 K. This makes it possible to determine the value of supercooling (taking the melting point of the bulk material as a reference point) of 21 K (η = 0.05). The behavior of samples containing larger inclusions is similar to that observed in [11]. Thus, indium in such nanocomposites undergoes overheating up to 435 K and crystallizes at a slightly higher temperature of 410 K, which corresponds to the values obtained by the authors of [11].

The melting and crystallization of fusible metals (In, Sn, Bi, Cd, Pb) embedded into the aluminum matrix was studied in the work [13]. Samples for the study were obtained by ball milling, which, according to [13], is a convenient way to create nanocomposite structures in contact pairs whose components are not dissolved in the solid state. The sizes of the embedded nanoparticles were determined by the data of XRD and TEM studies. The thermal properties of the samples were studied using differential scanning calorimetry.

Due to the ability to change the average size of particles of the nanocomposite by varying the ball milling time, the authors of [13] were able to establish that the transition from micro (of the size of 10–20 µm) to nanosized particles (less than 20 nm) causes a noticeable increase in the amount of supercooling.

The size dependence of the melting and crystallization temperature of bismuth embedded in a glass matrix was studied in the work [14]. The following scheme was used to create nanocomposite structures: first, industrially available samples of $72B_2O_3$–$28Na_2CO_3$–$3SnO$–$Bi_2O_3$ were heated to the temperature of 1318 K, after which they were cooled to room temperature. This caused the release of free bismuth, which was thus embedded into a more refractory matrix. The samples were then annealed at the temperature of 773 and 838 K at different times, which provided the possibility of the formation of bismuth particles of a given size.

The authors of [14] found that the melting-crystallization process has hysteresis, and the melting and crystallization temperatures decrease with the decrease in the characteristic size of particles according to a law close to $1/R$.

Ultrasound to determine the limiting value of supercooling of nanocomponents of a composite material was used in the work [15], the authors of which studied the melting and crystallization processes of gallium embedded into the matrix of opal. The study used artificial opal, which represents itself as a structure consisting of $SiO_2$ spheres with a characteristic size of about 250 nm. Such spheres in opal form a densely packed structure in which there are two types of pores: octahedral and tetrahedral; their sizes are 100 and 50 nm, respectively. In the work [15], a "gallium-opal" nanocomposite structure was created by saturating opal with gallium by applying external pressure. After the formation of the nanocomposite, the authors studied the temperature dependence of the propagation speed and the absorption coefficient of ultrasonic waves in the sample.

The authors found that both during heating and cooling, on the temperature dependencies of the speed of ultrasonic waves, there are particularities that can be correlated to phase transitions. At the same time, both melting and crystallization in the studied system are multistage processes. Thus, the jumps observed during the heating of the sample occur at temperatures of about 250 and 290 K. The observed jump temperatures correspond to the melting of the β and α gallium modifications characteristic for this nanocomposite [16]. During cooling of the sample heated above 310 K, on the temperature dependencies of the ultrasound speed, two particularities are also observed, whose termination temperatures are 260 and 185 K. Thus, for the β and α modifications of gallium, the authors of [15] determined the values of supercooling to be 30 and 65 K. That is, the relative supercooling is $\eta = 0.1$ and 0.26 for α and β modifications, respectively. It should be noted that in the case of heating the sample to a temperature below 310 K, its behavior was somewhat different. Thus, the value of supercooling of the low-temperature β modification in these samples does not change, and this modification still completely crystallizes at 185 K. At the same time, the temperature of the jump, which corresponds to the crystallization of the α phase, increases, and its supercooling in this case is about 10 K. Moreover, both melting and crystallization have a diffuse nature, and crystallization begins at a temperature very close to the melting termination temperature. The authors of [15] explained this phenomenon by the possible existence of an even more refractory crystalline modification of gallium in the sample, which melts at the temperature of 310 K but is not detected by the ultrasound method. The remaining part of the crystalline phase may be a catalyst for the crystallization of the supercooled melt. At the same time, the observed effect can also be explained by the fact that a small part of the gallium β modification remained in the crystalline state due to overheating, and it can cause accelerated crystallization of the β phase. It should also be noted that since, according to [17], ultrasound influence stimulates crystallization, the supercooling values obtained in the work [15] may be somewhat underestimated.

### 3.2 Liquid-crystal phase transitions in multilayered films

The particularity of the works reviewed in the previous subsection is that the authors studied nanocomposite systems in which nanoparticles were embedded into defects of the polycrystalline matrix (primarily in the grain boundaries), directly into crystallites, or into nanopores of various glasses and opals. However, due to the fact that the matrix itself is a bulk object, the study of phenomena caused by the presence of nanocomponents in such structures turns out to be quite a difficult task, and the effects observed in them are often masked by many external factors.

Therefore, it seems advisable to replace such nanocomposites with model systems when conducting basic scientific research. For example, when studying the stability limits of the liquid phase, a convenient model of composite material can be multilayer films in which a layer of material that models the inclusion under study is located between layers that act as a matrix. Similar model systems have a number of particularities that make them extremely convenient for studying supercooling in composite materials.

Thus, in contrast to the composites considered in the previous section, multilayer films are obtained using a highly reproducible vacuum condensation technique consisting of essentially a single step. A sample, suitable for research, can be only obtained by performing a set number of deposition cycles, and it does not require additional sample preparation (such as polishing, grinding or etching). In turn, the formation of samples under high vacuum conditions allows reducing the amount of outside impurities that can enter the sample both during its obtaining and during studying to a negligible level. In addition, layer-by-layer vacuum condensation provides atomic contact between the layers. This is necessary to eliminate side effects related to the quality of the interface. Finally, the ability to examine samples in a wide range of characteristic sizes by changing the thickness of the layers seems to be convenient.

While the method of changing the condensation, mechanism turned out to be extremely promising for studying the supercooling of fusible particles located on a more refractory substrate, other *in situ* techniques turned out to be more effective for studying the supercooling of the fusible component in layered film systems.

Thus, in the work [18], the use of the temperature dependence of the electrical resistance of film samples was proposed to determine the phase transition temperatures. As an object of study, the Bi-Al contact pair was chosen, which is characterized by a phase diagram of the eutectic type with low solubility of components in the solid state. Two series of experiments were carried out. In the first of them, two-layer Bi/Al films were studied. Supercooling during the crystallization of the bismuth-based eutectic in this case was determined with the help of the previously considered method of changing the condensation mechanism. The two-layer films were condensed onto the prolonged steel substrate with a previously deposited sublayer of amorphous carbon. Before condensation of the system under study, a gradient of temperatures was created along the substrate, which was chosen in such a way that at one end of the substrate, its temperature slightly exceeded the melting point of bismuth, while at the other end, it remained at the level of the room temperature. After the condensation was completed, the substrate with the film was cooled in a vacuum to room temperature, removed from the vacuum chamber, and examined by electron microscopy and visual inspection.

It was found that visually the boundary is observed on the substrate below which (by temperature) the film shows no signs of melting, and above which it consists of individual particles in the form of a spherical segment (Fig. 3.1). This clearly indicates that the film was in a liquid state. The boundary of change in the condensation mechanism in the samples was observed at a temperature of 453 K. That is, in these samples, the supercooling of bismuth, which is in contact with aluminum, was 91 K ($\eta = 0.17$). The wetting angle in this contact pair was found to be equal to 120°. It should be noted that the value of the wetting angle and the obtained supercooling value fit within the framework of the empirical dependence obtained from the results of the works [19, 20, 21].

In the second series of experiments, three-layer Al/Bi/Al films were studied, and for recording the melting and crystallization of the fusible eutectic, which is in contact with aluminum, the method of measuring electrical resistance in "heating-cooling" cycles was used. To realize this technique, the films were condensed onto glass substrates with a pre-deposited system of electrical contacts. Before condensation, the glass substrate was placed into a copper block equipped with a resistive heater and a K-type thermocouple. The substrate was pressed against the copper block using a special mask with a system of contacts that provided the electrical connection of the substrate with the measuring equipment.

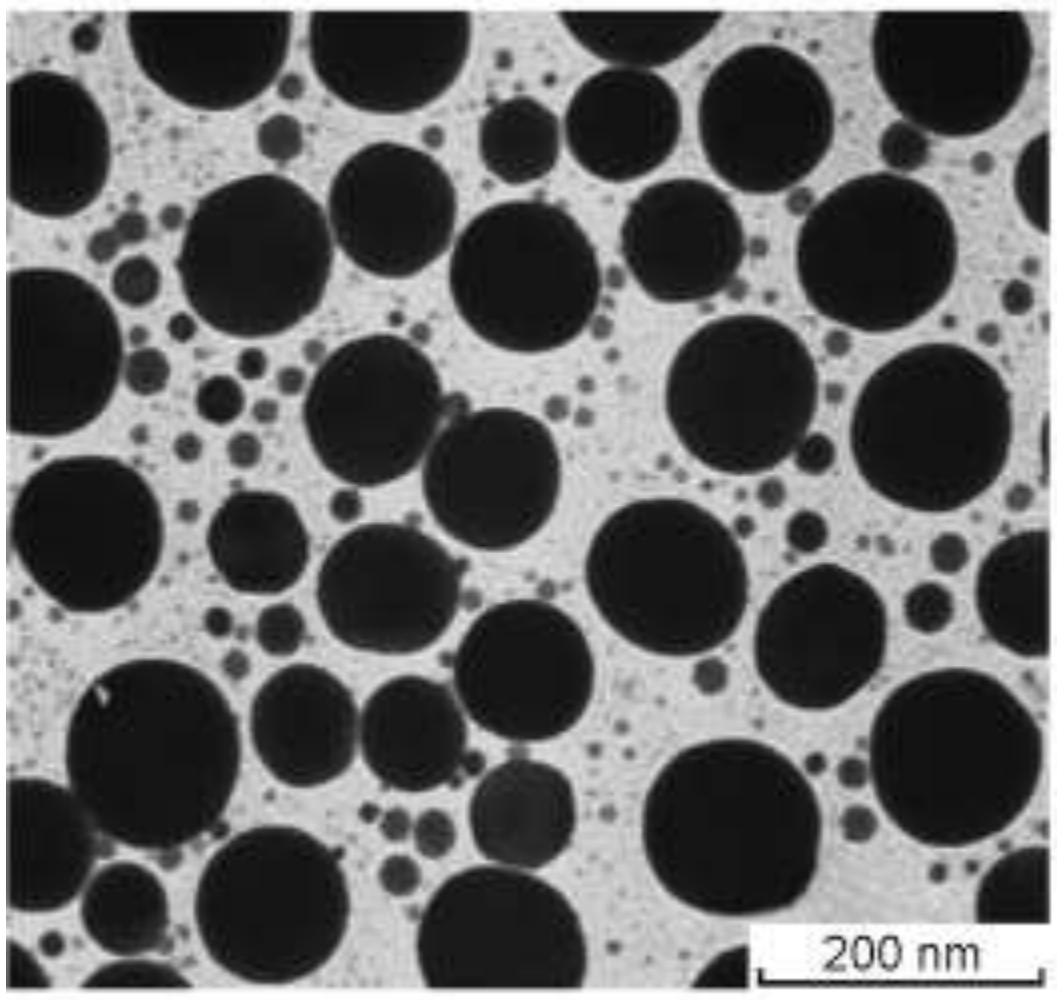

**Fig. 3.1** TEM image of a bismuth film on an aluminum substrate. The image corresponds to the sample, the temperature of which exceeded a little bit the temperature of the maximum supercooling [18]

Fig. 3.2 shows the temperature dependencies of the resistance obtained in this way, which correspond to several heating-cooling cycles. It can be seen that both the heating and cooling curves have particularities. For example, near the table value of the melting point of bismuth, there is a rather sharp drop in resistance. The opposite to it increase in resistance was observed when the sample is cooled to a temperature close to the previously determined temperature of the change in the condensation mechanism. This makes it possible to compare the jumps in electrical resistance, which are observed in three-layer films with the melting and crystallization of the bismuth layer.

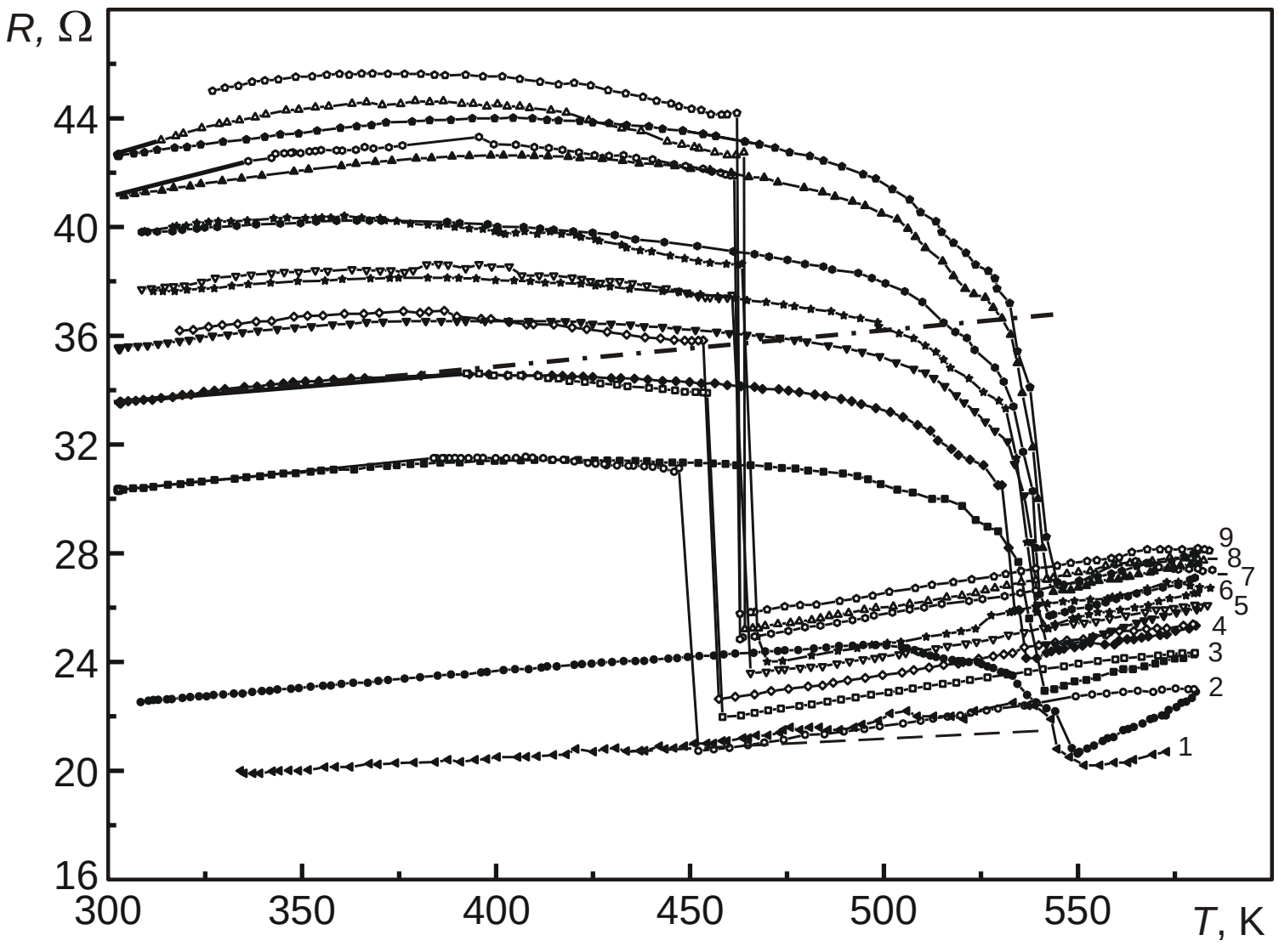

**Fig. 3.2** Resistance dependence of Al/Bi/Al three-layer films on the temperature. Immediately after the termination of the condensation (curve 1) and in the process of the following heating–cooling cycles performed after 8 (2), 9 (3), 10 (4), 14 (5, 6, 7), and 16 (8, 9) days in vacuum holding. In each cycle filled and unfilled symbols correspond to heating and cooling, respectively [18]

The value of resistance jumps accompanying phase transitions turns out to be quite significant. Thus, for a three-layer system with a bismuth content of 65 wt. %, the electrical resistance during crystallization changes by about one and a half times. At the same time, it is worth noting that crystallization, contrary to melting, which is still extended within a noticeable temperature range, occurs almost immediately throughout the whole sample, i.e., it has an avalanche-like character. During the thermal cycling process, the temperature dependencies of the resistance do not change qualitatively. Only a gradual increase in resistance is observed, as well as some changes within the course of the curves corresponding to the heating of the sample. Thus, in the first heating cycle, the electrical resistance of the three-layer film increases almost linearly up to a temperature of 440 K, after reaching which the linear dependence is broken – a faster increase in resistance begins that is changed by a drop near the melting point. During subsequent heating-cooling cycles, performed on the samples, that were kept in a vacuum chamber at room temperature for 8 days, an increase in resistance is observed up to a temperature of about 500 K, starting from which a decrease in resistance occurs, culminating in a jump that identifies the melting of bismuth.

It should be noted that the increase in resistance that occurs from cycle to cycle has a non-linear nature (Fig. 3.3).

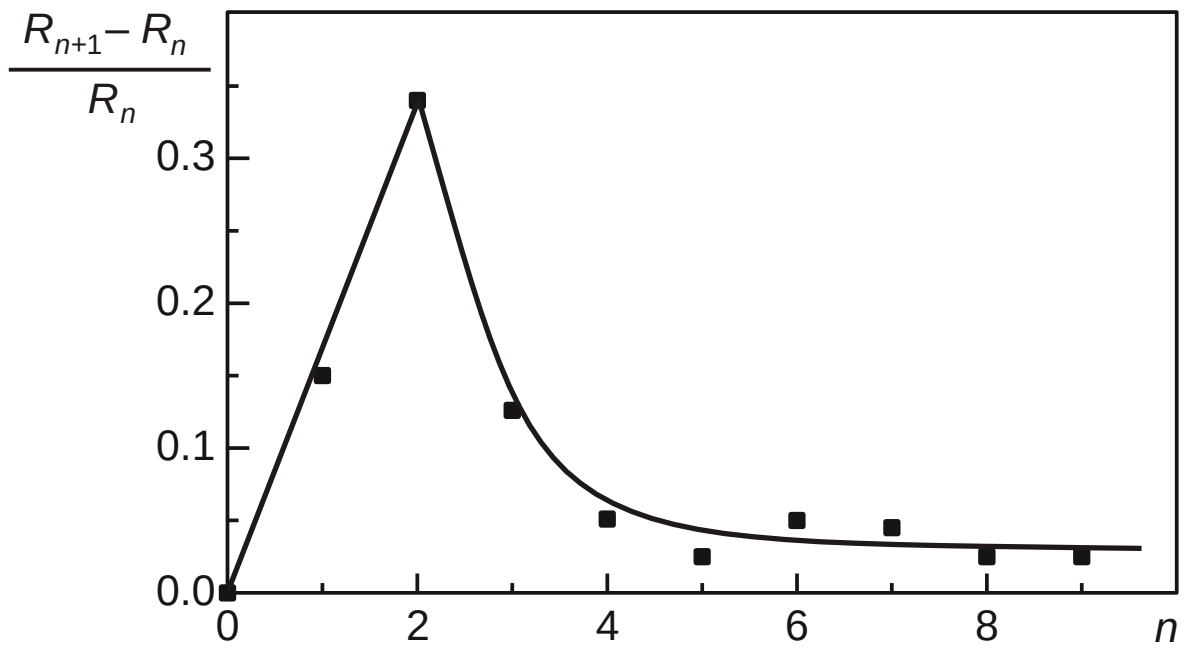

**Fig. 3.3** Resistance relative change versus the cycle number for Al/Bi/Al films during the phase transition [18]

Thus, the maximum increase in the electrical resistance of the sample is observed during the first three to four cycles, after which the relative increase in resistance practically stops to depend on the cycle number. At the same time, the relative change in resistance during crystallization becomes a constant already starting from the second cycle.

It is separately worth noting that starting from the second heating cycle, the deviation of the experimental resistance dependence ($\Delta R$) from the theoretical one, which was calculated based on the temperature dependence of the resistance (dotted line in Fig. 3.2), is observed. The temperature dependence of $\Delta R$ in the coordinates «$\ln\Delta R/R - 1/T$» has a complex nature (Fig. 3.4). However, four relatively linear sections can be highlighted in these graphs, each of which, according to the Arrhenius equation, can be assigned its own value of the activation energy of the process responsible for its appearance.

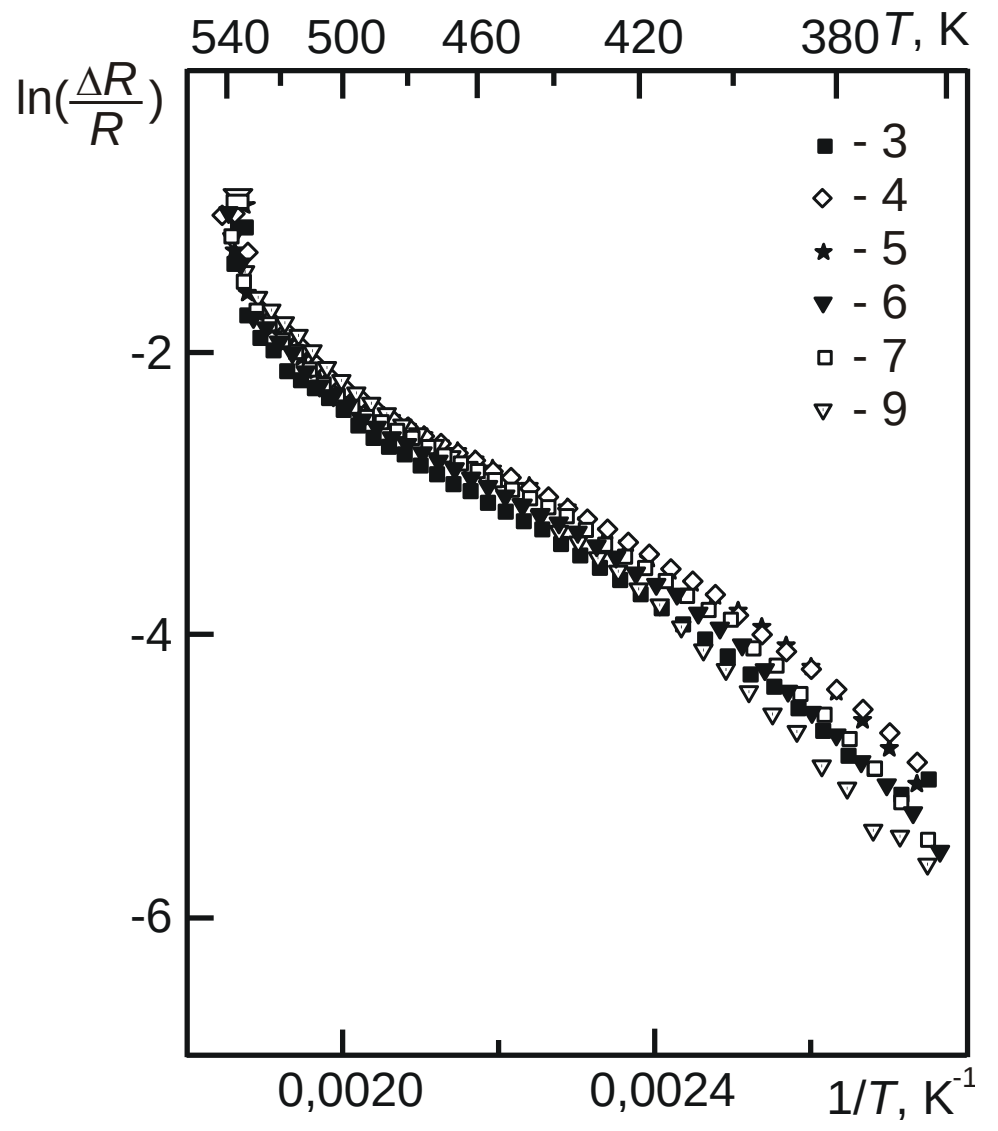

**Fig. 3.4** Relative decrement resistance versus inverse temperature for the Al/Bi/Al system (for 3–7 and 9 cycles of heating-cooling) [18]

Thus, the activation energy for the section corresponding to temperatures less than 400 K is 0.4 eV. It turns out to be somewhat less within the temperature range of 400-500 K, where the value obtained is 0.25 eV. The activation energy determined in the section, which immediately precedes the melting, is 0.65 eV, and the melting itself is determined by an energy in 1.3 eV.

It should be noted that the main features of the temperature dependence of the resistance are also observed in Bi/Al bilayer films. The temperature dependencies of the resistance in the first heating-cooling cycle show jumps in resistance, which indicate melting and crystallization of the supercooled melt. However, the thermal stability of such samples turns out to be significantly lower than that of three-layer structures. They lose their conductivity after the first heating-cooling cycle, which indicates the de-wetting of initially continuous films into separate islands. This decay is most intense after the crystallization of the supercooled melt.

The behavior of single-layer films obtained by simultaneous condensation of bismuth and aluminum turns out to be somewhat different. In these samples, the change in electrical resistance during crystallization turns out to be maximal. Thus, in Fig. 3.5, which corresponds to a film containing only 30 % bismuth, an almost threefold jump in resistance is observed. In addition, the graph does not show the particularities, which are characteristic of the first heating cycles of films obtained by sequential condensation.

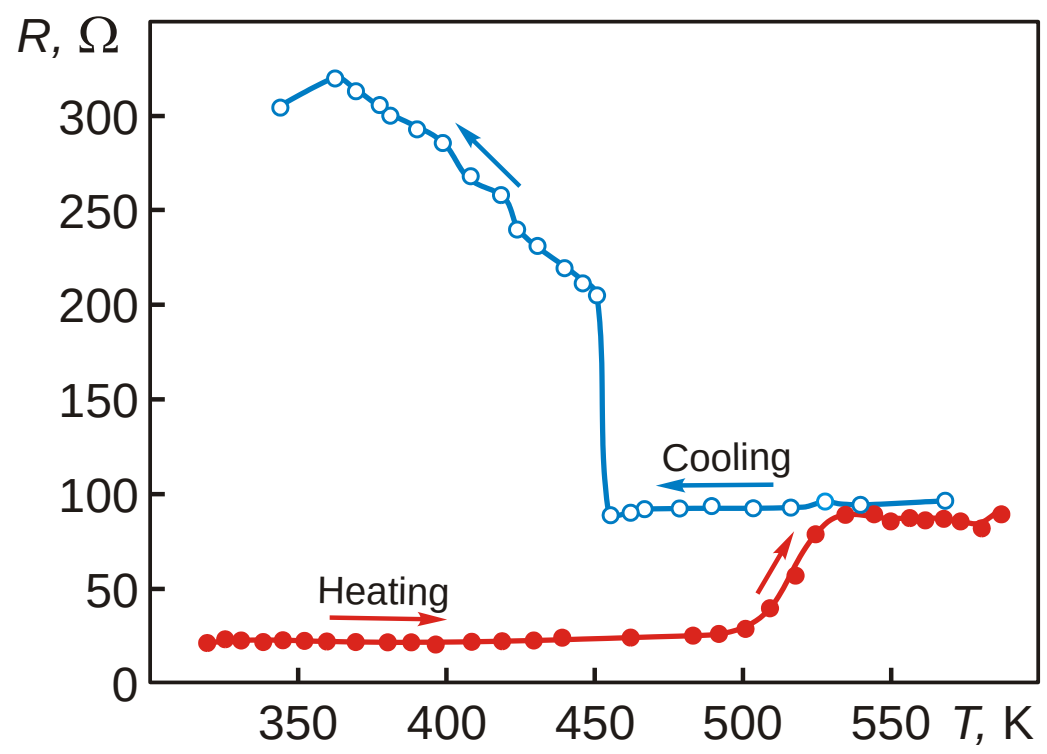

**Fig. 3.5** Electrical resistance of two-layer Bi/Al films versus temperature. The thicknesses of the layers of bismuth and aluminum are 30 and 50 nm, respectively [18]

It should be noted that in all three cases (three-layer, two-layer, and single-layer film obtained by simultaneous condensation of components), the limiting temperature of the existence of the liquid phase is close to the value obtained by using the method of changing the condensation mechanism.

The activation energy values determined in the work [18], as well as the comparison of the three types of films with each other, can provide important information about the kinetic processes occurring in layered film systems and determine the causes that lead to changes in resistance during the phase transition.

For example, in the case of the realization of the method of layer-by-layer condensation, the layers, which were formed, can be considered independent to some extent. Initially, only a small part of the bismuth atoms migrate through the accelerated diffusion pathways into the aluminum layers. During the first heating-cooling cycle, the part of such atoms increases due to the thermal activation of diffusion. The main pathway for accelerated diffusion for bismuth atoms is the boundaries of grains. The saturation of the grain boundaries of a polycrystalline aluminum film with bismuth atoms should cause a faster increase in the resistance of the sample than that expected, based on the temperature dependence of the resistance. This is what is observed in reality, and the Arrhenius plot constructed for the section of faster resistance increase in the first cycle (Fig. 3.6) gives a value of the activation energy of this process equal to 0.28 eV. The activation energy, which equals 0.26 eV and describes the resistance change processes observed in the films starting from the second heating cycle to a temperature of 500 K, is also attributed to bismuth diffusion by the authors of [18].

The low-temperature process with an activation energy of 0.4 eV in the work [18] is associated with the processes of annealing of defects that occurred, in the first turn, during the previous cooling cycle. The process that begins at a temperature of about 500 K (activation energy of 0.65 eV), according to the authors, is caused by the decomposition of a solid solution of aluminum in bismuth. Finally, the highest-energetic process, occurring near the melting point of bismuth, was attributed to the bulk diffusion of aluminum.

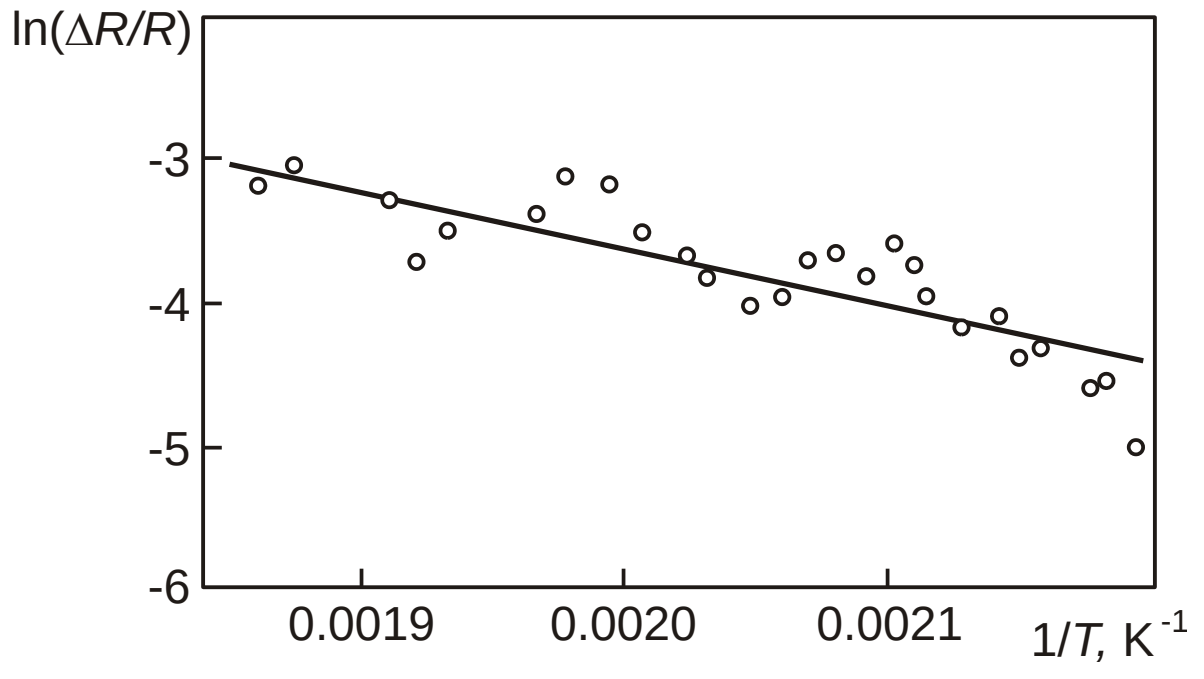

**Fig. 3.6** Arrhenius graphs which correspond to the irreversible increase of electrical resistance, which was observed in the first heating cycle of three-layer Al/Bi/Al films [18]

Thus, the authors of [18] not only showed the possibility of using the method of measuring electrical resistance to determine phase transition temperatures but also found out that this method is sensitive to many processes occurring in layered film systems and has important meaning for fundamental and applied science.

The method of measuring electrical resistance to study supercooling in germanium-based multilayer films has also been tested to establish the temperature limits of the stability of the liquid phase in Ge/Bi/Ge films. It is worth noting that this contact pair forms a phase diagram of the eutectic type, with insignificant mutual solubility of the components in the solid state. In addition to measuring the electrical resistance, the method of changing the condensation mechanism was used as a control technique for determining the phase temperatures in the study.

Similarly to other samples of binary systems obtained by the variable state method, two regions with different light scattering and a boundary separating them can be found on the substrate. The electron microscopic images shown in Fig. 3.7 confirm that in the higher temperature region, the film consists of particles in the form of a spherical segment.

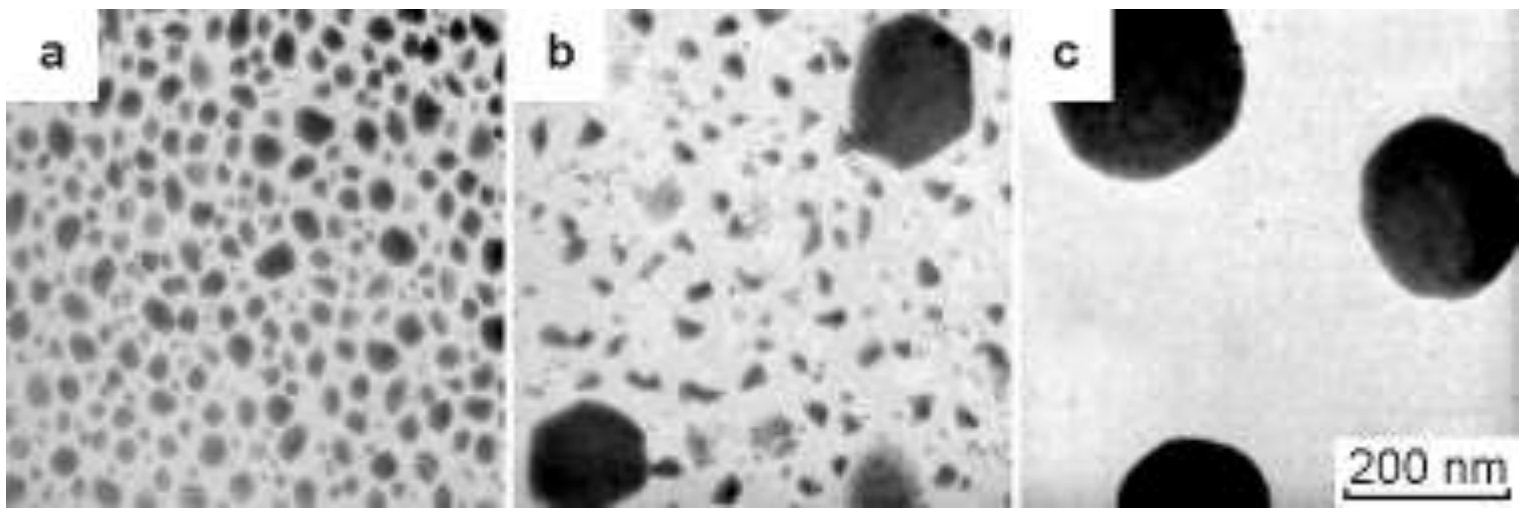

**Fig. 3.7** TEM images of bismuth films on a germanium substrate at the temperature of 434 K (**a**), 451 K (**b**), and 473 K (**c**)

This morphology of the sample is an indication that the "vapour-liquid" mechanism was implemented during its condensation. At the same time, in the low-temperature region, there are no features of the presence of a liquid phase. The boundary itself, which corresponds to the transition from the crystalline state to the metastable liquid state, has a finite width, i.e., there is a region of coexistence of solid and liquid phases on the substrate. This can be considered as an indication that crystallization in the system occurs diffusely. The minimum temperature for the existence of a supercooled melt determined in this way is 451 K. That is, the absolute value of supercooling in this case reaches 92 K, and the relative supercooling is η = 0.17. It should be noted that the electron diffraction studies carried out by the authors indicate that the germanium with which bismuth is in contact in these experiments has an amorphous structure (Fig. 3.8).

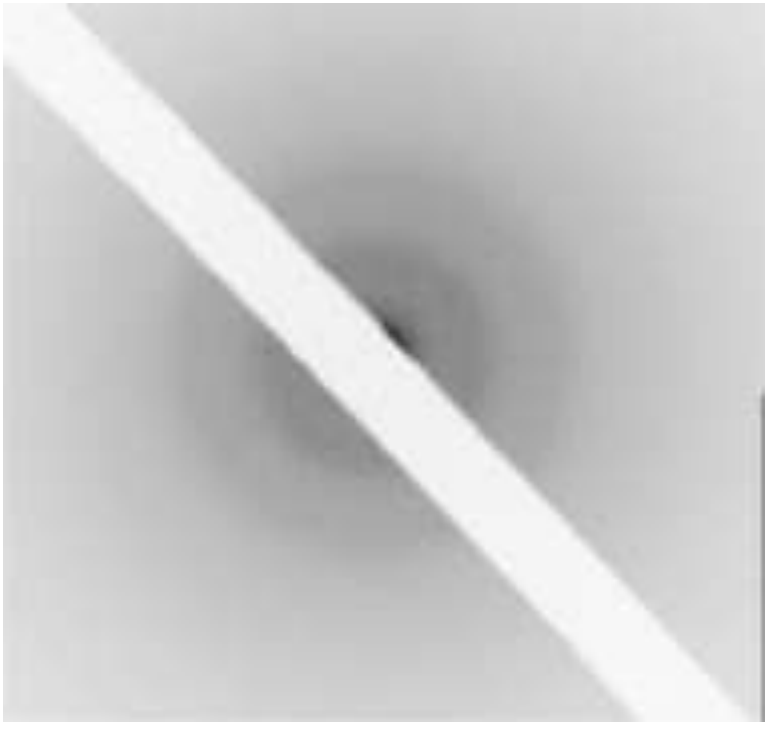

**Fig. 3.8** SAED pattern from as-deposited germanium films

In the second series of experiments, the temperature dependence of the resistance of three-layer Ge/Bi/Ge films was studied. As can be seen from Fig. 3.9, the dependencies of the electrical resistance of the three-layer films on the thickness have a clearly non-monotonic nature and undergo significant changes during the heating-cooling cycles.

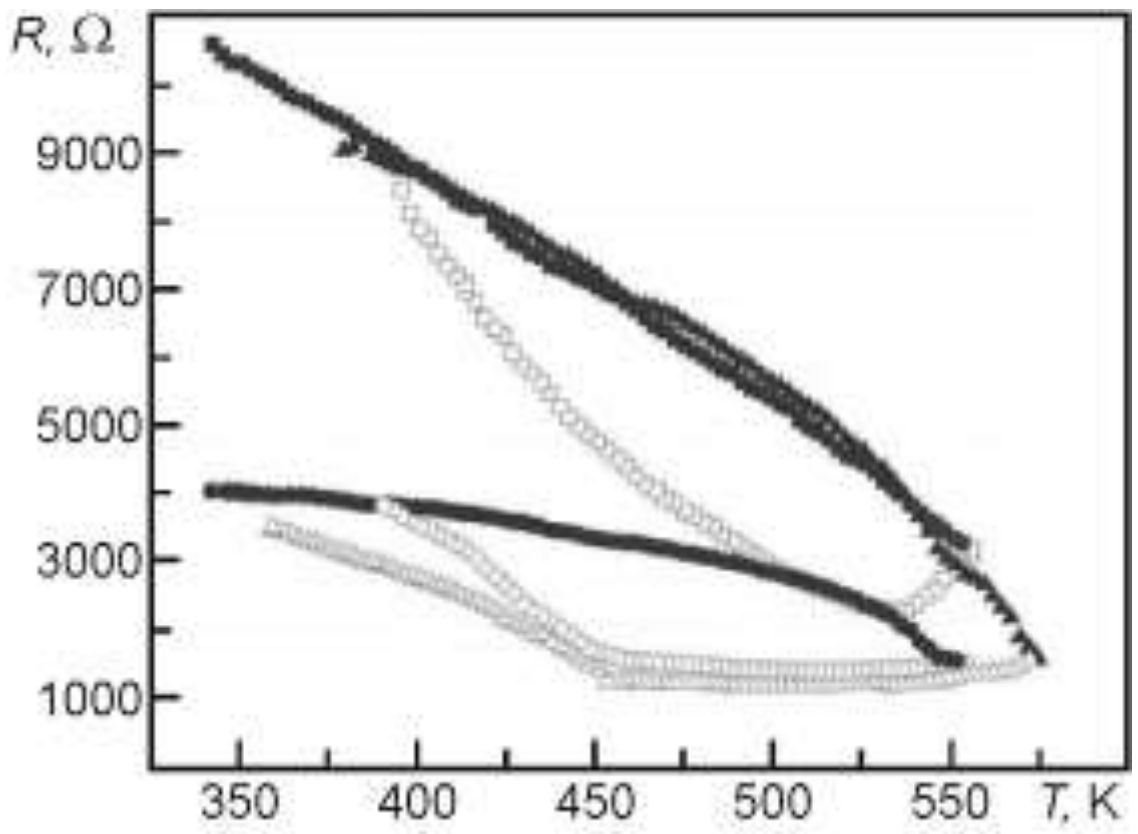

**Fig. 3.9** The electrical resistance change of the layered Ge/Bi/Ge film system for the first (■), second (▲), and third (●) heating-cooling cycles

Thus, in the first heating cycle, the nature of the temperature dependence of the resistance, typical for semiconductor systems, is observed, namely, its decrease with the temperature. At the same time, no particularities that correspond to the melting of a fusible component are observed. A similar situation occurs during cooling, in a process of which, apart from a small temperature region observed at the beginning of the cooling process, there are no other particularities that could be associated with crystallization.

At the same time, starting from the second heating-cooling cycle, resistance jumps appear and gradually become more obvious, which can be connected to the melting and crystallization of the eutectic. In the third cooling cycle, crystallization is completed at a temperature of about 410 K, i.e. the supercooling is 134 K, which corresponds to η = 0.25. Note that crystallization in this case has a diffuse nature and occurs in the temperature range of about 30 K.

The evolution of the studied contact pair continues up to the fifth heating-cooling cycle. Afterwards, the behavior of the system practically does not change, and the temperature dependence of the electrical resistance undergoes not only quantitative but also qualitative changes compared to the first cycles. Thus, starting from the fifth heating, the graphs take the form shown in Fig. 3.10. They show, along with the typical semiconductor dependence, distinct jumps corresponding to the melting and crystallization of bismuth. In this case, as in the first heating cycles, melting occurs diffusely, while crystallization becomes of avalanche-like nature. In addition to the change in the nature of the crystallization, the temperature at which this phase transition is observed also changes. Now the crystallization is fully completed at a temperature of 465 K, i.e. the supercooling is only 94 K (η = 0.17).

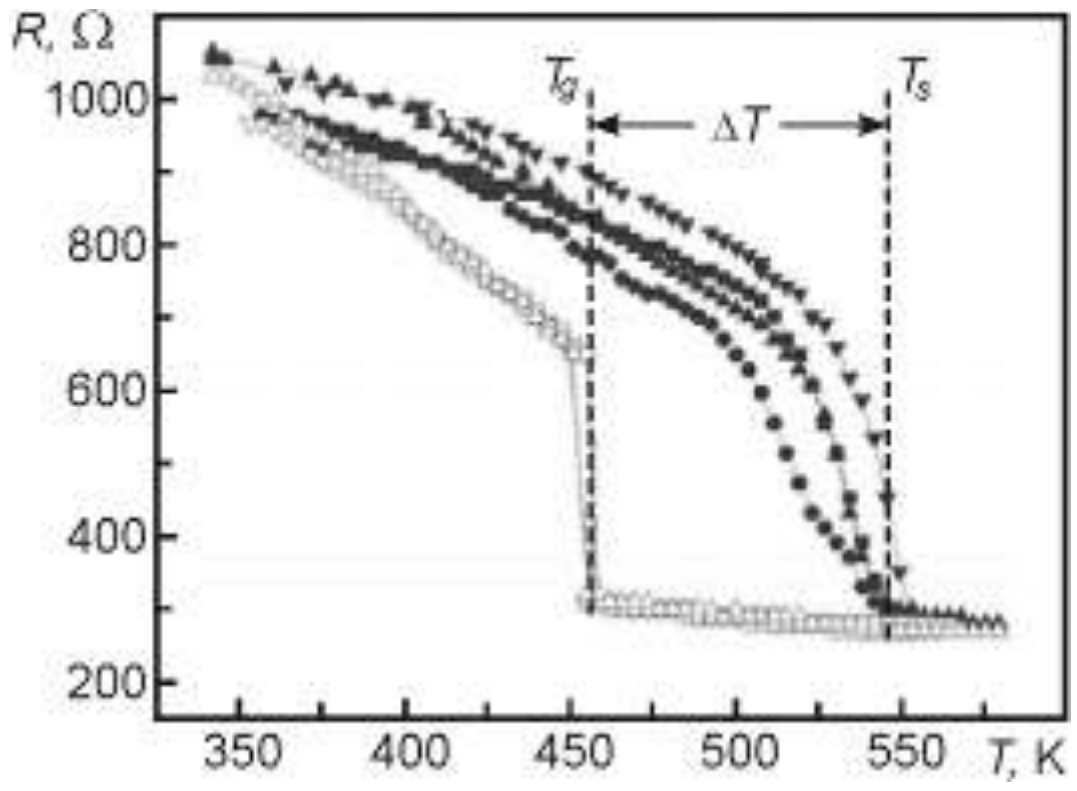

**Fig. 3.10** Electrical resistance dependences on the temperature for three-layered Ge/Bi/Ge films, which correspond to the fifth and to the next heating-cooling cycles

With the help of XRD studies, it was found that the thermal cycling of Ge/Bi/Ge films causes the crystallization of amorphous germanium (Fig. 3.11). At the same time, special studies have shown that only a diffuse halo is present on the electron diffraction patterns obtained from films, which were annealed for 15 hours at 610 K, indicating an amorphous germanium structure. Together with this, since the effective annealing time of three-layer Ge/Bi/Ge films that occurs in the study of melting-crystallization of the fusible component is much smaller, it can be argued that the crystallization of amorphous germanium is not directly connected to annealing. It is likely that in this case, we are dealing with metal-induced crystallization, which occurs in heating-cooling cycles. This phenomenon is caused by the presence of a thermodynamically stable or supercooled liquid phase of the second metal on the surface of amorphous germanium.

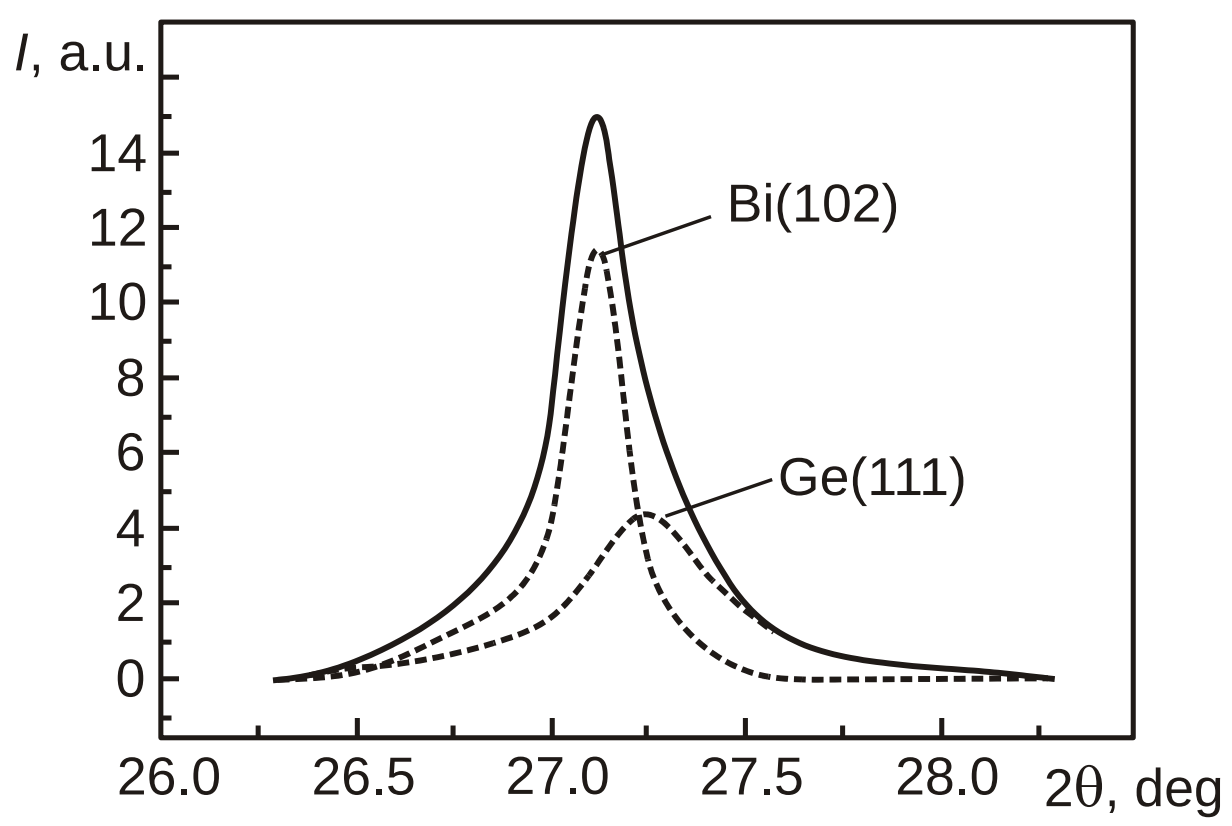

**Fig. 3.11** Part of a diffraction pattern obtained from Ge/Bi/Ge films, which shows the crystallization of initially amorphous germanium layers

The supercooling of bismuth located between layers of amorphous carbon was studied in the work [22]. To determine the temperatures of the "melting-crystallization" phase transitions, there were used methods of measuring electrical resistance in heating-cooling cycles and *in situ* electron diffraction studies.

In contrast to an earlier implementation of the electrical resistance measurement technique, which involved the use of clamping electrical contacts and a bulk inertial heater [18], in the work [22] special low-inertia cells with film contact pads were used to measure the resistance. Such cells represented themselves as two plates, fused or glued together, between which a thermocouple was fixed. The plates were made of dielectric material (glass, alumina ceramic). They were rectangular, 15 × 10 mm in size. The thickness of the plates was 1.2 mm for glass and 1.0 mm for ceramic. Prior to gluing, four through holes with a diameter of 1–2 mm were drilled in one of the plates, which was the working plate and was used for the deposition of the film systems under study. A copper wire was repeatedly threaded through these holes, which formed the basis of the contact pads required for resistance measurements. One of the ends of the wound wire was left between the plates to provide an electrical connection between the contact pads and the measuring system. After gluing or sintering, the working side of the measuring cell was carefully polished, after which the remains of wires on the working side represented themselves as good contact pads for providing electrical connection with the deposited layers. Then, a set of electrical contacts was applied to the surface of the polished plate by vacuum condensation through a special mask, on which the film system under study was subsequently deposited. The electrical resistance was measured using the four-point probe method. The measuring cell was suspended on the current-carrying and voltage-sensing electrodes in a special holder through electrical insulators. A furnace of radiant heating, which represented itself as an open spiral tungsten filament, was installed above its backside.

During the study of three-layer C/Bi/C films, it was found that in the first heating cycle, their resistance increases sharply and irreversibly near the melting point. This probably indicates the decay of the initially continuous film into separate islands.

This process is a well-known phenomenon and is typical of both contact pairs with non-wetting and systems with fairly good wetting. During subsequent heating-cooling cycles, characteristic jumps are observed in the temperature dependencies, but their small relative value makes it difficult to register their corresponding phase transitions. Therefore, in the work [22], instead of three-layer films, multilayer structures were used, in which up to five layers of bismuth were alternated with layers of amorphous carbon.

Fig. 3.12 shows the graphs of the resistance dependence on temperature obtained for such a multilayer system. It can be seen that there are jumps in the electrical resistance on the graphs of heating and cooling. At the same time, both melting and crystallization of bismuth occur diffusely and crystallization in the system is completed at a temperature of 175 °C, which corresponds to a supercooling of 96 K; η = 0.17.

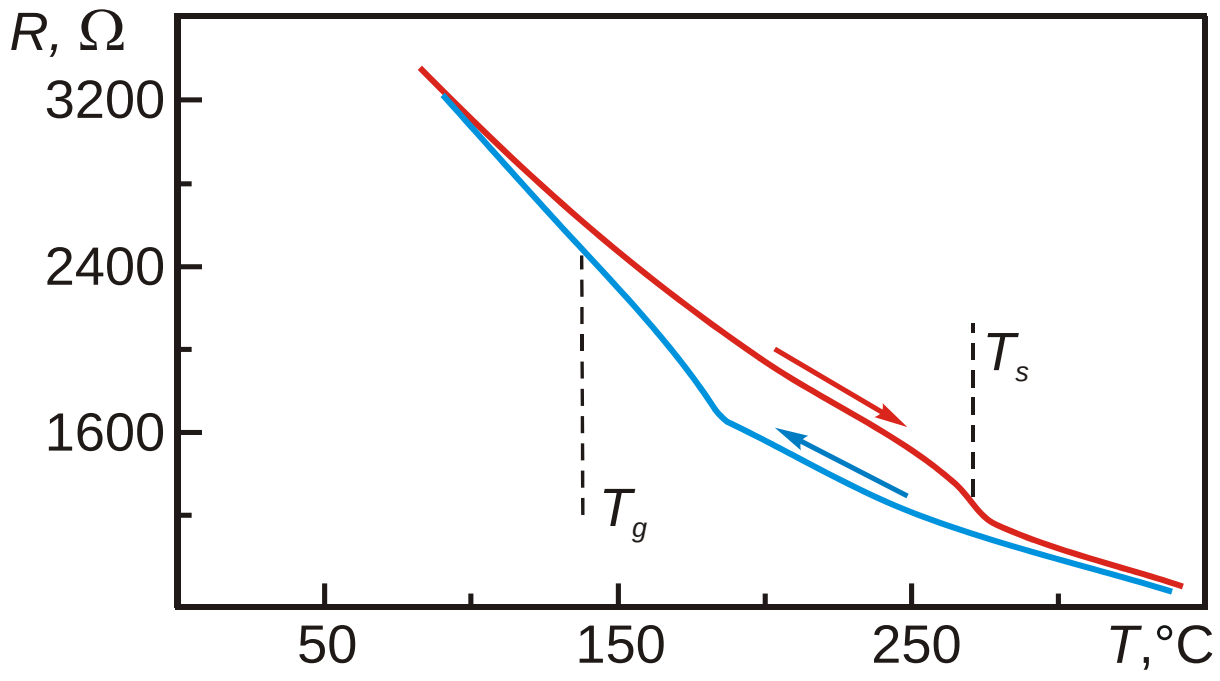

**Fig. 3.12** Electrical resistance dependence of C/Bi/C films on the temperature. The dependence was obtained from multilayered structures containing five layers of fusible component [22]

The position of phase transitions can be more accurately recorded if, instead of the electrical resistance of the films, the temperature dependencies of the rate of its change, i.e. the first derivatives of the experimental curves, are used. Fig. 3.13 shows such a graph obtained by numerical differentiation of the experimental data using the Savitzky–Golay filter. It can be seen that the differentiation made it possible to make the peaks more obvious. However, their prolonged nature (bismuth crystallization occurs in the temperature range of about 30–50 K), as well as in the case of using the temperature dependencies of resistance, somewhat complicates the exact determination of phase transition temperatures.

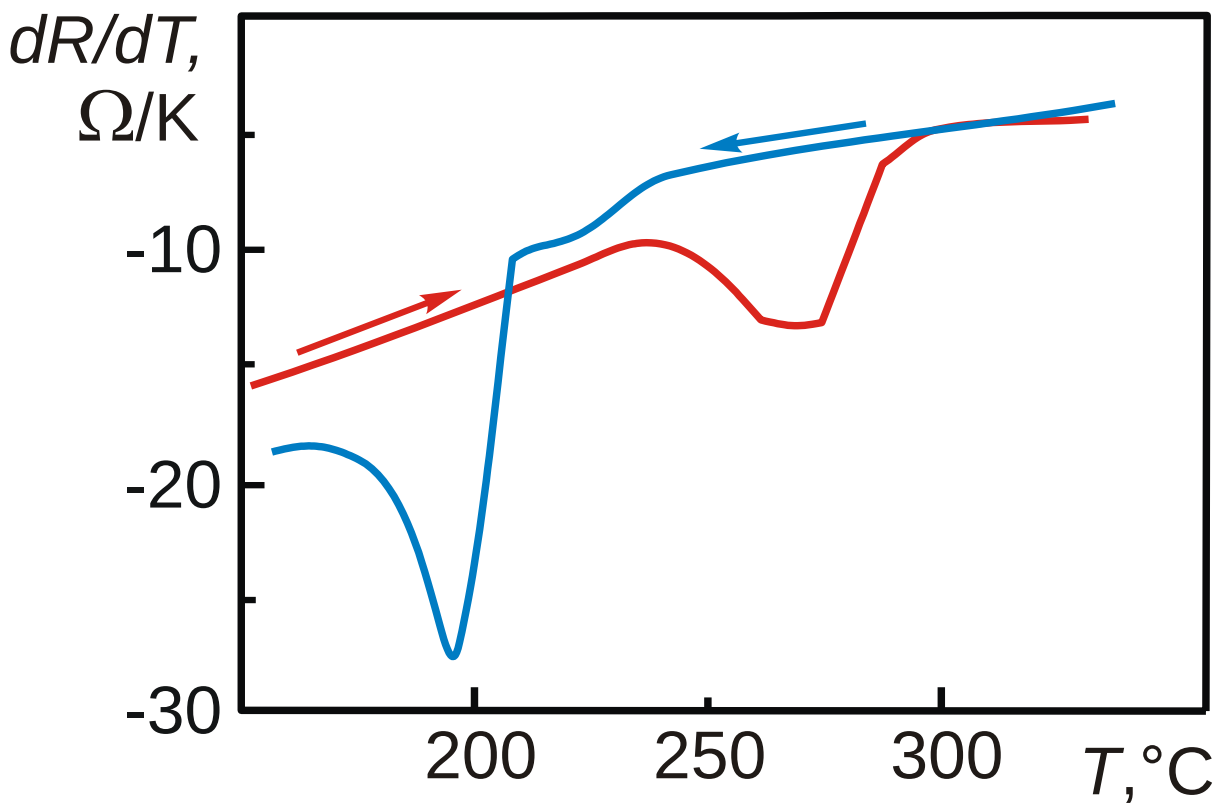

**Fig. 3.13** Numerical differentiation results of the temperature dependence of the electrical resistance for the C/Bi/C films [22]

As an alternative method for recording supercooling, *in situ* electron diffraction studies were used in the work [22]. During the *in situ* electron diffraction studies (Fig. 3.14), it was found that the electron diffraction lines corresponding to reflections from the crystallographic planes of bismuth completely disappear when the sample is heated to a temperature above 271 °C. During cooling, at a sample temperature below the melting point, reflections from the bismuth crystal structure are still absent and only begin to appear at significant supercoolings. They acquire their initial brightness near the temperature determined as the limiting temperature of the liquid phase existence according to the data of resistive studies.

It should be noted that the obtained supercooling values occupy an intermediate position between the previously reported values ($\eta$ = 0.18 and $\eta$ = 0.28) [23], which correspond to samples obtained under very different vacuum conditions ($10^{-5}$ mm Hg in a vacuum chamber, that is pumped out by an oil vapor jet pump and $5 \cdot 10^{-8}$ mm Hg under oil-free pumping conditions, respectively). This seems quite natural, since the films in the work [22] were condensed at a residual gases pressure at the level of $10^{-6}$ mm Hg, and the vacuum chamber was pumped out by a turbomolecular pump, i.e., the vacuum conditions were close to oil-free. Thus, the results of [22, 23] show that the regularities of the influence of the residual atmosphere on the supercooling of bismuth, located on the surface of a carbon film [19], also occur in systems of the "particle in a matrix" type.

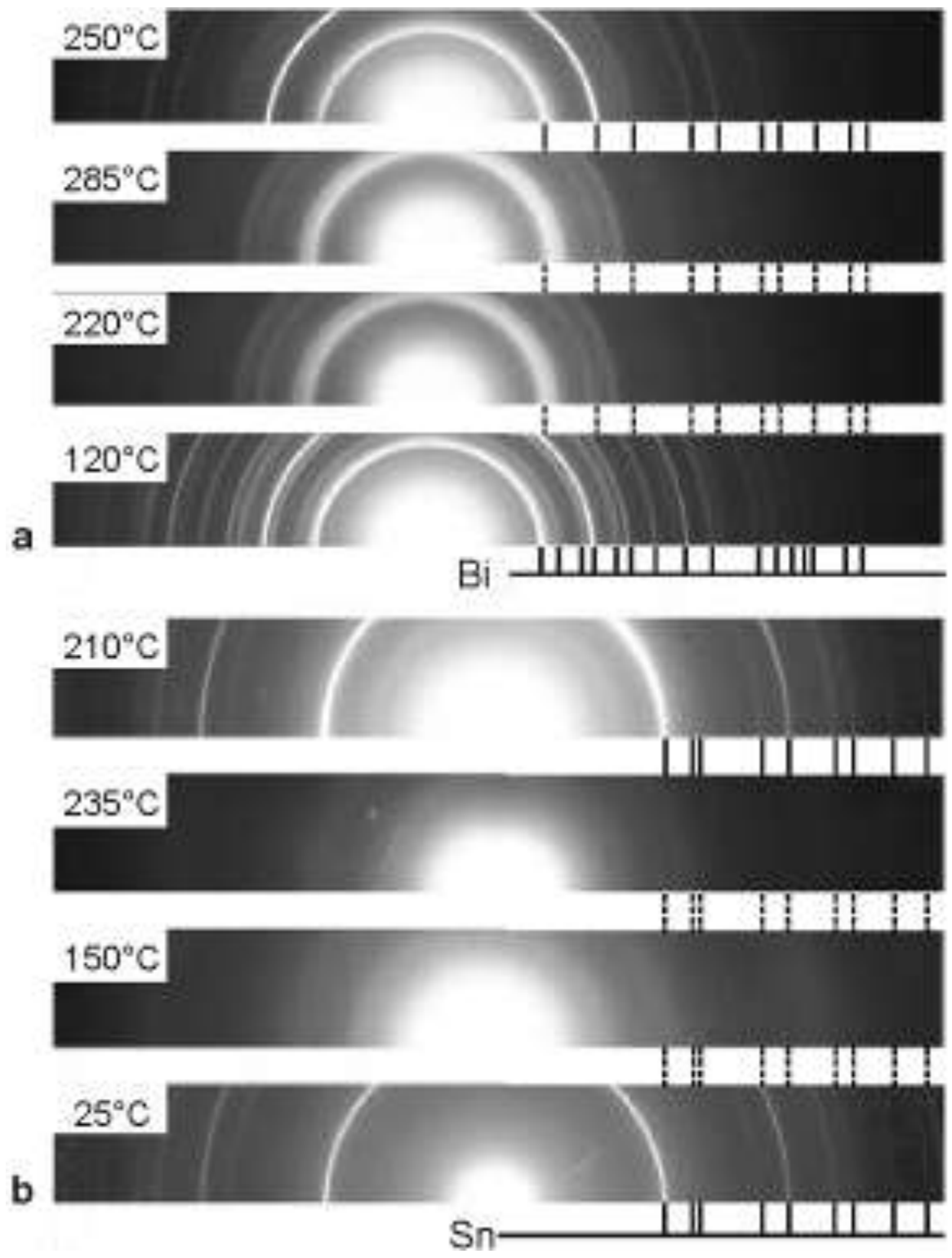

**Fig. 3.14** Electron diffraction patterns obtained from C/Bi/C (**a**) and C/Sn/C (**b**) films in the heating-cooling cycle (from up to down). The temperatures of the samples are shown in the pictures [22].

Supercooling of bismuth in Cu/Bi/Cu three-layer films was studied in [24]. Fig. 3.15 shows the dependence of the resistance of the Cu/Bi/Cu film on temperature. It should be noted that these results, as well as the previously reviewed works that used the method of measuring electrical resistance and a quartz resonator, were obtained on samples that had been condensed onto a substrate at room temperature.

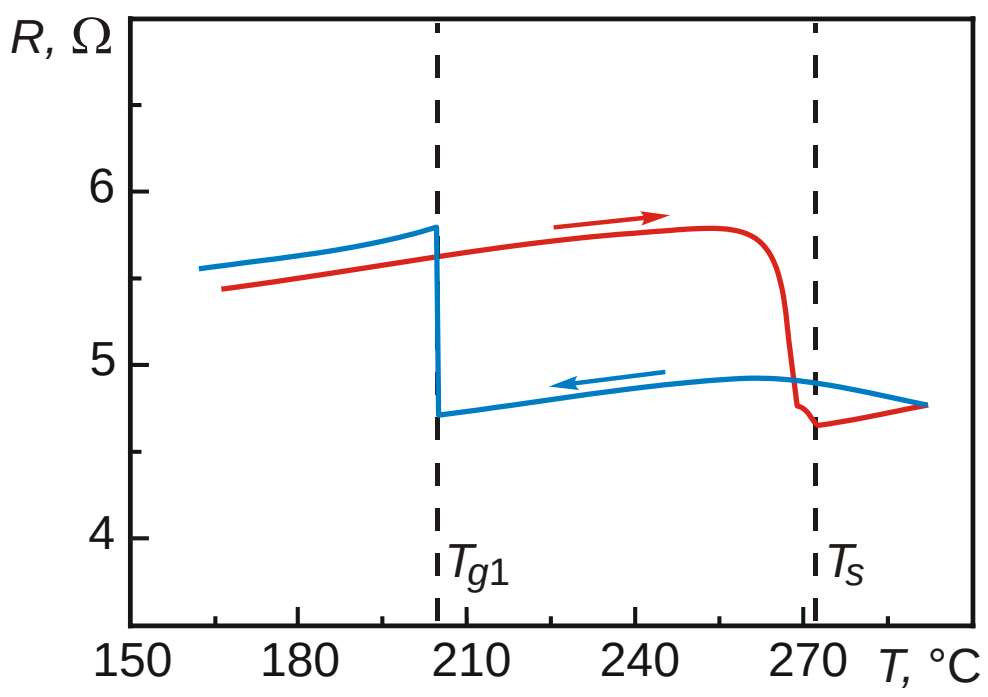

**Fig. 3.15** Electrical resistance dependence of Cu/Bi/Cu films on the temperature. The graph was obtained for the samples, which were subjected to several cycles of heating-cooling [24]

It can be seen that melting and crystallization are accompanied by jumps in electrical resistance. In this case, the crystallization has an avalanche-like nature and occurs at a temperature of 205 °C. This means that the observed supercooling is 65

K (η = 0.12). The obtained supercooling values were confirmed independently by *in situ* electron diffraction studies (Fig. 3.16).

It is worth noting an interesting regularity that follows from the analysis of the results presented in works [22, 23, 24]. In cases where bismuth is in contact with amorphous layers (carbon, as-deposited germanium films), its crystallization is diffuse, i.e. it stretches over a certain temperature range, the value of which can reach tens of degrees. In the case of a polycrystalline matrix (aluminum; copper; germanium, which has undergone metal-induced crystallization), the crystallization nature changes and bismuth hardens almost immediately throughout the sample. Moreover, from the comparison of Fig. 3.9 and Fig. 3.10, it can be concluded that the change in the nature of crystallization is also accompanied by a change in the value of supercooling.

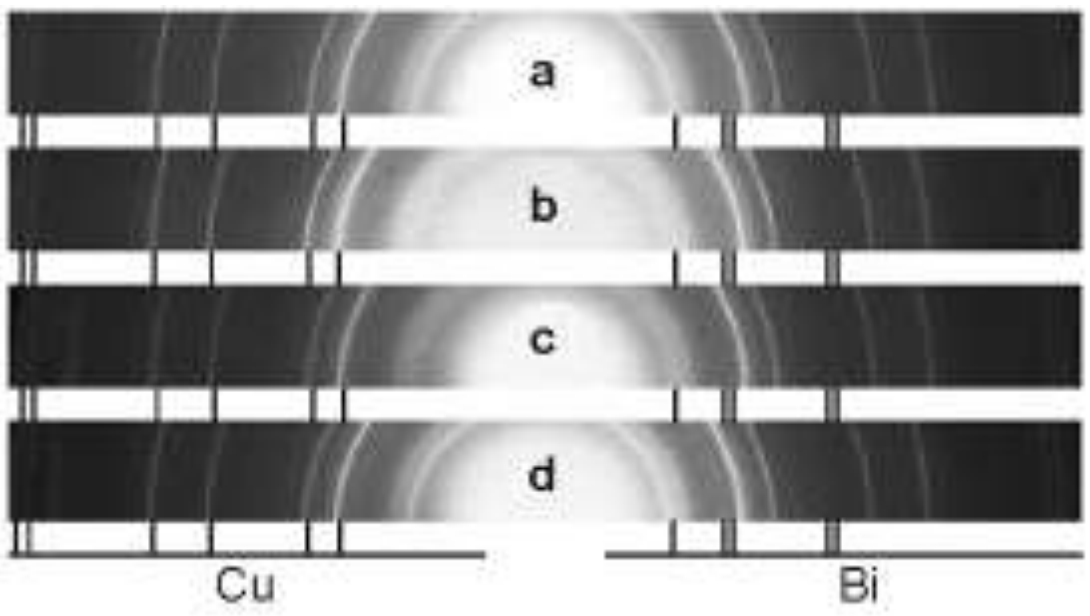

**Fig. 3.16** Electron diffraction patterns obtained from Cu/Bi/Cu films. Sample temperatures are 265 °C (**a**), 280 °C (**b**), 220 °C (**c**) and 200 °C (**d**) [24]

Avalanche-like crystallization, which generally seems to be an atypical phenomenon, may be due to the particularities of the morphological structure of the films, which, in turn, is determined by the technology of sample preparation. The results of the study of the effect of condensation conditions on the nature of crystallization confirmed the importance of this factor.

Thus, Fig. 3.17 presents a set of graphs showing the temperature dependence of the resistance for samples in which bismuth was condensed on substrates with different temperature.

As can be seen from the dynamics of changes in the shape of the curves in Fig. 3.17, the temperature of the substrate during the condensation of bismuth in Cu/Bi/Cu films really has a significant influence on the temperature and nature of the crystallization of the eutectic based on it. Thus, for the samples in which the substrate was maintained at a temperature of 220 °C during the deposition process, diffuse crystallization is observed, which is fully completed at a temperature of about 150 °C, i.e., at supercooling of 121 K (η = 0.22).

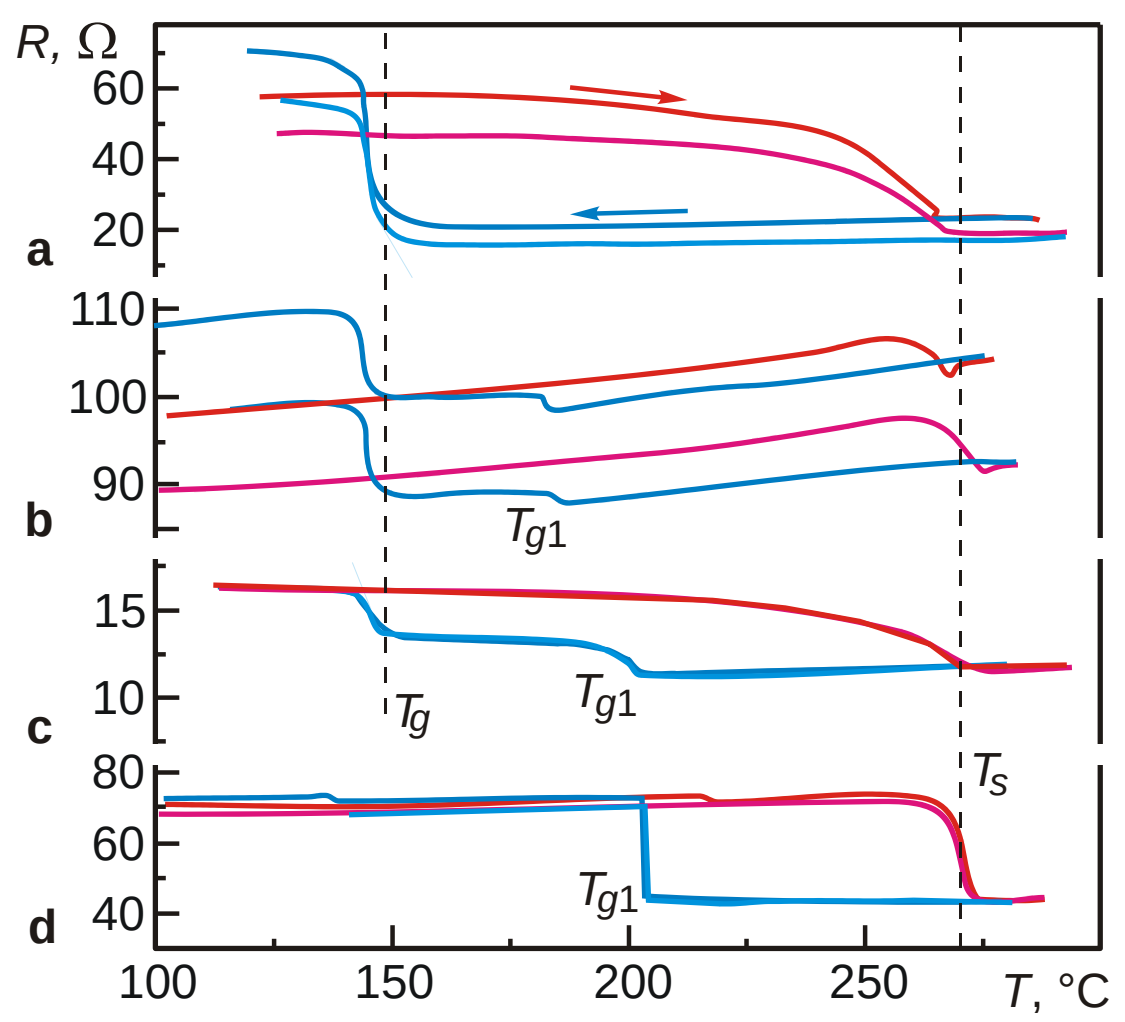

**Fig. 3.17** Electrical resistance dependence on the temperature for Cu/Bi/Cu films, in which Bi was condensed onto a substrate with a temperature of 220 °C (**a**), 170 °C (**b**), 155 °C (**c**), 140 °C (**d**) [24]

As the temperature of the substrate decreases, some evolution of the graphs can be seen. Thus, despite the fact that the temperature of the resistance jump caused by crystallization remains constant, in addition to this jump, another one appears on the curves, located near the crystallization temperature of samples obtained by the condensation of bismuth onto the substrate at room temperature. The relative contribution of this jump to the overall change in electrical resistance gradually increases. In a sense, the critical temperature of the substrate during bismuth condensation is 150 °C. If bismuth is condensed onto a substrate, the temperature of which is lower than this value, the temperature dependence of the resistance repeats exactly that, which is typical of samples, obtained at room temperature of the substrate.

It should be regarded, that the method of changing the condensation mechanism, which has been used to obtain a large amount of experimental data on the supercooling of fusible particles on the surface of a more refractory substrate, is based on the fact that the temperature of change of the condensation mechanism from the "vapor-liquid" to the "vapor-crystal" mechanism corresponds to the maximum value of supercooling possible in a particular contact system and under certain preparation conditions. The minimum value of the temperature of the existence of the supercooled liquid phase in Cu/Bi/Cu films obtained in [24] is 150 °C, and this is the temperature of the substrate during bismuth condensation that is the limiting temperature, by the transition through which, the temperature and nature of crystallization changes. Therefore, it can be assumed, that it is the change in the condensation mechanism from the "vapor-crystal" mechanism, which is realized at substrate temperatures below 150 °C, to the "vapor-liquid" mechanism, under which bismuth condenses at higher substrate temperatures, provides the change in crystallization conditions that is observed in Fig. 3.17.

As a result of electron microscopic studies (Fig. 3.18) of two-layer Bi/Cu films in which bismuth was deposited by the "vapor-liquid" and "vapor-crystal" mechanisms, it was found that even after several heating-cooling cycles, their morphology differs significantly. Thus, in the films deposited into the solid phase, even after several heatings, it is not possible to find separate bismuth particles. In these samples, bismuth uniformly covers the surface of the copper layer and thus forms a single connected system of fusible inclusions in the Cu/Bi/Cu three-layer films. In contrast to these samples, in the films obtained by condensation of bismuth into the liquid phase, scattered bismuth particles of 100–1000 nm in size are observed. In turn, as follows from thermodynamic considerations and numerous observations of the crystallization process, if a nucleus of the crystal phase appears in the melt, the size of which exceeds the critical value, it causes extremely rapid crystallization of the entire supercooled melt which is in contact with it.

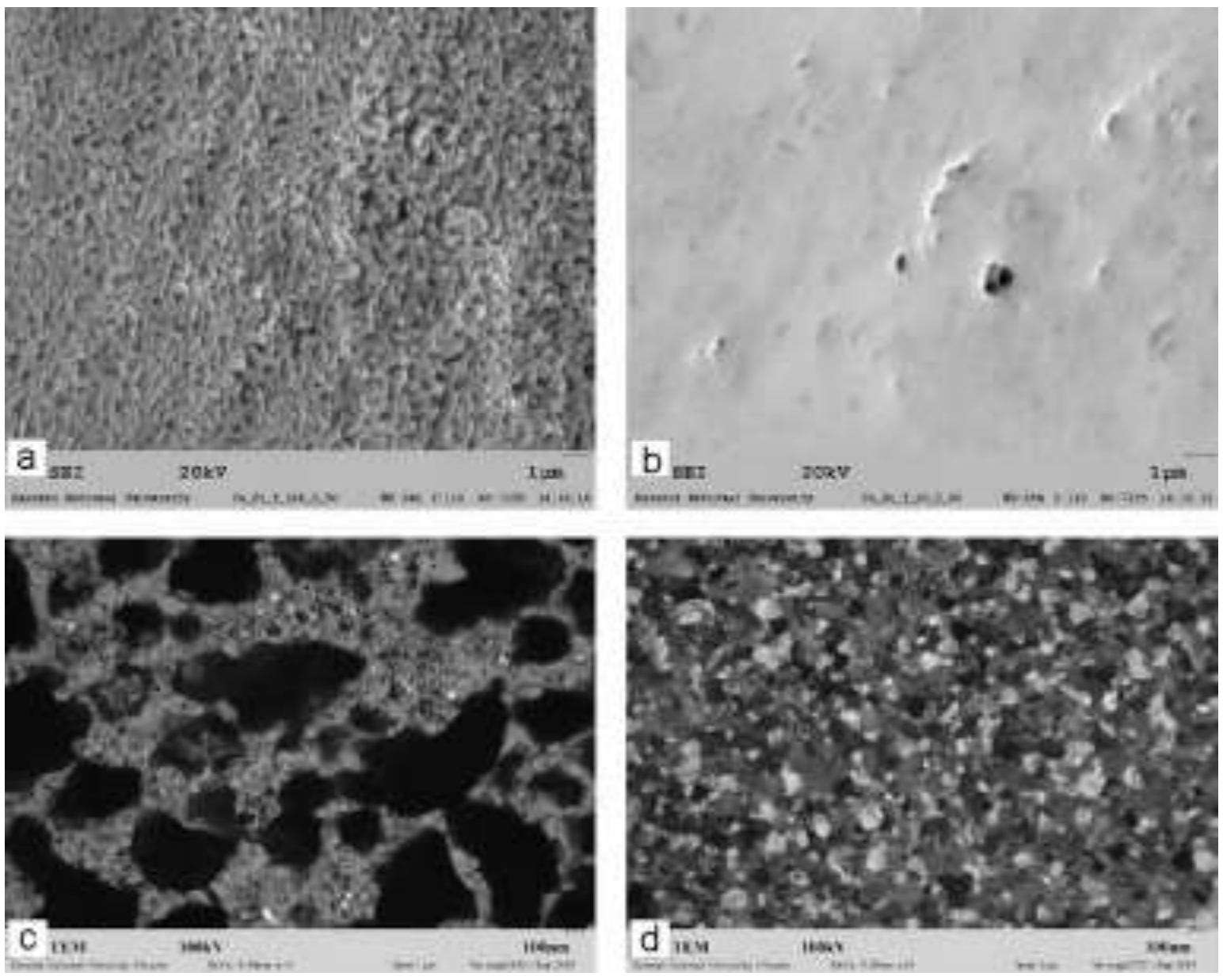

Fig. 3.18 SEM (**a**, **b**; the shooting angle of 60°) and TEM (**c**, **d**) images of Cu/Bi films in which bismuth was deposited on copper at the temperature of 180 °C (**a**, **c**) and 20 °C (**b**, **d**). The images correspond to the samples that were subjected to several cycles of heating-cooling [24]

Thus, a connected system of inclusions observed in Bi/Cu films condensed into the solid phase explains the avalanche-like nature of crystallization in these samples. The large size of such a system, which is comparable to the size of the entire sample, provides a high probability of crystallization of the supercooled melt on the external (e.g., impurity) nucleation centre long before the maximum supercooling are reached.

The structure formed in the samples obtained by the "vapor-liquid" condensation mechanism turns out to be much more favourable for the realization of deep supercooling. Thus, the accidental crystallization of one of these scattered particles will not affect its neighbour. That is, in this case, the conditions of the micro-volumes method are fulfilled in the sample, and the obtained supercooling, which is $\Delta T = 121$ K ($\eta$ = 0,22), is maximum and corresponds to the energy characteristics of the interface of the studied contact pair.

The influence of preparation conditions on the temperature and nature of bismuth crystallization has been observed in other works [25]. Fig. 3.19 shows the dependence of the resistance on temperature for three-layer Ag/Bi/Ag films obtained by sequential condensation of components on a substrate at room temperature. As can be seen, melting and crystallization are accompanied by jumps in electrical resistance and avalanche-like crystallization is observed at a temperature of 205 °C under a supercooling of 66 K ($\eta$ = 0.12).

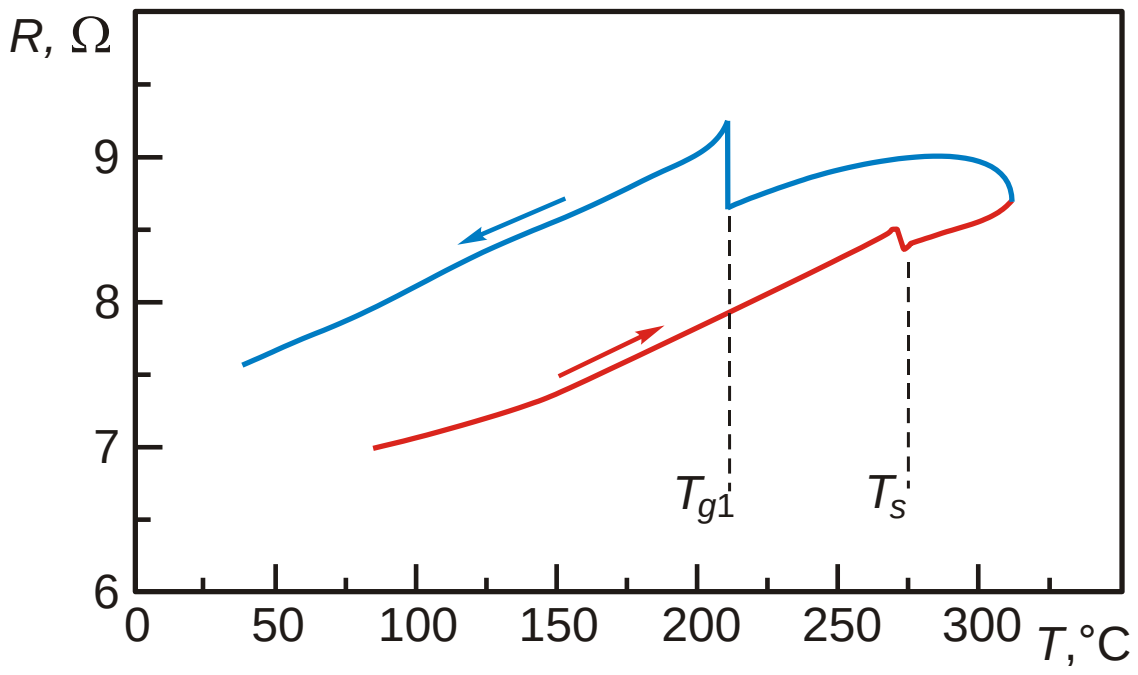

**Fig. 3.19** Electrical resistance dependence on the temperature in Ag/Bi/Ag films, all components of which were deposited into the solid phase [25]

It is worth noting that, unlike layered film structures based on copper, and even more so carbon, the thermal stability of Ag/Bi/Ag samples is very low, and their electrical conductivity tends to zero after several heating-cooling cycles. This indicates the thermal activation of the process of decay of the initially continuous films into separate islands, which naturally causes degradation of electrical conductivity. At the same time, as in the case of Al/Bi/Al films, the main contribution to the irreversible increase in resistance is made by the crystallization stage of supercooled bismuth. The rapid destruction of the samples under study probably indicates a high defectiveness of the silver layers, which, accordingly, have numerous accelerated diffusion pathways. As a result, there is an active diffusion of bismuth, which is embedded in the defects of the polycrystalline film, which probably causes the low thermal stability of the layered film system. Ag/Bi/Ag films in which bismuth was condensed through the liquid phase turn out to be even less stable (Fig. 3.20).

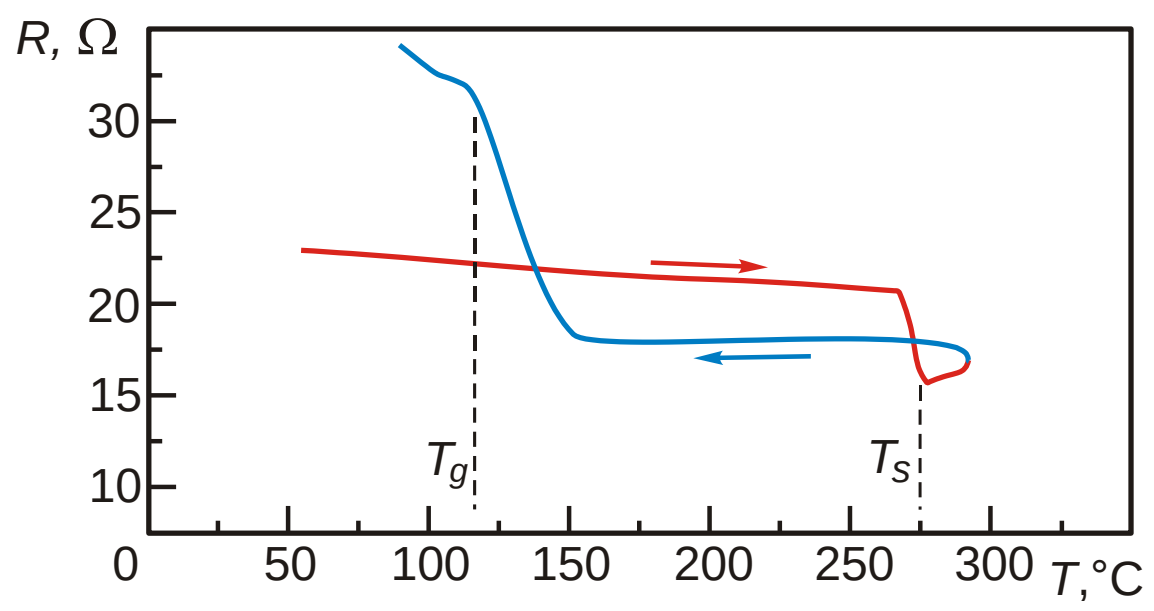

**Fig. 3.20** Electrical resistance dependence on the temperature in Ag/Bi/Ag films, in which bismuth was deposited onto the substrate with a temperature of 220 °C [25]

As in Cu/Bi/Cu films, the crystallization of the bismuth layer condensed by the "vapour-liquid" mechanism occurs diffusely and is completed at a much higher supercooling, than the one, which is happening in films, in which all components were deposited into the solid phase. The supercooling in these films is very large and reaches 150 K, η = 0.28.

The results of *in situ* electron diffraction studies of Ag/Bi/Ag samples, in which bismuth was deposited on a substrate at a temperature of 220 °C, are shown in Fig. 3.21. Similar to the Cu/Bi/Cu system discussed above, electron diffraction studies confirm the results of electrical resistance measurements. The bismuth lines are present in the sample when it is heated up to the melting point, and when cooled, they reappear at lower temperatures and achieve their original brightness at the temperature corresponding to the end of the sharp increase in resistance on its temperature dependence (Fig. 3.20).

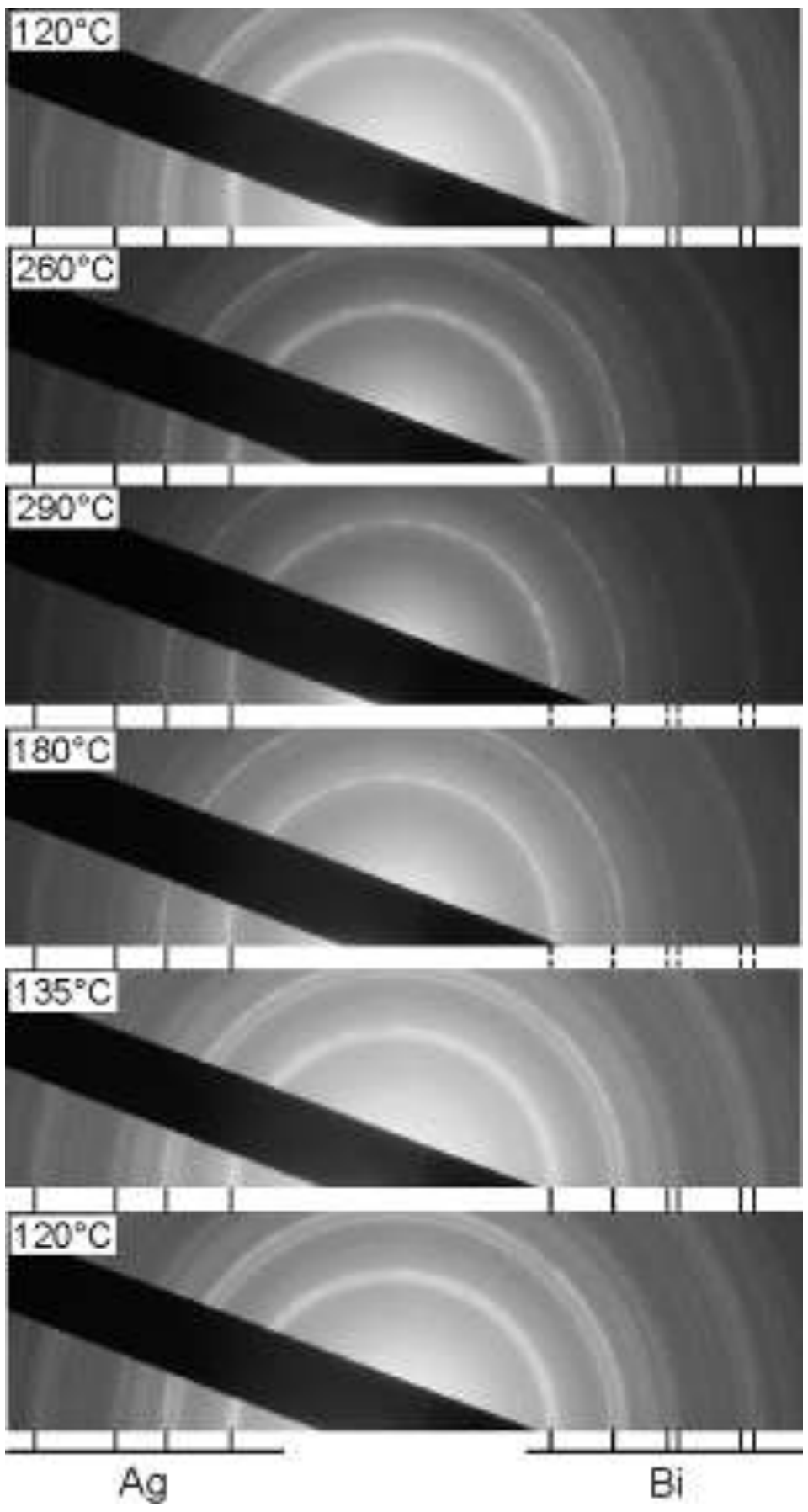

**Fig. 3.21** Electron diffraction patterns of Ag/Bi films, which correspond to the heating-cooling cycle (from up to down). The temperatures of the samples are shown in the pictures. Bismuth was deposited by the vapor-liquid mechanism [25]

The considered contact pairs, except for Bi-C, correspond to systems in which phase diagrams of eutectic type are realized. The C/Bi/C films are characterized by a complete absence of component interaction, and it is in this system that avalanche-like crystallization of the fusible component is not observed. In addition, the carbon layers in the C/Bi/C samples are amorphous with a smooth surface, and they lack morphological and structural defects characteristic of condensed metal layers, in particular, pores, a developed surface, and grain boundaries. In this regard, to find out the causes that cause avalanche-like crystallization, it is necessary to study contact pairs that are also characterized by the absence of interaction, but in which the layer modelling the matrix is a metal. One such system is Mo/Bi/Mo films, the curves of temperature dependence of electrical resistance for which are shown in Fig. 3.22.

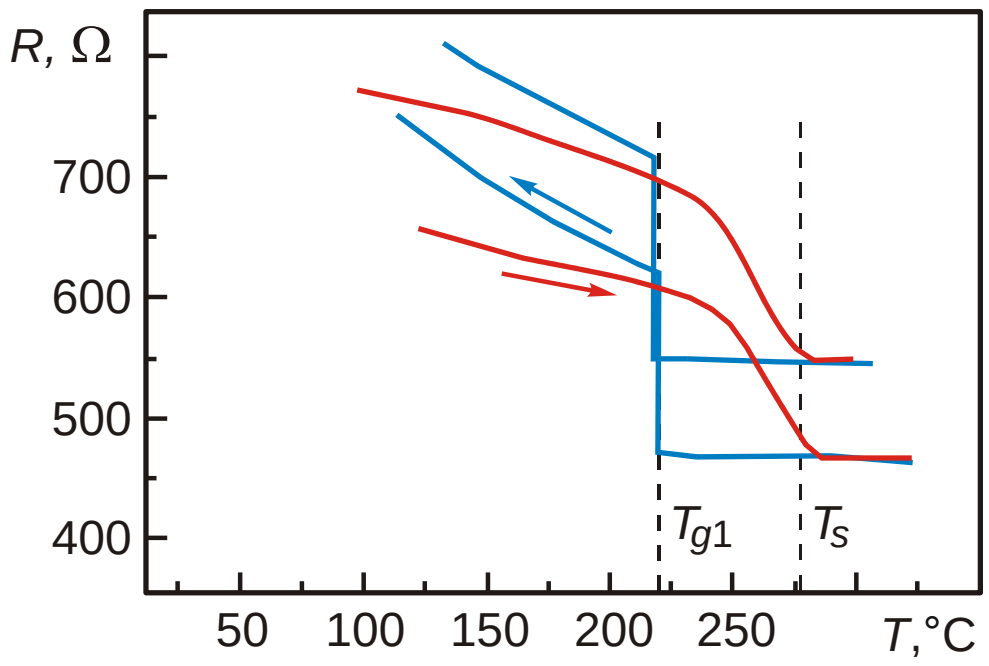

**Fig. 3.22** Electrical resistance dependence on the temperature in Mo/Bi/Mo films. Bismuth was condensed by the vapor-crystal mechanism [27]

As in other systems in which bismuth is placed between polycrystalline layers, in samples condensed by the "vapor-crystal" mechanism, bismuth crystallization is avalanche-like and occurs at a temperature of $T_{g1} \approx 200$ °C.

By analogy with the systems described above, it should be assumed that the supercooling during bismuth crystallization in such samples is less than the value, which is limiting for this contact pair.

Indeed, if bismuth was deposited on a molybdenum layer the temperature of which was 230 °C (i.e., by the "vapor-liquid" mechanism), the temperature dependence of the electrical resistance of the three-layer films took the form shown in Fig. 3.23.

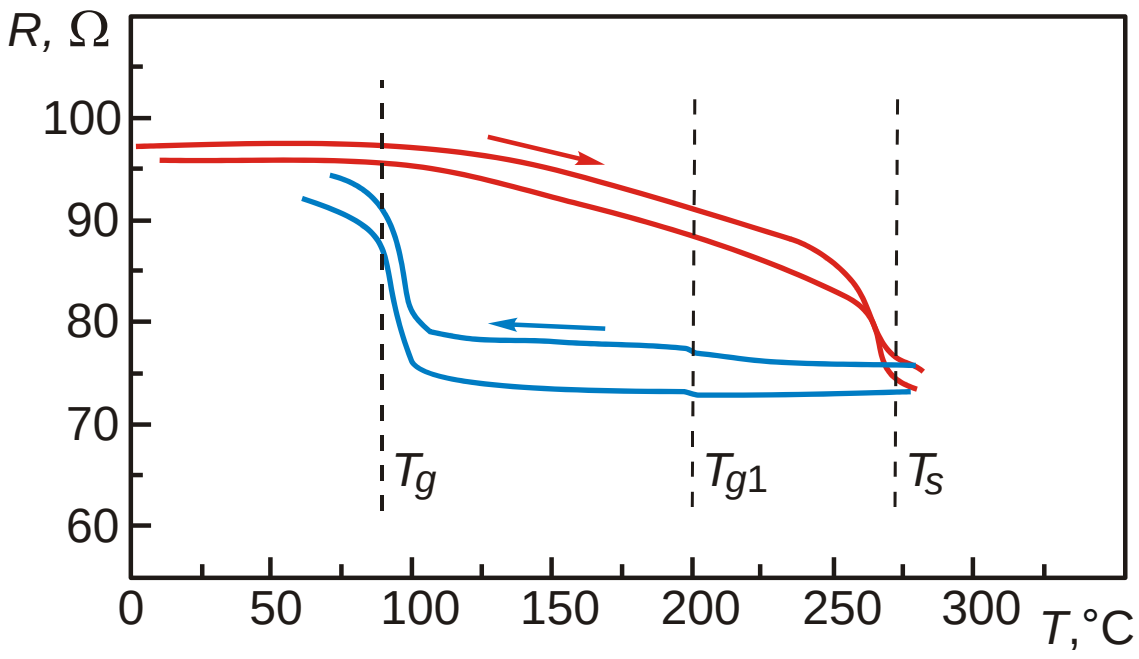

Fig. 3.23 Temperature dependence of the electrical resistance in Mo/Bi/Mo films. Bismuth was condensed by the vapor-liquid mechanism [27]

As in aluminum-based films [18], a slight increase in electrical resistance is at the beginning observed with increasing temperature, which is probably primarily due to the temperature dependence of the resistance of molybdenum and bismuth. With a further increase in temperature, somewhat a slight decrease in resistance is observed, which can be connected with the annealing of defects created by mechanical stresses that occurred on the previous cooling cycle during bismuth crystallization. The fastest change in resistance during heating occurs near the melting point of bismuth, and this change is used to identify the phase transition temperature.

During cooling, an increase in the resistance of the films is observed (relatively weak in the region of the existence of the supercooled melt and more intense immediately after crystallization), which is overlaid by jumps in electrical resistance connected with the crystallization of supercooled bismuth. The crystallization of the main part of bismuth in these films occurs diffusely in the temperature range of about 10 K and is completed at a temperature of about $T_g \approx 90$ °C. Thus, the supercooling in Mo/Bi/Mo films in which bismuth was condensed through the liquid phase is 180 K, and the value of the relative supercooling reaches η = 0,33.

It should be noted that, in addition to the jump in electrical resistance corresponding to the maximum supercooling, the graphs show another low particularity whose position is close to the value of the crystallization temperature of samples condensed by the "vapor-crystal" mechanism. That is, a small fraction of the supercooled substance in films, which were condensed into a liquid phase and contain scattered inclusions of a fusible component, crystallizes at a slight supercooling. However, unlike films in which a connected system of inclusions is observed, this process does not cause the entire sample to crystallize. Moreover, the sample crystallization process turns out to be stretched over a certain temperature range. It is reasonable to assume that early crystallization in both types of samples is due to the same causes.

According to the results of electron microscopic studies, the different nature of crystallization of films in which bismuth was deposited by different mechanisms can be explained by differences in the morphology of the films, just as in the Cu/Bi/Cu system. At the same time, the mechanisms by which bismuth condensed through the solid phase creates a connected system of inclusions that is preserved for several heating-cooling cycles, are not fully understood.

As for the reasons that cause early crystallization of films deposited on a cold substrate, the following considerations can be made. The most obvious factor that causes small supercooling values in such samples is impurities that may have been embedded into the melt either during condensation or during heating-cooling cycles. In addition, the polycrystalline structure of the matrix containing many defects, such as grain boundaries, triple junctions, and pores, can contribute to the reduction of supercooling [26]. In particular, in such systems, a decrease in supercooling can be observed [26] if the size of the critical nucleus is larger than the matrix cavities. If the size of the critical nucleus under these conditions is significantly smaller than the matrix pores, then its role is limited to reducing the energy of new phase formation, which is typical for particles on a given substrate.

The size of the critical bismuth nucleus, which provides the supercooling values, which are observed in experiments [25, 27], can be estimated within the framework of the classical nucleation theory from the relation (1.4). Such an estimate, using the value of $\sigma_{sl}T_s/\lambda$ presented in [26], gives a value of the radius of the critical nucleus of about 1 nm. The obtained value is significantly less than the thickness of the bismuth layer in the studied films. Probably, the thickness of the bismuth layer in the studied three-layer films, in which bismuth was located between polycrystalline layers, is already large enough, so that the presence of films of another substance on both sides of it does not have a direct effect on the crystallization particularities. However, vacuum condensates are usually very non-equilibrium structures and contain a significant number of nucleation pores and grain boundaries. It can be assumed that the melt particles that have entered the defects with a size smaller than the value of the critical nucleus determined in accordance with (1.4) crystallize first. This will provoke a rapid crystallization of a connected system of inclusions of the fusible component, which is formed in the samples obtained by the "vapor-crystal" condensation mechanism.

When analyzing these results, it is noteworthy that the crystallization temperature of samples condensed into a solid phase is weakly dependent on the material of the layers modeling the matrix and on the vacuum conditions during the preparation of layered film systems. Such samples crystallize at a temperature of about 200 °C. Special studies have shown that impurities that can enter the sample from the residual atmosphere should cause crystallization at a different supercooling value.

Fig. 3.24 shows graphs of the temperature dependence of the electrical resistance obtained on samples condensed by the "vapor-liquid" mechanism and kept in a vacuum chamber at a residual gases pressure of about 10 Pa for 10 to 15 days. It can be seen that during a sufficiently long holding time, crystallization takes on a distinctly expressed stepwise nature. Thus, the jump in resistance corresponding to the limiting supercooling gradually smoothes out, but at the same time, another particularity $T_{g2} \approx 130$ °C appears on the graphs, which probably corresponds to crystallization caused by impurities from the residual atmosphere (e.g., due to the formation of oxides) accumulated in the particles during the holding process. At the same time, the value of the particularity observed near 200 °C does not change during the holding process, i.e., it is unlikely that this particularity is due to crystallization on those impurity centers connected with the influence of the residual atmosphere. It can be assumed that the appearance of the given jump is due to the bismuth particles that, due to their location in the defects of the polycrystalline matrix, initiate crystallization at a lower supercooling.

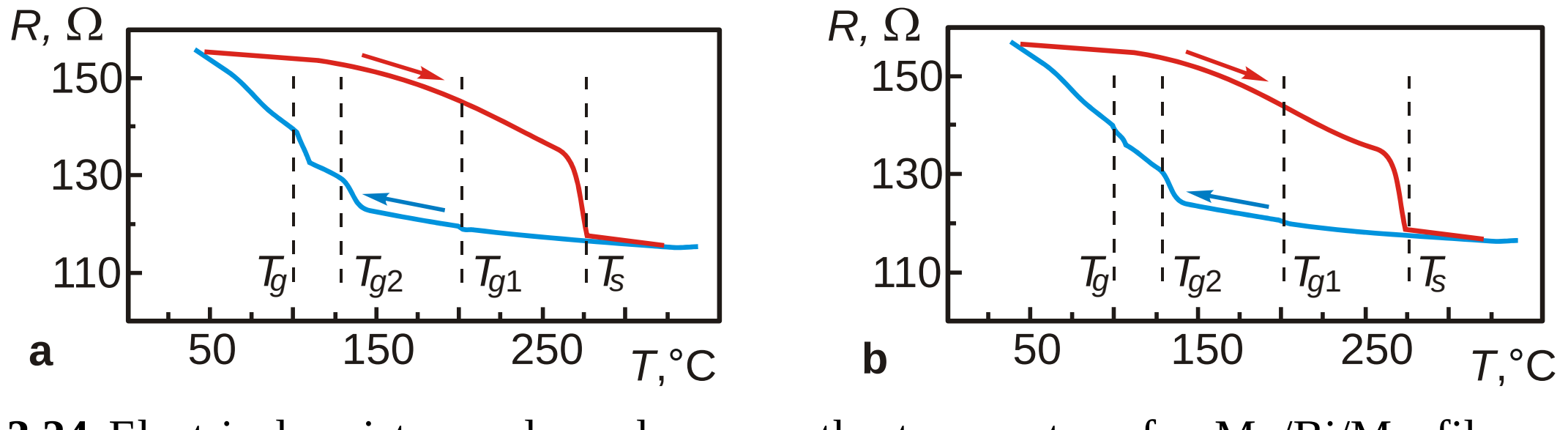

**Fig. 3.24** Electrical resistance dependence on the temperature for Mo/Bi/Mo films after holding in conditions of the residual atmosphere pressure of 10 Pa during 10 (**a**) and 15 (**b**) days [27]

As follows from the results presented, avalanche-like crystallization of the melt is typical for films in which bismuth is condensed into a solid phase. Such crystallization occurs at relatively low supercooling and is explained by the formation in the sample of the connected system of fusible inclusions (Fig. 3.18). Taking into account the nature of crystallization, bismuth condensed by the vapor-crystal mechanism creates a connected system of inclusions of the fusible component in Bi/Al, Bi/Cu, Bi/Ag, and Bi/Mo films. A special place among the studied contact pairs is occupied by Bi/Ge films. The nature of bismuth crystallization in these samples changes during thermal cycling. During the first heating-cooling cycles, bismuth in Ge/Bi/Ge films crystallizes diffusely (Fig. 3.9). However, after several successive cycles, the situation changes and the crystallization becomes avalanche-like (Fig. 3.10). The change in the type of crystallization of supercooled bismuth is accompanied by metal-induced crystallization of germanium (Fig. 3.11). Thus, the results of the study of supercooling in the Bi-Ge contact pair indicate an important role of the crystal structure of the more refractory layer in the process of the formation of a continuous layer of bismuth.

At the same time, vacuum condensates are characterized by a nanocrystalline structure. Typically, the size of crystallites forming vacuum condensates decreases with an increase in the ratio of their melting point to the temperature of the substrate. That is, when the films are condensed onto the substrate at room temperature, samples with a higher melting point will be more dispersed. In some cases, the size of the crystallites can be so small, that it will give the films signs of an amorphous state.

Copper and silver have similar melting points. Accordingly, the crystal structure of these layers should also be similar. Aluminum, which has a lower melting point, should form a structure with a slightly larger crystallite size. This probably explains why film systems based on these layers allow reliable observation of avalanche-like crystallization. Fig. 3.25 shows a TEM image of copper and silver films condensed on a substrate at room temperature. After condensation, the samples were subjected to a short annealing. This was to emulate the heating that such layers experience during thermal cycling (Figs. 3.15, 3.19).

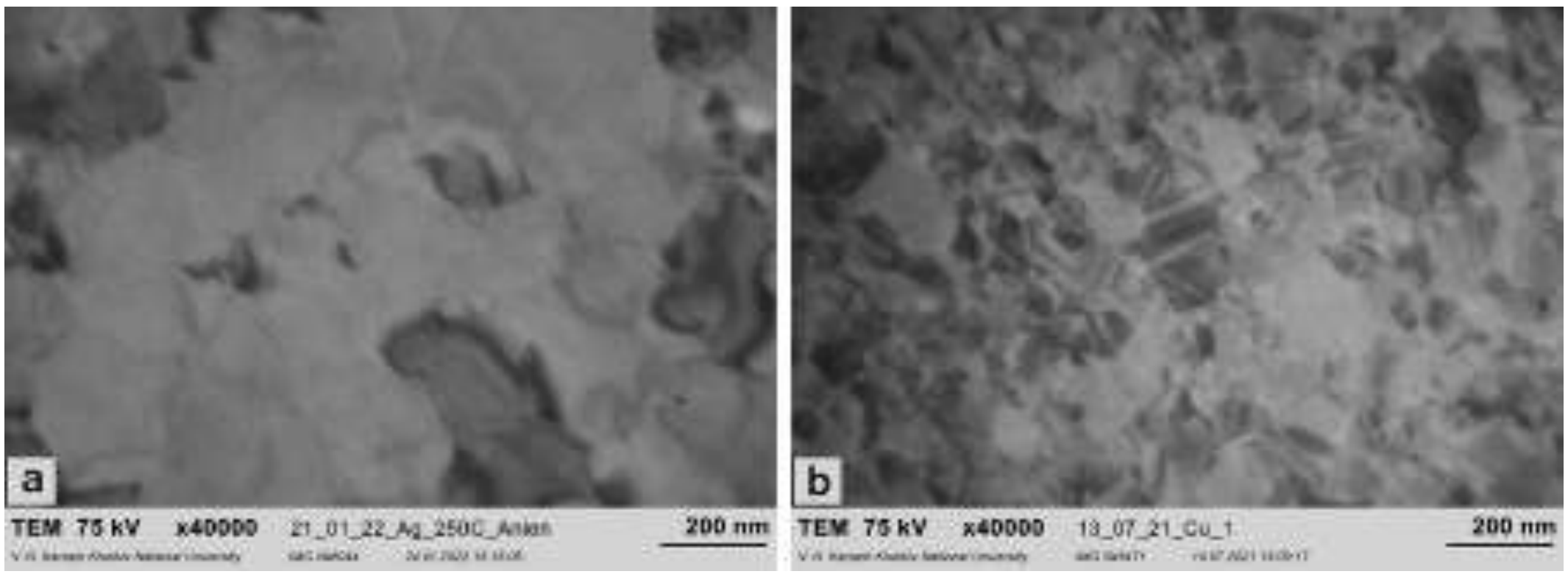

**Fig. 3.25** TEM images of Ag(a) and Cu(b) films obtained after short-term annealing of the samples at 250 °C

According to the results of TEM studies, the characteristic grains size in shortly annealed copper and silver films is 50-70 nm. This allows us to consider them as "true" crystalline systems that do not contain features of an amorphous state. However, more refractory vanadium films, even after annealing, consist of crystallites with the most likely size of about 8 nm (Fig. 3.26).

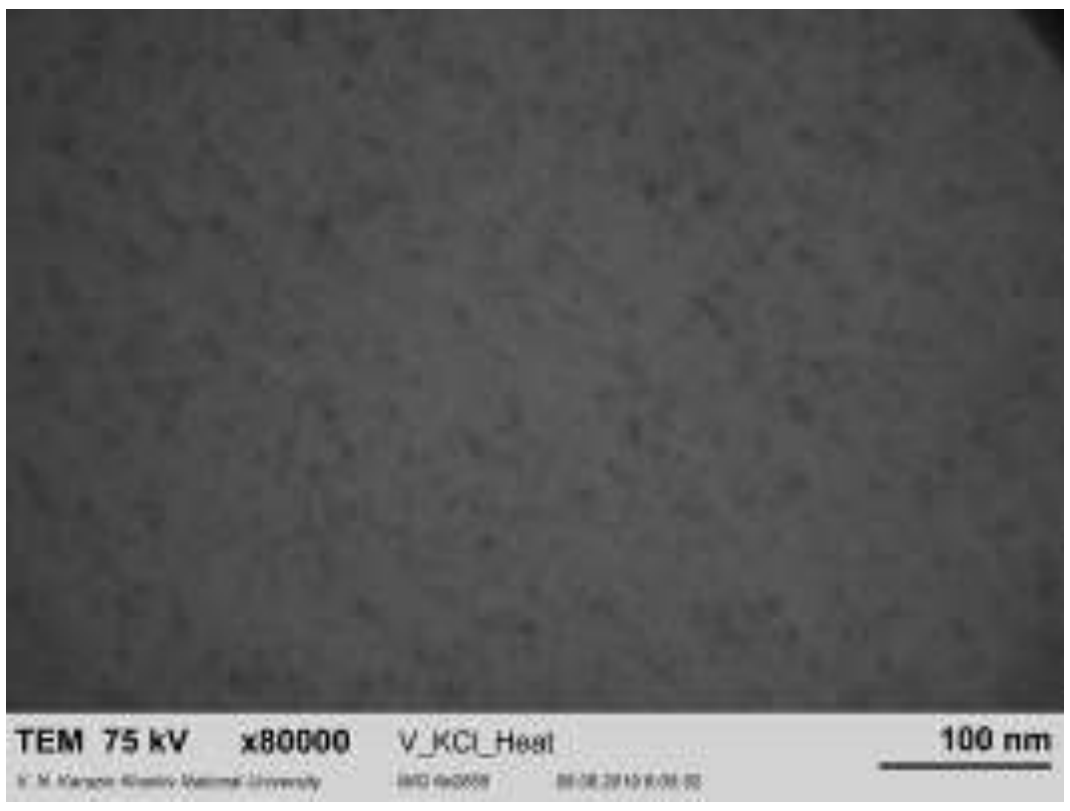

**Fig. 3.26** TEM image of a shortly annealed vanadium film

The small grains size in vanadium films (Fig. 3.26) gives them some features of the amorphous state. In particular, broad diffraction lines are observed in the electron diffraction patterns obtained from such films (Fig. 3.27). The width of such lines approaches the width of the diffuse halo, which is typical for amorphous samples. Thus, taking into account the crystallization particularities of Ge/Bi/Ge films, it can be expected that the transition from copper and silver to more dispersed vanadium may affect the particularities of crystallization of supercooled bismuth.

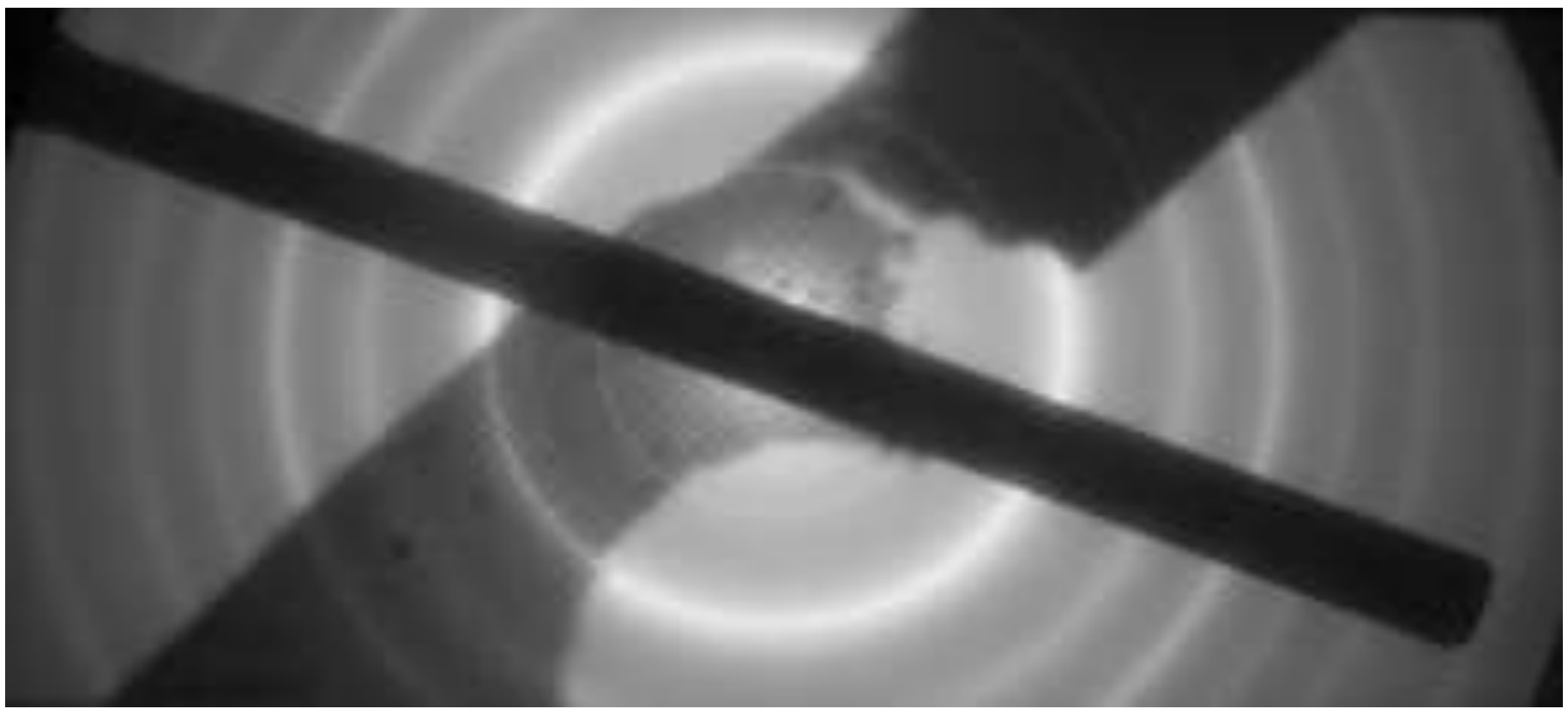

**Fig. 3.27** Electron diffraction patterns of Bi/V films. Thin lines correspond to the standard placed below the sample [28]

Fig. 3.28 shows the temperature dependence of the electrical resistance of Bi/V films obtained for samples in which bismuth was condensed into solid *(a)* and liquid *(b)* phases. It can be seen that in both cases the crystallization is diffuse. Moreover, it turns out that the crystallization range in the samples deposited by the vapor-crystal mechanism is larger than in the films condensed into the liquid phase. Thus, the observed kinetics of crystallization indicates that in this case there is no connected system of fusible inclusions, which is formed in other contact pairs with bismuth.

Fig. 3.28 shows that the temperature of maximum supercooling in Bi/V films does not depend on the condensation mechanism. However, crystallization in the samples deposited by the vapor-crystal mechanism begins much earlier than in the films in which bismuth was deposited into the liquid phase.

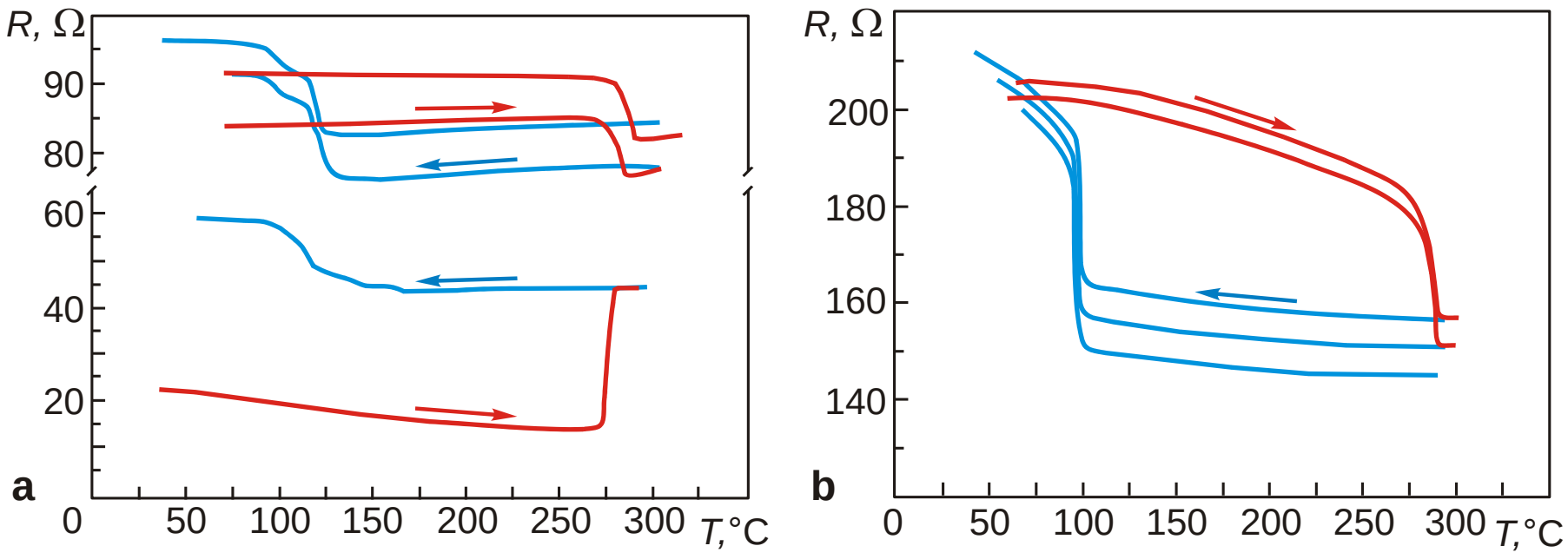

**Fig 3.28** Temperature dependence of the electrical resistance in Bi/V films in which bismuth was deposited by the vapor-crystal (**a**) and vapor-liquid (**b**) mechanisms [28]

The microstructure of Bi/V films is shown in Fig. 3.29. It can be seen that, regardless of the condensation mechanism, bismuth forms an island film after melting. This is the microstructure which is expected based on the results of resistive studies (Fig. 3.28). The transformation of initially continuous layers into island structures explains the irreversible increase in electrical resistance observed in Fig. 3.28a.

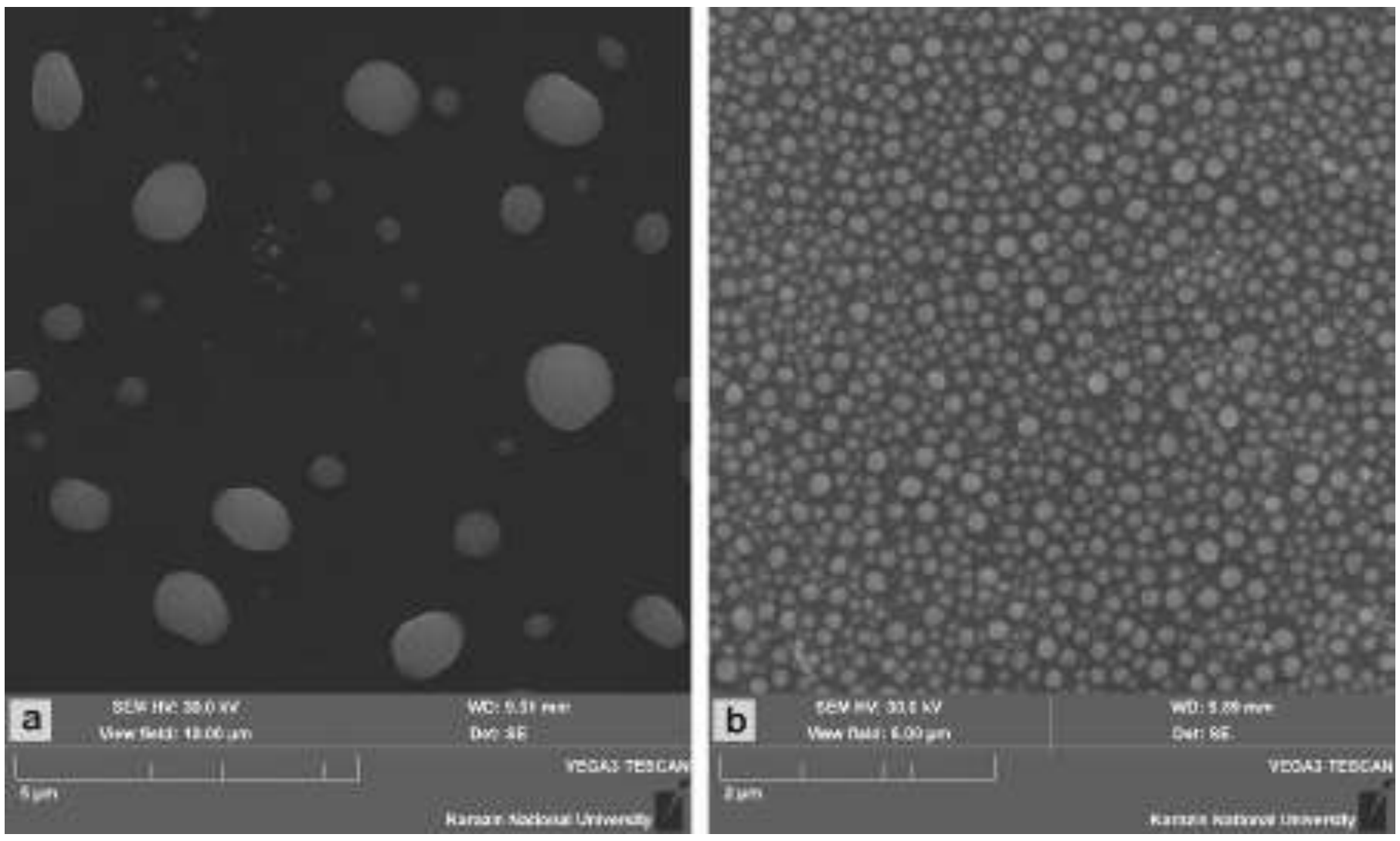

**Fig. 3.29** SEM images of Bi/V films obtained by the vapor-crystal (**a**) and vapor-liquid (**b**) mechanisms. [28]

To interpret the results of the study of supercooling in the Bi-V contact pair, it is worth referring to the histograms of the size distribution (Fig. 3.30) of the particles observed in the SEM images. It can be seen that the most probable size of particles in the films deposited into the liquid phase is about 5-10 times smaller than the size of particles formed during the melting of films deposited by the vapor-crystal mechanism.

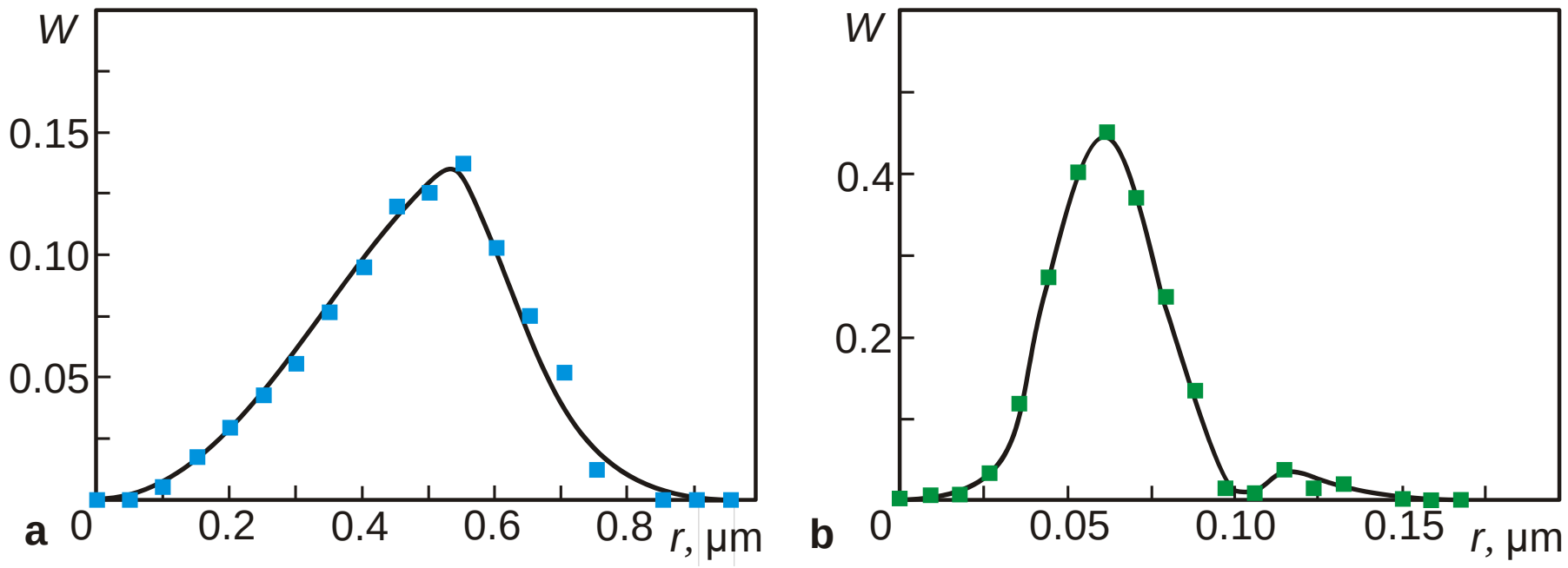

**Fig. 3.30** Histograms of the particles size distribution observed in Bi/V films condensed by the vapor-crystal (**a**) and vapor-liquid (**b**) mechanisms [28]

It is the difference in the sizes of the particles in which the fusible component of Bi/V films is concentrated that causes differences between the crystallization of samples deposited by different mechanisms (Fig. 3.28). Thus, in the films deposited by the vapor-liquid mechanism, the size of particles is already small enough to reliably satisfy the conditions of the micro-volumes method. Due to the influence of random factors, individual particles of such samples can crystallize at low supercooling. However, the crystallization of single particles in an array will have little effect on its overall state. Most of the bismuth particles in such samples will remain in a metastable state up to the temperature determined by thermodynamic considerations.

In contrast to the samples obtained by the vapor-liquid mechanism, the size of bismuth particles, which were formed during the melting of crystalline layers, is much larger. It can be said that such particles occupy an intermediate position between samples in which bismuth forms a unified fusible system (Fig. 3.18b, d) and island structures (Fig. 3.18a, c). Such particles do not fully satisfy the conditions of the method of micro-volumes. Consequently, the amount of particles whose crystallization is caused by random factors rather than interaction at the bismuth-vanadium interface increases significantly. As a result, a significant mass of the metastable melt begins to crystallize at low supercooling. However, due to the isolated nature of the supercooled particles, such crystallization is localized in separate islands. Thus, despite the relatively large size of the particles, many of them remain in a supercooled state up to the maximum possible supercooling.

Comparing the graphs shown in Figs. 3.10, 3.15, 3.19, and Fig. 3.28, it is clear that crystallization in Bi/V films condensed into the solid phase begins much later than in other contact pairs. This can be viewed as an indication of the significant role of the grain boundaries of the more refractory layer in promoting the crystallization of the supercooled melt. It can be assumed that precisely places of concentration of a large number of grain boundaries are the defects, that locally reduce the energy barrier of phase formation. In turn, due to the small size of the crystallites, the excess energy localized in the boundary is smaller than in structures with a larger grain size and with a more clearly formed boundary. Thus, the crystallization effect of vanadium films is also less than that of copper, silver, and aluminum layers.

Studies of supercooling during the crystallization of lead layers located between more refractory films were carried out in works [22, 27]. Fig. 3.31 shows the temperature dependencies of the resistance for Pb-C contact couple films which contain five layers of lead alternating with carbon layers [22]. This contact pair is characterized by almost complete non-wetting of the outer layers by the melt of a fusible metal ($\theta = 142^{\circ}$ [10, 25, 26]) and a complete absence of interaction between the components in the temperature range under study. It can be seen that the obtained curves contain jumps in electrical resistance, while, as in all previously studied contact pairs, lead melting is accompanied by a decrease in resistance, and its crystallization – by an increase in resistance. The corresponding differential curves constructed from the experimental data are shown in Fig. 3.32.

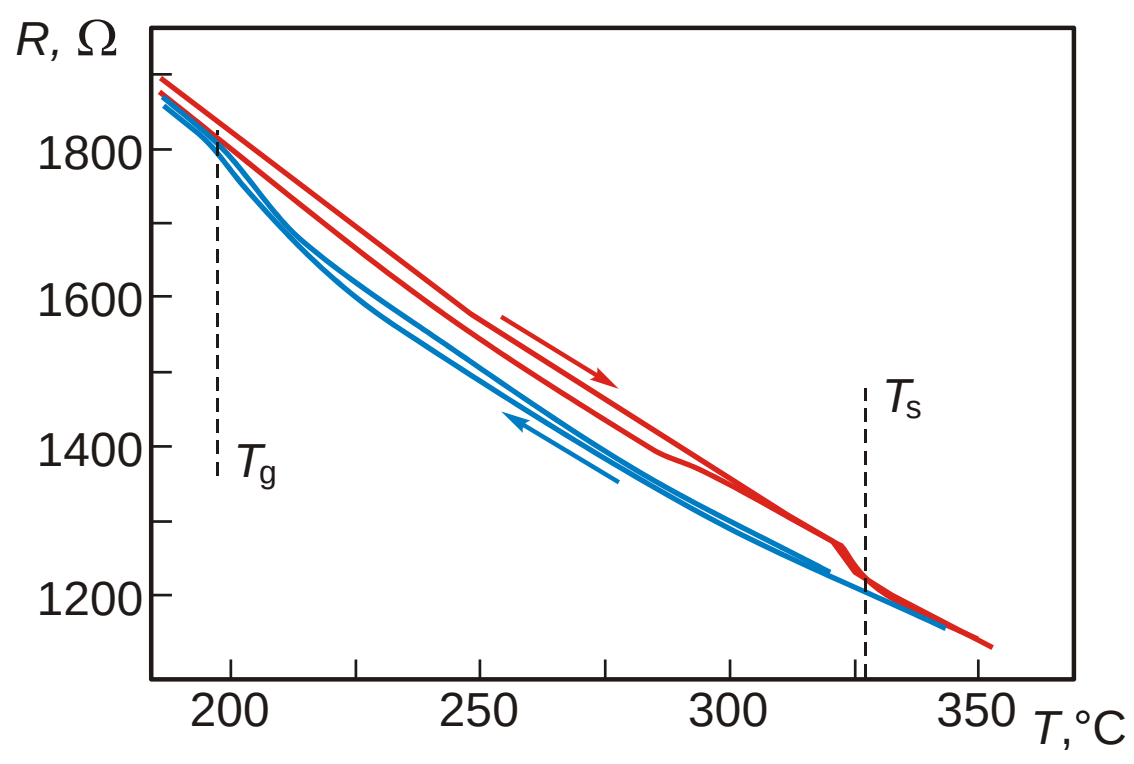

**Fig. 3.31** Electrical resistance dependence on the temperature for C/Pb/C layered systems. The samples contain five layers of the fusible component [22]

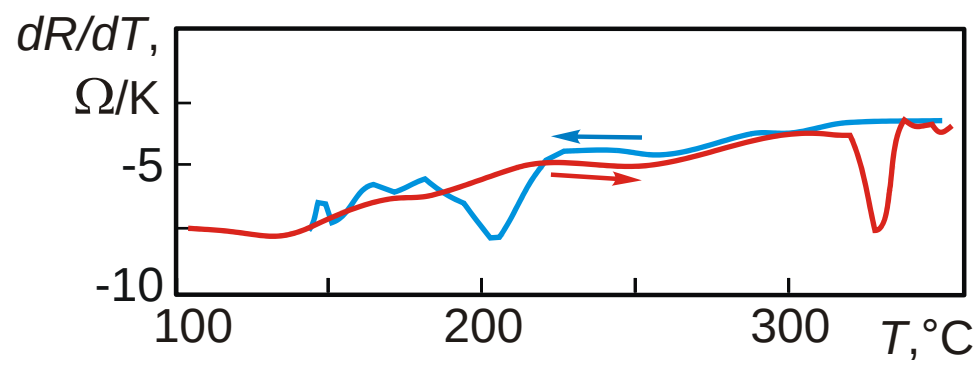

**Fig. 3.32** Numerical differentiation results of the temperature dependence of C/Pb/C films resistance [22]

The crystallization of the fusible component in the layered films of the Pb-C contact pair is completed at 192 °C, i.e. the supercooling observed in this system is 135 K (η = 0.23).

A different nature of the temperature dependence of the electrical resistance occurs in three-layer Cu/Pb/Cu films. The results of the studies performed in the work [24] are shown in Fig. 3.33. The resistance of copper films containing lead increases rapidly and irreversibly when they are heated to the melting point of lead. This behavior indicates the decay of initially continuous films into separate islands.

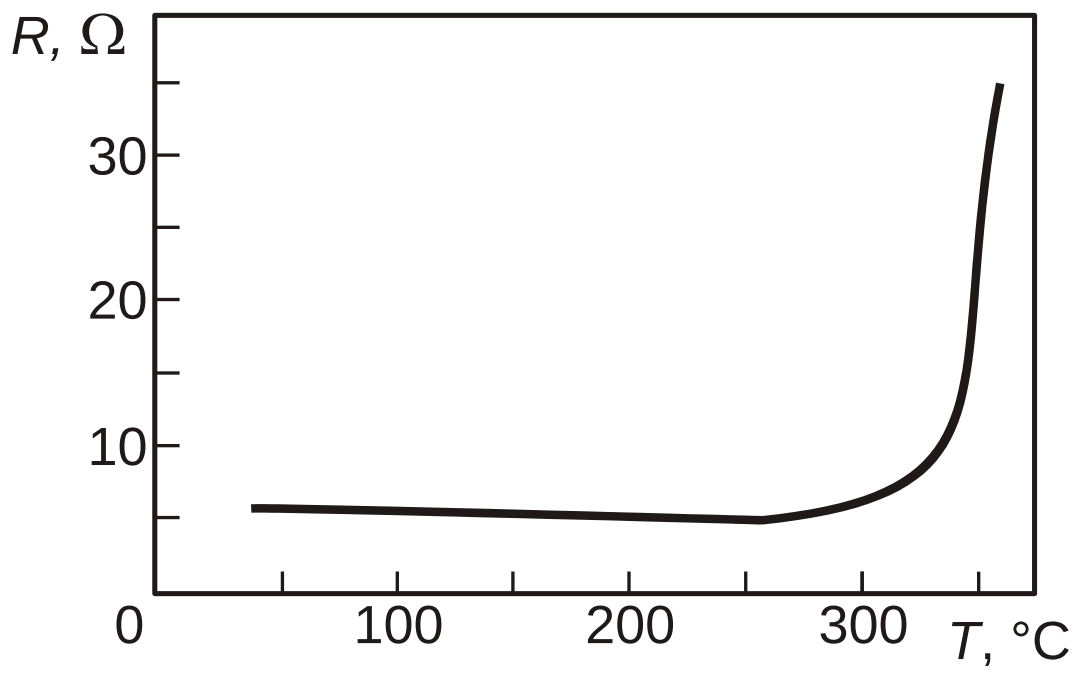

**Fig. 3.33** Pb/Cu two-layer films electrical resistance dependence on the temperature

To slow down this process, a 1-2 nm thick molybdenum surfactant layer was used in the work [24], which was deposited onto a glass substrate before condensation of the studied layered film system. According to the results of [31, 32, 33], the presence of a thin wettable layer should not only reduce the thickness at which the film becomes continuous, but also complicate its thermal decay.

Fig. 3.34 shows the temperature dependence of the electrical resistance of three-layer Cu/Pb/Cu films condensed on a molybdenum sublayer. Indeed, as can be seen from the figure, the dewetting of the films deposited on the molybdenum sublayer slowed down significantly, which made it possible to determine the crystallization temperature of the fusible component. The graphs show repeated jumps in electrical resistance, which, according to in situ electron diffraction studies, can be attributed to the melting and crystallization of the lead-based eutectic. Despite the fact that the resistance of the films increases from cycle to cycle, their thermal stability turns out to be sufficient to allow carrying out several heating-cooling cycles and recording phase transition temperatures. Crystallization in this system ends around 240 °C, i.e. the supercooling of lead in contact with copper is approximately 90 K (η = 0.15).

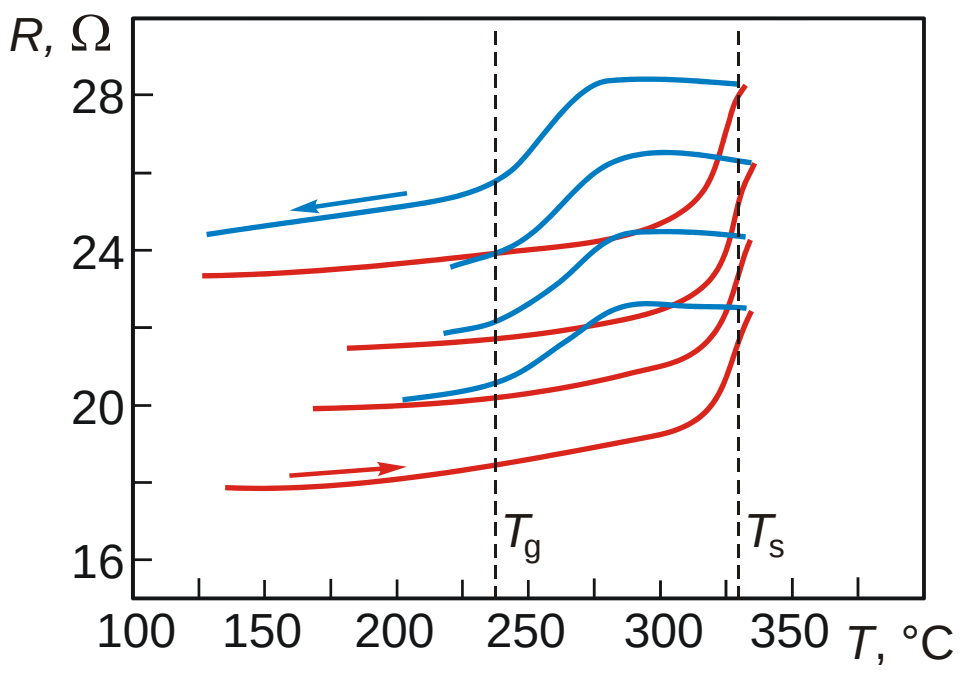

**Fig. 3.34** Electrical resistance dependence on the temperature in Cu/Pb/Cu three-layer films condensed on a molybdenum sublayer [24]

The relationship of electrical resistance jumps with phase transitions in Cu/Pb/Cu three-layer films is also confirmed by *in situ* electron diffraction studies. Fig. 3.35 shows the electron diffraction patterns obtained during heating (a, b) and cooling of the samples (c, d, e). The disappearance of reflexes from the lead crystal lattice during heating of the sample due to melting of the fusible component and their recovery in the process of cooling during its crystallization occurs at different temperatures and is fully consistent with the course of change in electrical resistance which is observed in the experiments (Fig. 3.34).

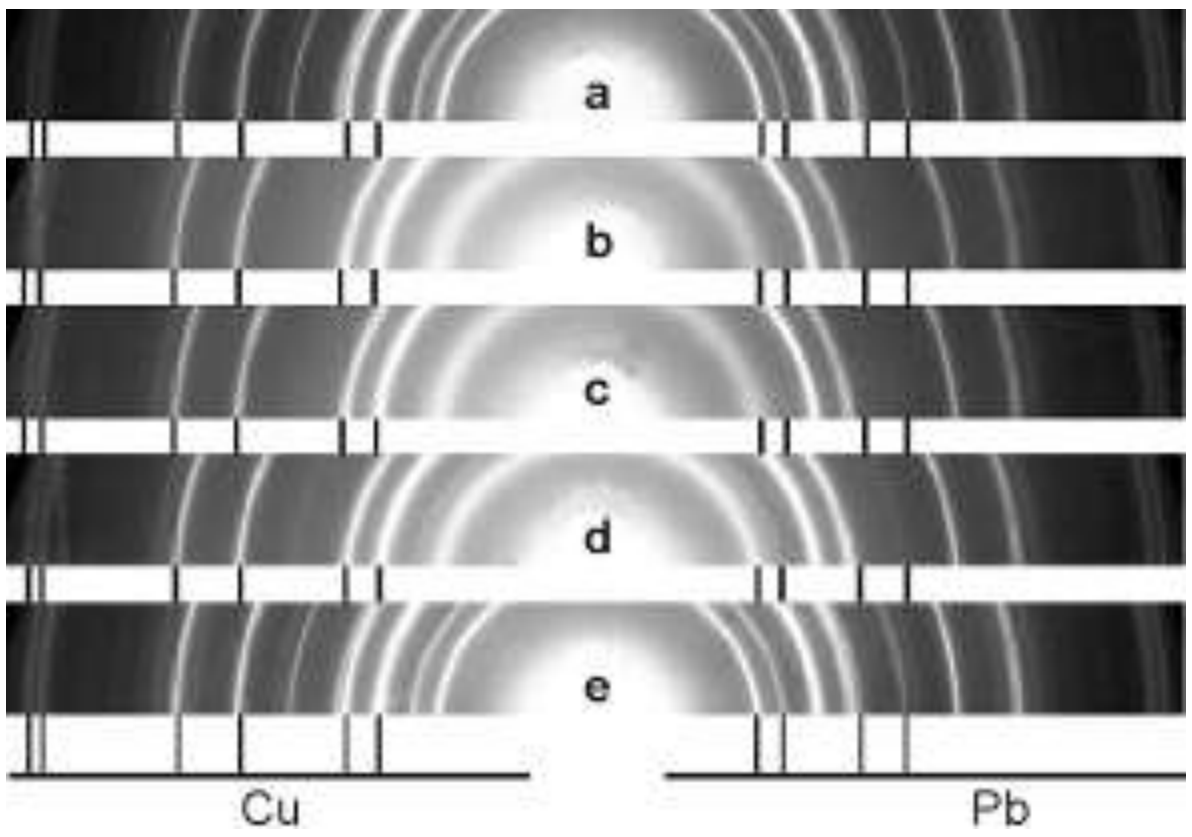

**Fig. 3.35** Electron diffraction patterns of Pb/Cu films. Images were taken at different temperatures: **a** – 30 °C, **b** – 335 °C, **c** – 300 °C, **d** – 260 °C, **e** – 200 °C (**a**–**b** images correspond to heating, **c**–**e** – to cooling)

It should be noted that, unlike films with a bismuth located between the polycrystalline layers, lead crystallization in the studied contact pair always occurs diffusely, regardless of the temperature at which condensation was carried out. Another important difference from the previously discussed films is that the electrical resistance of Cu/Pb/Cu samples increases during melting and decreases during the crystallization of the fusible eutectic. In addition, it was found that liquid lead destroys continuous copper films, even in the presence of a surfactant sublayer. This phenomenon can be of independent interest both for the creation of ordered nanoarrays and in developments that require predicting the stability of modern electronic devices.

An even greater influence of lead on the thermal stability of metal layers is observed in Ag/Pb/Ag samples [25]. From Fig. 3.36 it can be seen that during the first heating of silver films, an irreversible decrease in their electrical resistance due to annealing is observed (curve a). It should be noted that the same behavior is characteristic of all the metal layers discussed earlier. In the subsequent heating-cooling cycles, the temperature dependence of the electrical resistance of the silver film is almost linear (Fig. 3.36, curve b). At the same time, the addition of lead, as well as in the case of Cu/Pb/Cu films, causes rapid degradation of the layered film system (Fig. 3.36, curve c). In this case, the use of a surfactant layer has a limited effect and usually only allows observing a few heating-cooling cycles at one sample.

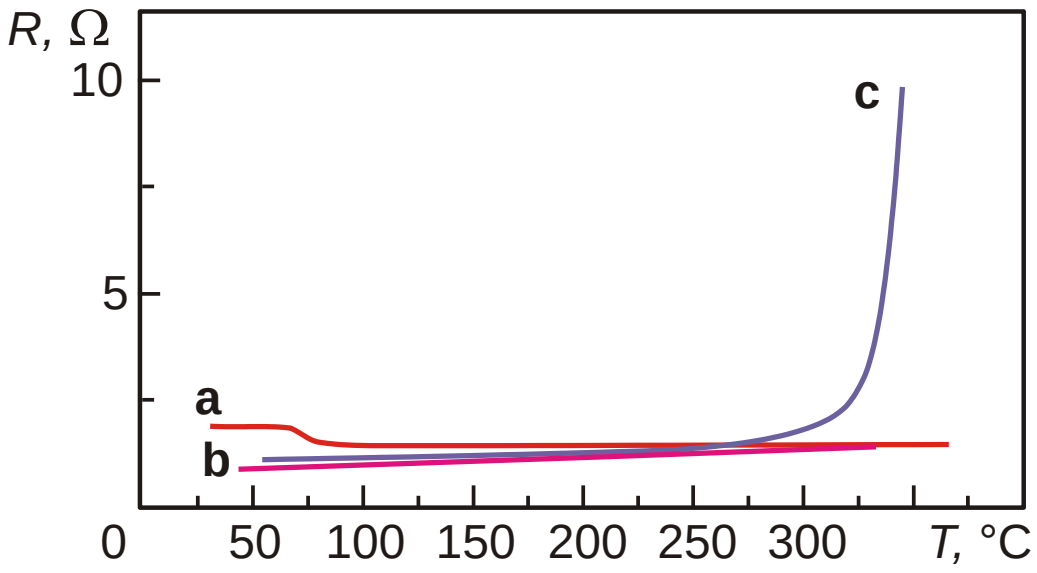

**Fig. 3.36** Electrical resistance dependence on the temperature for Ag and Pb/Ag films. The curve (**a**) corresponds to the first, and (**b**) – to the second (and to the next ones) heating of the silver film. The graph (**c**) demonstrates the behavior of the Pb/Ag film during the first heating [25]

One such cycle is shown in Fig. 3.37. However, observation of even one act of melting and crystallization was sufficient to determine the supercooling in the given contact pair. The obtained value was $\Delta T = 60$ K, i.e., the relative supercooling is equal to $\eta = 0.1$.

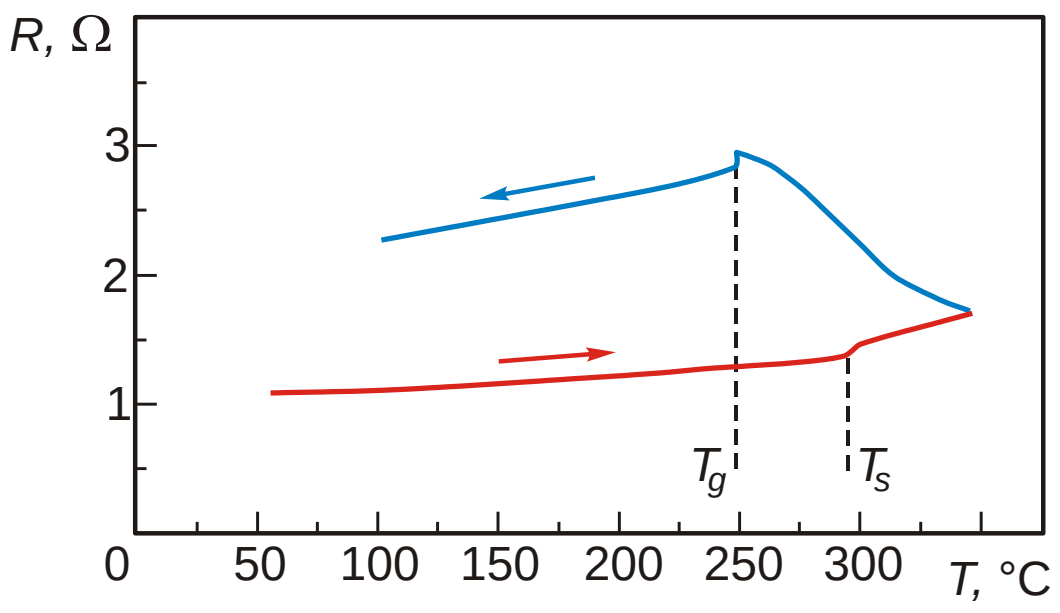

**Fig. 3.37** Electrical resistance dependence on the temperature in Ag/Pb/Ag films condensed onto the surfactant molybdenum layer [25]

As in Cu/Pb/Cu films, the melting of lead in Ag/Pb/Ag samples is accompanied by an increase in the electrical resistance, and crystallization - by a decrease in the electrical resistance. It should be noted that an irreversible increase in the resistance of Ag/Pb/Ag samples is observed in the temperature range in which lead is in a thermodynamically stable or supercooled liquid state. In this way, the samples in which lead was the fusible component differ from the previously discussed films containing bismuth, whose irreversible degradation occurred mainly during the crystallization of the supercooled melt.

The results of the study of the temperature dependence of the electrical resistance in Mo/Pb/Mo films are shown in Fig. 3.38. Despite the fact that lead also causes thermal degradation of molybdenum layers, this effect is rather slow, and the stability of the films turns out to be sufficient to endure several heating-cooling cycles. The figure shows that this contact pair is also characterized by the presence of electrical resistance jumps that identify phase transitions. Similarly to Cu/Pb/Cu and Ag/Pb/Ag films, the resistance of the films increases during melting and decreases during lead crystallization. The crystallization of lead is diffuse, and the minimum temperature of the existence of the liquid phase is 217 °C, i.e. lead in contact with molybdenum is supercooled by 110 K, and the relative supercooling is $\eta = 0.18$.

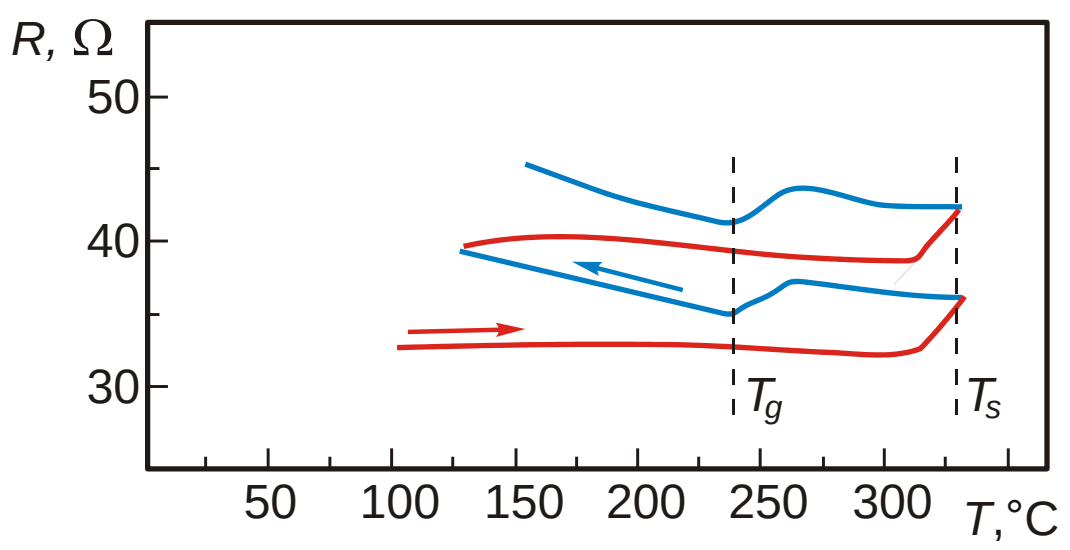

**Fig. 3.38** Electrical resistance dependence on the temperature for Mo/Pb/Mo films [27]

The particularity of the contact pairs under consideration, which are mostly described by the phase diagram of the eutectic type, is that the temperature of the eutectic transition in them turns out to be very close to the melting point of the fusible component. At the same time, for many eutectic pairs, the eutectic temperature turns out to be significantly lower than the melting point of both components. One of these contact pairs is the Au-Ge system, where the eutectic melting point is 577 K below the melting point of germanium. In addition, the study of the particularities of the formation and existence of the liquid phase in this contact pair is of great applied importance. For example, crystalline germanium is widely used in modern technologies. At the same time, germanium films deposited by the methods of vacuum condensation are amorphous, and their crystallization can be achieved by sufficiently high-temperature heating. However, this thermal effect is often undesirable and researchers are seeking to reduce the required annealing temperature. One of the ways to solve this problem is based on metal-induced crystallization, which is based on the fact that the presence of a liquid eutectic on the surface of amorphous germanium stimulates its transition to the crystalline state [34, 35, 36, 37]. Due to its low eutectic temperature, the Au-Ge contact pair is promising for the practical application of metal-induced crystallization.

Fig. 3.39 shows the temperature dependence of the quality factor of a quartz resonator with a condensed Au/Ge film for the first three heating-cooling cycles [23]. The thicknesses of the films were 80 and 40 nm for gold and germanium, respectively, which provided a eutectic concentration of components. As can be seen, the studied system shows melting-crystallization hysteresis, and the observed supercooling value is 165 K, i.e., the relative supercooling is η = 0.26.

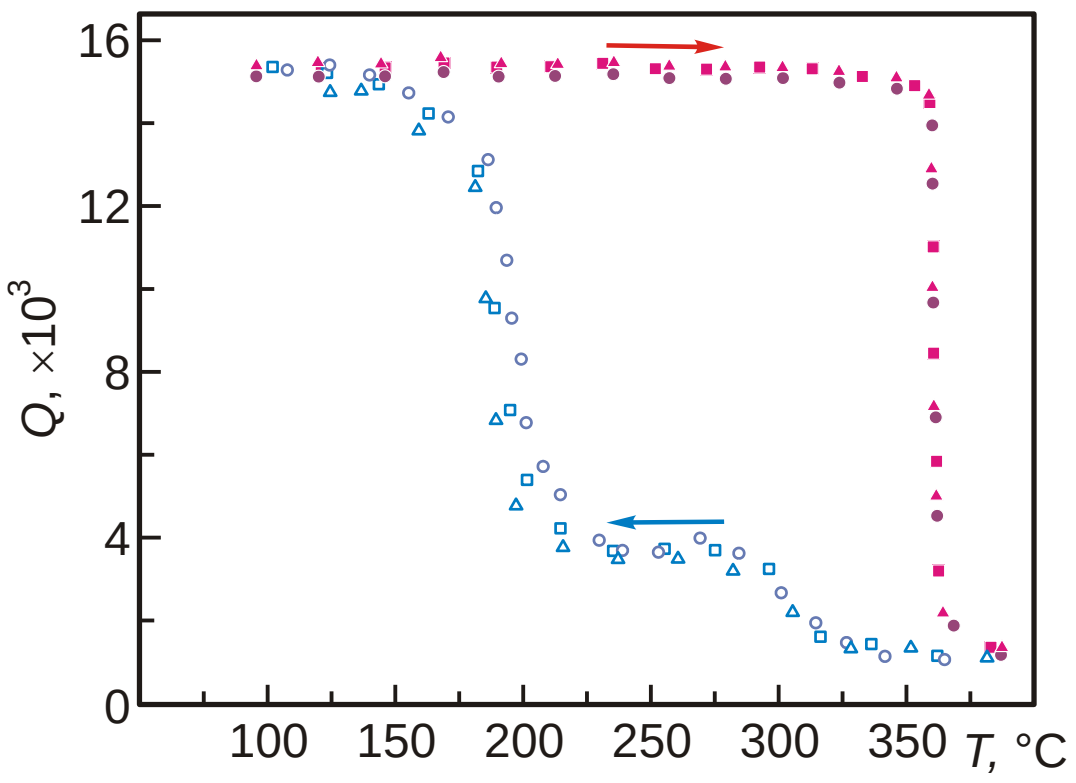

**Fig. 3.39** Quality factor temperature dependence of a quartz resonator with a Ge/Au condensed film [60]

The complex nature of crystallization observed in this system is noteworthy. This behavior can probably be explained by the heterogeneity of the system. Thus, initially, the sample obtained by the sequential condensation method consists of two continuous films, and its average concentration corresponds to the eutectic value. During heating, due to contact melting, a liquid phase appears at the interface of the films, the part of which gradually increases. This eventually causes the sample to break up into separate islands. At the same time, due to the action of random factors, the homogeneity of the system is disturbed and, in addition to particles with a eutectic concentration of components, there will be particles with a different concentration of components in the sample.

These statements are confirmed by the data of scanning electron microscopy (SEM) (Fig. 3.40). In the sample that was heated to 400 °C, both particles of spherical shape (indicating their complete melting) and shapeless objects, that obviously have not been completely melted, are observed. Microanalysis using energy-dispersive X-ray spectroscopy showed that, while the composition of spherical particles is close to eutectic, as the "sphericity" of the particle decreases, the concentration of components in it moves away from the eutectic.

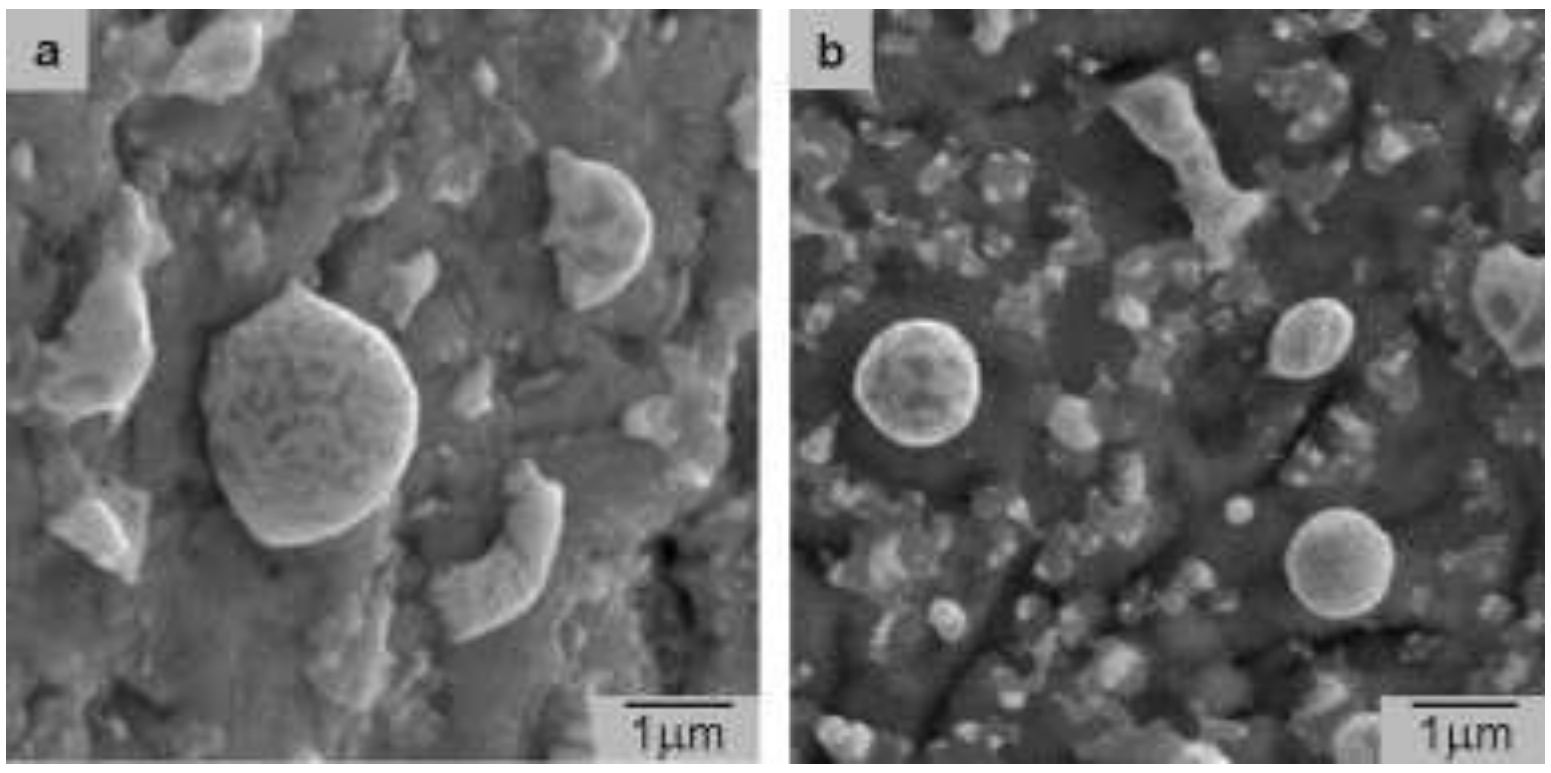

**Fig. 3.40** SEM image of the Au/Ge/C film system with a mass thickness of 120 nm on a quartz plate after its heating to 400 °C. Image (**a**) shows the micrographs of Au-Ge eutectic droplet profiles on a germanium substrate at an angle of 46° to the optical axis of the microscope [60]

Melting of a continuous film in the initial stages mainly occurs at interfaces, causing heterogeneous dewetting of films into islands that have a different ratio of components. Meanwhile, the prevailing majority of the islands had a ratio of masses of gold and germanium close to eutectic. As a result, almost all of the substance was transformed into a liquid state, collecting in droplets that do not wet the carbon sublayer. In other islands, the excess component, according to the phase diagram, remained in the form of a solid phase on its base. Thus, on the surface of the quartz plate, there were located islands, which were in contact only with amorphous carbon and islands that, apart from this, were also in contact with the excess element.

The melting of most part of the substance of the eutectic composition of the binary film system explains the sharp drop in the quartz quality factor observed in Fig. 3.39. It is obvious that the crystallization temperatures for fully melted particles, in which the liquid eutectic is in contact with the carbon sublayer, and partially melted particles, in which it interacts with the corresponding components, will be different. Thus, the two-stage crystallization observed in Fig. 3.39, can be explained by the fact that first the partially melted particles crystallize, and only then those whose concentration of components corresponded to the eutectic value.

A two-layer Bi/Ge film system with layer thicknesses of components of 50 nm and 100 nm, respectively, was also studied by the quartz resonator method. Fig. 3.41 shows a typical temperature dependence of the oscillations amplitude of a quartz resonator loaded with a Bi/Ge film system.

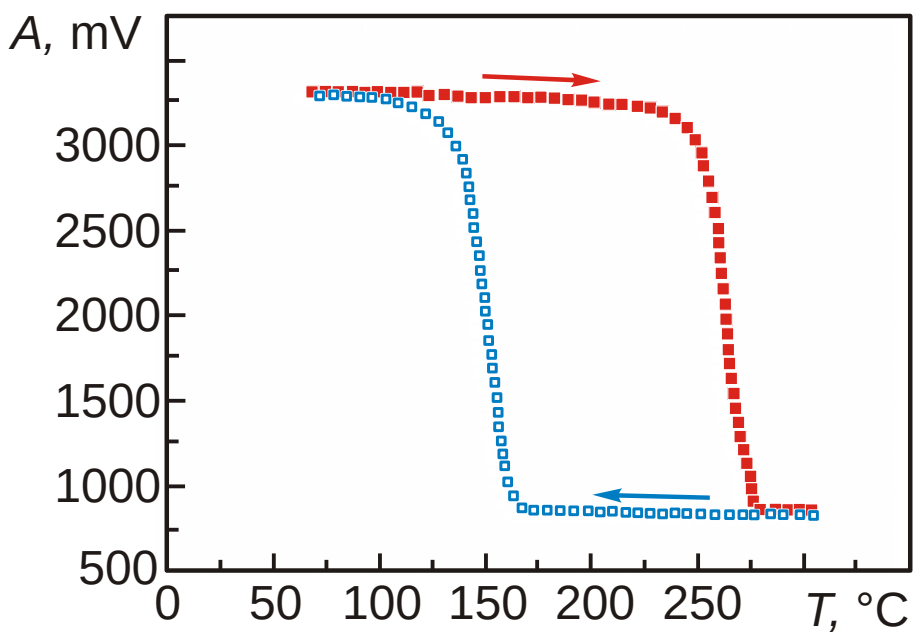

**Fig. 3.41** Typical temperature dependence of the quartz resonator oscillations amplitude for Bi(50 nm)/Ge(100 nm) film system; ▪ – heating, ▫ – cooling [60]

As for the Au/Ge system, the presence of a reproducible temperature hysteresis is common to all dependencies, which allows observing the fact of formation and crystallization of the liquid phase in the system. The melting and crystallization temperatures determined at the points with the maximum rate of change in the quality factor were 270 °C and 150 °C, respectively.

Thus, the value of supercooling during the crystallization of the Bi-Ge eutectic on a germanium substrate turned out to be 120 K. At the same time, the wetting angle measured by the oblique observation method [29, 30, 38] was approximately 80°.

SEM image of the eutectic particles that were crystallized on amorphous germanium, is shown in Fig. 3.42.

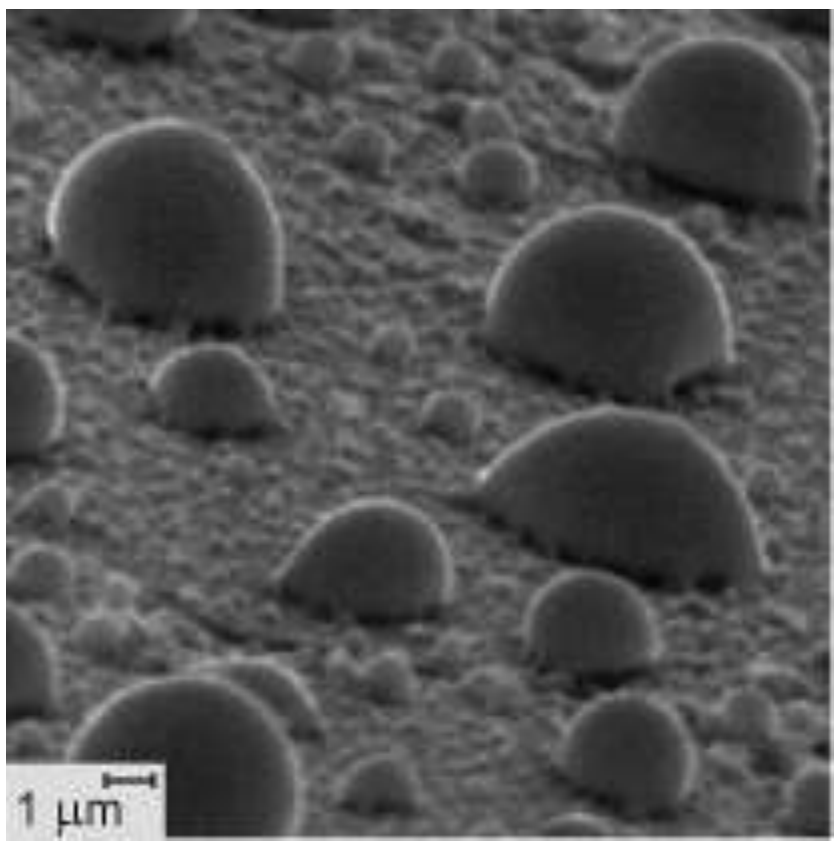

**Fig. 3.42** SEM image of Au-Ge eutectic droplet profiles on the amorphous germanium substrate, which was taken at an angle of 75° to the optical axis of the microscope; the wetting angle is $\theta = 80°$ [60]

The large number of results, obtained for contact pairs, containing bismuth and lead is due not only to their applied significance but also to methodological considerations, namely, the fact that these metals with many substances form simple phase diagrams that do not contain new phases and chemical compounds. This makes it possible to study supercooling itself in contact systems without being distracted by external factors.

At the same time, many contact pairs that have prospects for applied use are characterized by more complex phase diagrams. The behavior of such systems can have a number of particularities that complicate research and should be taken into account in technological developments.

In particular, in the Cu-In binary system, the phase diagram of which can be found, for example, in [39, 40], a large number of new phases and chemical compounds can be formed.

Phase transitions in this system were studied in [41]with the help of a technique based on the study of the temperature dependence of electrical resistance. Fig. 3.43 (curve a) shows the temperature dependence of the resistance in Cu/In/Cu films. In this graph, it is not possible to find any particularities that could be connected with phase transitions in the studied system. However, after a single heating of the samples to a temperature of 350 °C, the pattern changes and the melting-crystallization hysteresis is clearly observed on the temperature dependencies of the resistance, indicating the existence of a supercooled liquid phase in the films (Fig. 3.43, curve b). It should be noted that according to electron diffraction studies (Fig. 3.44), such annealing leads to the formation of the chemical compound $Cu_{11}In_9$ in the system, the formation of which above 320 °C follows from the phase diagram.

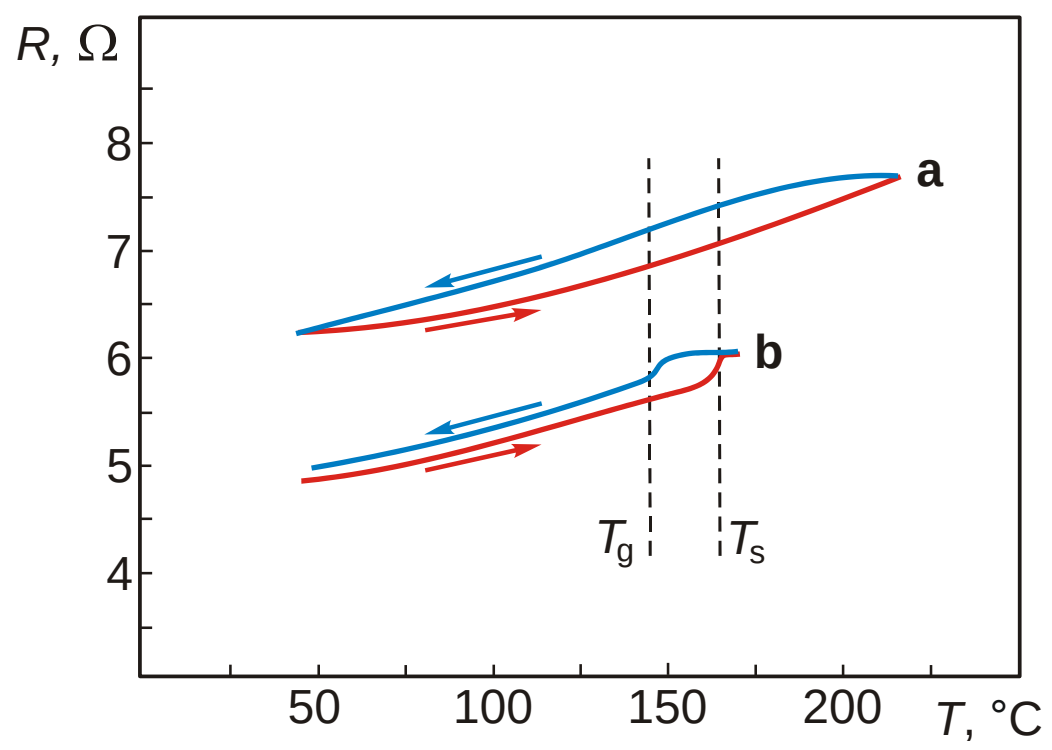

**Fig. 3.43** Temperature dependence of the resistance of Cu/In/Cu films before (**a**) and after (**b**) their heating to the temperature of 350 °C

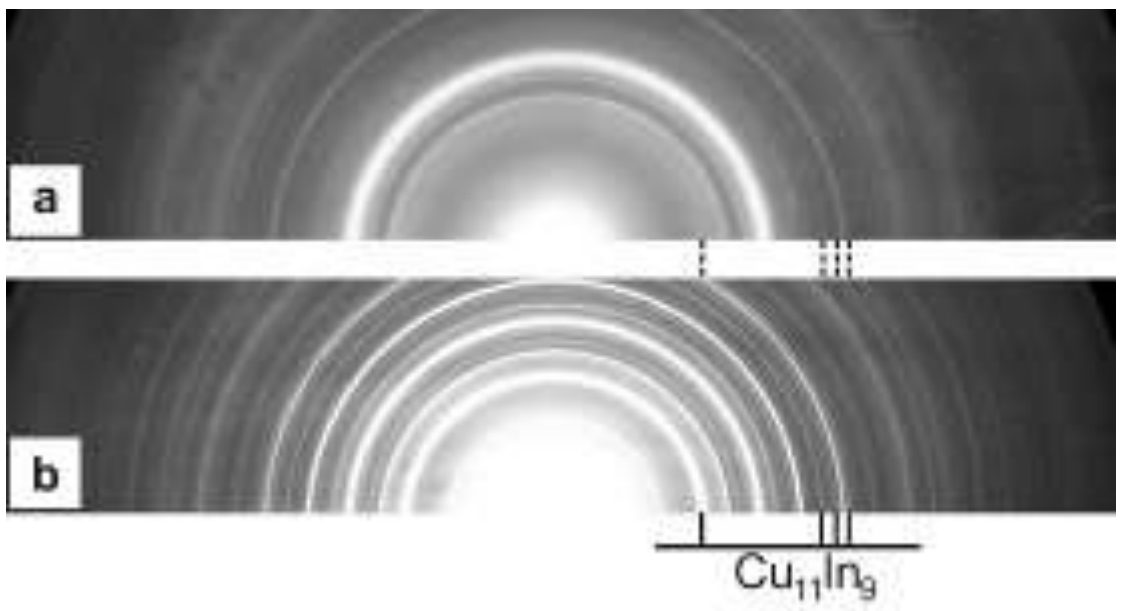

**Fig. 3.44** Electron diffraction patterns of Cu/In/Cu films before (**a**) and after (**b**) their heating to the temperature of 350 °C

The value of supercooling during the crystallization of indium in a layered Cu/In/Cu film system obtained in the work [41] is 20 K (the corresponding relative supercooling is $\eta$ = 0.05). The interpretation of the obtained value looks rather complicated. It is reasonable to assume that the $Cu_{11}In_9$ intermetallic, from which reflexes are observed through electron diffraction studies, must appear in films on the copper-indium interface. Probably, after heating to 350 °C, the indium particles will be covered with the intermetallic. Thus, the recorded value of supercooling should be attributed not to the In-Cu contact pair, but to the In–$Cu_{11}In_9$ pair.

A simpler situation takes place in Mo/In/Mo films. According to the phase diagram, there are no new phases or chemical compounds for this contact pair in the temperature range under study. Fig. 3.45 shows the resistance versus temperature for these films. As can be seen, it qualitatively repeats the main features of the curves characteristic of Mo/Pb/Mo films, except that indium practically does not accelerate the thermal dispersion of the samples. The supercooling in this contact pair is 75 K, which corresponds to a relative supercooling of $\eta$ = 0.17.

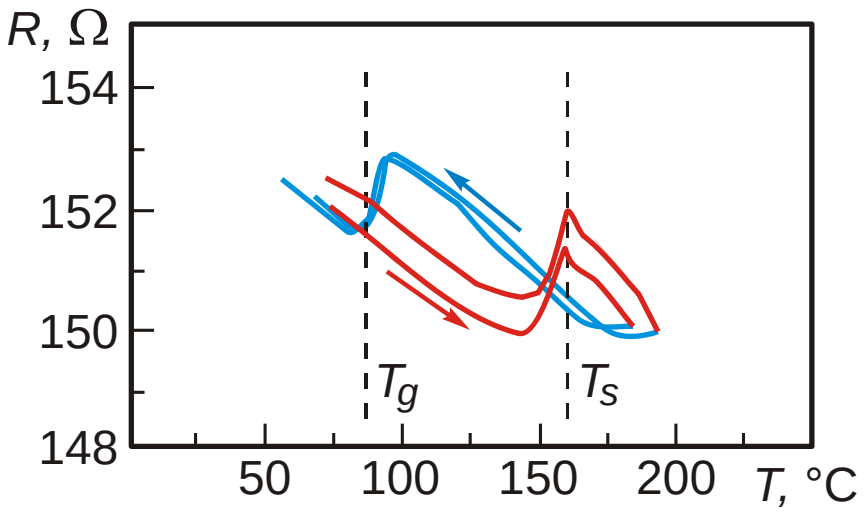

**Fig. 3.45** Electrical resistance dependence on the temperature for Mo/In/Mo films

In works [22, 41], layered film systems in which tin was used as a fusible component were investigated. Among the considered contact pairs, only the Sn-C pair is non-interacting, while in the Sn-Cu and Sn-Mo systems, the formation of various chemical compounds is possible. The results obtained from the temperature dependence of the electrical resistance of the layered films are generally consistent with the above regularities for similar contact pairs.

The Sn-C system is characterized by almost complete non-wetting and the absence of additional phases and chemical compounds. Fig. 3.46 shows the temperature dependence of the electrical resistance of multilayer films in which five layers of tin alternated with layers of carbon. The graphs of the numerical differentiation of the experimental curves using the Savitzky–Golay filter are shown in Fig. 3.47.

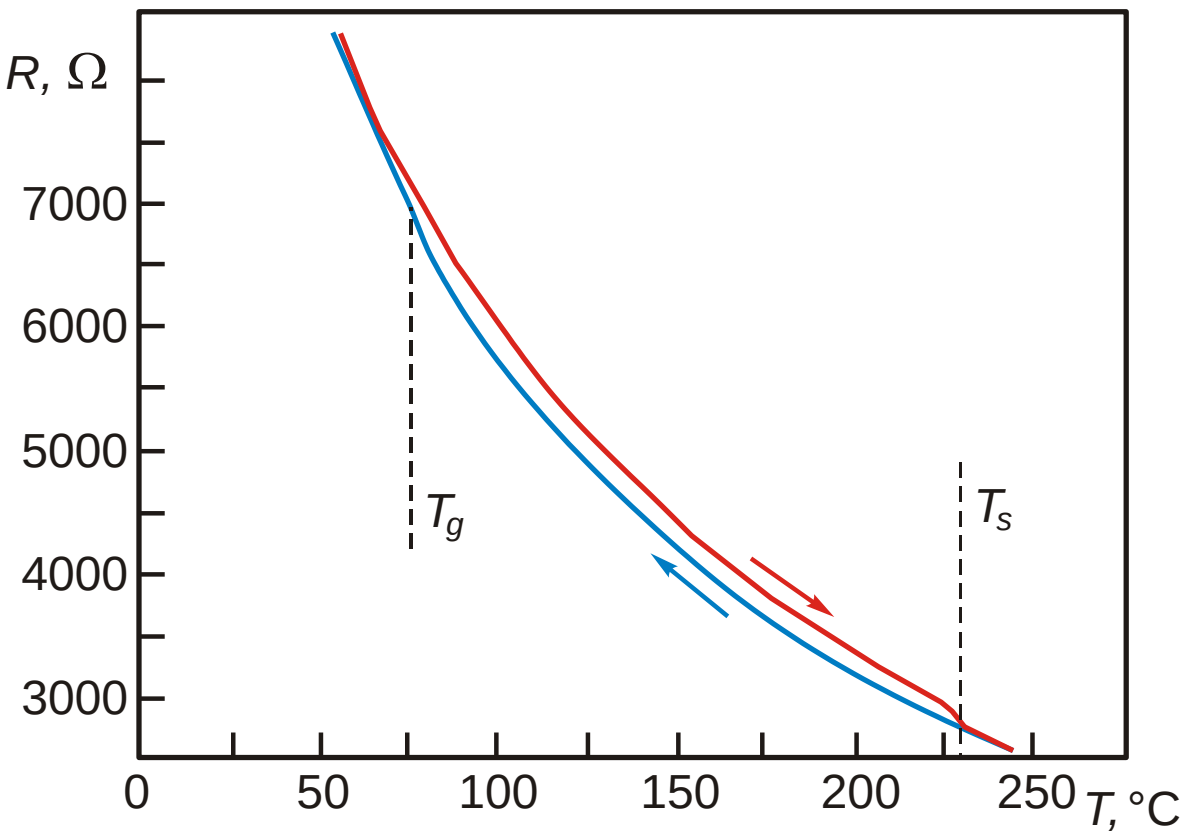

**Fig. 3.46** Temperature dependence of the electrical resistance of C/Sn/C layered film systems, which contain five tin layers between carbon layers [22]

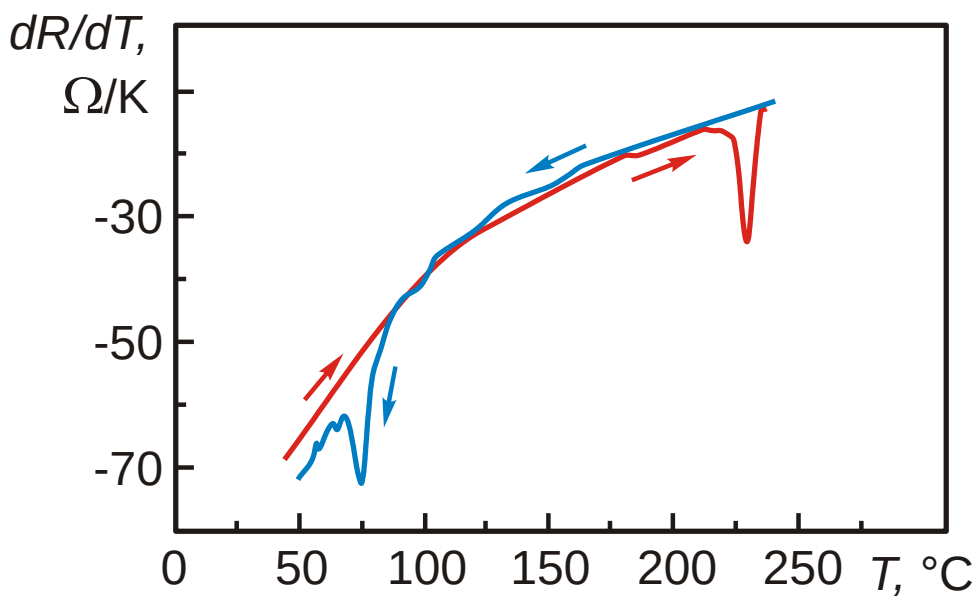

**Fig. 3.47** Graphs of numerical differentiation of the temperature dependence of C/Sn/C film electrical resistance [22]

As can be seen from Fig. 3.46 and Fig. 3.47, the melting and crystallization of tin are accompanied by jumps in electrical resistance. At the same time, as in the case of C/Bi/C and C/Pb/C samples, the resistance of the films decreases during melting and increases during the crystallization of the fusible component. The supercooling recorded in these samples is 160 K, which corresponds to a relative supercooling of η = 0.32.

The situation in Cu/Sn/Cu films looks much more complicated. The phase diagram of this contact pair is marked by a large number of phases and intermetallic compounds [40, 42]. At relatively low (less than 20 wt%) tin concentrations, the temperature dependence of the resistance of the layered film system is similar to copper annealing: the graphs show an irreversible decrease in electrical resistance. During subsequent heating-cooling cycles, the dependence of the resistance of the films on temperature has a linear nature and does not show any particularities (Fig. 3.48).

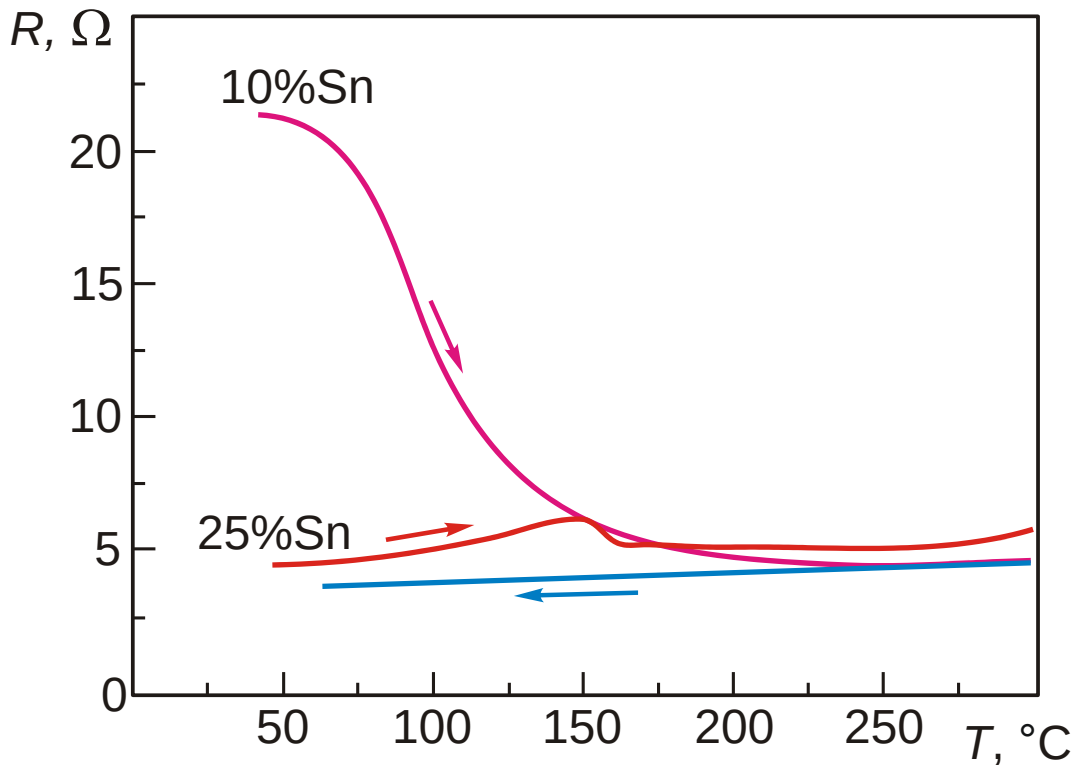

**Fig. 3.48** Cu/Sn/Cu films electrical resistance dependence on the temperature in the first heating cycle. The mass concentration of tin in the sample is indicated near the curve

A slightly different nature of the dependence of electrical resistance on temperature in the first heating cycle is observed in Cu/Sn/Cu samples containing more than 20 wt. % of tin. In such films, during the first heating cycle, the resistance initially increases slightly, after which a sharp drop in resistance occurs in the temperature range of 150–160 °C, which probably indicates a change in the phase composition of the sample.

During subsequent heating-cooling cycles, the electrical resistance dependence on temperature is almost linear but contains small and repeated particularities: during heating near the melting point of tin, a faster increase in resistance can be seen than would be expected based on the linear nature of its change (Fig. 3.49). During cooling, a rapid decrease in resistance is observed, which extends over a temperature range of about 10 K and ends at the temperature of 186 K. Thus, the supercooling in this contact pair is 45 K, and the relative supercooling is η = 0.09. It should be noted that, like lead, tin contributes to the dispersion of copper layers. However, this requires significant overheating above the melting point. The effect of dewetting becomes noticeable only under the condition when the samples are heated to a temperature above 300 °C.

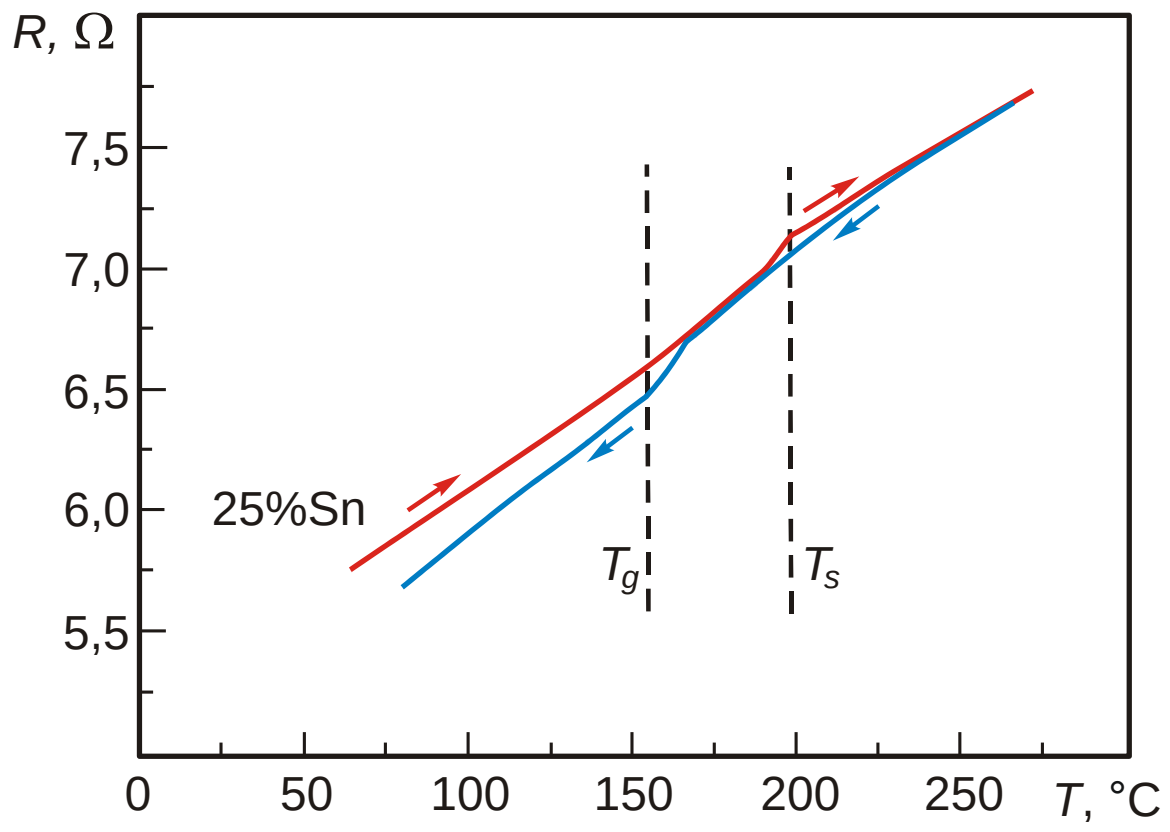

**Fig. 3.49** Temperature dependence of Cu/Sn/Cu films electrical resistance in the fifth heating cycle. The mass concentration of tin in the sample is 25 %

Similarly to Cu/In/Cu films, the complex nature of the phase diagram of the Cu-Sn contact pair requires consideration of the processes at the interfaces whose formation is possible in this system. As follows from the results of the work [43], during the deposition of Sn/Cu films by the method of layer-by-layer vacuum condensation on a substrate at room temperature, the $Cu_6Sn_5$ intermetallic appears at their interface. Consequently, in the obtained bilayer films, tin will be in contact with $Cu_6Sn_5$ intermetallic and, possibly, with pure copper.

During heating, the decomposition of $Cu_6Sn_5$ occurs with the formation of a new chemical compound $Cu_3Sn$ [43]. The jump in electrical resistance in the first heating cycle (Fig. 3.48) can be attributed to the decomposition of the $Cu_6Sn_5$ intermetallic. The appearance of particularity in the temperature dependence of the resistance can be explained by the fact that, according to [44], the decomposition of $Cu_6Sn_5$, which occurs near a temperature of 150 °C, is accompanied by a change in the coefficient of thermal expansion which should affect the mechanical stresses in the sample. The relaxation of such stresses usually changes the concentration of crystal lattice defects, which, in turn, affects the electrical resistance. An additional contribution to the change in the resistance of the layered film system should be made by the difference in the specific resistivity of $Cu_6Sn_5$ and $Cu_3Sn$ intermetallics. During subsequent heating-cooling cycles, due to the active formation of $Cu_6Sn_5$ at the copper-tin interface, competition between the two compounds is observed [45, 46, 47], which are likely to be present in the sample simultaneously. Thus, the value of supercooling, determined in this work, should be attributed to the supercooling of tin which is in contact with two intermetallic compounds and, possibly, with pure copper.

The results of the study of the temperature dependence of the electrical resistance in the films of the Mo-Sn contact pair are shown in Fig. 3.50 [41]. It can be seen that in these samples there is a clearly visible evolution of the nature of the temperature dependence of the electrical resistance as the heating-cooling cycles are carried out. In the first cycle, both during heating and cooling, resistance jumps are observed, each of which causes an increase in the total resistance of the sample. Starting from the second heating cycle, the graphs also show jumps whose directions are opposite to the jumps of the specific resistance of tin during the "liquid-crystal" phase transitions. And only starting from the fifth heating-cooling cycle, the curves show generally reversible particularities, the direction of which coincides with the direction of the jump in the specific resistance of tin. Such particularities can already be compared to the melting and crystallization of a fusible component. At the same time, the low-temperature jump in electrical resistance has a complex nature: after an initial increase in resistance, a decrease is observed, which probably corresponds to the end of tin crystallization. The obtained value of supercooling in such films is 115 K, with a relative supercooling of η = 0.23. It should be noted that the graphs presented in Fig. 3.50 correspond to two-layer Sn/Mo films in which the molybdenum layer was subjected to a single heating cycle to a temperature of 300 °C before tin condensation. This was necessary so that the changes in electrical resistance during the heating-cooling cycles were not affected by the annealing of the molybdenum layers, which usually causes an irreversible decrease in the resistance of the samples.

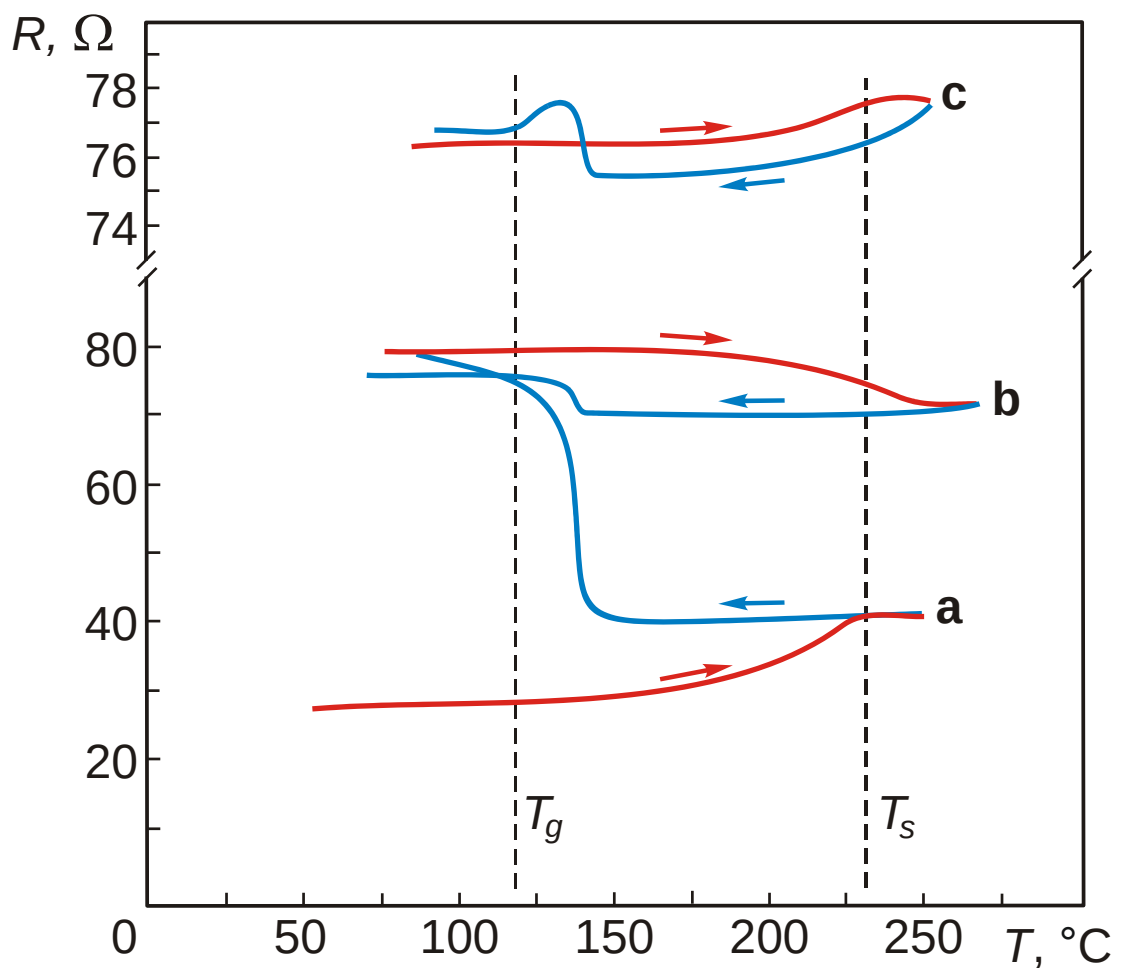

**Fig. 3.50** Temperature dependence of the electrical resistance of Sn/Mo films which corresponds to the first (**a**), second (**b**), and fifth (**c**) heating-cooling cycles

It can be assumed that some increase in resistance observed in two- and three-layer films of the Sn-Mo contact pair is due to the decomposition of $Mo_3Sn$ intermetallic, which, according to the works [48, 49], appears in this system at a temperature of about 300 °C.

The results of the study of the temperature dependence of the resistance in three-layer Mo/Sn/Mo films corresponding to samples, that have already undergone several heating-cooling cycles, are shown in Fig. 3.51. These dependencies also show jumps in the electrical resistance, the temperatures of which coincide with the values obtained in two-layer samples.

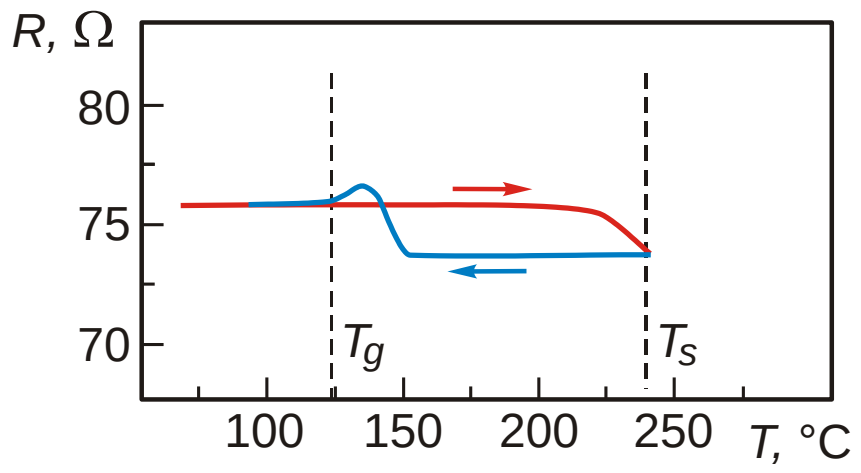

**Fig. 3.51** Temperature dependence of the electrical resistance of Mo/Sn/Mo films after several heating-cooling cycles

## 3.3 Physical causes of electrical resistance jumps in layered film systems

The results presented above show that the study of the temperature dependence of electrical resistance is a powerful tool that allows not only to determine the temperature of phase transitions *in situ* but also to obtain information about the processes of diffusion and phase formation in layered film systems. However, the physical causes that lead to sharp changes in the resistance of films during phase transitions are not yet well understood today.

The particularity of the described contact pairs of the type "metal-to-metal" is that the directions of resistance jumps in the multilayer film systems created from them coincide with the jump in the specific resistance of the fusible component during the phase transition. As a rule, an increase in electrical resistance is observed during the melting of indium, tin, and lead layers in the studied film systems. The only metal for which a decrease in resistance was observed in these studies during its melting in layered film systems based on more refractory metals or crystalline germanium is bismuth. Unlike most metals, its specific resistance decreases when it passes into the liquid phase. This circumstance gives reason to believe that the main cause for the sharp change in electrical resistance during phase transitions is the jumps in the resistance of the fusible component.

In order to quantify the contribution that these jumps provide to the overall resistance of the three-layer structure, we assume that the samples under study can be represented as three conductors connected in parallel. Under this assumption, and taking into account the fact that the resistance of the films modeling the matrix is usually much lower than the resistance of the fusible component layer, the following can be obtained:

$$\frac{\Delta R}{R_l} \approx \frac{h_m \rho_u^s}{h_u}\left(\frac{1}{\rho_m^l} - \frac{1}{\rho_m^s}\right), \tag{3.1}$$

where $\Delta R = R_s - R_l$ ($R_s$ is the total electrical resistance of the multilayer film in the case when the fusible metal is in the solid phase and $R_l$ is in the liquid phase; the index *m* indicates the fusible component of the layered system; the index *u* is used to indicate the material of the layers that model the matrix); $\rho^l$ and $\rho^s$ are the specific electrical resistance of the substance of the layer, indicated in the lower index, in the crystalline and liquid state, respectively; *h* is the thickness of the layers.

The application of expression (3.1) to the studied film systems gives values that correspond in order of values to those obtained in experiments for films in which layers of lead, indium, or tin are located between layers of copper, molybdenum, or silver. At the same time, the jumps in electrical resistance in films containing bismuth turn out to be much larger. This indicates the existence of additional factors that affect the resistance of the studied film systems. It should be noted that bismuth, unlike most metals, is characterized not only by a sharp decrease in specific resistance during the phase transition, but also by an increase in its specific volume during crystallization. Obviously, under conditions of limited volume, the process of bismuth crystallization causes mechanical stresses to occur in the layered film system. The relaxation of these stresses by the appearance of crystal structure defects, in turn, causes an increase in the electrical resistance of the films.

It is well known that the electrical resistance of polycrystalline films is determined in the first turn by the contribution of grain boundaries. According to the results of [18], thermal cycling of layered film systems initiates a gradual diffusion of bismuth atoms along the grain boundaries, which is accompanied by an increase in the resistance jumps of Al/Bi/Al films during crystallization. Characteristically, the largest jumps in resistance occur in films that were obtained by simultaneous condensation of the components. In such samples, bismuth turns out to be embedded in various defects in the aluminum layer from the very beginning, which in fact determine the resistance of the film system. During the crystallization of bismuth inclusions located in defects in the polycrystalline external layers, mechanical stresses, and new defects occur. This is probably the main cause for the sharp jumps in the electrical resistance of layered films containing bismuth as a fusible component.

Defects that occurred during the relaxation of mechanical stresses play a crucial role in the jumps in the resistance of Al/Bi/Al, Cu/Bi/Cu, Ag/Bi/Ag, Mo/Bi/Mo, and Ge/Bi/Ge films (after metal-induced crystallization of amorphous germanium layers), as evidenced by their behavior in heating-cooling cycles. Thus, in these contact pairs, even before the melting of the fusible component begins, instead of a linear increase, which should be expected based on the temperature dependence of the specific resistance of the components, a rather strong decrease in resistance is observed (Fig. 3.52). The most natural explanation for this phenomenon is the annealing of various defects that appeared in the films during the previous crystallization. The activation energy of the process that provides such annealing is 0.4 eV for Al/Bi/Al films, which is a typical value for annealing and relaxation of nonequilibrium defects. For example, during the annealing of freshly deposited copper films, which are extremely non-equilibrium structures, the value of the activation energy of 0.35 eV was obtained (Fig. 3.52, inset). Thus, the anomalously large value of the electrical resistance jump that occurs in samples in which the bismuth layer is located between more refractory polycrystalline films is probably due to the part of bismuth that is embedded into defects in the external polycrystalline layers.

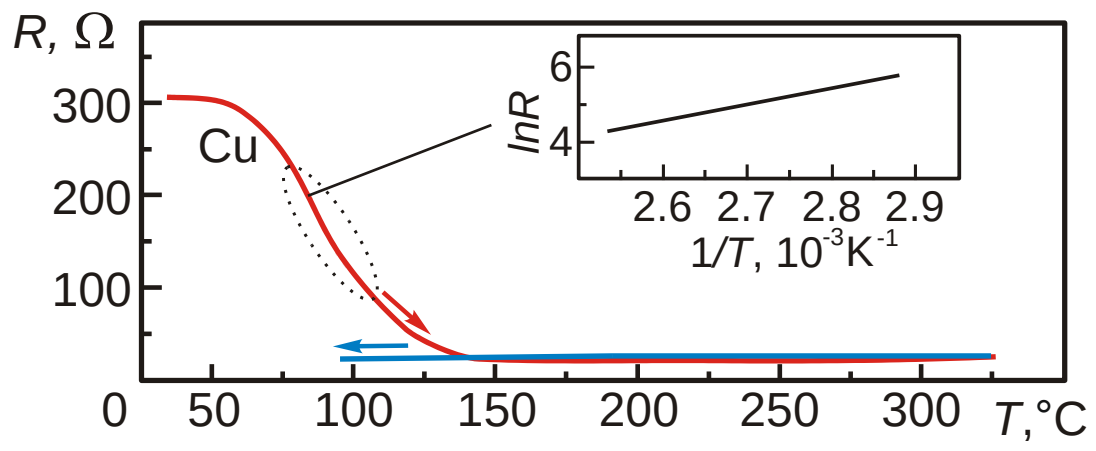

**Fig. 3.52** Dependence of the electrical resistance of copper films on the temperature in the first heating cycle. Inset is an Arrhenius plot that corresponds to a sharp decrease in sample resistance in the first heating cycle

For the other investigated fusible metals (In, Sn, Pb) in contact pairs based on copper, silver, and molybdenum, the resistance jumps are satisfactorily described by the model of parallel conductors. This indicates that in these cases, the main cause of the particularities observed on the temperature dependencies of electrical resistance is a change in the specific resistance during melting or crystallization of a layer of a fusible component located between layers of a more refractory one.

Among all the studied contact pairs, the carbon-based films are distinguished, the directions of resistance jumps in which do not correlate with the direction of jumps in the specific resistance of the fusible component during phase transitions. To explain the effects in such systems, it is necessary to take into account that the specific resistance of carbon in amorphous layers is much higher than that of the studied fusible metals. Due to poor wetting, the metal layer will have an island structure after the first melting. The electric current flows through the sample mainly via the carbon film, and the metal islands shunt some of its sections. Thus, the overall resistance of multilayer structures will be largely determined by the metal-carbon interface, whose resistance changes significantly during the phase transition. Liquid particles provide a virtually perfect electrical contact that has minimal resistance. At the same time, the electrical contact between the carbon film and the solid particles due to the volume jump during crystallization, mismatch of crystal structures, etc. will be more defective, and a significantly higher resistance should be expected for it.

In addition to the change in films resistance caused by the improvement in conductivity due to the replacement of the solid interface with a liquid one, a significant contribution to the jump in films resistance, as earlier, will be due to the change in the specific resistance of the fusible component during its melting or crystallization.

The contribution of each of these mechanisms, which are responsible for the presence of electrical resistance jumps accompanying phase transformations, can be estimated by comparing the values of the relative change in electrical resistance in layered films in which the embedded components have opposite directions of jumps of specific electrical resistance during the phase transition. This estimate done in this way allows stating that the contribution of the change in the electrical resistance of the interface to the total electrical resistance of the sample is about 60 %, and the changes in the electrical resistance of the fusible particles themselves are 40 %. As can be seen, the change in interface quality and the jump in specific resistance of fusible particles have almost the same contribution to the overall electrical resistance of the sample. This opens up the possibility, by changing the morphological structure of the embedded fusible particles located in the layered films between the carbon layers (for example, by adding small amounts of a third component that will change the wetting parameters in the required way), to influence the value of the electrical resistance jump in a controlled manner and even change the sign of the effect.

Taking into account the above factors that cause jumps in electrical resistance in C/Pb/C, C/Bi/C, and C/Sn/C films, it looks natural that it is bismuth that provides the largest change in electrical resistance among all the studied fusible components during phase transformations.

Unlike most metals, the specific resistance of bismuth decreases after melting. This effect is summed up with a reduction in the electrical resistance of the contact interfaces. At the same time, for tin and lead, the reverse change in electrical resistance is observed during melting, which partially compensates for the improvement in the electrical conductivity of the interface, which is provided by the liquid phase. Therefore, the changes in electrical resistance, which accompany phase transformations in films with particles of these fusible metals, turn out to be much smaller.

# Chapter 4 Supercooling during the crystallization of alloys in condensed films

**Abstract** Compared to supercooling of pure metals, the study of supercooling of alloys is complicated by a number of factors. Thus, the difference in temperatures of solidus and liquidus, which is natural for alloys, leads to some ambiguity in the numerical determination of the supercooling value. In addition, local segregation of components can be observed in real alloys. Due to the concentration dependence of the melting point, violation of the film homogeneity leads to a local change in the phase transition temperatures. In this connection, for studying the concentration dependence of supercooling of alloys, integral methods that allow obtaining average information about the processes in the sample become very effective. In particular, these are the methods based on the use of electrical resistance and the resonant frequency of a quartz oscillator. The chapter presents the results of the study of supercooling in various contact pairs. It is shown that in the contact pair (Bi-Sn)/Cu the concentration dependence of the crystallization temperature of the supercooled melt generally repeats the liquidus line, but lies significantly below it and the solidus line. This indirectly indicates the approximate constancy of wetting in the system. The results of the study of supercooling in (In-Pb)/Mo samples indicate that the relative supercooling reaches a maximum for alloys containing equal amounts of lead and indium. In the Bi-Sn system, minimal supercoolings are observed during the crystallization of the eutectic, located on one of its components. This confirms the general trend according to which the achievable supercooling decreases as the interaction in the system improves.

## 4.1 Phase transitions in particles of binary alloys embedded in a more refractory matrix

Compared to the information on supercooling during the crystallization of pure metal particles that are in contact with another substance (substrate or matrix), there are significantly fewer scientific papers on supercooling of alloys in the matrixes. At the same time, phase transitions in such structures are characterized by unique particularities associated with the influence of the matrix on the formation of various phases, including metastable ones. The study of In-Sn alloys of eutectic concentration (46 at.% indium) embedded into an aluminum matrix was carried out in the work [1]. The samples were obtained by rapid cooling of an Al-In-Sn melt containing the components in a given concentration. It should be noted that this method of sample preparation is possible due to the very low solubility of indium and tin in solid aluminum.

During X-ray diffraction studies, it was found that the obtained samples contain crystalline aluminum, as well as β- and γ-phases of the In-Sn alloy [2]. In electron-microscopic images, it was found that in the aluminum matrix, particles with a size of 150–250 nm are observed, within the image of which there are zones of different diffraction contrast. This indicates that such particles consist of two crystalline modifications separated by an interface.

Differential scanning calorimetry (DSC) was used to study melting and crystallization in the work [1]. Using this technique, it was found that both melting and crystallization in this system are multistage processes. One of the peaks which are detected during heating occurs at a temperature of 119 °C. This temperature is close to the melting point of the eutectic of the In-Sn binary system. At the same time, one more peak is observed in the system during heating near the temperature of 154 °C. The first melting peak has a significant temperature width, which is probably explained by the superposition of the size dependence of the melting point with a wide distribution of particles present in the sample by sizes. The peak, which appears at 154 °C in the first heating, is blurred during subsequent cycles, its intensity decreases, and it shifts to a region of lower temperatures. The authors explained the presence of this particularity on the DSC curves by the formation of metastable structures which have heterogeneities of composition.

Crystallization in this system is also a complex process that starts at a temperature of 115 °C, ends at 95 °C, and consists of three stages. According to the conclusions drawn in the work [1], the first crystallization peak is due to grain boundary crystallization, and the other two correspond to the crystallization of the γ- and β-phases, which make up the fusible particles in the aluminum matrix.

The study of particles of Pb-Sn alloys embedded into an aluminum matrix was carried out in the work [3]. The samples for the studies were obtained by mechanical milling of Pb, Sn, and Al microparticles with an initial size of 45, 45, and 35 μm, respectively. The milling process was carried out in two stages. First, the mixtures of Pb-Al and Sn-Al particles were separately milled. Then the obtained structures were mixed and subjected to further milling. To prevent oxidation and coalescence of the particles, all dispersion processes were carried out in the environment of toluene.

The results of X-ray diffraction studies were used to determine the crystal structure of the particles and their characteristic size. It was found that the milling process causes a slight shift in the positions of the diffraction maxima, and the average size of lead and tin particles after milling is 17 nm and 21 nm, respectively.

Using TEM and SEM studies, the authors of [3] found that the method used, consisting of sequential mechanical milling, allows the formation of structures in which particles of the fusible Pb-Sn alloy turn out to be embedded into the aluminum matrix. The studies of the thermal properties of the obtained structures, performed by the method of differential scanning calorimetry, allowed us to establish that during the heating of the initial samples, which represent themselves a mechanical mixture of Pb-Al and Sn-Al structures, two peaks are observed, corresponding to the melting of tin and lead. With an increase in the milling time to which the obtained mechanical mixture is subjected, the intensity of the peaks corresponding to the pure components decreases, and they are slightly shifted to the region of lower temperatures. For example, for samples that have been milled for 20 hours, the peak, which identifies the melting of tin, shifts by 14 K compared to its initial position.

In addition to the jumps caused by the pure components, another peak appears on the DSC curves and grows with increasing milling time, the temperature of which gradually decreases from 190.6 to 179.4 °C. The authors of [3] attribute the presence of this particularity on the DSC graphs to the formation of two-phase particles in the system.

During the study of the processes occurring during the cooling of Pb-Sn-Al structures, it was found that these processes are fundamentally different from those observed in Pb-Al and Sn-Al. The DSC graphs of Pb-Al and Sn-Al samples show clear, rather narrow, and reproducible peaks in heating-cooling cycles, which correspond to the crystallization of the supercooled melt. The melting of lead and tin occurs near the table value and only slightly decreases with increasing milling time, which in turn determines the size of the particles and provides their embedding into the aluminum matrix. It has been found that these contact pairs crystallize with supercooling, which, unlike melting, strongly depends on the degree of milling. Thus, in the initial samples, which represent pure particles of lead or tin with a size of 45 µm, relatively small supercoolings are observed, equal to 25 and 71 K, respectively.

At the same time, already after three hours of milling, which forms Pb-Al and Sn-Al composite structures, the supercoolings increase to 127.5 and 132.6 K (relative supercoolings $\eta_{Pb}$ = 0,21, $\eta_{Sn}$ = 0,26), respectively. At the same time, despite the fact that the crystallization has a diffuse nature, the peaks on the DSC curves turn out to be quite clear and actually do not contain any particularities that could indicate a multi-stage crystallization process.

The crystallization of Pb-Sn-Al samples is observed in the temperature range of 160–250 °C. The authors observe that many weakly expressed peaks, each of which obviously corresponds to a phase transition observed in several or even in one particle. The authors of the work [3] explain this behavior by the fact that the aluminum matrix has a weak orientation effect on the embedded lead and tin particles. Consequently, crystallization will be determined by local orientation relations that are characteristic of a particular particle and a given point of the matrix and will consist of many stages, which temperatures distributed over a wide range of values.

The particularities of melting and crystallization of particles of Pb-Sn alloys in an aluminum matrix were studied in the work [4] on samples obtained by the rapid solidification method. The authors studied alloys of three concentrations: $Sn_{82}$–$Pb_{18}$, $Sn_{64}$–$Pb_{36}$, and $Sn_{54}$–$Pb_{46}$, which correspond to the hypereutectic and hypoeutectic compositions on the Pb-Sn phase diagram.

Using electron microscopy, it was found that the investigated fusible particles turned out to be embedded into the aluminum matrix. At the same time, some of them are embedded in aluminum grains, while others remain at the grain boundaries. The size of particles located at the boundaries turns out to be somewhat larger than that of particles embedded into grains. The microstructure which is characteristic of fusible particles has a complex nature and is determined by their composition. For example, in the hypereutectic $Sn_{82}$–$Pb_{18}$ samples, fusible nanoparticles have a diffraction contrast and represent a binary structure, the larger part of which is enriched in tin and the smaller one in lead. At the same time, in samples whose composition corresponds to the hypoeutectic concentration ($Sn_{54}$–$Pb_{46}$), in addition to acorn-shaped particles, a certain number of lamellar particles are observed.

The study of the obtained nanocomposites by the method of differential scanning calorimetry revealed that starting from the second heating cycle, melting peaks are observed in all samples, the temperature of which corresponds to the eutectic values. The melting occurs in the range of 20-30 K, which grows slightly with increasing tin content. In addition, an additional peak is observed during the cooling of the hypereutectic alloys at a temperature higher than the eutectic one. Its existence is probably due to the heterogeneity of the composition.

Thus, despite the fact that the microstructure of the particles turns out to be significantly dependent on the composition, this has virtually no effect on their melting point. At the same time, studies conducted in the work [4] showed that the crystallization temperature and some particularities of this process are determined by the composition of the particles. In samples with the composition of the fusible phase $Sn_{82}$–$Pb_{18}$, three crystallization peaks with temperatures of 445, 408, and 381 K are observed on the curves of differential scanning calorimetry. The first of these peaks corresponds to a very small supercooling. There is no thermodynamic driving force in the phase diagram at this temperature and concentration that can cause lead crystallization. In this case, the driving force for the phase transition comes from the crystallization of tin. According to the conclusions of the authors of the work [4], under such supercooling, particles concentrated at grain boundaries crystallize. The second crystallization peak observed in $Sn_{82}$–$Pb_{18}$ alloys at the temperature of 408 K must also be initiated by the heterogeneous crystallization of tin that has contact with aluminum. Only the third peak, the temperature of which corresponds to 381 K, is located in that region of the phase diagram, in which there are thermodynamic causes for the crystallization of both tin and lead. And it is under these conditions that the crystallization of the fusible alloy, as a whole, occurs. At the same time, the eutectic nature of melting observed in subsequent heating cycles indicates that this phase transition is caused by phase interfaces rich in fusible components.

During the cooling of the $Sn_{54}$–$Pb_{46}$ alloy, a single broad crystallization peak is recorded, the maximum of which occurs at a temperature of 379 K. The blurred nature of this peak on the DSC curves is consistent with the fact that the tin which initiates this phase transition does not have clear orientation relationships with the aluminum matrix. The crystallization of $Sn_{64}$–$Pb_{36}$ alloys starts at a temperature of 433 K, also occurs in several stages and is accompanied by several peaks, the temperature position of which changes from cycle to cycle. This behavior can be explained by the fact that crystallization in these samples must have been initiated by lead, which is located in a certain orientation relationship with the aluminum matrix. The presence of two peaks on the DSC curves indicates that there are two preferred orientations for lead particles, which are responsible for crystallization.

The study of $Bi_{55}Pb_{44}$ alloys which are embedded in the aluminum matrix was carried out in the work [5]. Samples for the studies were obtained through the rapid cooling of alloys of appropriate concentrations. During TEM studies of nanocomposites, fusible particles are observed which are located in the body of the grains and also at the grain boundaries. Similarly to the samples studied by the authors of the work [4], the size of the particles concentrated at the grain boundaries turns out to be larger than those contained in the body of the grain.

At the same time, the embedded particles observed in the samples reveal a phase contrast and consist of bismuth and β-phase, the presence of which is confirmed by X-ray diffraction studies. The most probable size of the fusible particles present in the composite is 40 nm. Two types of particles are found in the samples: “acorn-type” particles, which consist of a β-phase covered with a bismuth cap, and lamellar particles. It is important to note that the aluminum matrix has an orientation effect on the embedded particles, and the interfaces formed are highly energetic.

The study of the obtained structures using differential scanning calorimetry revealed that the melting of fusible particles occurs at 16 K above the eutectic temperature, i.e., overheating is observed in the system. The crystallization of $Bi_{55}Pb_{44}$ particles embedded into the aluminum matrix occurs with supercooling, the peak of the DSC curves occurs at 332.8 K, and the phase transition is completed at a temperature of 314 K. The temperature range in which crystallization occurs is 30 K. The value of supercooling obtained by the authors of the work [5] is 111 K, and the relative supercooling is $\eta = 0.26$.

In the work [5], using *in situ* heating of samples directly in an electron microscope, it was shown that the melting of the embedded particles begins from their surface, which is, actually, in contact with aluminum. The melt front moves deeper, and at a temperature of 429 K, the particle completely turns into a liquid state. Crystallization occurs under supercooling at 40 K, and begins with the formation of crystalline bismuth or β-phase, depending on the particle composition. At the same time, mechanical deformations occur in the matrix, which cause the appearance of dislocations near the particle that is crystallizing. Upon further cooling, another crystalline phase is formed, after which eutectic growth of the two-phase particle is observed.

### 4.2 The method of changing the condensation mechanism in studies of supercooling of alloys

The method of changing the condensation mechanism, which is widely used to determine the supercooling of free single-component particles [6, 7, 8], turns out to be effective in studying the supercooling of two-component alloys. The main difficulty, in this case, is obtaining the alloy of the required concentration and preventing its fractionation. Since the components of binary systems usually have different pressures of saturated vapor at a certain temperature, the deposition of films by thermal evaporation of weighed portions of pre-prepared alloys usually does not provide reproducible results. One of the solutions to this problem is to use the flash evaporation technique to obtain samples under study.

For example, in the work [9], flash evaporation was used to prepare films during the study of supercooling at the crystallization of eutectic particles of the Bi-Pb alloy on an amorphous carbon substrate. To obtain the samples, small weighed portions (of 10-50 mg) of the pre-prepared alloy were placed on a tungsten strip, where they were almost instantly heated to a temperature of about 2000 °C. Due to the small size of the evaporated weighed portions, very rapid heating to a temperature significantly higher than the evaporation temperatures of the components is achieved (the vapor pressure of Bi and Pb is $10^{-2}$ mm Hg at temperatures of 670 and 715 °C, respectively). This, despite a slight difference in the pressure of the saturated vapor of bismuth and lead, made it possible to almost completely prevent the fractionation of the alloy. A certain disadvantage of this implementation of the flash evaporation method is that even the 10-50 mg particles, used to load the evaporator, are still quite massive, and it takes some time to heat them to a temperature, that provides the possibility of condensation of unfractionated samples. In addition, it is difficult to provide good thermal contact between the loaded alloy particles and the superheated surface when using practically realizable supply options for evaporators.

These difficulties can be overcome if the alloy components are placed on the evaporator by their preliminary deposition. In practice, this technique is implemented in two stages. In the first stage, a film with a given concentration of components is applied to a refractory (usually tungsten) strip by deposition from one or two independent evaporators. For this purpose, either a simultaneous or sequential deposition technique can be used. The required concentration of the components is either provided by complete evaporation of the respective weighed portions or determined *in situ*, e.g. using the quartz resonator method. Afterwards, the evaporator with the condensed film is kept at an elevated temperature, which provides diffusion homogenization of the condensate but is not yet sufficient for active evaporation of the components. Only after homogenization, the flash evaporation of the alloy under study is performed. The low mass of the condensate and the perfect thermal contact of the deposited layers with the evaporator surface allow for providing almost immediate heating of the alloy under study to the required temperature [9].

To determine the temperature of maximum supercooling, the methodology, tested for single-component samples, can be used: eutectic alloys were condensed onto an amorphous carbon substrate, along which a gradient of temperatures was created [9]. After the condensation was completed, the substrate with the film was removed from the vacuum chamber and examined by electron microscopy and visually. The visible boundary corresponding to the temperature of change of the condensation mechanism in the samples of eutectic composition obtained using both variants of the realization of flash evaporation technique is observed at a temperature of 268 K, which corresponds to a relative supercooling of $\eta = 0.33$. The value of relative supercooling found in this way is in good agreement with the value of the wetting angle, which, according to TEM studies for eutectic particles on an amorphous carbon substrate, is 135°. It should be noted that the supercooling value obtained in the work [9] is significantly higher than the values, known for massive samples, and this probably indicates the homogeneous nature of crystallization in the system under study.

It is also worth noting that the data of the work [9] indicate the presence of a dependence of the crystallization temperature on the concentration. Thus, with a decrease in the condensation rate, the temperature of this phase transition increases and gradually reaches a temperature of 363 K, which is close to the crystallization temperature of supercooled bismuth on an amorphous carbon substrate. This is explained by the fact that at low evaporation rates, the role of the difference within the saturated vapor pressures of the condensed components grows, which increases the degree of alloy fractionation. Since bismuth evaporates more easily than lead, this causes, that in the initial moments of condensation, the first to form on the substrate are the nuclei of the condensed phase enriched in bismuth, whose crystallization temperature is significantly higher than that of the eutectic.

The method of changing the condensation mechanism to study the concentration dependence of the value of supercooling during the crystallization of alloys of the Bi-Sb and Bi-Pb binary systems was used in works [10, 11, 12]. Two approaches were used to prepare the samples under study. In the first, the method of samples of constant content was used, i.e., several samples with different concentrations of components were obtained in a few experiments using the previously tested flash evaporation technique. In practice, the samples were obtained in a step of 10 mass percent of the concentration of the components. In the second series, the method of samples of variable content and variable state was used [6], in which the alloy components were condensed simultaneously from separated evaporators. Due to the special relative positioning of the evaporators, screens, and extended substrate, a set of samples with a composition, which continuously changes over a wide range of component concentrations, could be obtained in one vacuum cycle. Despite the slightly higher implementation complexity, it was the method of samples of variable content and variable state that proved to be effective for studying supercooling in binary systems.

As in the case of single-component films, in samples obtained by the method of variable content and variable state, a border is visually observed on the substrate (Fig. 4.1), which, according to TEM studies, corresponds to the concentration dependence of the temperature of the change of the condensation mechanism. Fig. 4.1 clearly shows that, as should be expected from the results obtained in the work [9], an increase in the concentration of bismuth in the binary alloy is accompanied by an increase in the crystallization temperature of the supercooled melt. At the same time, in a significant range of characteristic sizes (at a vacuum condensate thickness of approximately $10^2$–$10^4$ nm), the crystallization temperature of the supercooled melt does not show a size dependence but is determined only by the concentration of components.

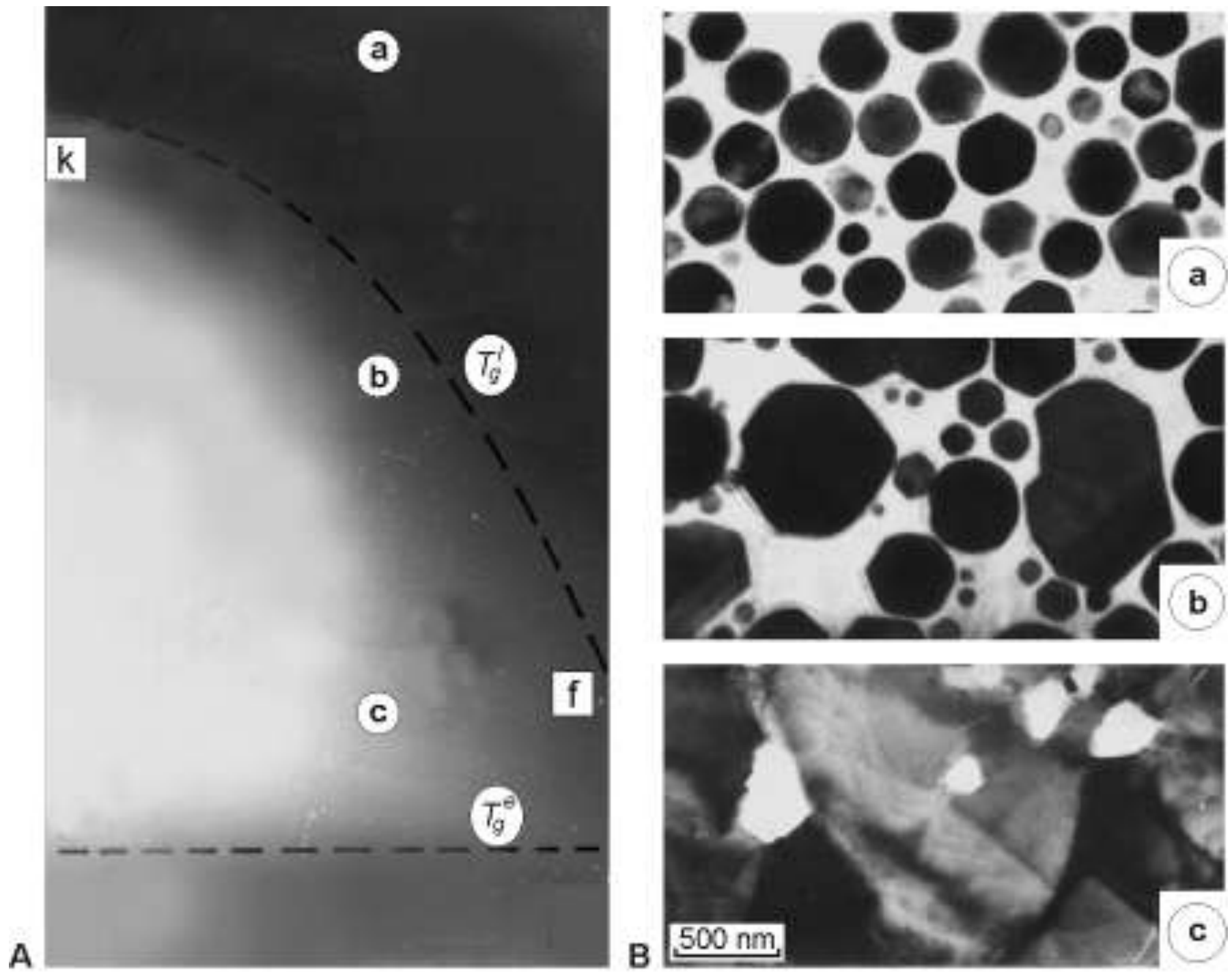

**Fig. 4.1** Photo (×1) of Bi-Pb film of variable content on a substrate with a temperature gradient (**A**) and TEM images of its various sections (**B**)

The concentration course of the crystallization temperature generally repeats the course of the liquidus line of the studied binary systems, but lies below it (Fig. 4.2). At the same time, the value of supercooling changes slightly when moving from the $Al_2O_3$ substrate to the amorphous carbon substrate, but the nature of the dependence remains unchanged. It should be noted that, according to TEM studies, the temperature range in which the crystallization of binary alloys occurs, is significantly higher than the values that occur in single-component systems. The value of such a blurring of the crystallization temperature in samples with a constant concentration of components turns out to be somewhat larger than in those samples for which the variable content method was used for condensation. Thus, it is likely that the fractionation of the alloy, the degree of which is largely determined by the particularities of the sample condensation method, plays a significant role in increasing the temperature range in which crystallization occurs.

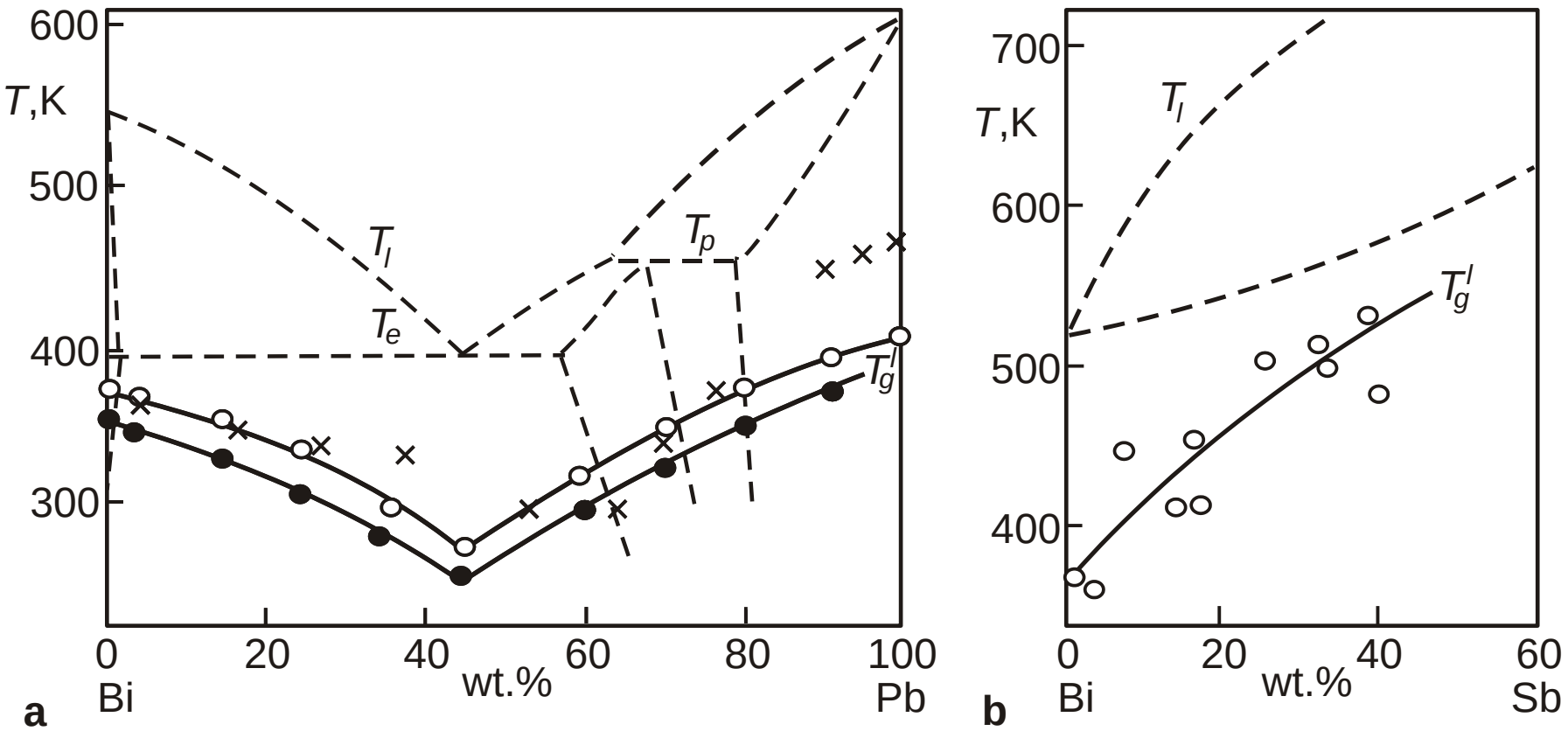

**Fig. 4.2** Dependence of the boundary temperature $T_g^l$ on the composition of Bi-Pb (**a**) and Bi-Sb (**b**) films on carbon (O) and $Al_2O_3$ (●) substrates [12] (× – data of [13]; dotted line – "*T* – *C*" diagrams for bulk samples)

Despite the fact that the crystallization temperature and, with it, the absolute value of supercooling of the metastable melt strongly depend on the concentration of the components, the value of relative supercooling turns out to be almost constant throughout the studied concentration range (Fig. 4.3). A somewhat decrease in relative supercooling, indicated by the data of the work [13], represented in Fig. 4.2 by the sign "×", is probably due to the low condensation rate used by the authors of [13] for the preparation of films. For this reason, their samples turn out to be not completely free of oxide impurities, which are known to reduce the achievable supercooling. Since lead oxidizes somewhat more actively than bismuth, the decrease in supercooling with increasing lead concentration seems natural but is not explained by the presence of a concentration dependence, but by external factors. The stability of the value of relative supercooling is also indicated by the authors of the work [14], who studied the crystallization of macroscopic (about 50 μm in diameter) particles of the Ag-Au alloy. The values of relative supercooling obtained in the work [14] are practically independent of the alloy concentration and are in the range of 0.170–0.183.

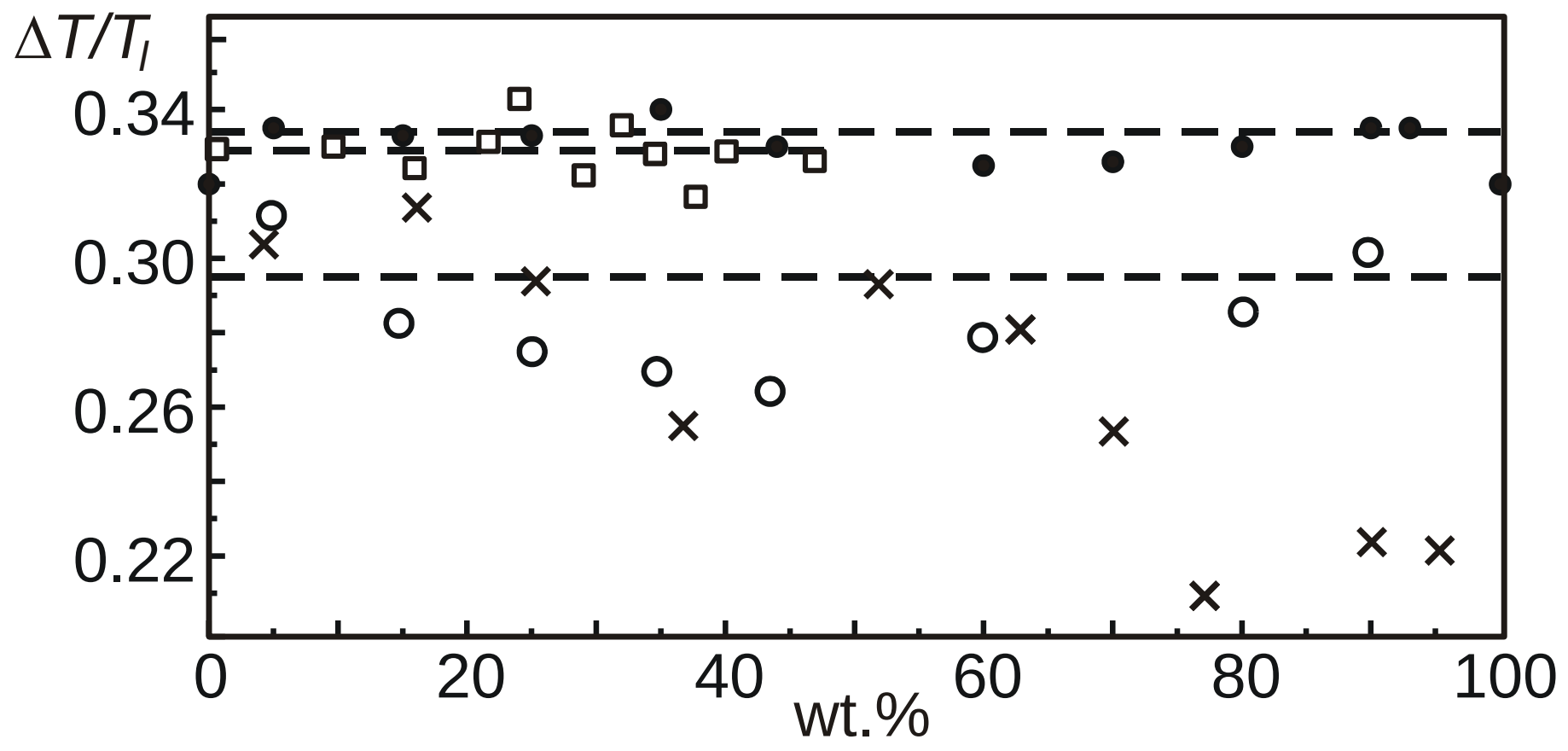

**Fig. 4.3** Relative supercooling dependence during the crystallization on the composition of the films Bi-Pb (●, ○) and Bi-Sb (□) (substrate material: ● – $Al_2O_3$, ○ and □ – amorphous carbon), × – data of the work [13]

The results of the studies of the effect of the substrate material on the values of relative supercooling, performed in the work [12], are shown in Fig. 4.4. Bi-Pb binary alloys of eutectic composition and samples containing 10 wt% lead were used as the fusible component under study. As can be seen from the graph, the behavior of such alloys generally coincides with that of single-component samples: relative supercooling increases with poorer wetting and, for contact systems with a wetting angle of more than 130°, reaches maximum values close to those, which are characteristic of homogeneous crystallization.

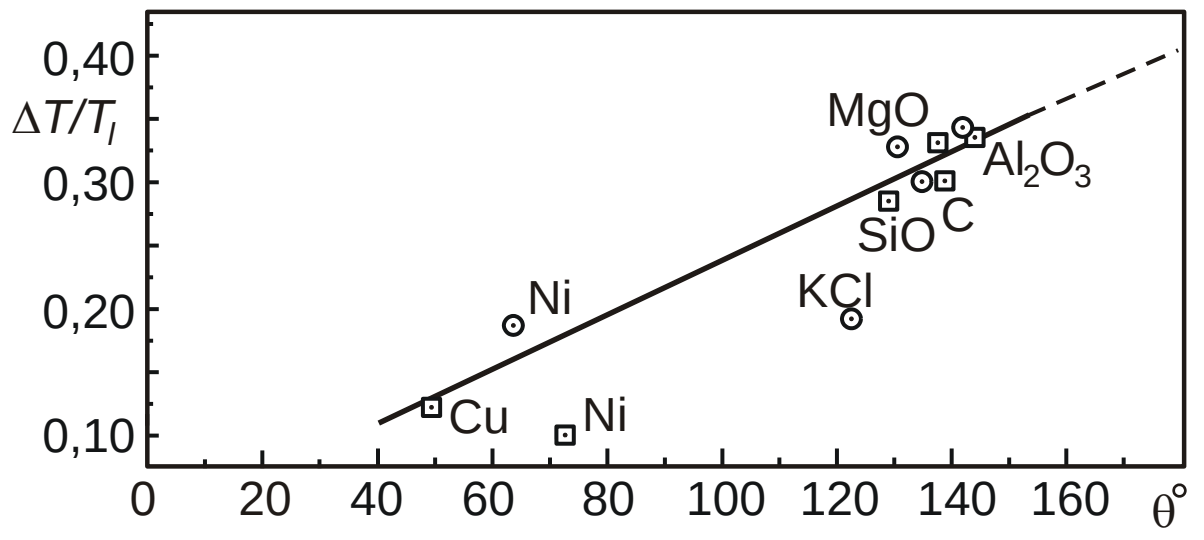

**Fig. 4.4** Relative supercooling during the crystallization of Bi-Pb films of the eutectic composition (⊙) and with 10 wt% Pb (⊡) versus the contact angle of wetting (substrate material is indicated on the graph) [12]

## 4.3 Application of the quartz resonator method for measuring the supercooling during alloy crystallization

Work [15] studied vacuum condensates of Bi-Sn alloys that are in contact with an amorphous carbon layer or one of their own components. To investigate phase transitions in alloys of eutectic composition, layers of bismuth and tin with a thickness of 100 nm were condensed onto an amorphous carbon film that had previously been deposited on the surface of a quartz resonator. After that, the quartz resonator with the deposited film was subjected to a series of heating-cooling cycles with simultaneous recording of temperature, resonant frequency, and quality factor. According to the results of earlier studies, the melting and crystallization of films of pure metals deposited on the working surface of a quartz resonator cause a sharp change in its frequency and quality factor [16]. The same effect is expected to be observed for films of alloys.

Fig. 4.5 shows, obtained in the work [15], temperature dependencies of the quality factor of a quartz sensor loaded with a two-layer Bi-Sn film on a carbon sublayer. The thicknesses of the bismuth and tin layers were 100 nm each. According to these data, melting in this contact pair is observed at a temperature of 139–143 °C. This range is close to the value of the eutectic temperature in the Bi-Sn contact pair.

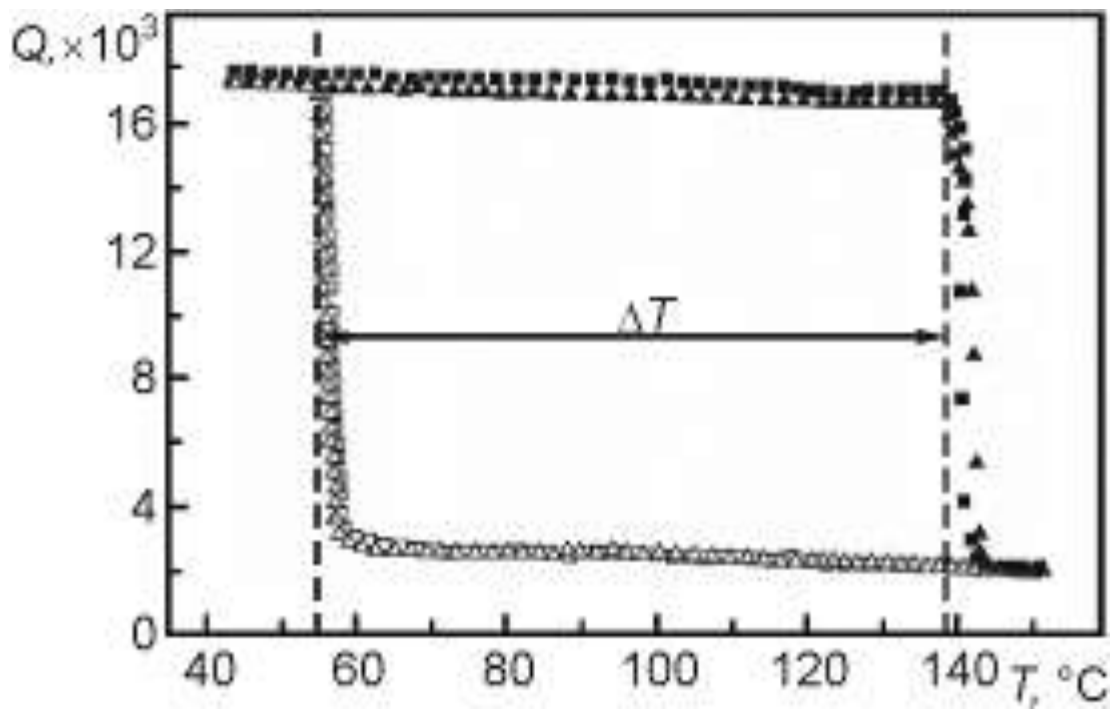

Fig. 4.5 Quality factor temperature dependence of the quartz resonator with a condensed Bi/Sn film. The thickness of each layer is 100 nm [15]

The fact of melting of Bi/Sn films after their heating to a temperature of 143 °C is confirmed by SEM studies (Fig. 4.6). After such heating, the films consist of particles of spherical shape, which clearly indicates that these particles were in a liquid state for some time.

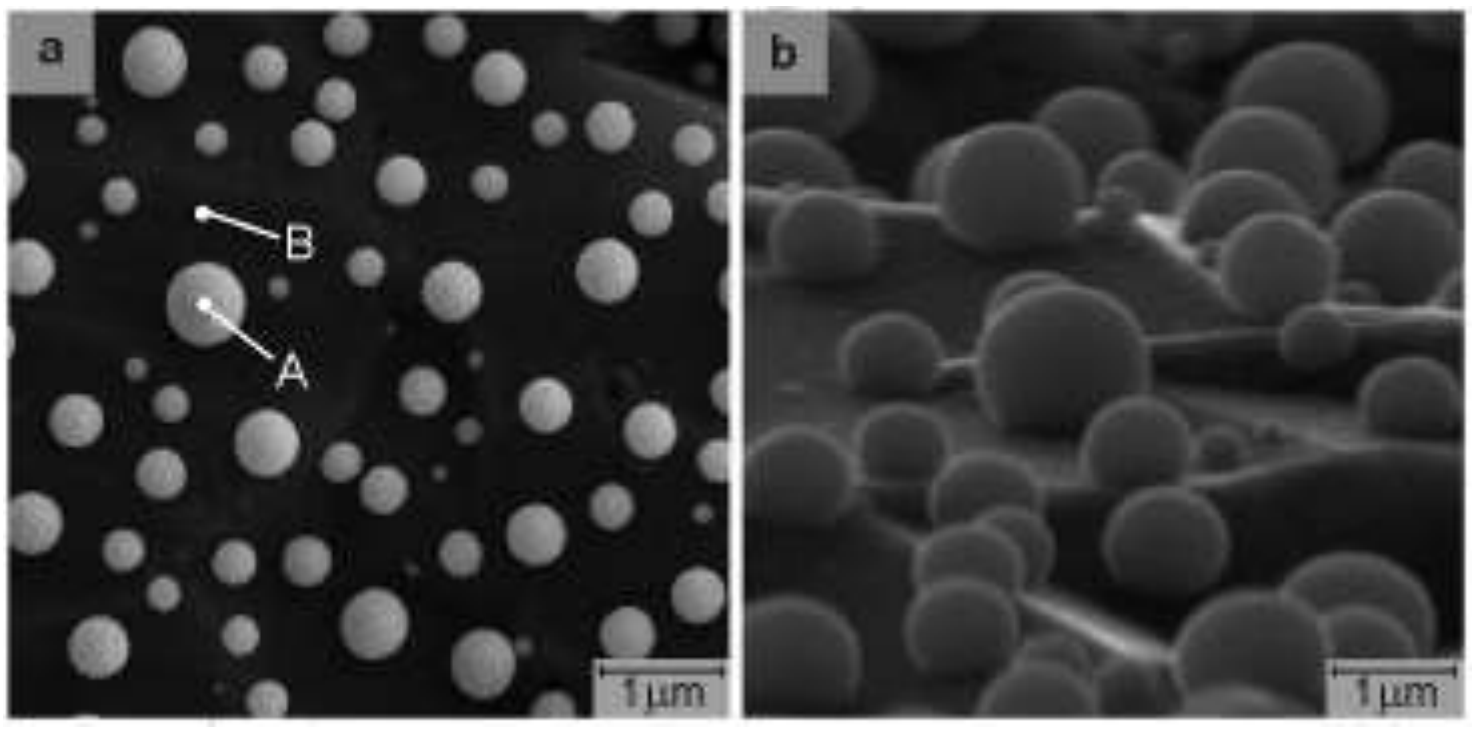

**Fig. 4.6** SEM images of Bi/Sn films (the thickness of each layer is 100 nm) after their melting on the amorphous carbon substrate. Image (**b**) was obtained at an angle of 75° to the substrate [15]

The crystallization of the supercooled eutectic alloy on an amorphous carbon substrate is observed in the range of temperatures 55–58 °C. Thus, the supercooling value in the studied system reaches 84 K, and the relative supercooling is η = 0.2.

It is important to note that the temperature dependencies of the quality factor shown in Fig. 4.5, already starting from the first heating cycle, have a reproducible nature and are repeated from cycle to cycle. Thus, all the processes connected with the formation of the eutectic alloy in the samples are completely finished during the first heating.

The results of the study of the temperature dependence of the quality factor in Sn/Bi films with thickness of layers of 20 and 200 nm, respectively, are shown in Fig. 4.7. Since in these samples tin is significantly less than is required in order for all bismuth to participate in the formation of the eutectic alloy, after melting the eutectic alloy will be in contact with crystalline bismuth.

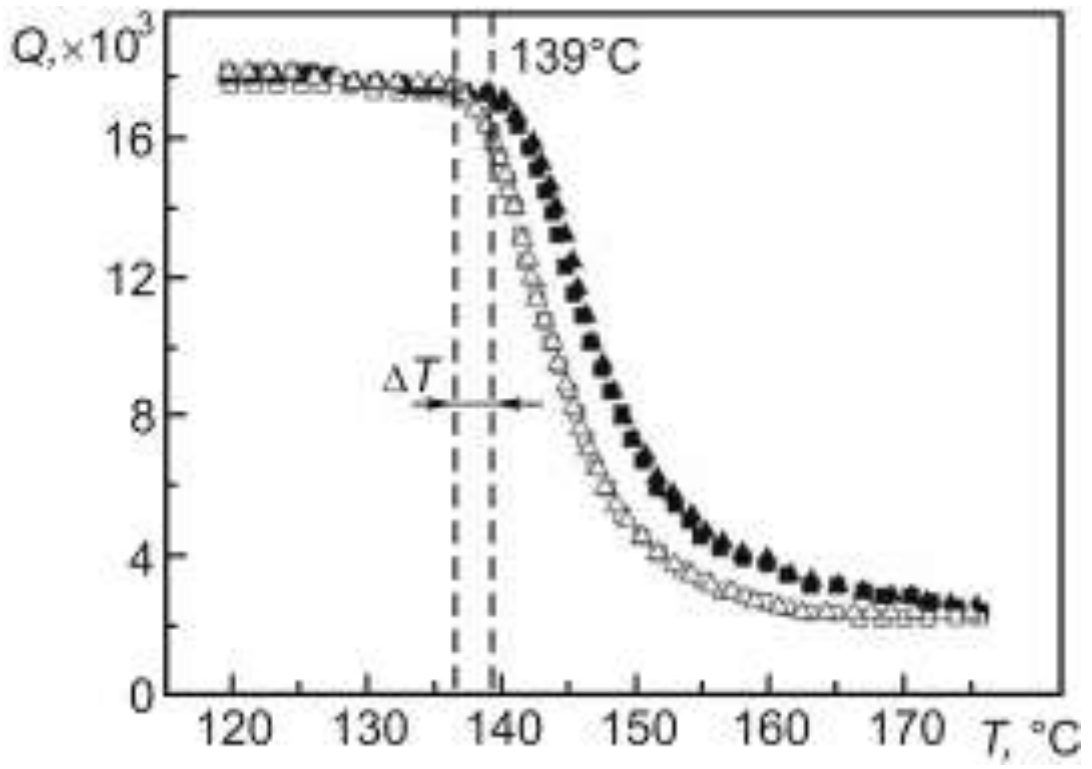

**Fig. 4.7** Quality factor temperature dependence of the quartz resonator on which Sn/Bi films were condensed with a thickness of 20 and 200 nm, respectively [15]

As can be seen from Fig. 4.7, melting-crystallization hysteresis is also observed in this contact pair, but the course of the curves of the quartz resonator's quality factor during its heating and cooling differs significantly from that, which occurs in the case of contact between the eutectic alloy with a carbon film. If we accept the obvious assumption that the decrease in the quality factor of the quartz sensor correlates with the liquid phase content in the film, then the behavior of the heating curve in Fig. 4.7 fully corresponds to the melting dynamics of an alloy that is far from the eutectic of the composition. The melting in these samples begins at a temperature of 139 °C and continues during further heating. The amount of liquid grows with increasing temperature, and at 170 °C it turns out to be sufficient to almost completely stop the oscillations of the resonator. At this temperature, the composition of the melt is determined by the phase diagram and is very different from the eutectic value.

During cooling, the liquid phase turns out to be thermodynamically unstable and crystallizes with the release of excess bismuth. The melt of the eutectic composition crystallizes last. The obtained supercooling value corresponding to the crystallization of the eutectic Sn-Bi alloy, that is in contact with bismuth, is 3 K, with a relative supercooling of η = 0.007.

As a result of TEM studies (Fig. 4.8), it was found that in the process of the above-described thermal effect, particles of rounded shape appear in the sample, the composition of which is close to the eutectic value (point A in Fig. 4.8). These particles are located on the surface of the layer containing 91% Bi (point B in Fig. 4.8).

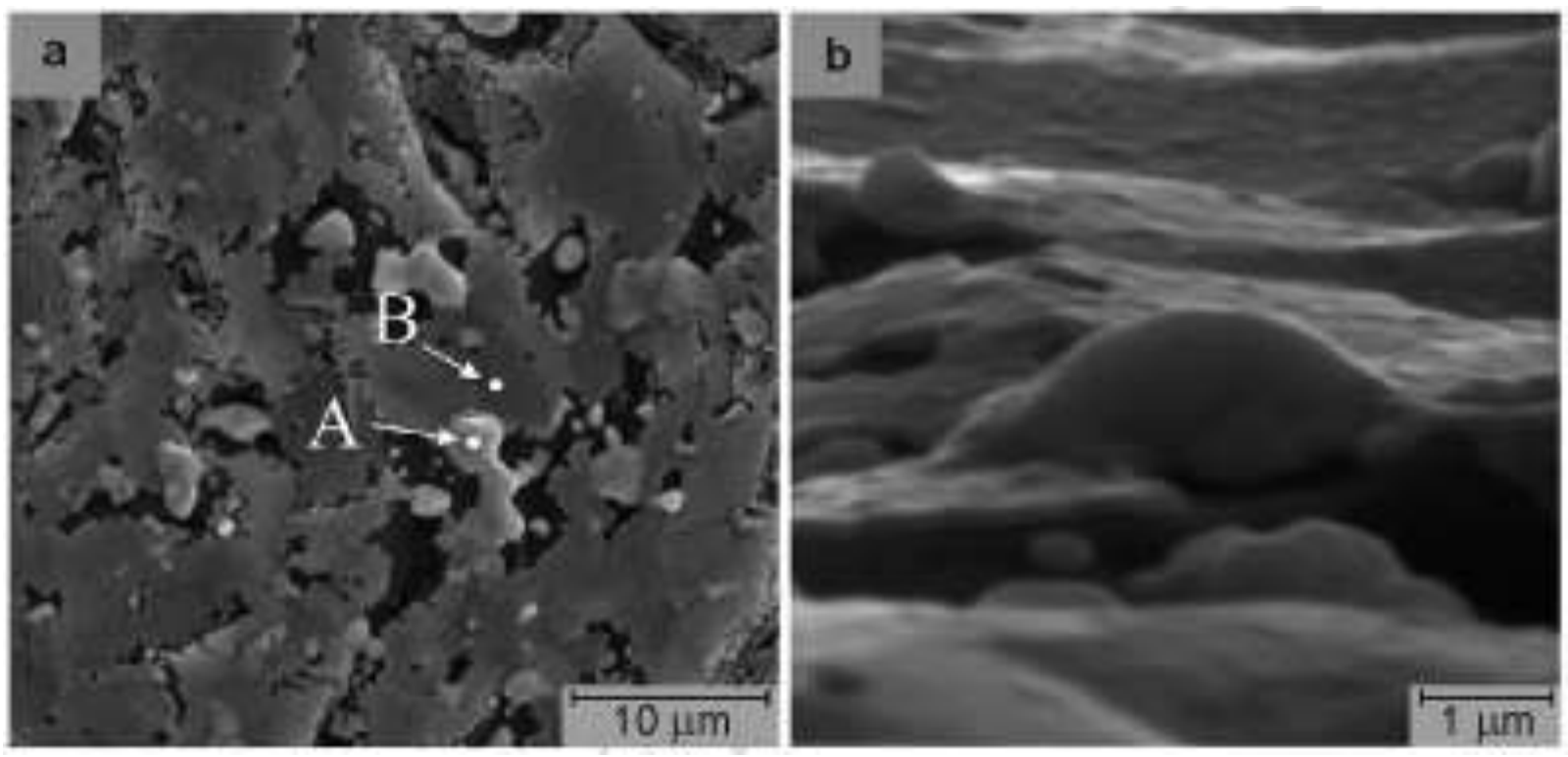

**Fig. 4.8** SEM images of Sn/Bi films (the mass thicknesses of component layers are 20/200 nm, respectively). Image (**b**) was obtained at an angle of 75° to the optical axis of the microscope [15]

It should be noted that SEM studies show that the particles observed in the sample can be divided into two types. They differ in size and in the wetting angle that the rare islands form with the substrate. Particles, the size of which is 2–3 µm, have a contact angle of 39°. Particles, the average size of which turns out to be much less – around 0.3 µm, have a contact angle which is 15°. The presence of two types of particles can be explained by the partial dispersion of the film. Indeed, when the sample is heated (Fig. 4.9a) in accordance with the mechanism described in the work [17], tubercles are formed on the film surface, which are the "nuclei" of large particles (Fig. 4.9b). According to the concept presented in the article [17], a further increase in temperature causes the redistribution of the substance of the upper layer (Sn), which leads to an increase in tubercles and thinning of the tin layer (Fig. 4.9c). As a result of the melting of such a two-layer structure, it becomes possible to form large and small particles of eutectic composition on the surface of the bismuth film (Fig. 4.9d).

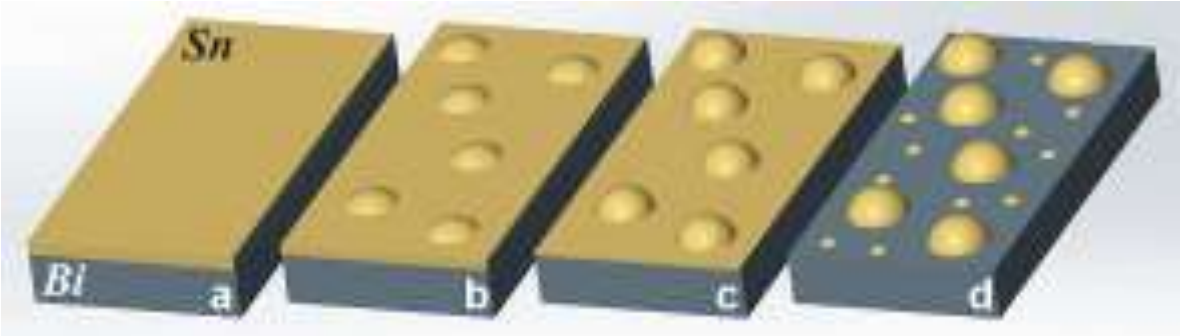

**Fig. 4.9** Formation model of particles of two types on the surface of the sample during its annealing

Large particles containing a significant amount of bismuth are likely to interact not only with the remains of the bismuth layer but also with the amorphous carbon substrate. At the same time, small particles use only a small part of the bismuth contained in the lower sublayer for their formation. Therefore, they are in contact exceptionally with the bismuth layer, which is wetted by the melt much better than carbon. Thus, only those particles of the eutectic alloy that have formed on the surface of a continuous bismuth layer and for which a contact angle of 15° is characteristic, allow us to observe supercooling of the eutectic Bi-Sn alloy on the bismuth layer.

Similar studies were performed for bilayer Bi/Sn films containing excess tin on a carbon sublayer. The thicknesses of the bismuth and tin layers were 40 and 170 nm, respectively. In this case, the tin layer condensed first, so that the melt of the eutectic concentration was in contact with the tin-based solid solution. The ratio of components was chosen so that the concentration of Bi (24 wt%) exceeded the maximum solubility of bismuth in tin. In such a contact system, supercooling is also observed during the crystallization of eutectic particles in contact with tin, which is 7 K. As shown by SEM studies (Fig. 4.10), in samples, the particles whose sizes are distributed in the range of values from 0.5 to 5 µm, are observed, and the contact angles, which can be determined by the method of oblique observation, are from 10 to 30°. The above considerations allow us to conclude that the case of "eutectic Bi-Sn melt on the surface of a solid solution" corresponds to contact angles equal to 10°. This is close to the value obtained in [18].

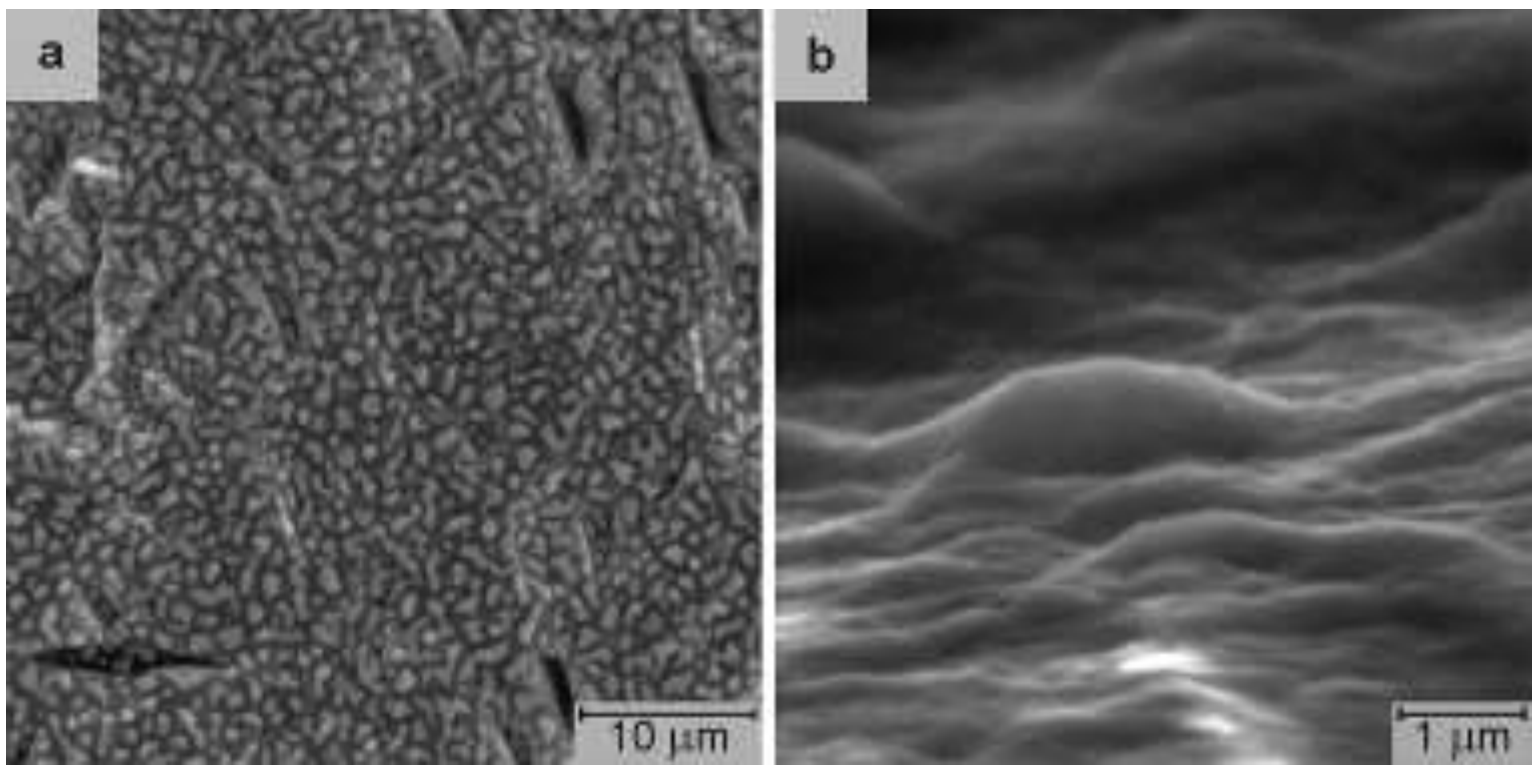

**Fig. 4.10** SEM images of Bi/Sn films (the thicknesses of component layers are 40/170 nm, respectively). Image (**b**) was obtained at an angle of 75° to the optical axis of the microscope

Based on the experimental data obtained in [15], assuming that the crystallization of the eutectic melt on an amorphous carbon film is homogeneous, the interfacial energy of this contact system was estimated. For this purpose, expression (1.7) [7] was used in the work [15], transformed into the following form:

$$(\sigma_{sl})^3 = \frac{3k \ln N}{16\pi} \left( \frac{\Delta T}{T_s} \right)^2 \lambda^2 T_g, \tag{4.1}$$

where *k* is the Boltzmann constant; *N* is the number of atoms in a liquid particle; *λ* is the latent heat of phase transition normalized per unit volume.

Substituting the supercooling value determined in the study and the latent heat of melting of $5 \cdot 10^6$ J/m$^3$ (49 J/g [19]) into the expression (4.1), the authors of the work [15] obtained a value of the crystal-melt interfacial energy, which is equal to 46.5 mJ/m$^2$, which expectedly turned to be lower than the values for pure components (54.4 and 59 mJ/m$^2$ for bismuth and tin, respectively).

The knowledge of the interfacial energy of the crystal-melt allows, using expression (1.4), which is valid for both homogeneous and heterogeneous crystallization, to estimate the value of the radius of the critical nucleus of the new phase. For Bi/Sn films of eutectic composition on a carbon substrate, the value $r^*$ = 1.1 nm was obtained.

The values of the critical radii estimated using the relation (1.4) for the cases of crystallization of the eutectic melt which has contact with a layer of bismuth and tin are 30 and 13 nm, respectively. In the work [15], on the basis of the experimental data obtained, the energy of the "eutectic melt - crystalline bismuth" interphase boundary was also estimated. For this purpose, based on the consideration of the balance of surface forces on the lines of triple contact "nucleus – melt – substrate" and "melt drop – vacuum – substrate" (see Fig. 1.4), the following expression was obtained:

$$\sigma_s^{Bi-Sn/Bi} = \sigma_s^{Bi} - \sigma_{sl} \cos\psi - \sigma_l^{Bi-Sn} \cos\theta, \tag{4.2}$$

where ψ is the contact angle between the crystal nucleus of the eutectic phase and the solid substrate (in this case, a bismuth layer); $\sigma_l^{Bi\text{-}Sn}$ is the surface energy of the melt under study; $\sigma_s^{Bi}$ is the surface energy of the solid substrate; θ is the wetting angle of the substrate by the melt of the eutectic composition.

It should be noted that to date, there are no methods that allow direct measurement of the angle ψ. At the same time, this angle, as well as the wetting angle θ, which is reliably determined in experiments, is a quantitative measure of the influence, that the substrate on the crystallization particularities has. Therefore, to estimate the interfacial energy $\sigma_s^{Bi\text{-}Sn/Bi}$ in the work[15], it was assumed that ψ = θ as a first approximation. The value of $\sigma_l^{Bi\text{-}Sn}$ was obtained by extrapolating the data from the works [20, 21] to the eutectic temperature. The surface energy of the bismuth layer, according to [22], was assumed to be $\sigma_s^{Bi}$ = 521 mJ/m$^2$. The value of the interfacial energy of the interface "eutectic melt – crystalline bismuth" estimated in this way was $\sigma_s^{Bi\text{-}Sn/Bi} \approx 50$ mJ/m$^2$.

The experimental study of the stability of the liquid phase in bismuth films with a thickness of 3-100 nm, which were located between germanium layers, was carried out in the work [23] by *in situ* TEM using the method of selected area electron diffraction (SAED). The studied three-layer Ge/Bi/Ge film system was formed by sequential deposition of components in a high vacuum and acted as a model of the "nanoscale object – matrix" system.

The study found that the nature of the temperature evolution of the crystal structure was the same for all Ge/Bi/Ge samples with bismuth film thicknesses in the range from 40 to 5 nm. The sequence of electron diffraction patterns of the Ge/Bi/Ge film with a thickness of each layer of 10 nm at different temperatures is shown as an example in Fig. 4.11. A set of diffraction rings corresponding to polycrystalline bismuth as well as to the diffuse halo of amorphous germanium was clearly observed in the diffraction pattern of the sample at room temperature (Fig. 4.11a). The intensity of the diffraction lines of the bismuth was uniform throughout the ring, and no signs of texture were observed, which allowed us to state that the bismuth layer consists of small grains randomly oriented with respect to the electron beam. The most intense diffraction lines (102) of Bi and (111) of Ge overlap in the SAED patterns, so lines (220) and (311) with relative intensities of 0.57 and 0.39, respectively, were used to unambiguously monitor the phase state of Ge, and line (110) was used for Bi.

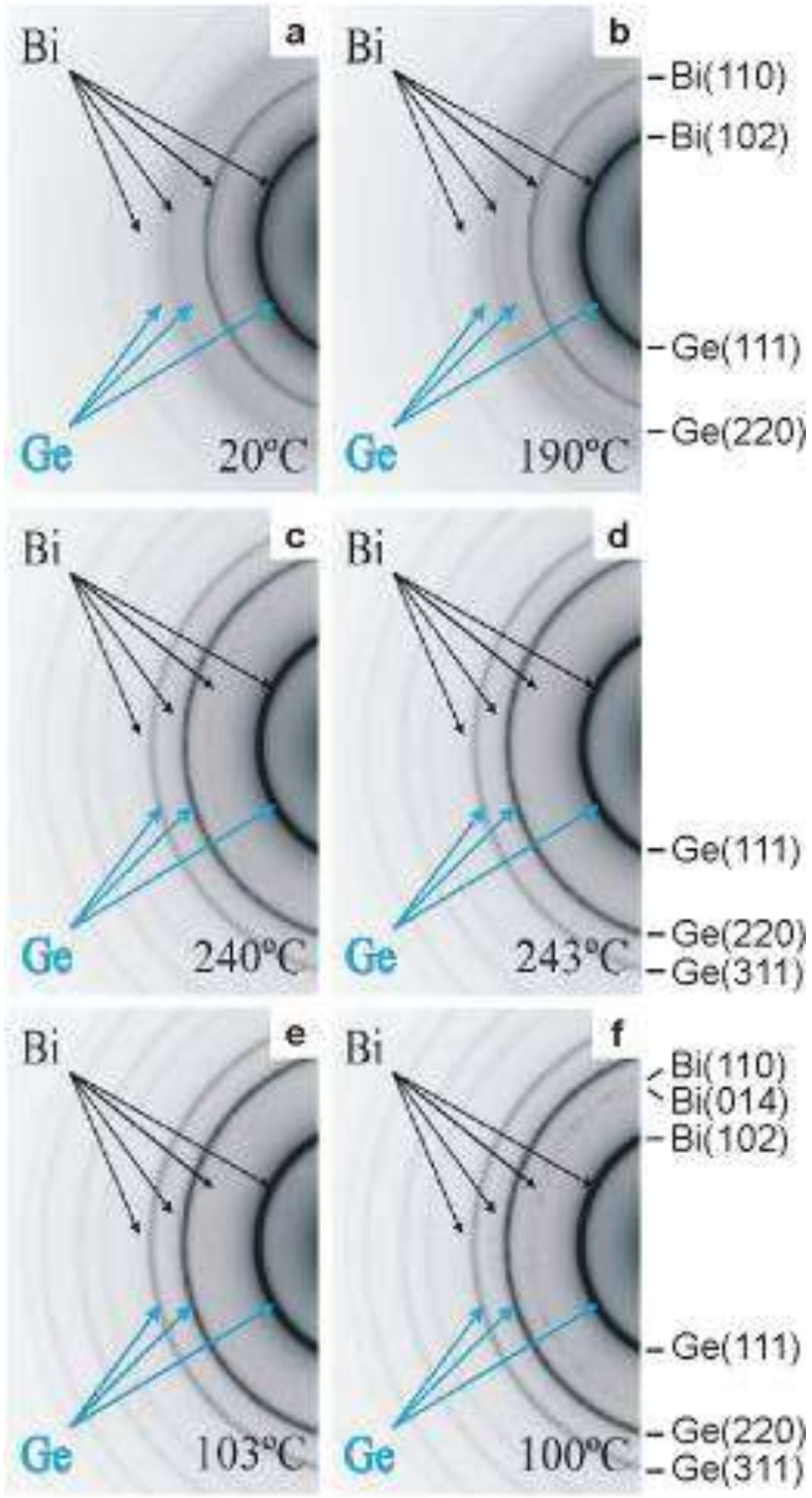

**Fig. 4.11** SAED images of Ge/Bi/Ge films with a thickness of each layer of 10 nm at different temperatures [23]

In the process of the temperature increase, the diffraction pattern remained unchanged up to about 190 °C, when the diffraction rings of crystalline germanium began to appear along with a-Ge halo (Fig. 4.11b). When approaching the melting point of Bi, the intensity of the germanium rings increased and that of the bismuth rings decreased, and at a temperature of 243 °C, the remains of crystalline bismuth completely disappeared. Thus, at a temperature of 240 °C (Fig. 4.11c), the diffraction pattern barely shows rings of crystalline bismuth, and already at 243 °C, no reflexes from crystalline bismuth are detected in the diffraction pattern at all (Fig. 4.11d). It should be emphasized that despite the fact that melting in the system occurred in a certain temperature range, when determining the eutectic temperature, it was assumed that the entire bismuth film participated in the formation of the liquid phase, i.e., the temperature $T_e$ of eutectic melting in a system with a characteristic layer thickness of 10 nm was assumed to be equal to 243 °C. When the melting point was reached, the heating was stopped and the sample was cooled to room temperature in a controlled manner. The crystallization of the bismuth-based eutectic was monitored by the appearance of the first reflexes of crystalline bismuth (Fig. 4.11e). The crystallization process occurred in the temperature range of 3–10 K and led to complete crystallization in this system (Fig.

4.11f). During a single heating-cooling cycle of a sample with a characteristic size of 10 nm, the melting points of the eutectic $T_e \approx 243$ °C and its solidification $T_g \approx 103$ °C were determined. These temperatures were determined in a similar manner for Ge/Bi/Ge samples with bismuth layer thicknesses ranging from 3 to 40 nm. The measurement results are shown in Fig. 4.12.

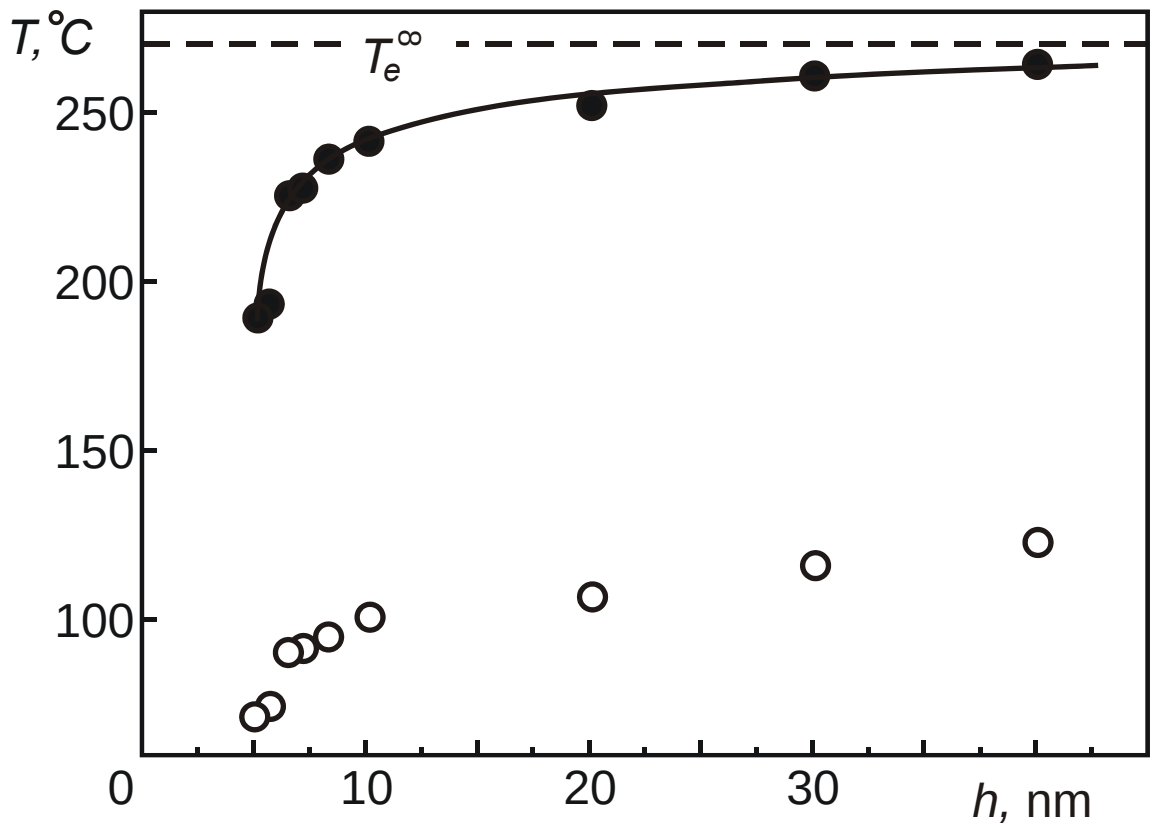

**Fig. 4.12** Size dependencies of the eutectic temperature $T_e$ (●) and crystallization temperature $T_g$ (○) in Ge/Bi/Ge films (*h* is the characteristic size of the system during its studying, i.e. the thickness of the bismuth film)

The study of the size dependence of the melting and crystallization temperatures in the Ge/Bi/Ge system was also carried out using the method of a quartz resonator [24]. Initially, the melting and crystallization temperatures were determined in a three-layer Ge/Bi/Ge film system with a thickness of 50 nm for each layer. This configuration of the film system at the specified layer thickness was sufficient to study the stability of the liquid phase without taking into account the influence of the size factor [6].

Fig. 4.13 shows the temperature dependencies of the quality factor of the resonator for the case of an unloaded quartz plate and for the first 5 heating-cooling cycles of quartz with a sample.

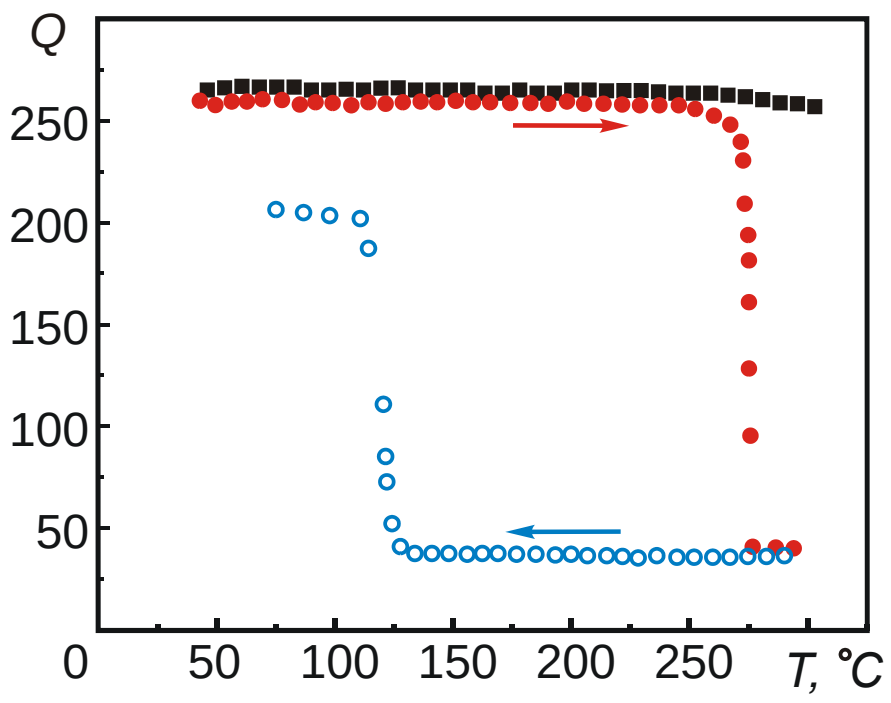

**Fig. 4.13** Temperature dependencies of the resonator quality factor for the unloaded quartz plate (■) and for the first 5 cycles of heating-cooling of quartz with Ge(50 nm)/Bi(50 nm)/Ge(50 nm) sample. Filled points correspond to heating, unfilled – to cooling [24]

The formation of the liquid phase occurred at a temperature of about 270 °C (Fig. 4.13). This corresponded to the eutectic temperature in the system. When the system was cooled after reaching a temperature of 120 °C, a sharp increase in the amplitude of the resonator oscillations was observed due to the crystallization of the liquid phase in the system. It should be noted that the value of supercooling of the liquid phase in three-layer systems significantly exceeds (by almost 30 K) the value characteristic of two-layer samples. This effect can be explained by the presence of additional mechanical stresses that arise during the crystallization of liquid particles of bismuth with an increase in their volume in conditions of limited space between two continuous germanium films. Already in the second heating-cooling cycle, melting occurs in a much narrower temperature range compared to the first cycle (Fig. 4.13). This is due to the enlargement of bismuth particles to a macroscopic size during the first heating-cooling cycle. At the same time, the crystallization of the liquid phase occurs at the same temperature as in the first heating-cooling cycle, i.e., at 120 °C. Thus, the value of supercooling of the liquid phase in the macroscopic three-layer Ge/Bi/Ge system was approximately 150 K.

Next, the main attention was paid to the study of the effect of the size factor on the melting-crystallization phase transitions in this binary system. For this purpose, a series of three-layer samples were formed with component layer thicknesses, varying from 5 to 50 nm. Fig. 4.14 shows typical temperature dependencies of the quartz quality factor for some thicknesses of nanodispersed systems. The general view of the corresponding temperature dependencies coincides with the dependence characteristic of macroscopic samples (Fig. 4.13), which also present regions of sharp drop and increase in the quality factor, which are the result of melting and crystallization of the system.

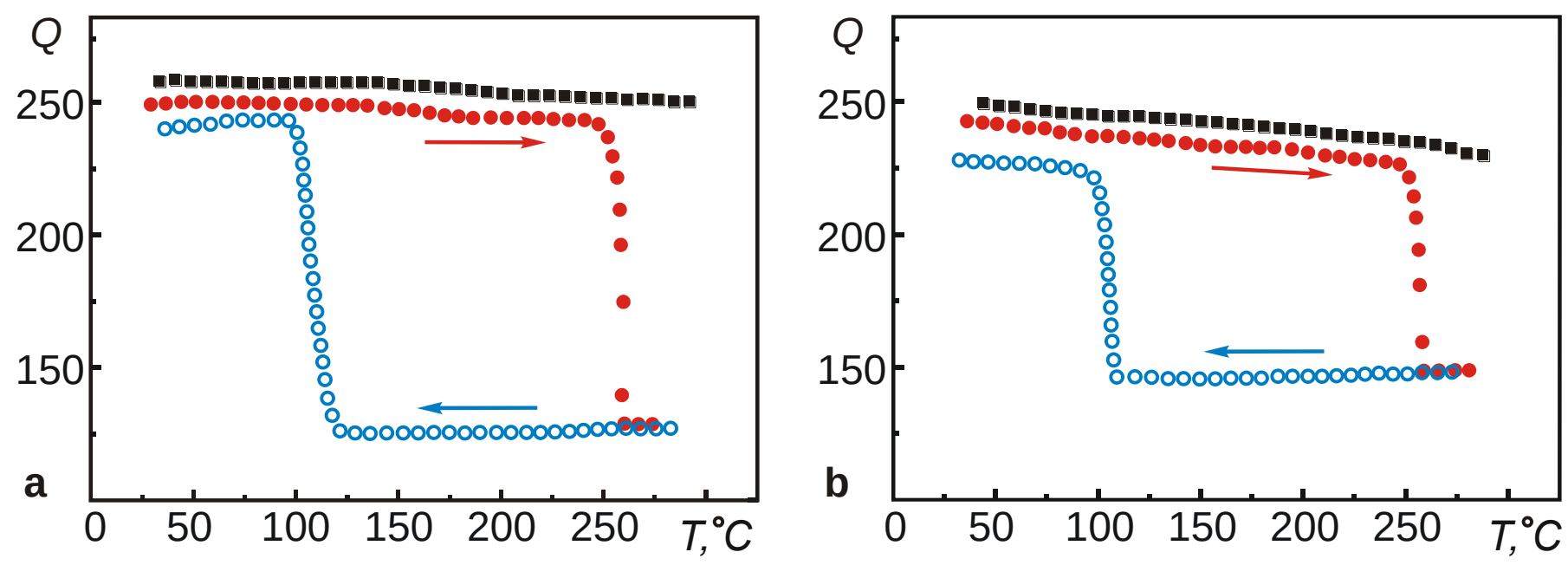

**Fig. 4.14** Temperature dependencies of the resonator quality factor for the unloaded quartz plate (■) and for the heating-cooling cycle of quartz with a sample: **a** – Ge(20 nm)/Bi(20 nm)/Ge(20 nm), **b** – Ge(15 nm)/Bi(15 nm)/Ge(15 nm) [24]

It can be clearly seen that the eutectic temperature in the system is a size-dependent value and decreases with decreasing thickness of the film system layers. In particular, the melting point in the system with a characteristic thickness of 20 nm was approximately 157 °C (Fig. 4.14a), which is 13 K lower than for macroscopic samples. In addition, with a decrease in the characteristic size of the samples, a decrease in the crystallization temperature in these systems was also observed. For a system with a bismuth film thickness of 20 nm, the eutectic crystallization occurred at a temperature of ≈106 °C, while for samples with a thickness of layers of 50 nm, it was 120 °C (Fig. 4.13). The data obtained for some samples are summarized in Table 4.1.

**Table 4.1** $T_e$ melting and $T_g$ crystallization temperatures in selected samples

| Ge/Bi/Ge system, nm | $T_g$, °C | $T_e$, °C |
|---|---|---|
| 50/50/50 | 120 | 270 |
| 25/25/25 | 108 | 265 |
| 20/20/20 | 105 | 258 |
| 15/15/15 | 100 | 255 |
| 10/10/10 | 97 | 244 |

When studying layered systems with the thickness of layers of components less than 10 nm, certain difficulties arose. First, a decrease in the thickness of the germanium film led to a loss of continuity of samples when heated to the melting point. Therefore, in subsequent experiments, the thickness of germanium films was kept constant and amounted to 10 nm. Secondly, at a thickness of bismuth films less than 10 nm, the amount of liquid phase that is formed during the eutectic melting of the system was insufficient to cause noticeable changes in the amplitude or quality factor of the quartz resonator. Taking into account that as a result of melting, the bismuth film shatters and collects into droplets the size of which is approximately 12 times larger than the initial film thickness, the indicated "insufficiency of the liquid phase" is consistent with numerical estimates, according to which the minimum thickness of the liquid layer that can affect the characteristics of quartz is about 100 nm. Therefore, in subsequent experiments, the samples represented multilayer systems in which films of the fusible component alternated with 10 nm germanium films, with their total thickness exceeding 10 nm.

At this stage, the samples were studied, which represented a layered system in which bismuth films were located between 10 nm thick germanium films, as shown in Fig. 4.15.

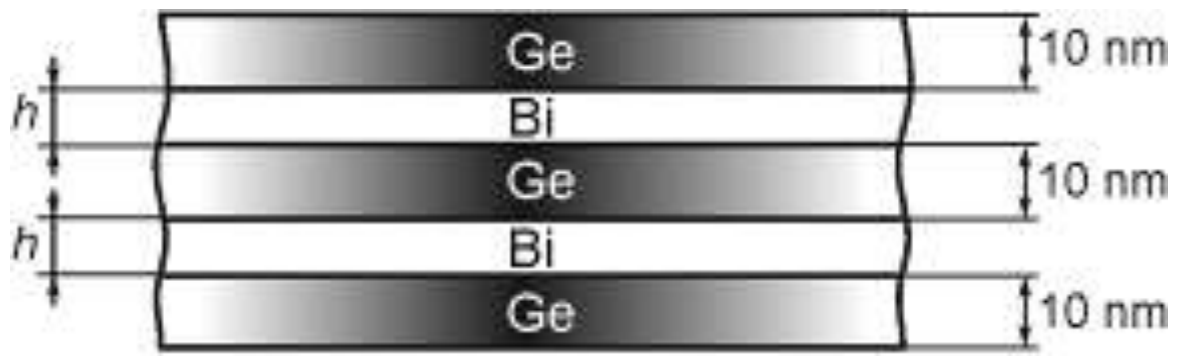

**Fig. 4.15** Schematic representation of the multi-layered Bi-Ge film system [24]

Since the thickness of the germanium films was unchanged, in the following, when describing the object of study, we will operate with the value of the thickness of one bismuth layer (the characteristic size of the system – $h$), as well as the number of bismuth layers. For example, when we talk about two layers of bismuth with a thickness of *h* – 5 nm, we mean the Ge/Bi/Ge/Bi/Ge system.

The initial system was chosen with two bismuth layers of 7 nm. This choice was due to the fact that when reducing the characteristic size of the system, the total thickness of bismuth should be greater than 10 nm, in accordance with the considerations given above. However, in the course of research, it turned out that the reaction of quartz to melting-crystallization phase transitions in this system is at the level of measurement error. This is probably due to the fact that the attenuation of a sound wave on a periodic structure (multilayer film) is not a linear function of the number of layers. Therefore, it was decided to increase the number of layers until the changes in the characteristics of quartz become the ones that can be unambiguously identified. Below, in Fig. 4.16, the temperature dependence of the quality factor of a quartz resonator loaded with a system with 10 layers of bismuth of 7 nm each is presented.

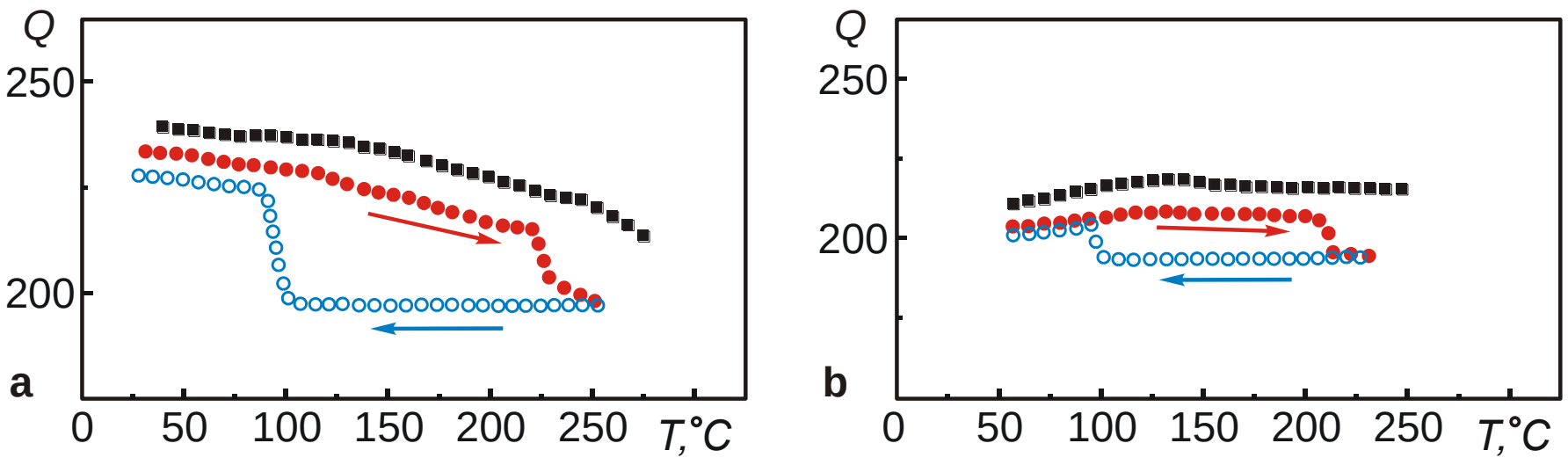

**Fig. 4.16** Temperature dependence of the quality factor for the unloaded quartz resonator (■) and for loaded with a film system: **a** – (10nmGe/7nmBi)×10, **b** – (10nmGe/5,7nmBi)×20 [24]

It has been shown that melting and crystallization in the given system are clearly identified by the change in the quartz quality factor. Thus, the melting point amounted to 225 °C, and crystallization was observed at 95 °C. During the study of samples with a characteristic thickness of the fusible component *h* = 5.7 nm, the number of bismuth layers was increased to 20. The graph of the corresponding temperature dependence of the quartz quality factor is shown in Fig. 4.16b. The melting point in this system amounted to approximately 212 °C, and crystallization was observed at a temperature of 85 °C.

The results of the study of the size dependence of the temperatures of melting-crystallization phase transitions in layered Bi-Ge systems are shown in Fig. 4.17.

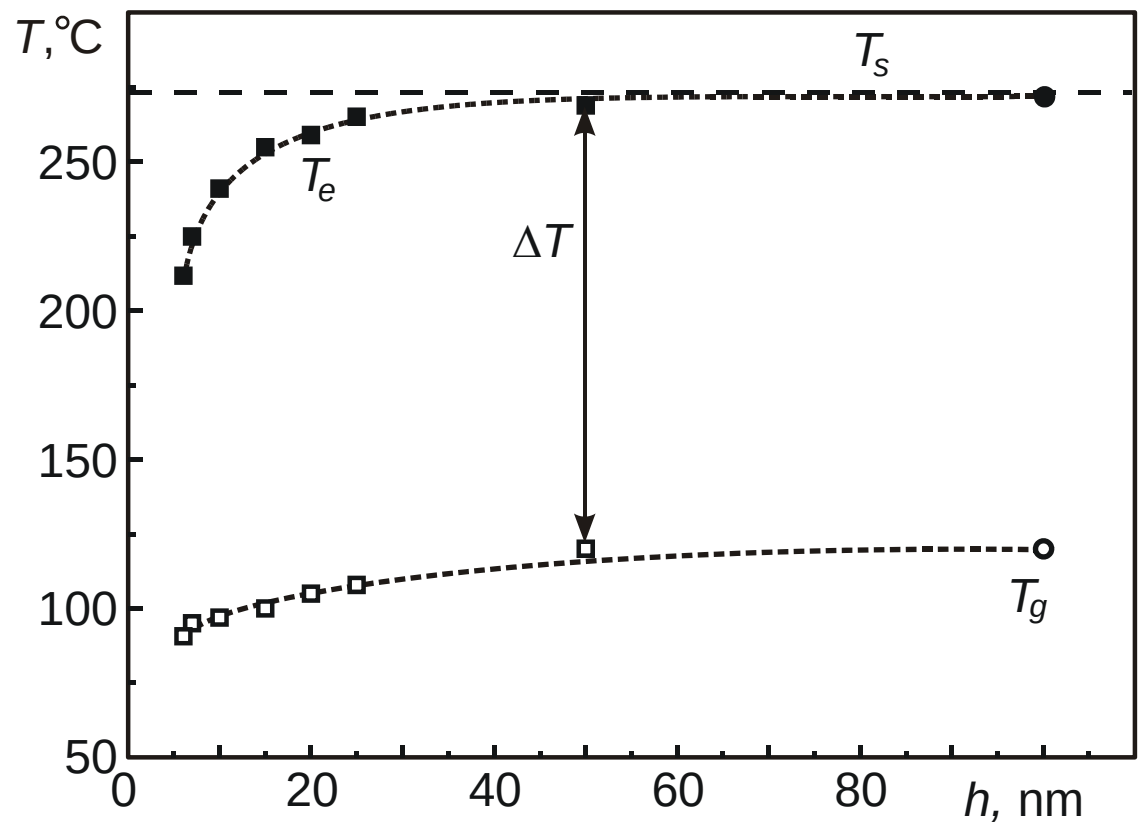

**Fig. 4.17** Size dependence of the melting points of $T_e$ (■) and crystallization of $T_g$ (□) in a Bi-Ge layered film system (•, ○ – correspond to the melting and crystallization temperatures of Bi-Ge bulk samples according to the data of [6]) [24]

## 4.4 Supercooling during crystallization of alloys in layered film structures

One of the results of the above study of supercooling of eutectic alloys [15] is that fusible particles of different sizes are observed in the samples, forming different contact angles with more refractory layers. Thus, the differences in the morphology of the samples should certainly affect the particularities of their crystallization. Below are the results of the study [25, 26] of supercooling in (Bi-Sn)/Cu films, which were subjected to different thermal effect, which largely determines the morphology of the samples.

Samples for the studies were obtained by sequential condensation of the components in a vacuum of $10^{-6}$–$10^{-7}$ mm Hg. First, a layer of amorphous carbon was applied to the chips of KCl single crystals, on which a 30 nm thick copper film was deposited. After that, the condensation of bismuth and tin was performed sequentially. The deposited bilayer Sn/Bi film contained 7 wt. % tin and the layer thicknesses were chosen from the condition that the total copper content in the sample was 40 wt. %.

Two series of experiments were conducted to evaluate the effect of preliminary heat treatment on the particularities of supercooling in binary film systems. In the first of them, all components of multilayer films were deposited on the substrate at room temperature, right immediately after the deposition of the previous layer without the implementation of any additional annealing. The second series of experiments was performed on samples that, after condensation of the Bi/Cu film, were subjected to heating to provide melting of bismuth. In this case, tin was deposited only after the crystals with the films had completely cooled to room temperature. The melting and crystallization temperatures were determined by *in situ* electron diffraction studies during heating-cooling cycles carried out directly in the column of the electron microscope.

Electron diffraction patterns obtained from samples that were not annealed during the obtaining process are shown in Fig. 4.18. As expected, reflections from the crystallographic planes of bismuth are present in the electron diffraction patterns up to a temperature of about 270 °C, at which the melting of the fusible alloy is completed. Due to its low concentration, tin lines are present only as traces in the electron diffraction patterns. The crystallization of the fusible alloy begins at a significant supercooling: bismuth lines appear only at a temperature of about 200 °C. The value of supercooling in this contact system amounts to 70 K (relative supercooling η = 0.13), which is close to the values obtained in Cu/Bi/Cu, Ag/Bi/Ag, and Mo/Bi/Mo films in which bismuth was deposited on a substrate at room temperature [27, 28, 29]. It should be noted that, just as in these systems, the crystallization of Bi + 7% Sn alloys, which are in contact with copper also has an avalanche-like nature.

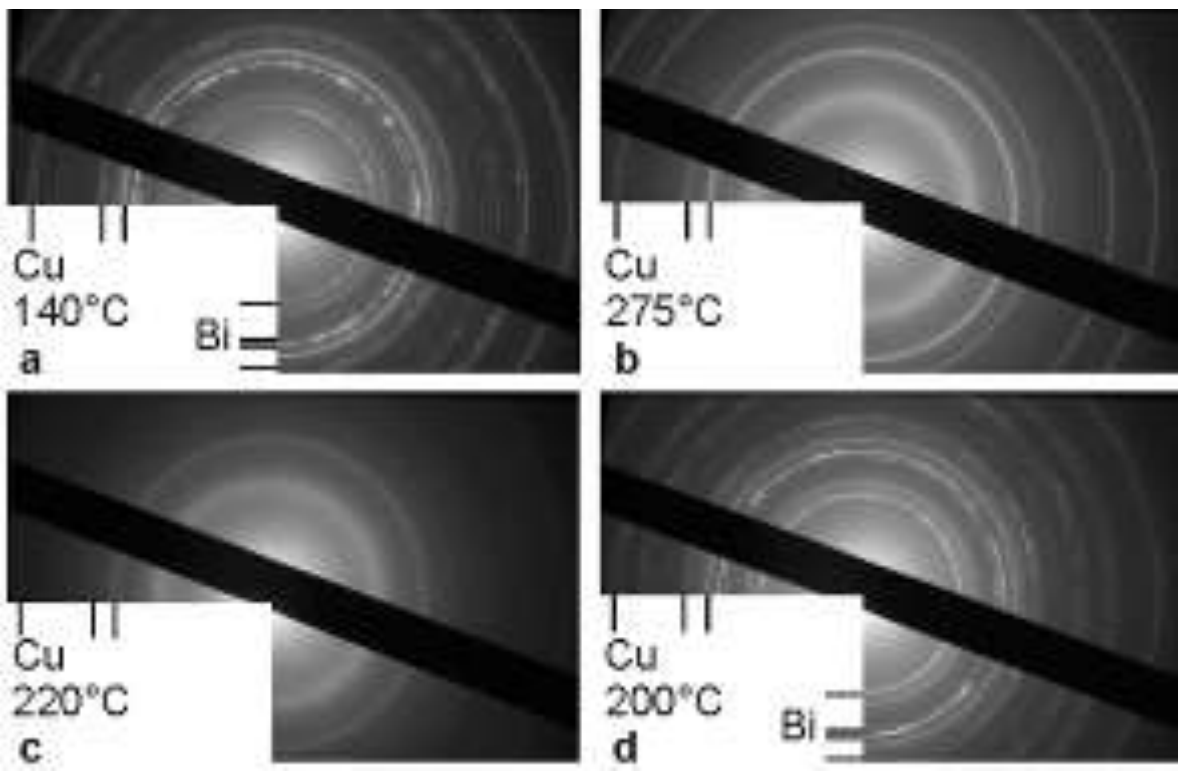

**Fig. 4.18** Electron diffraction patterns of films (Bi + 7% wt. Sn)/Cu, which were not annealed until the condensation termination. Sample temperatures are shown in the images. Patterns (**a**) and (**b**) correspond to heating, (**c**) and (**d**) – to cooling. The images were taken in the third cycle of heating-cooling of the samples [25]

It should be noted that the diffraction rings obtained from the crystallographic planes of bismuth in the Bi + 7% Sn alloy have a dotted nature. This indicates the formation of a fusible component in the samples of large-crystal structures. Comparison of the electron diffraction patterns of Figs. 4.19a and 4.19b indicates that the appearance of such objects occurs after the first heating cycle during the subsequent crystallization. For example, the electron diffraction pattern obtained near the melting point during the first heating of the films is typical of polycrystalline samples and indicates their nanosized structure. At the same time, after melting and subsequent crystallization, the look of the electron diffraction pattern changes qualitatively, and it already has a look that indicates the large-crystal nature of the samples.

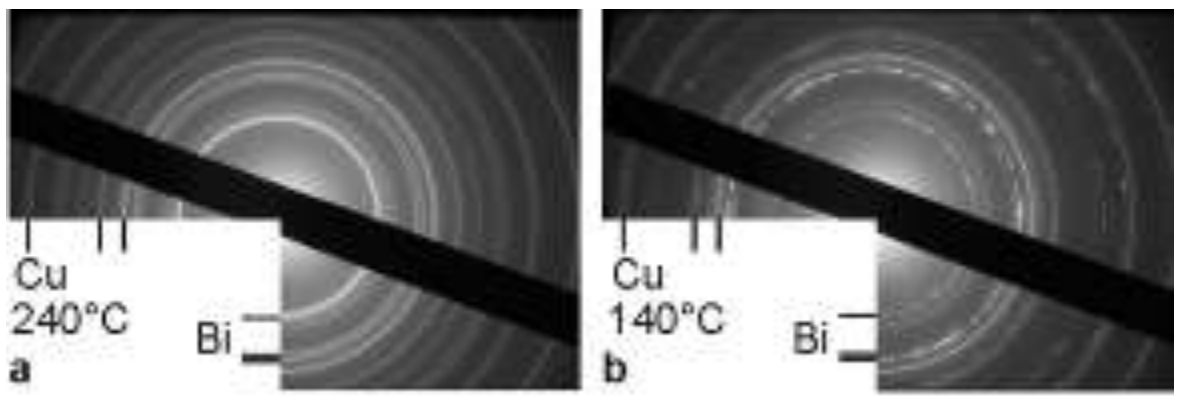

**Fig. 4.19** Electron diffraction patterns obtained in the first heating-cooling cycle of films (Bi + 7% wt. Sn)/Cu, which were not annealed during the condensation process; image (**a**)

corresponds to the first heating: (**b**) – represents cooling of the sample after the alloy melting [25]

The change in the microstructure of the films as a result of five heating-cooling cycles is shown by the TEM images shown in Fig. 4.20. While in the initial state of the sample, a finely dispersed structure is typical, after thermal cycling the region under study appears to be completely covered with a layer of bismuth through which copper crystallites are visible.

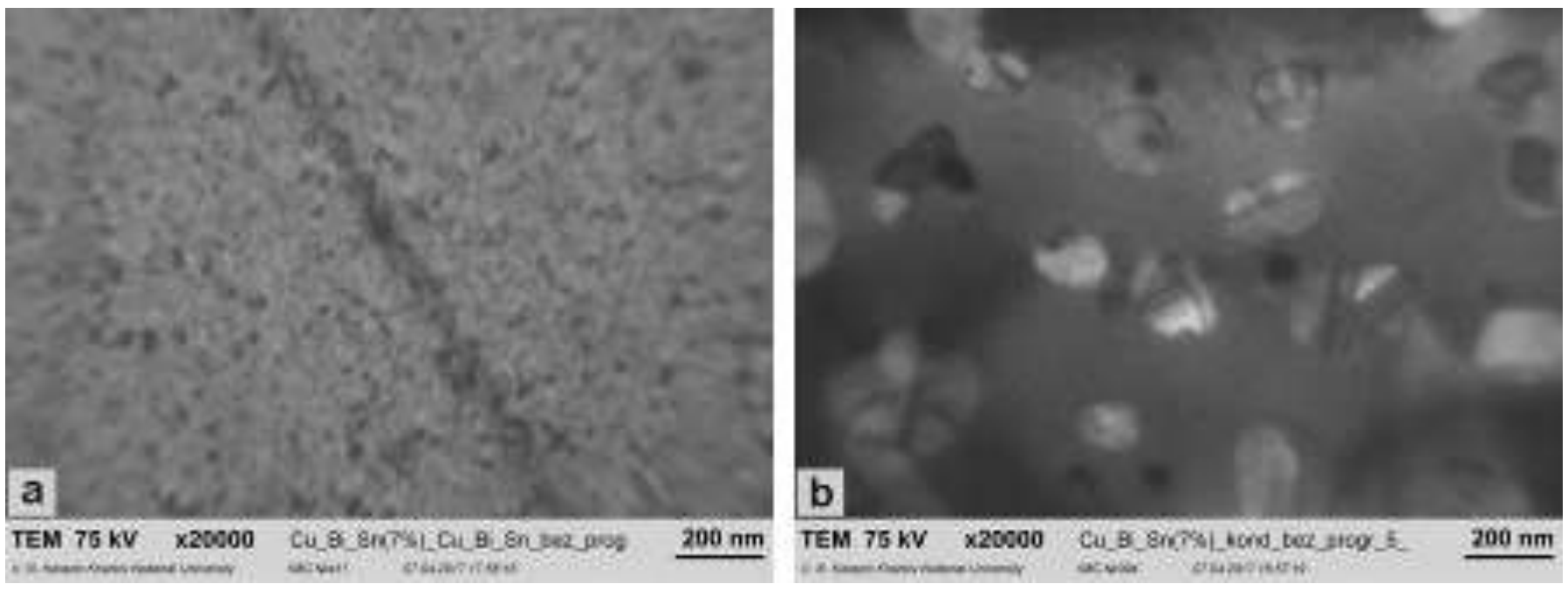

**Fig. 4.20** TEM images of the (Bi + 7% wt. Sn)/Cu film before (**a**) and after (**b**) five heating – cooling cycles. The content of the (Bi + 7% wt. Sn) alloy in the film (Bi + 7% wt. Sn)/Cu is 60% wt. The sample was not annealed during the condensation [25]

The results of the study of supercooling in films that were subjected to heating before tin deposition to provide melting of bismuth are shown in Fig. 4.21. In this case, the behavior of the samples differs significantly from that of the films of the first series.

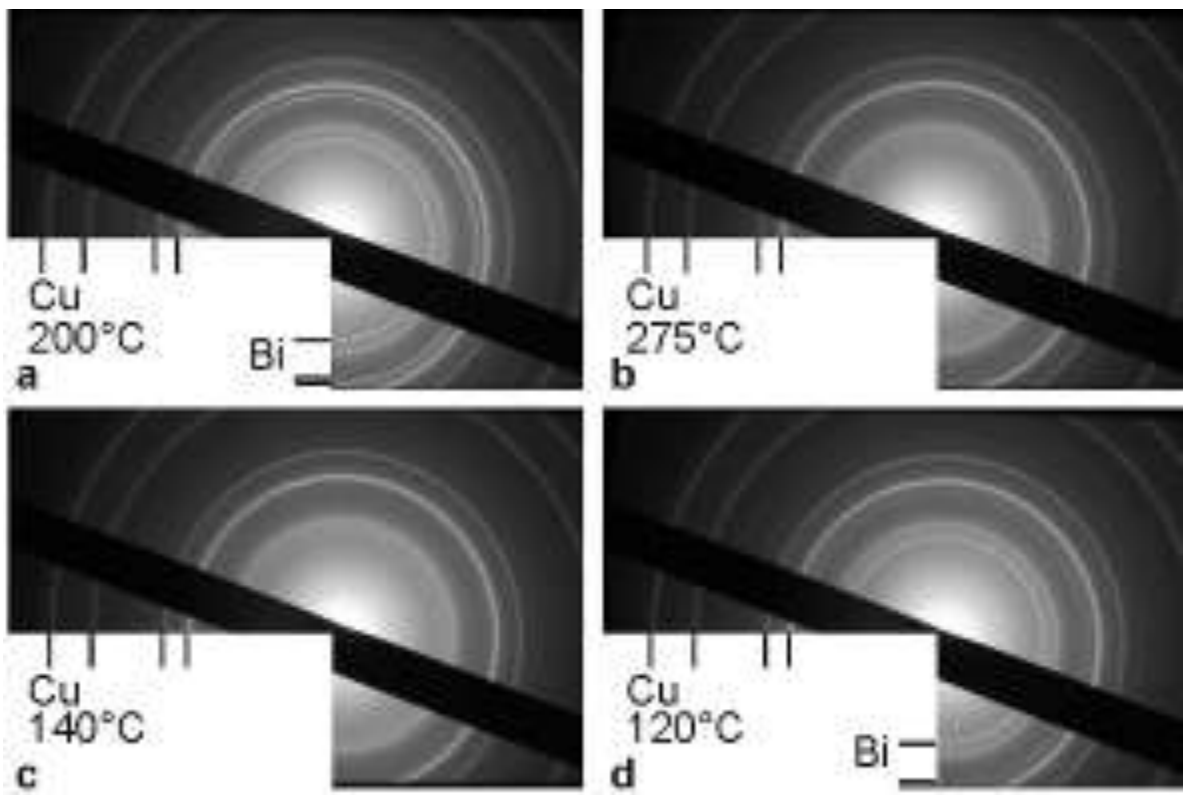

**Fig. 4.21** Electron diffraction patterns of films (Bi + 7% wt. of Sn)/Cu. Samples were annealed after the condensation of the Cu and Bi layers. Images (**a**), (**b**) correspond to heating; (**c**), (**d**) – to cooling [25]

The crystallization of the fusible component in the films of this series occurs diffusely, which causes the gradual appearance of diffraction reflexes that acquire their initial brightness when the sample is cooled in the temperature range of 150–130 °C. Thus, the supercooling recorded in these samples increases to 140 K ($\eta$ = 0.25). In addition, unlike the layered film systems that were not subjected to annealing during the condensation process, heating and cooling of the films of the second series does not cause a qualitative change in the diffraction pattern. That is, the samples preserve their finely dispersed structure even after crystallization. TEM images of the films of the second series before and after thermal cycling are shown in Fig. 4.22.

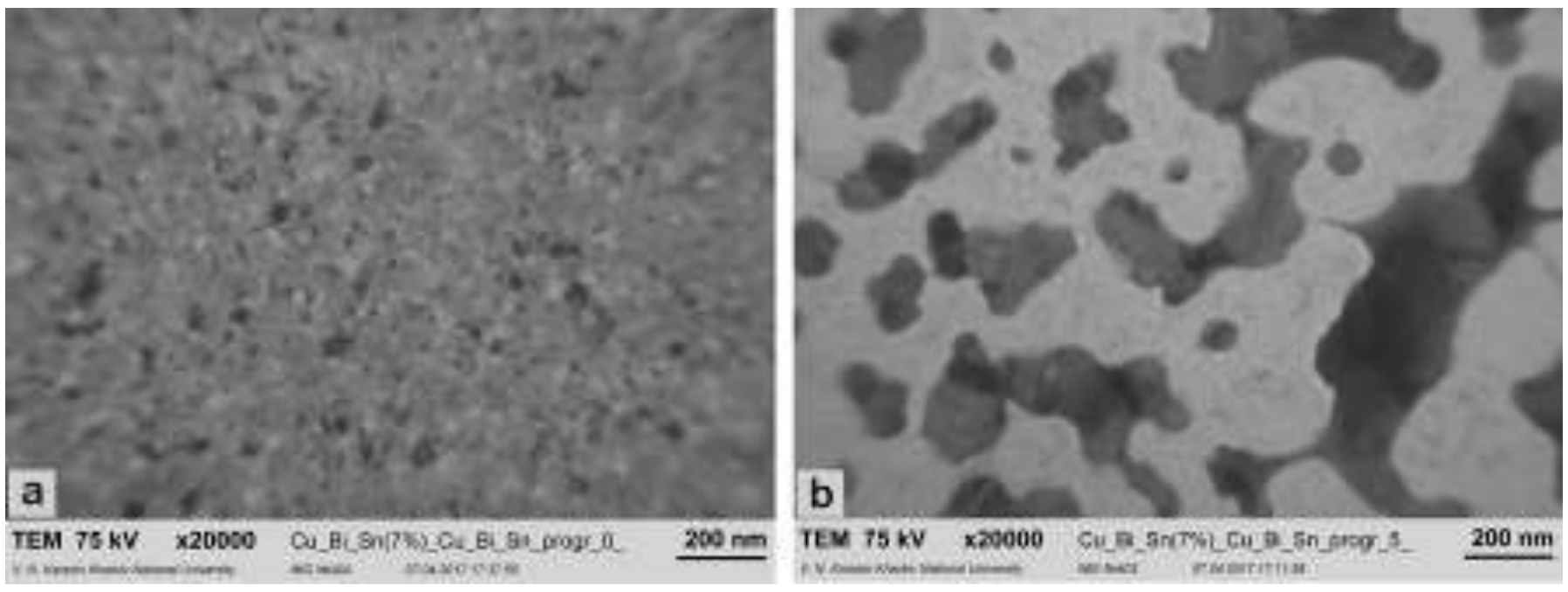

**Fig. 4.22** TEM images of the (Bi + 7% wt. Sn)/Cu films before (**a**) and after (**b**) five heating-cooling cycles. The content of the (Bi + 7% wt. Sn) alloy in the (Bi + 7% wt. Sn)/Cu film is 60% wt. Samples were annealed before tin depositing. [25]

The microstructure of the as-deposited films is generally similar to that of the films of the first series. There is only a slight enlargement of its characteristic elements. At the same time, the structure of the annealed films of both series differs dramatically. After heating, the films of the second series decay into separate islands of irregular shape, the sizes of which are distributed over a wide range of values: the smallest particles are about 20 nm in size, and the largest ones reach several microns.

The morphological differences in the films of the two series explain the differences in the nature of their crystallization and the values of supercooling observed. For example, the large-crystal structure that appears in the samples of the first series does not provide for the fulfillment of the conditions of the micro-volume method required to achieve the limiting supercooling. In this case, as well as in the Cu/Bi/Cu, Ag/Bi/Ag, and Mo/Bi/Mo systems, in which all components were deposited into the solid phase, early crystallization can be caused by impurities or defects in the polycrystalline copper layer. At the same time, the thermal dispersion of the films of the second series forms structures that are optimal for observing deep supercooling.

The question of the causes of morphological differences in the samples of both series is much more complicated. First, it should be recalled that tin in Sn/Cu binary film systems stimulates their dispersion at temperatures around 300 °C. In the case of the layered (Bi-Sn)/Cu system, if the fusible alloy contains more than 20 wt.% tin, an irreversible increase in electrical resistance is observed already at a temperature close to the eutectic melting point (Fig. 4.23). The value of the resistance jump and, consequently, the intensity of dispersion naturally decreases with a decrease in the tin content in the alloy.

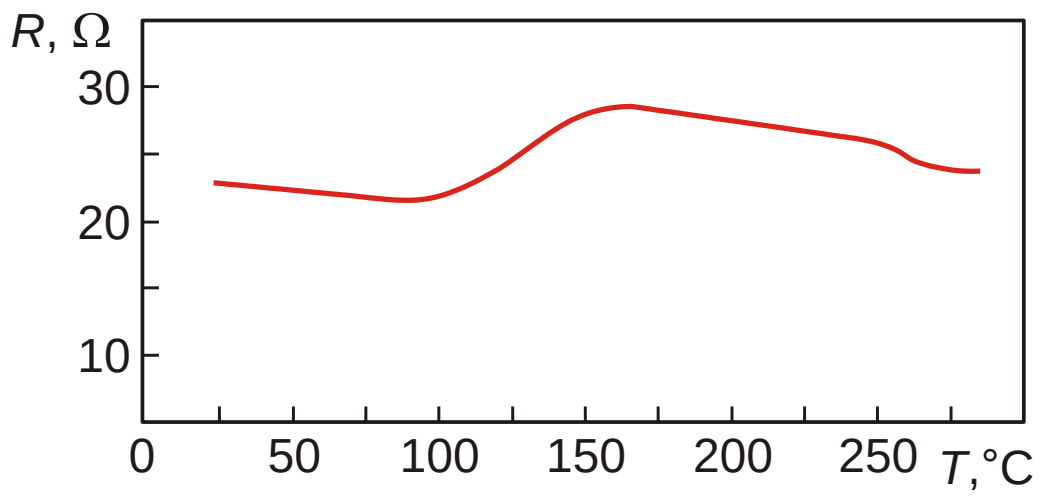

**Fig. 4.23** Electrical resistance dependence on the temperature of (Bi + 20% wt. Sn)/Cu films during the first heating [26]

It should be noted that the results of many studies (e.g., [30, 31, 32]) indicate that fusible inclusion particles in composite systems of the "particle-in-matrix" type are largely concentrated at grain boundaries and other defects typical of the polycrystalline structure. According to [30, 31, 32], conducting heating-cooling cycles increases the amount of bismuth located in the inter-boundary space. The first heating cycle contributes the greatest effect to this process. Thus, in the (Bi + 7% Sn)/Cu samples, which were subjected to a single heating-cooling cycle before tin condensation, part of the bismuth diffuses deep into the copper film. This leads to a slight increase in the relative content of tin in the alloy layer that is formed on the surface of the copper film after its condensation. It is the increased concentration of tin that stimulates faster dispersion of the samples.

It should be noted that information on the concentration dependence of supercooling values is necessary for solving many fundamental and applied problems. A convenient object for determining the concentration dependence of supercooling during the crystallization of alloys is the Bi-Sn binary system, which not only has a simple phase diagram of eutectic type but also looks promising as a basis for lead-free solders. Below are the results of studying supercooling in layered (Bi-Sn)/Cu film systems obtained using the methods of electrical resistance measurement and *in situ* electron diffraction studies [26].

Samples for studying the temperature dependence of electrical resistance were obtained by the method of sequential vacuum deposition. An important particularity of the preparation of the studied layered film systems was that after the condensation of each layer, the sample was subjected to a heating-cooling cycle. Short-term annealing within the framework of realization of such cycles does not cause significant morphological changes in the samples, but by reducing the amount of non-equilibrium defects concentrated mainly at the boundaries of crystalline grains of the copper film, it significantly and irreversibly reduces the resistance of the films. This procedure for obtaining samples made it possible to get rid of changes in electrical resistance caused by copper annealing and to study the behavior of Bi-Sn binary films starting from the first heating cycle.

During the study of the temperature dependence of the resistance, it was found that, unlike Cu and Bi/Cu films, the first heating of (Bi-Sn)/Cu samples is accompanied by an increase in resistance rather than a decrease. An increase in resistance at the first heating of the three-layer (Bi-Sn)/Cu films obtained by sequential condensation is observed in that case if the two-layer Bi/Cu films were subjected to short-term heating before tin deposition.

According to [27], the first heating of such structures provides annealing of the copper layer, which is accompanied by an irreversible decrease in its electrical resistance. At the same time, the resistance of the films can decrease tenfold, and the temperature at which the most intense resistance decrease is observed is close to the eutectic temperature of the Bi-Sn system. That is, the first heating of (Bi-Sn)/Cu samples that have not undergone preliminary annealing is accompanied by two oppositely directed processes, which makes it difficult to highlight the effects caused by the presence of the Bi-Sn alloy. Therefore, in studies [26], films subjected to annealing after condensation of each layer were used.

The irreversible increase in the resistance of (Bi-Sn)/Cu samples, which occurs near the eutectic transition temperature, is probably due to contact melting, which takes place in a binary system of fusible components [33, 34]. Contact melting, which begins in places determined by random factors, from the beginning of continuous layers of bismuth and tin, causes a violation of their continuity, which is naturally accompanied by an increase in the resistance of the sample.

However, starting from the second heating cycle, the temperature dependencies of the resistance take on a generally reversible nature. The temperature dependencies of the electrical resistance of (Bi-Sn)/Cu films corresponding to the subsequent heating-cooling cycles are shown in Fig. 4.24. They contain jumps that can be correlated with the melting and crystallization of a fusible alloy. A certain irreversible increase in resistance, which occurs mainly during alloy crystallization, is obviously associated with the gradual de-wetting of the films.

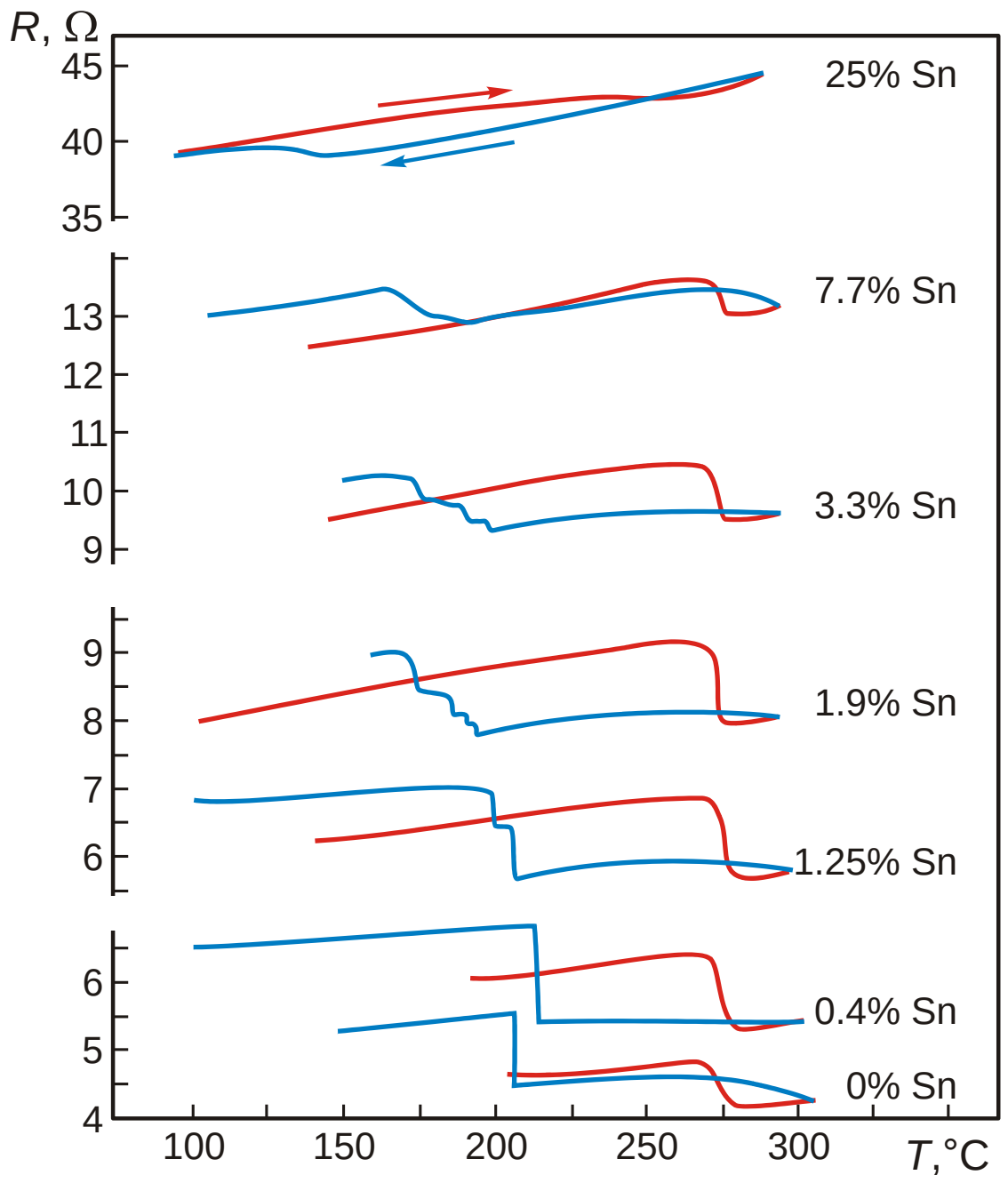

**Fig. 4.24** Resistance dependence on the temperature of (Bi-Sn)/Cu films with different content of tin (indicated near the graphs) [26]

As can be seen from the graphs presented in Fig. 4.24, an increase in tin content changes not only the temperature but also the nature of crystallization. In films containing less than one mass percent of tin in the fusible alloy, crystallization has an avalanche-like nature. As the tin content increases, crystallization first becomes step-like, i.e., several discrete jumps can be distinguished in the temperature dependencies of the resistance, each of which corresponds to the crystallization of a significant part of the sample. The step-like crystallization occurs until the tin content in the alloy is less than 3–4 wt. %. It should be noted that this value corresponds to the limiting solubility of tin in solid bismuth. With a further increase in its concentration, crystallization acquires a clearly defined diffuse nature.

The concentration dependence of the temperature width of the range in which crystallization occurs is shown in Fig. 4.25. Initially, with increasing tin concentration, the temperature range of crystallization increases, but starting from about 4 wt. % it almost stops changing.

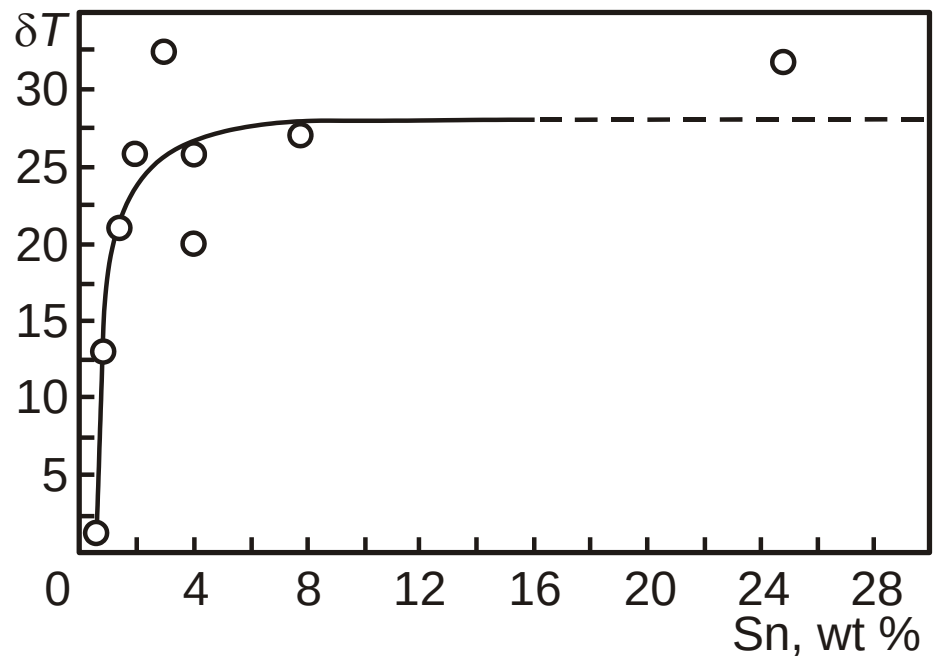

**Fig. 4.25** Temperature range crystallization versus tin concentration in a supercooled Bi – Sn melt, which contacts with copper [26]

The observed transition from avalanche-like crystallization to diffuse crystallization can be explained as following. According to the results of the work [25], an increase in the tin content in the alloy intensifies the dispersion process. At the same time, at its low concentrations, the destruction of the films does not have a global nature, and the layer of the fusible component in the sample breaks up into several rather large regions, each of which crystallizes independently. The sufficiently large size of such regions causes their crystallization to occur at external centers, even before reaching the limiting supercooling for this contact pair.

As the tin content increases, the decay intensity of the connected system of inclusions formed by bismuth grows. In a dispersed system of non-connected fusible inclusions, each of which crystallizes independently, the conditions of the micro-volume method are already met, and the supercooling observed in this case is determined by the energy particularities of the interfaces in this contact system.

The change in the morphology of the films after the first heating cycle for samples with different concentrations of components is shown by the SEM images shown in Fig. 4.26. The images clearly show that melting in samples containing even a small amount of tin in the fusible layer causes the destruction of a connected system of inclusions characteristic of tin-free samples. This process is most likely to occur near the eutectic temperature and is provided by contact melting.

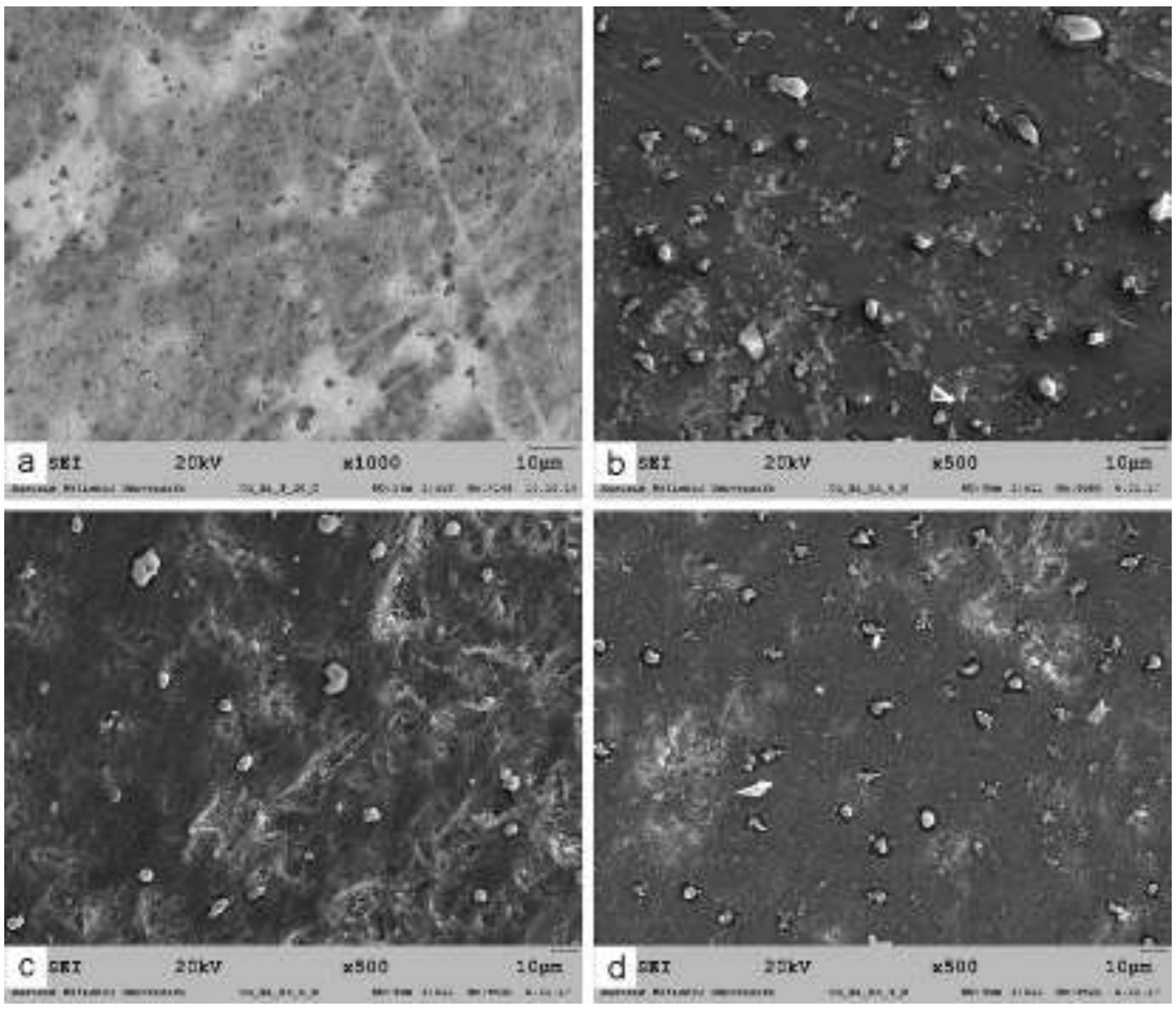

**Fig. 4.26** SEM images of (Bi – Sn)/Cu bilayer films: **a**) (Bi + 0% wt. Sn)/Cu, **b**) (Bi + 1% wt. Sn)/Cu, **c**) (Bi + 4% wt. Sn)/Cu, **d**) (Bi + 8% wt. Sn)/Cu [26]

It should also be noted that variations in the composition of fusible particles can make a significant contribution to the expansion of the crystallization temperature range. However, since the temperature width of the range in which the phase transition in (Bi-Sn)/Cu films is observed is close to that, which occurs in layered systems with pure fusible metals, it can be argued that variations in the composition of fusible inclusions are insignificant.

As already mentioned, tin contributes to the de-wetting of layered film systems. Therefore, (Bi-Sn)/Cu films, as its content in fusible alloys increases, lose stability and break up into separate islands. The use of the electrical resistance measurement method for such structures is inefficient or even impossible. Therefore, in addition to resistive studies, (Bi-Sn)/Cu films were studied using *in situ* electron diffraction studies.

The concentration dependence of the relative supercooling during the crystallization of the liquid phase of the Bi-Sn alloy, obtained using two independent *in situ* methods, is shown in Fig. 4.27. As can be seen, at first there is a rapid increase in relative supercooling, which ends at a tin concentration of about 5 wt. %. After that, the relative supercooling η remains practically constant over a wide range of concentrations. It is worth noting that an increase in the relative supercooling, as well as an increase in the crystallization range δ*T*, are observed up to the concentration, that corresponds to the maximum solubility of tin in solid bismuth, i.e., in the concentration range in which only a solid bismuth-based solution is present in the studied films. For this reason, their behavior turns out to be partly similar to that of Bi/Cu samples, i.e. in this case, the fusible component can form a connected system of inclusions for which a significant reduction in supercooling is observed. The appearance of a tin-based phase containing up to 20 wt. % bismuth according to the phase diagram, causes the decay of the fusible alloy film into separate islands, and thus leads to a significant change in the conditions of the existence of the supercooled melt. A decrease in relative supercooling at high tin concentrations is observed in the single-phase region on the phase diagram, in which only the tin-based solid solution is equilibrium.

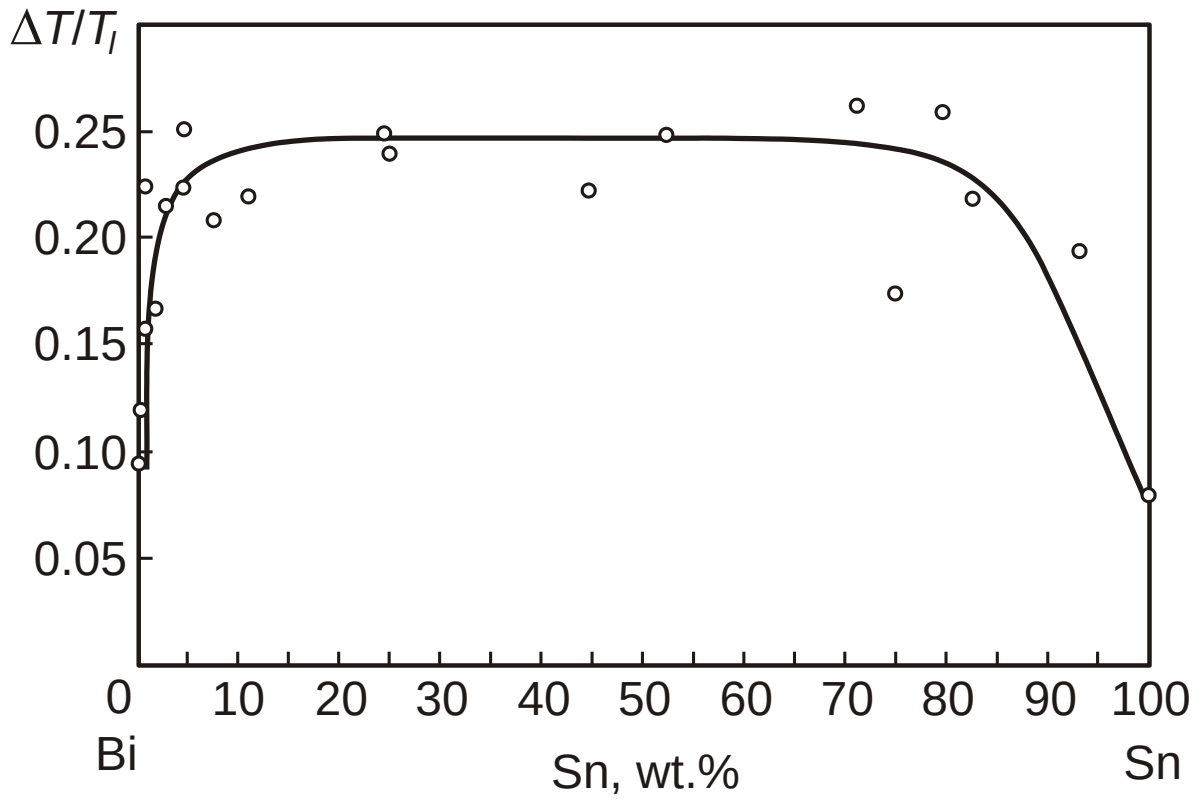

**Fig. 4.27** Concentration dependence of relative supercooling of Bi-Sn layers, which contact with copper [26]

The constancy of the limiting supercooling in a wide range of concentrations has been observed in other works [6, 35]. During the study of supercooling in (Bi-Pb)/C and (Bi-Sb)/C films [6], it was found, that the relative supercooling of these alloys is practically independent of concentration and generally corresponds to the values characteristic of pure components (Fig. 4.2). For alloys of fusible metals, work [35] shows that the crystallization temperature of particles with a radius of about 30 nm depends on the concentration of the components. At the same time, it follows from the results of the work [35] that the value of relative supercooling for Pb-Bi and Pb-Sn films (on condition of normalization to the value of the corresponding liquidus temperature) has an almost constant value.

The concentration dependence of the crystallization temperature in the studied films, put on the phase diagram of the Bi-Sn binary system, is shown in Fig. 4.28. The curve corresponding to the crystallization of the supercooled melt is located below the liquidus line and generally follows its course.

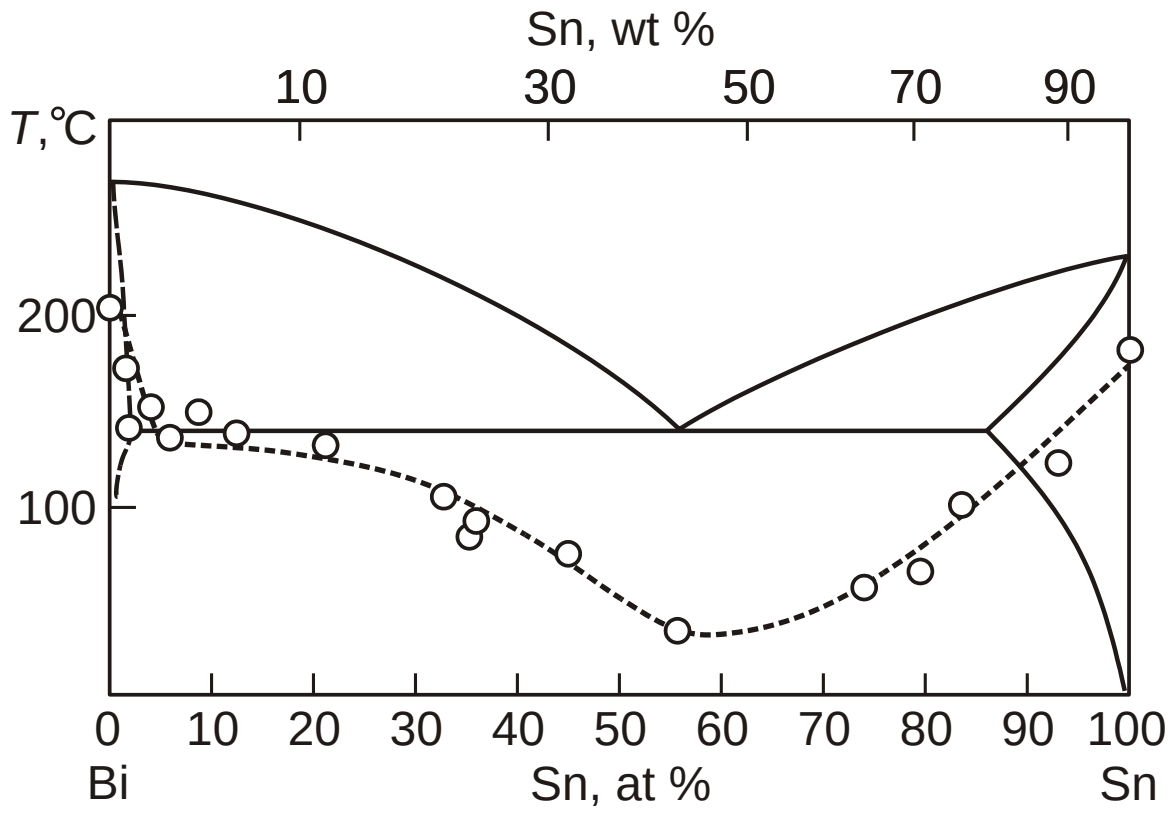

**Fig. 4.28** Crystallization temperature dependence of the supercooled melt of the fusible component in (Bi-Sn)/Cu films on Bi-Sn alloy composition [26]

The Cu-Sn binary system is characterized by a complex phase diagram, according to which various phases and chemical compounds can be formed. Thus, a typical result in the study of the film structure based on this contact pair is the detection of the intermetallic compound $Cu_6Sn_5$, which is formed at the condensation stage. During heating, $Cu_6Sn_5$ decomposes with the formation of Cu3Sn intermetallic. Thus, in films subjected to heating-cooling cycles, a constant competition between the two intermetallic compounds $Cu_6Sn_5$ and $Cu_3Sn$ is observed [36, 37, 38, 39].

At the same time, the presence of the third component can significantly affect the conditions of phase formation in the studied systems. According to the results of [40], the only intermetallic observed in the ternary (Bi-Sn)/Cu system is $Cu_6Sn_5$. This intermetallic occurs after annealing the system at a temperature of 120 °C for 30–100 days. Similar results were obtained in the work [41], where it was shown that the only intermetallic copper-based compound, observed in the (Ag-Sn)/Cu system after a three-day annealing, was $Cu_6Sn_5$. The formation of the $Cu_3Sn$ compound, which is generally typical for the Cu-Sn binary system [36, 37, 38, 39], was observed only for samples, that were subjected to even longer annealing. According to [42], for the ternary (Ag-Sn)/Cu system, the appearance of a thin $Cu_6Sn_5$ intermetallic layer occurs 1 minute after the system is heated to 250 °C. At the same time, its expansion into the internal part of the copper layer occurs rather slowly. Annealing the samples for about 50 hours at the temperature of 180 °C increases the thickness of the intermetallic layer by only 3–5 times. It should be noted that this time is significantly longer than the effective annealing time to which the three-layer (Bi-Sn)/Cu films were subjected during the study of the temperature limits of liquid phase stability. Therefore, the obtained supercoolings probably correspond to the crystallization of Bi-Sn alloys which are in contact with a thin $Cu_6Sn_5$ layer.

In contrast to (Bi-Sn)/Cu films, in which one of the components of the fusible alloy interacts significantly with the copper layer, (In-Pb)/Mo samples are a more convenient object of study. In fact, the complete absence of interaction between the fusible melt and molybdenum allows us to focus on the thermodynamic aspects of crystallization. In such a system, phase transformations are not complicated by the formation of intermetallics and phase interaction of the fusible layer with the matrix.

It is worth noting that molybdenum films obtained by the thermal condensation method are amorphous, as evidenced by the electron diffraction patterns obtained from these samples. As can be seen from Fig. 4.29, only a diffuse halo is observed in the electron diffraction patterns obtained from as-deposited and even briefly annealed molybdenum films.

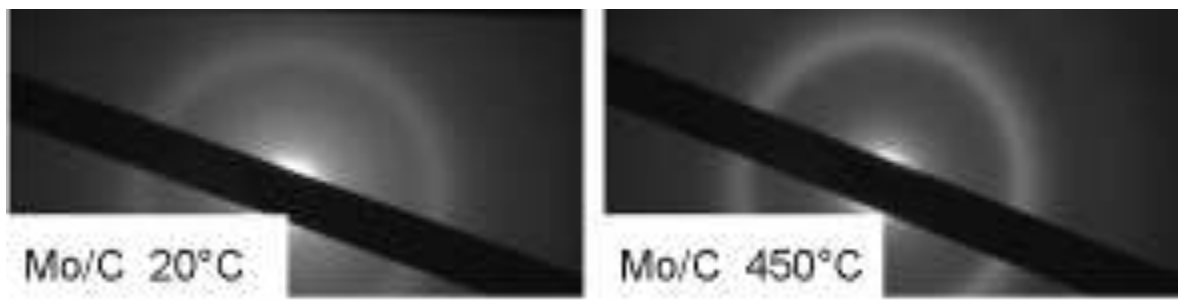

**Fig. 4.29** Electron diffraction patterns obtained from as-deposited and annealed molybdenum films

However, despite their amorphous state, in the context of supercooling studies, molybdenum layers behave similarly to other metal films. For example, bismuth condensed on a molybdenum layer by the vapor-crystal mechanism crystallizes in an avalanche-like manner. The first annealing of as-deposited Mo films is accompanied by an irreversible decrease in their electrical resistance (Fig. 4.30). After the first heating, the temperature dependence of the resistance approaches to linear and stops changing.

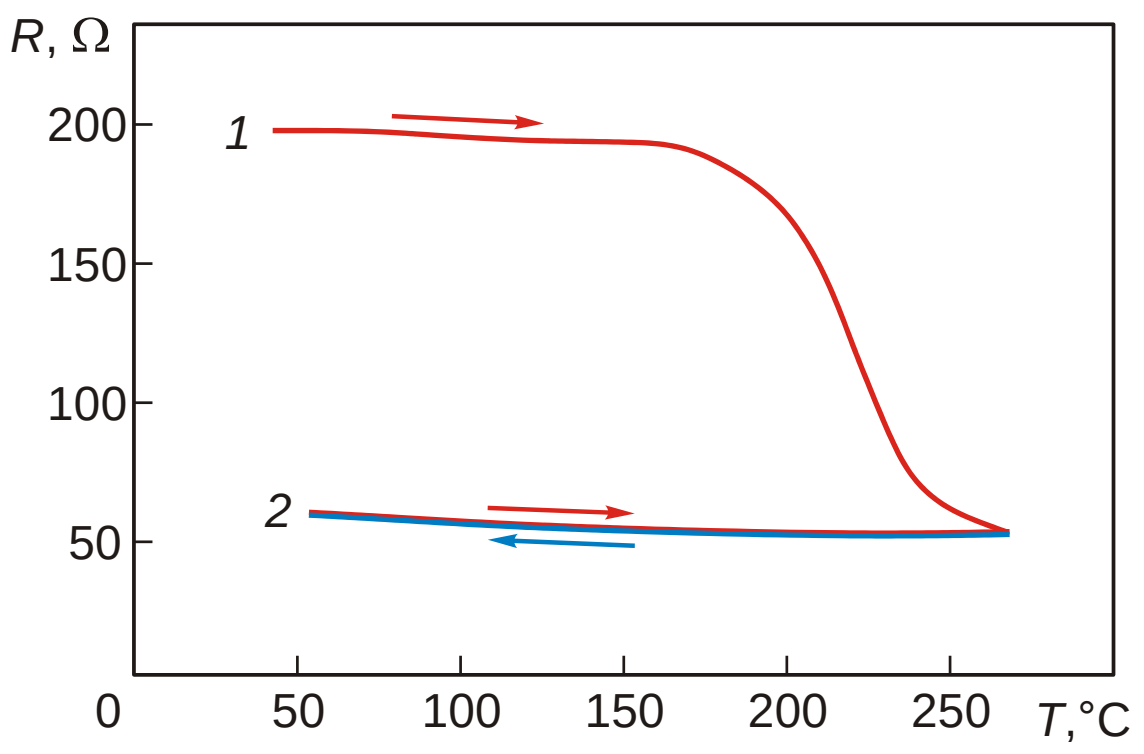

**Fig. 4.30** Temperature dependence of the electrical resistance of molybdenum films obtained in the first (1) and subsequent (2) heating-cooling cycles

Given the amorphous state of molybdenum, the irreversible decrease in electrical resistance (Fig. 4.30) can no longer be attributed to the recrystallization of the films. It can be explained by the relaxation of other defects characteristic of vacuum condensates. The activation energy of such processes can be obtained [43, 44] from the Arrhenius plot for the region of irreversible decrease in electrical resistance. The values obtained in this way are 0.2–0.5 eV and correspond to typical energies of activation of diffusion of vacancies. Thus, the first heating of the molybdenum layer reduces the number of vacancies in the film and allows obtaining a stable structure.

Another particularity of molybdenum films is the negative temperature coefficient of electrical resistance. That is, the electrical resistance of such layers decreases with temperature (Fig. 4.30). The observed phenomenon is a manifestation of the inner size effect, the study of which has been devoted to in works [45, 46]. At the same time, since only jumps in electrical resistance are of importance in the study of supercooling, molybdenum films stabilized during the first annealing become a convenient object for studying the supercooling of binary alloys.

Fig. 4.31 shows the temperature dependence of the electrical resistance of the Mo/(In + 20 wt% Pb)/Mo film corresponding to the first heating of the sample. It can be seen that during the first heating there is a gradual, irreversible increase in the electrical resistance of the layered structure. This can be explained by the optimization of the shape of the fusible layer, due to which the initially continuous fusible layers (Fig. 4.32a) turn into island layers (Fig. 4.32b). Due to these processes, the moment of melting is not well observed in the first heating cycle (Fig. 4.31) In contrast to melting, the crystallization of the melt becomes clearly visible already during the first cooling (Fig. 4.31).

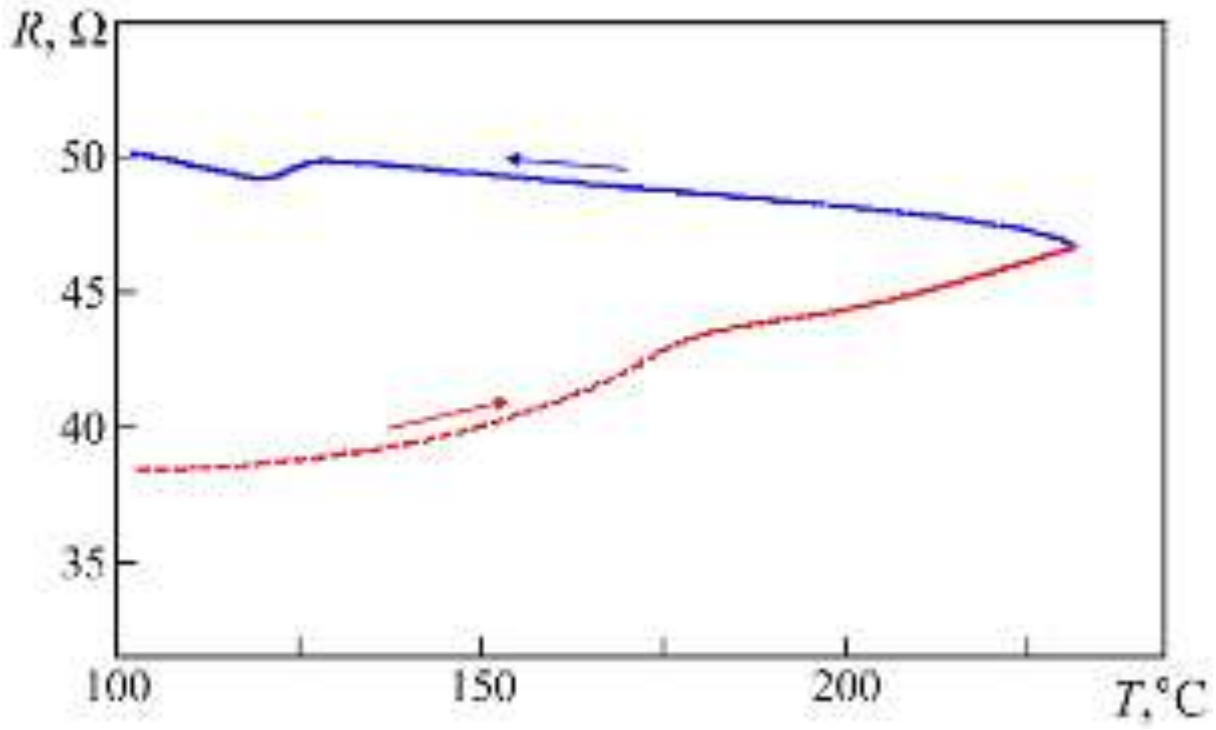

**Fig. 4.31** Temperature dependence of the electrical resistance of Mo/(In + 20 wt% Pb)/Mo films obtained in the first heating cycle

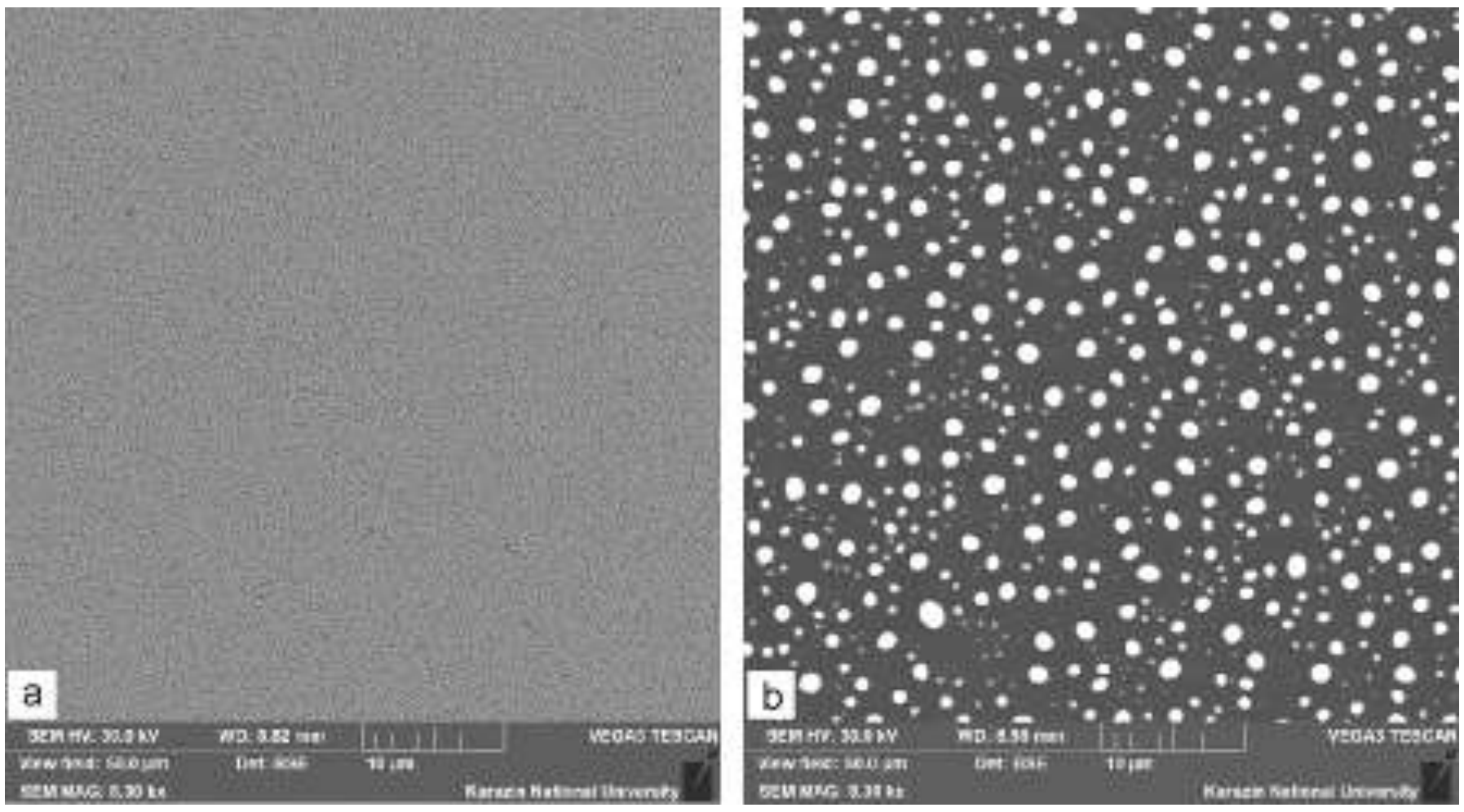

**Fig. 4.32** SEM images of (In + 80 wt.% Pb)/Mo films obtained before (**a**) and after (**b**) melting of the fusible layer

An important aspect of the study of binary structures is the homogeneity of the alloys. The formation and stabilization of binary alloys in films obtained by sequential condensation of components was confirmed by electron diffraction studies (Fig. 4.33). Thus, the diffraction pattern observed in the electron diffraction patterns of Mo/(In + 70 wt% Pb)/Mo corresponds to the FCC structure. The parameter of crystal lattice determined by electron diffraction studies is 0.485 nm. This is somewhat lower than the parameter of the crystal lattice of pure lead and corresponds to the solid solution of InPb of the studied concentration [47]. The same situation occurs in Mo/(In + 20 wt.% Pb)/Mo films, in which the electron diffraction patterns correspond to a body-centered tetragonal crystal lattice. The crystal lattice parameters are 0.327 and 0.499 nm and indicate the formation of an indium-based solid solution [48]. Since the electron diffraction patterns shown in Fig. 4.33 were obtained at room temperature, they indicate the stabilization of alloys that do not decompose upon cooling.

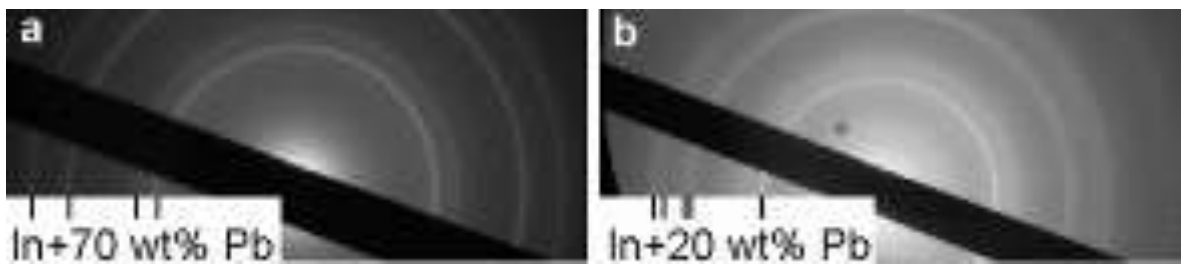

**Fig. 4.33** Electron diffraction patterns obtained from Mo/(In + 70 wt.% Pb)/Mo (**a**) and Mo/(In + 20 wt.% Pb)/Mo (**b**) films

It is interesting to note that, according to electron diffraction studies, the homogenization of the fusible layer occurs even before heating. As can be seen from Fig. 4.34, even in as-deposited Mo/(In + 70 wt.% Pb)/Mo films, no indium lines are observed. The existing crystal structure probably corresponds to a lead-based solid solution. Up to the melting point, heating does not cause qualitative changes in the diffraction pattern, and the parameter of the crystal lattice increases linearly with temperature (Fig. 4.34). This growth is natural given the thermal expansion of the layers. The coefficient of linear thermal expansion of the alloy is $6{\cdot}10^{-5}$ $K^{-1}$. This value is significantly lower than the expansion coefficient of both lead and indium. Although the decrease in thermal expansion is typical for alloys, the value obtained may indicate the size effect of thermal expansion observed for single-component nanocrystalline films [49].

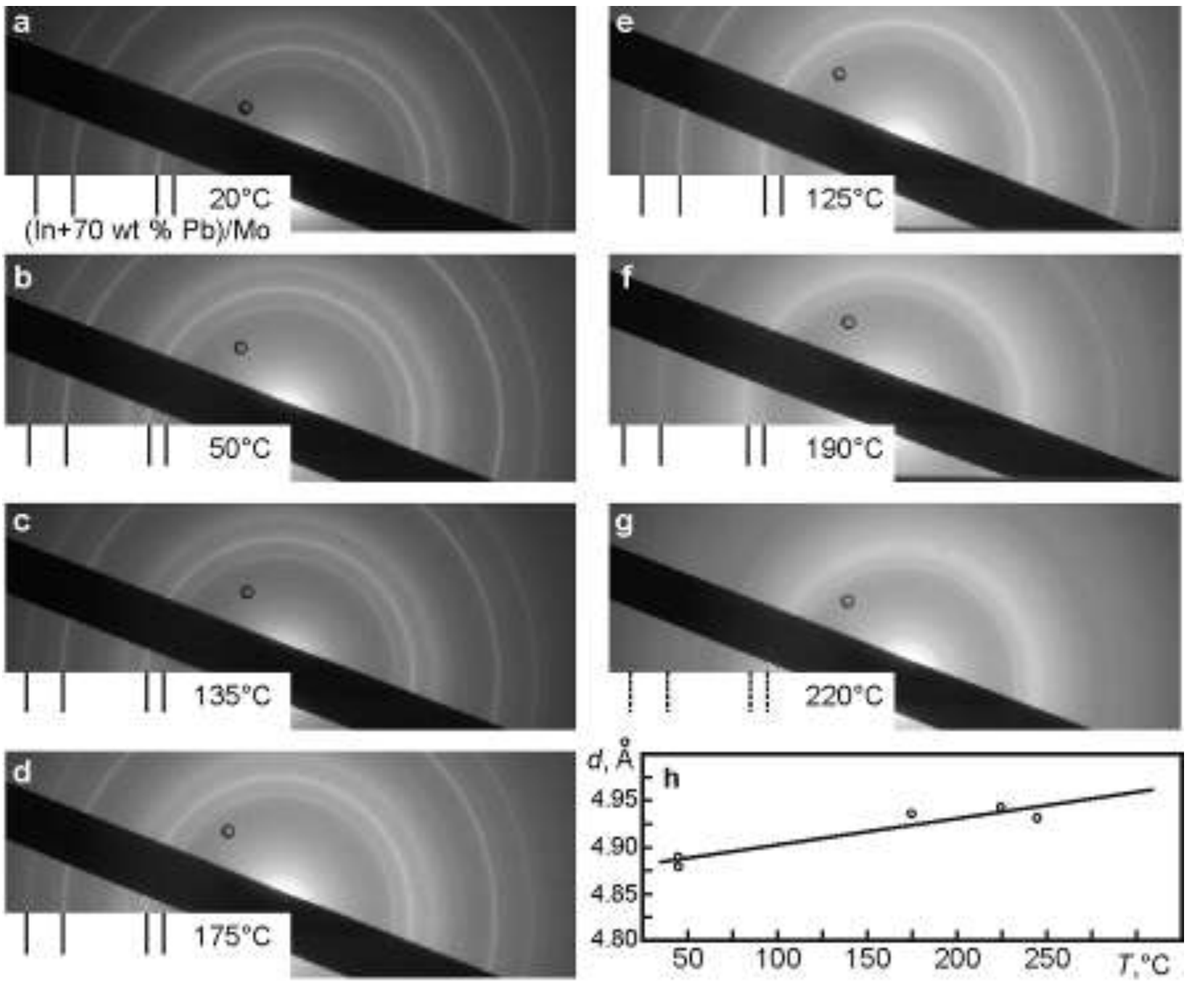

**Fig. 4.34** Electron diffraction patterns of Mo/(In + 70 wt.% Pb)/Mo films obtained in the first heating cycle. The lower part shows the temperature dependence of the parameter of the crystal lattice of the solid solution of In + 70 wt.% Pb

Taking into account the results of the electron diffraction study, it can be assumed that the formation of a solid solution in Mo/(In-Pb)/Mo films occurs due to condensation-stimulated diffusion even at the deposition stage. This is due to the fact that the intensity of such diffusion processes is much higher than the intensity of diffusion. Thus, the irreversible increase in electrical resistance in Fig. 4.31 should be attributed only to the morphological evolution of the samples, which is not necessarily accompanied by a change in their phase composition. However, it should be borne in mind that the thickness of the fusible layer in samples intended for electron diffraction studies is at least five times less than the thickness of the samples used for the study of electrical resistance. According to TEM studies (Fig. 4.35), such samples are island even before heating. It is interesting to note that the layered structure affects the morphological evolution of Mo/(In-Pb)/Mo thin films. Due to the fact that the fusible layer is located between the molybdenum thin films, melting does not cause the formation of spherical particles typical of systems with poor wettability. The conditions of the limited geometry provide for the overall preservation of the shape of the initial particles, which are only slightly rounded, but do not acquire a clearly defined spherical shape.

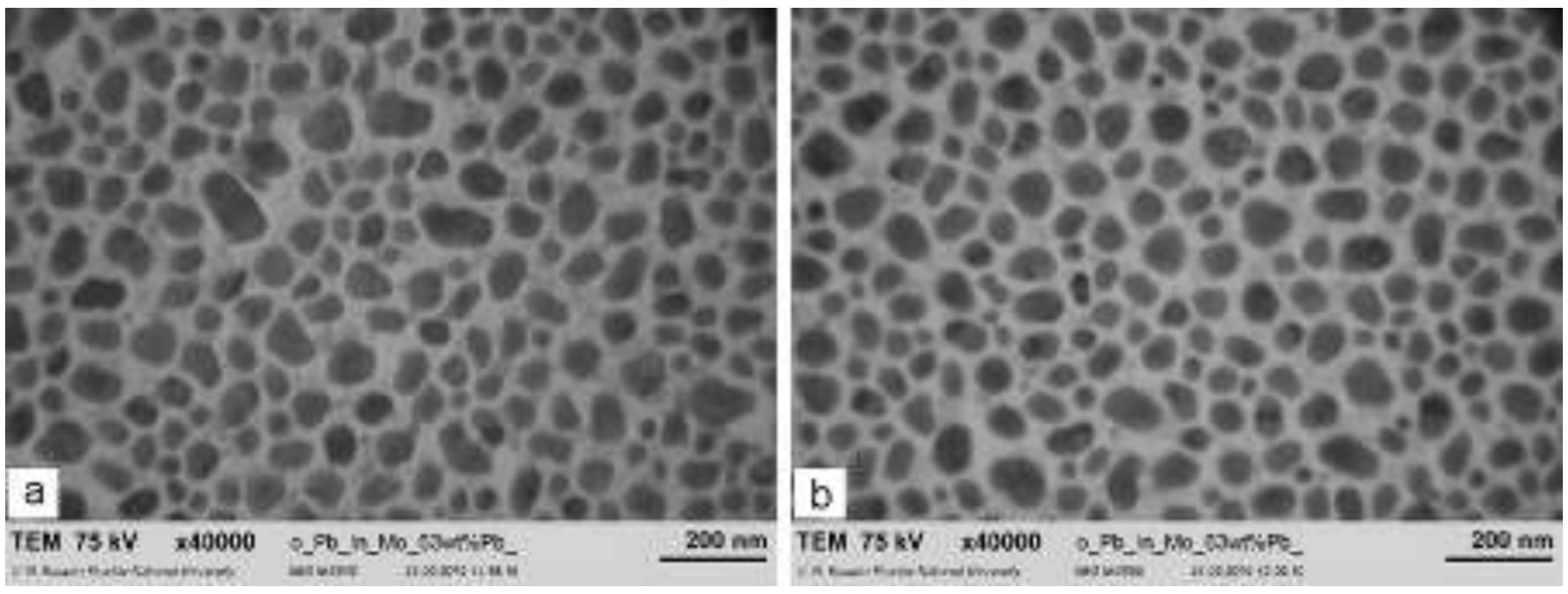

**Fig. 4.35** TEM-images of Mo/(In+70 % Pb)/Mo films before (**a**) and after (**b**) melting

Increasing the thickness of the fusible layers in itself increases the time required to homogenize the sample in the vertical direction. Another factor that can slow down the formation of solid solutions is the size dependence of the diffusion coefficient [50, 51, 52]. The growth in the diffusion coefficient, which can reach several orders of magnitude in thin films [50, 51, 53, 54], accelerates the homogenization of thin film structures. However, films of greater thickness will homogenize at a rate typical of the bulk state. Thus, the irreversible increase in electrical resistance (Fig. 4.31) that occurs during the first heating may be due not only to the morphological evolution of the film systems (Fig. 4.32). It may also indicate the formation of a solid solution that could not form during condensation due to the relatively large thickness of the layers.

Fig. 4.36 shows the temperature dependencies of the electrical resistance of Mo/(In + 9.2 wt.% Pb)/Mo films obtained during the second and subsequent heating cycles. Both melting and crystallization of the fusible alloy are clearly observed in the graphs, and the corresponding temperatures are reproducible from cycle to cycle. A gradual increase in the electrical resistance of the films (Fig. 4.36) probably indicates a partial decay of the film systems. This can be explained by the influence of liquid lead, which contributes to the thermal dispersion of more refractory layers [55, 56].

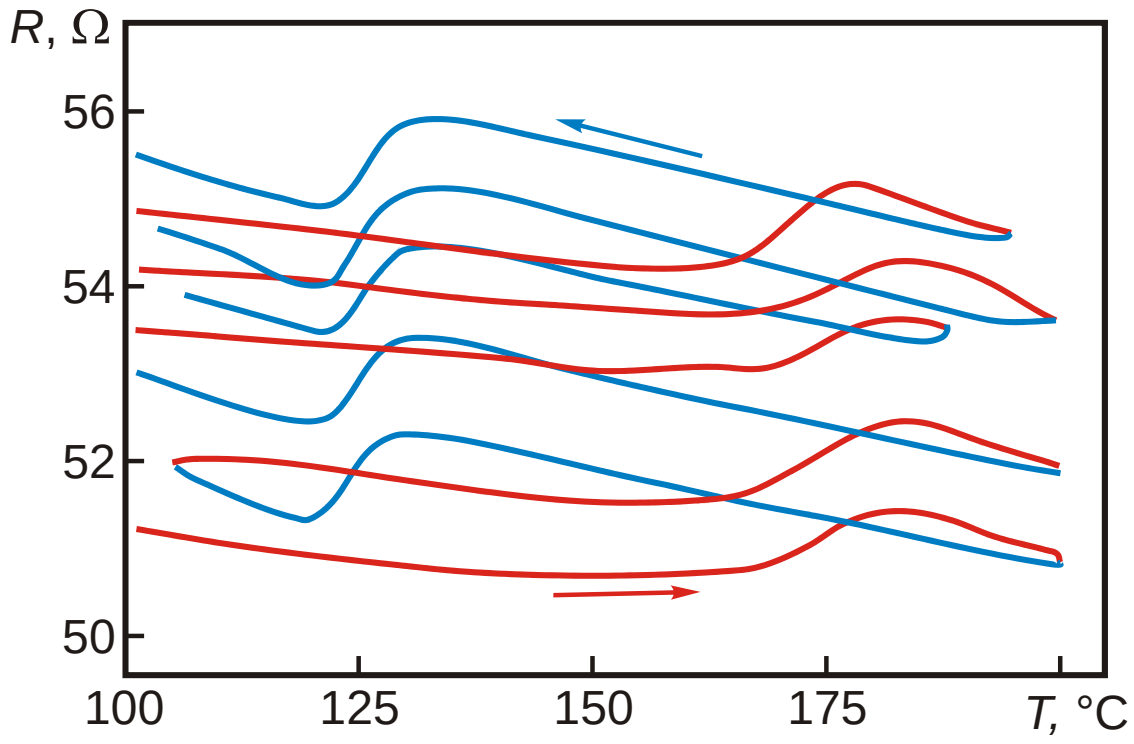

**Fig. 4.36** Temperature dependencies of the electrical resistance of Mo/(In + 9.2 wt.% Pb)/Mo films obtained in sequential heating-cooling cycles [57]

Fig. 4.36 clearly shows that both melting and crystallization of InPb alloys that contact with molybdenum occur in a certain temperature range. The concentration dependence of the melting and crystallization ranges is shown in Fig. 4.37. It can be seen that the range in which phase transformations occur varies smoothly with concentration. It reaches a maximum for samples containing approximately 50% of each of the components. Such dependencies generally correspond to the phase diagrams and can be explained by the coexistence of the solid and liquid phases, which is common for alloys. However, the significant scatter of experimental points (Fig. 4.37), which significantly exceeds the experimental errors, indicates additional mechanisms for expanding the temperature range of phase transformations. This is probably due to the significant influence of random factors on phase transformations in the island structures, which are the studied samples (Fig. 4.32). Such factors do not actually affect the maximum values of melting and crystallization temperatures, which are determined by thermodynamic considerations. However, random factors have a decisive influence on the start of phase transformations, causing early melting and especially early crystallization of individual particles. This is what in an uncontrollable way increases the observed ranges of phase transformation processes.

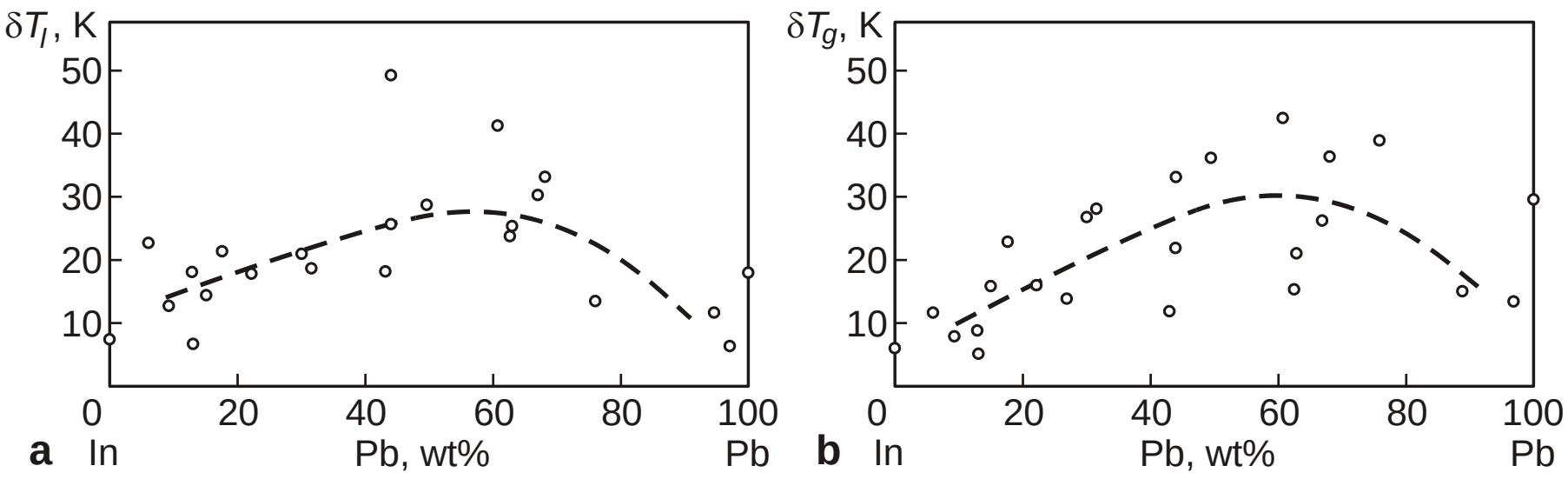

**Fig. 4.37** Concentration dependence of the temperature range of melting (a) and crystallization (b) of InPb alloys that are in contact with molybdenum layers [57]

The concentration dependence of relative supercooling in the Mo/(In-Pb)/Mo system is shown in Fig. 4.38. The graph shape in Fig. 4.38 is similar to the concentration dependencies of the melting and crystallization ranges. The maximum supercooling is observed for alloys containing approximately 50% of each component and decreases when approaching pure metals. The smooth course of the dependence is disturbed in Mo/Pb/Mo and Mo/In/Mo films, for which a significant increase in supercooling is observed. This can probably be explained by the fact that binary liquid solutions are less equilibrium systems compared to single-component melts. It is worth noting that despite the fact that the layered structure of the samples affects the morphology of the films after melting (Fig. 4.35), its effect on the crystallization temperature has not been established. That is, (In-Pb)/Mo films and Mo/(In-Pb)/Mo samples crystallize at the same temperature.

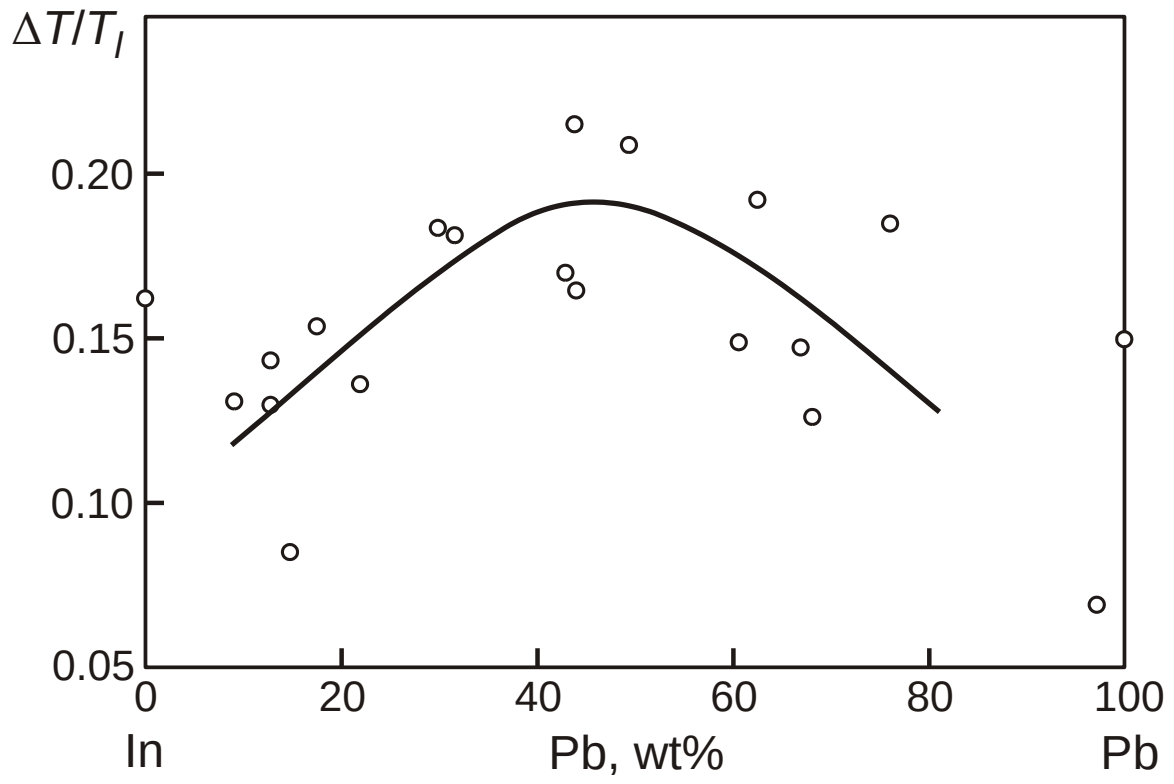

**Fig. 4.38** Concentration dependence of relative supercooling in Mo/(In+Pb)/Mo films [57]

To quantitatively describe the supercooling of alloys, one can use the theory of nucleation of two-component melts, which is presented in [58]. According to the results of [58], the radius *(r*)* and the work of formation *(A)* of a critical nucleus during the crystallization of a two-component melt are determined by the equations

$$r^* = \frac{2\sigma_{sl}\nu}{f_l(x_l) - f_s(x_s) + (x_l - x_s)df_l/dx_l}, \tag{4.3}$$

$$A = \frac{1}{3}\sigma_{sl} \cdot S_k + \frac{1}{2}\frac{V_k^2}{Nv^2}(x_l - x_s)^2 \frac{d^2 f_l}{dx_l^2}, \tag{4.4}$$

where $f_l$ and $f_s$ are the chemical potentials of the liquid and solid phases, respectively; $V_k$ and $S_k$ are the volume and surface area of the crystal nucleus; $v$ is the atomic volume.

The first term in equation (4.4), which defines the work of the nucleus formation, is due to the excess energy associated with the appearance of the phase interface, and the second describes the variation in the composition of the solid phase nucleus compared to the composition of the melt that gives rise to it. For small supercooling, the second term of Eq. (4.4) has a decisive contribution to the energy of nucleation of the crystalline phase. This determines that in this case, the composition of the critical nucleus of the solid phase will differ from the composition of the liquid from which it arises.

At the same time, according to [58], the role of the first term of equation (4.4) increases with the increase in the achieved supercooling. As a result, the composition of the nucleus, which arises, will approach the concentration of components in the initial melt. Starting from a certain temperature $T_0$, which corresponds to the temperature at which the specific free energies of liquid and solid solutions of the same composition become equal between themselves, the concentration of components in the nucleus can correspond to their content in the melt. On the condition of further growth of supercooling, the appearance of nuclei of the original composition becomes the predominant process. It should be noted that expressions (4.3) and (4.4) were obtained for homogeneous nucleation in the melt volume.

It is obvious that during the crystallization of binary films in contact with more refractory layers, the conditions of homogeneous crystallization are not realized, and crystal nuclei are formed at the point of contact of the melt with solid layers. If a liquid is in contact with a solid, the volume of the crystal phase nucleus that appears on its surface (see Fig. 1.4) decreases according to the following equation:

$$V_k^* = V_k \Phi(\psi), \tag{4.5}$$

where $\Phi(\psi) = \frac{1}{4}\left(2 - 3\cos\psi + \cos^3\psi\right)$.

It should be noted that since the volume is included in the second term of equation (4.4) squared, the heterogeneous nature of crystallization will provide a reduction of the contribution of this term to the overall work of the formation of the nucleus. In addition, significant supercoolings (0.2–0.25 of the liquidus temperature $T_l$) were obtained in the studied system, which can already be considered close to the values corresponding to homogeneous crystallization. Thus, it can be assumed that the crystallization of the supercooled melt in the Mo/(In-Pb)/Mo and (Bi-Sn)/Cu systems occurs without a change in composition. In view of this, the second term of equation (4.4) can be neglected. Then it is easy to show that the work of formation of a crystal nucleus can be determined using the following equation:

$$A = \frac{1}{3}\sigma_{sl} \cdot 4\pi \left(\frac{2\sigma_{sl} T_l}{\lambda \Delta T}\right)^2 \Phi(\psi) = \frac{16\pi}{3} \cdot \frac{\sigma_{sl}^3}{\lambda^2} \cdot \left(\frac{T_l}{\Delta T}\right)^2 \Phi(\psi), \tag{4.6}$$

from which the following relation can be obtained for relative supercooling:

$$\left(\frac{\Delta T}{T_l}\right)^2 = \frac{16\pi}{3k\ln N}\left(\frac{\sigma_{sl}}{\lambda}\right)^3\left(\frac{\lambda}{T}\right)\Phi(\psi), \tag{4.7}$$

which coincides with a similar equation for single-component melts (1.13).

Numerous results obtained during the study of supercooling of pure metals indicate that the approximate constancy of relative supercooling is a consequence of the constancy of the combination of values:

$$\left(\frac{\sigma_{sl}}{\lambda}\right)^3\left(\frac{\lambda}{T_s}\right) \approx const, \tag{4.8}$$

where $T_s$ is the melting point of the metal or the liquidus temperature of the alloy.

Thus, the independence of relative supercooling from the content of components in a wide concentration range, obtained experimentally in the work [26], indicates that the empirical relation (4.8) is also valid for alloys. From condition (4.8), under independently determining the specific heat of melting of the alloy, the concentration dependence of the interfacial energy of the crystal-melt interface can be estimated. Estimates made using literature data on the latent heat of fusion [19] show that the values of the interfacial energy of the crystal-melt interface of Bi-Sn alloys are in the range of 70–90 mJ/m$^2$.

If we assume that the relation (4.8) is also valid for the In-Pb alloy in contact with molybdenum layers, then the concentration dependence of relative supercooling in the binary system can be explained by a change in the wetting of amorphous molybdenum by the In-Pb melt when the concentration of components in it changes. As follows from expression (4.8) and the dependence shown in Fig. 4.38, the contact angle formed by the melt of the fusible layer with the molybdenum layer should reach a maximum for alloys containing components with the same mass concentrations. It should decrease with the transition to pure components. This thesis is consistent with the results of SEM studies, which indicate an improvement in wetting for indium-rich alloys.

# Chapter 5. Summary

The paper systematizes the results of a study of supercooling during the crystallization of one- and two-component melts. The main object of the analyzed scientific works became melts remaining in the liquid state below the melting point under conditions of slow cooling. A quantitative description of supercooling processes was performed using the methods of phenomenological thermodynamics. Mutually complementary methods such as a method of changing the condensation mechanism, quartz resonator method, electron diffraction studies, and electrical resistivity measurement of multilayer films have been used to experimentally determine the crystallization temperature and phase transition kinetics.

The phenomenon of supercooling during crystallization is quite natural for the liquid phase, but sufficiently clean conditions are required for its observation. In fact, it is necessary to get rid of external crystallization centers as much as possible, and ideally to create conditions for homogeneous nucleation. It is shown that the physical cause of supercooling is the presence in the system of additional energy connected with the crystal-melt interface. Such a boundary naturally occurs when a nucleus of a crystalline phase is formed in the melt. Due to this, crystallization turns out to be a barrier process and for its start requires the emergence of a nucleus whose size exceeds a certain value called critical. According to phenomenological thermodynamics, the critical size turns to infinity at the melting point (see expression (1.4)) and decreases hyperbolically with increasing $\Delta T$, i.e., the difference between the melting point and the current temperature. Therefore, crystallization cannot occur directly at the melting point and requires some supercooling of the melt. The value of supercooling is determined by the temperature at which a nucleus of sufficient size will emerge in the liquid by fluctuation.

According to (1.4), the size of the critical nucleus is determined not only by supercooling but also by the crystal-melt interfacial energy. Consequently, the value of supercooling achieved in the system also depends on it. Thus, the difference between the melting and crystallization temperatures can be used to estimate the interfacial energy of the crystal-melt boundary. For quantitative estimation, it is possible to use (1.7) obtained within the framework of this model. It should be noted that experimental ways of determining this value are limited. However, it is the interfacial energy that determines the kinetics of many physical processes that have both applied and fundamental significance.

The expressions (1.4) and (1.7) were obtained for homogeneous crystallization and should be modified for the case when a wetted surface is present in the melt. In this case, due to wetting, the shape of the nucleus changes (Fig. 1.4), and the supercooling itself is already determined by (1.13). Quantitatively, the reduction of supercooling is determined by the shape factor (1.10). For practical applications, it is possible to use a linear approximation of the expressions (1.13) and (1.10), which agrees with the results of experiments in a wide range of contact angles.

It is worth noting that, although the supercooled melt preserves the basic properties of the liquid state, the decrease in temperature leads to some differences in the supercooled and equilibrium melt. In particular, in some cases, primarily in binary systems, intermediate metastable ordered structures are observed (Fig. 1.6) [1]. In addition, due to the temperature dependence of viscosity in highly supercooled liquids, the formation of ordered structures possessing long-range order elements is possible [2, 3, 4, 5]. That is, it can be expected that during cooling, even before crystallization, some phase transitions will occur in the liquid, which will change the properties of the system. Such phenomena are often observed in compounds containing Sb ($GeSb_2Te_4$, GeTe, $Ag_4In_3Sb_{67}Te_{26}$, $Ge_{15}Sb_{85}$) and consist in the formation of ordered regions of about 0.6 nm [3, 4, 5, 6, 7, 8]. This size is twice as large as the radius of the first coordination sphere, which exceeds the values characteristic of the short-range order and allows us to consider the formed structures as some intermediate state.

The sensitivity of melts to the presence of wetted surfaces, which can be not only vessel walls, but also various impurities, makes the task of maximum melt purification a key aspect of the study of supercooling. The classical method of minimizing the influence of impurities was proposed in works [9, 10] and consists of strong dispersion of the melt. Given that the amount of impurities in the melt is finite, if the degree of dispersion is sufficient, one can expect the appearance of particles that will contain no impurities at all. A natural way to realize such an approach is to use island films obtained by vacuum condensation methods [11, 12, 13]. In this case, due to the processes carried out in a vacuum, high purity of the investigated objects is achieved. In addition, in a number of cases, the use of vacuum condensation turns out to be a simple way to form island films, the size of islands which is much smaller than those, which are typical for ultrasonic dispersion [9, 10]. The combination of these factors significantly increases the probability of achieving supercooling that is due to thermodynamic rather than impurity factors. In the case of vacuum condensates, the main impurity is the substrate, which is chosen by the researcher. This simplifies the process of investigating the effect of wetting on supercooling. In particular, this approach allows the impurity factor limiting supercooling in technological applications to be studied under controlled conditions.

The method of changing the condensation mechanism allowed us to determine the temperatures of maximum supercooling in numerous contact pairs of "melt – substrate" obtained under high vacuum conditions. This method is based on the fact that the transition from the vapor-crystal condensation mechanism to the vapor-liquid mechanism occurs at the substrate temperature equal to the temperature of maximum supercooling of the condensed substance under the given conditions. This conclusion follows from the particularities of condensation of the substance on the substrate at temperatures below its bulk melting point. In the initial stages of film formation, due to the size dependence of the melting point, the condensed phase nuclei are thermodynamically equilibrium liquid. As they grow, their equilibrium state becomes crystalline, but crystallization itself will occur only if their temperature is below the temperature of limiting supercooling. Otherwise, they remain in the liquid, but already in the supercooled state. The morphological structure of films deposited through liquid and solid phases differs significantly. As a consequence, the light scattering by such films is also different. The mentioned difference can be easily observed on a substrate with a gradient of temperatures, on which a sharp boundary corresponding to the temperature of maximum supercooling is visible after film deposition (Fig. 2.15). Using this approach, an empirical dependence between relative supercooling and wetting angle in the investigated "melt – substrate" pair was obtained (Fig. 2.6).

It is found that the relative supercooling is approximately linearly related to the contact angle, at least up to angle values at the 130° level. The maximum possible supercooling is realized in the case when the melt poorly wets the substrate on which it is located. The limiting supercooling, which is observed when the liquid cools slowly, is almost the same for all metals and is 0.3–0.4 of the melting temperature. The experimental values are consistent with thermodynamic estimates, which are based on the assumption that the maximum supercooling value corresponds to the temperature at which the size of the critical nucleus is reduced to the size of a single atom. The influence of the pressure of the residual atmosphere in the vacuum chamber on the value of supercooling was found and the threshold nature of this factor was established (Fig. 2.4) [14, 15]. The role of the residual atmosphere composition in increasing the crystallization temperature of supercooled melt was determined (Fig. 2.5) [16]. Experimental data obtained using the method of changing the condensation mechanism indicate that supercooling decreases with decreasing size. For very small particles, crystallization occurs without supercooling [17]. In general, at this stage of the study of supercooling in vacuum condensates, it was obtained that particles located on a poorly wetted substrate in the context of supercooling can be considered as free, and crystallization itself turns out to be homogeneous in this case.

In addition to the method of changing the condensation mechanism, methods of quartz resonator and electrical resistance measurement can be named as convenient ways to investigate supercooling. The first one is based on a jump in quality factor, which occurs during the melting and crystallization of a film deposited on a piezoelectric crystal [18, 19]. In the second method, melting and crystallization are registered by the resistance jumps of the film system that accompany phase transitions [20, 21].

Using such methods, it is possible to observe melting-crystallization hysteresis (Figs. 2.23, 2.24, 3.2, 3.15) and to study supercooling in multilayer films. Multilayer films are a convenient model of nanocomposite structures [21, 22, 23]. In turn, nanocomposite systems are not only an important object of modern technologies [24, 25, 26, 27], but can also have unique physical particularities. In particular, superheating is often observed in such structures [30, 31, 32], which should be taken into account in the study of supercooling. The data obtained using resistive and dynamic studies complement the information obtained by the method of changing the condensation mechanism. In particular, they allow us to observe the kinetic effects of phase transitions. Using such methods, supercooling in various contact pairs was determined, and the results obtained earlier using the method of changing the condensation mechanism were confirmed. The ability to observe crystallization kinetics has allowed us to obtain new information on the temperature width of phase transitions and the influence of the morphology and composition of melts on this value.

By comparing single-component samples (Fig. 3.28), with significantly different sizes of islands (Figs. 3.29, 3.30), it is shown that decreasing the most probable size of liquid particles decreases the temperature width of the range in which their crystallization occurs. By studying the crystallization kinetics of as-deposited films and samples subjected to prolonged exposure under low vacuum conditions (Figs. 3.23, 3.24), it has been obtained an indication of the existence of other factors, different from impurity factors, which cause a decrease in the supercooling value. Such factors should probably include substrate defects, which could potentially become places of facilitated crystal nucleation.

The temperature width of phase transitions can also be influenced by the crystalline state of the substrate. Thus, the study of supercooling in Bi/Ge films shows that for as-deposited samples there is an extended and multistage crystallization. However, after several heating-cooling cycles, the picture changes and crystallization becomes one-stage and occurs in the temperature range of about 20 K (Fig. 3.41).

For alloys, it is shown that the temperature width of crystallization and melting of the fusible layer depends on the concentration (Figs. 4.25, 4.37). For example, the crystallization temperature range of Bi-Sn melt in contact with copper increases rapidly with growing tin content. This is explained by morphological changes occurring in the samples with increasing tin content in the alloy. In (In-Pb)/Mo films, the concentration dependence of the temperature ranges in which the phase transition occurs has a dome-shaped appearance and apparently correlates with the value of the maximum supercooling (Fig. 4.38).

The use of multilayer films, which is possible due to the methods of quartz resonator and electrical resistivity measurement, allows protection of the investigated substance from external factors. This, along with the possibility of thermal cycling, significantly increases the reproducibility of the results. Also, the use of the resistive technique allowed us to discover an interesting effect, which is that bismuth deposited by the vapor-crystal mechanism tends to form a connected system of inclusions in the samples. This system is preserved during thermal cycling and causes avalanche-like crystallization of samples (Figs. 3.2, 3.15, 3.19) [21, 22, 23]. Probably, the grain boundaries, where segregation of bismuth takes place, which is the basis of the system of inclusions, also play an important role in this process.

It is shown that the jump in resistivity of the fusible component, which occurs when the substance melts, can qualitatively explain the results of resistive studies for most bismuth-free contact pairs. For bismuth-based samples (Bi/Cu, Bi/Ag, Bi/V), the observed resistance jump exceeds the expected value [21, 22, 23]. The significant value of the electrical resistance jump at the phase transition can also be explained by the grain boundary segregation of bismuth, due to which the contribution of the fusible component to the total resistance of the film system increases significantly.

In the study of supercooling of binary alloys, it is shown that the general regularities established for single-component systems are also valid for them. Thus, the supercooling of a eutectic melt turns out to be minimal in the case, if it is located on one of its components [28]. This fact is explained by the fact that melts, as a rule, wet their components well. At the same time, the supercooling of alloys that are on the surface of metals not included in their composition is much greater [20, 29]. This is explained by a decrease in the interaction of components in the second case, and is consistent with general thermodynamic ideas.

The concentration dependencies of supercooling were obtained for (Bi-Sn)/Cu, (Bi-Pb)/C, (Bi-Pb)/Al2O3, (Bi-Sb)/C, and (In-Pb)/Mo films. It is shown that the concentration dependence of supercooling in (Bi-Sn)/Cu films generally repeats the liquidus line (Fig. 4.28). The same situation takes place in the contact pairs (Bi-Pb)/C, (Bi-Pb)/$Al_2O_3$ and, apparently, (Bi-Sb)/C. In these contact pairs, the relative supercooling is practically independent of concentration, indicating an approximate constancy of melt wetting of crystalline copper.

Nevertheless, in the (Bi-Sn)/Cu contact pair, the supercooling decreases when going to pure components (Fig. 4.27). In the case of tin, this indicates a strong interaction in the Sn-Cu contact pair. For pure bismuth, this can be explained by the formation of a connected system of inclusions, which crystallizes at small supercooling and collapses with growing tin content. The destruction of a connected system of inclusions is accompanied by a transition of crystallization from avalanche-like to diffuse type (Fig. 4.24) and is confirmed by the results of SEM studies (Fig. 4.26). The concentration dependence for the (In-Pb)/Mo system has a somewhat different view. In this case, the relative supercooling demonstrates a dome-shaped concentration dependence and reaches a maximum for samples containing equal amounts of indium and lead. It is shown that for deep supercooling, which is achieved in vacuum condensates, the crystallization process of binary melt proceeds without change of composition and in many respects is similar to the crystallization of single-component liquids.

As well as for single-component melts, at the crystallization of alloys the condition $\left(\frac{\sigma_{sl}}{\lambda}\right)^3\left(\frac{\lambda}{T_s}\right) \approx const$, the physical interpretation of which requires further investigations, is observed.

In general, the experimental data indicate that the empirical relations established in the study of the crystallization of pure metals remain valid for alloys. This indicates the commonality of physical principles conditioning supercooling of single- and multicomponent systems and confirms the effectiveness of using the methods of phenomenological thermodynamics for quantitative description of supercooling in fundamental research and applied developments.

**Acknowledgment**

This project has received funding through the MSCA4 Ukraine project, which is funded by the European Union. The project is also supported by the Ministry of Education and Science of Ukraine.